\documentclass[a4,12pt]{article}
\usepackage[dvips]{graphicx,color}
\usepackage{pslatex}
\usepackage[hyperindex]{hyperref}
\definecolor{darkblue}{rgb}{0,0,.5}
\hypersetup{pdftex=true, colorlinks=true, breaklinks=true, linkcolor=darkblue, menucolor=black, pagecolor=darkblue, urlcolor=darkblue}

\input epsf
\def\gs{\mathrel{\lower0.6ex\hbox{$\buildrel {\textstyle >}
 \over {\scriptstyle \sim}$}}}
\def\ls{\mathrel{\lower0.6ex\hbox{$\buildrel {\textstyle <}
 \over {\scriptstyle \sim}$}}}

\def\japitem#1{\smallskip\noindent\rlap{#1}\hglue2em\hangindent2em}

\def\kms{{\,\rm km\,s^{-1}}}
\newcommand{\lya}{{\mbox Ly$\alpha$}}

\def\hompc{{\,h\,\rm Mpc^{-1}}}
\def\mpcoh{{\,h^{-1}\,\rm Mpc}}
\def\blob{\japitem{$\bullet$}}
\def\blobb{\japitem{$-$}}
\def\key#1{#1}
\def\half{{\textstyle{\scriptstyle 1\over\scriptstyle 2}}}
\def\frac#1#2{{\textstyle{\scriptstyle #1\over\scriptstyle #2}}}
\def\japsub{\rm\scriptscriptstyle}

\def\bigstrut{\vrule width0pt height1.6em depth1.6em}
\def\topstrut{\vrule width0pt height1.1em depth0em}
\def\botstrut{\vrule width0pt height0em depth0.7em}

\def\be{\begin{equation}}
\def\ee{\end{equation}}

\def\m@th{\mathsurround=0pt }
\def\eqalign#1{\null\,\vcenter{\openup1\jot \m@th
 \ialign{\strut\hfil$\displaystyle{##}$&$\displaystyle{{}##}$\hfil
 \crcr#1\crcr}}\,}

\def\epsfigsimp#1#2#3#4{
\begin{figure}
\begin{center}
\epsfxsize = #2\hsize
\epsfbox{#1}
\end{center}
\caption{#3}
\label{#4}
\end{figure}
}

\def\epsfigsimptwo#1#2#3#4#5{
\begin{figure}
\begin{center}
\epsfxsize = #3\hsize
\epsfbox{#1}
\epsfxsize = #3\hsize
\epsfbox{#2}
\end{center}
\caption{#4}
\label{#5}
\end{figure}
}

\let\ssec = \subsection

\def\japsec#1{\newpage\section{#1}}

\def\sssec#1{\par\noindent{\bf #1}\quad}

\makeatletter
\renewcommand\section{\@startsection {section}{1}{\z@}%
                                   {-5.5ex \@plus -1ex \@minus -.2ex}%
                                   {0.8ex \@plus.2ex}%
                                   {\normalfont\Large\bfseries}}
\renewcommand\subsection{\@startsection{subsection}{2}{\z@}%
                                     {-3.25ex\@plus -1ex \@minus -.2ex}%
                                     {0.3ex \@plus .2ex}%
                                     {\normalfont\large\bfseries}}
\makeatother

\begin{document}

\renewcommand{\labelenumi}{\theenumi}
\renewcommand{\theenumi}{(\arabic{enumi})}

\pagenumbering{roman}
\pagestyle{empty}

\vglue 1em

\centerline{\Large\bf Report by the ESA-ESO Working Group on}
\medskip
\centerline{\Large\bf Fundamental Cosmology}

\vskip 3em

\section*{Abstract}

In September 2003, the executives of ESO and ESA agreed to establish a
number of working groups to explore possible synergies
between these two major European astronomical institutions on key
scientific issues. The first two working group reports (on Extrasolar
Planets and the Herschel--ALMA Synergies) were released in 2005
and 2006, and this third report covers the area of Fundamental Cosmology.

The Working Group's mandate was to concentrate on fundamental issues
in cosmology, as exemplified by the following questions:
(1) What are the essential questions in fundamental
cosmology?
(2) Which of these questions can be tackled, perhaps exclusively,
with astronomical techniques?
(3) What are the appropriate methods with which these key questions
can be answered?
(4) Which of these methods appear promising for realization within
Europe, or with strong European participation, over
the next $\sim 15$ years?
(5) Which of these methods has a broad range of applications and a
high degree of versatility even outside the field of fundamental
cosmology?

From the critical point
of view of synergy between ESA and ESO, one major resulting
recommendation concerns the provision of new generations of
imaging survey, where the image quality and near-IR sensitivity
that can be attained only in space are naturally matched by
ground-based imaging and spectroscopy to yield massive
datasets with well-understood photometric redshifts (photo-z's).
Such information is essential for a range of new cosmological
tests using gravitational lensing, large-scale structure,
clusters of galaxies, and supernovae. All these methods can
in principle deliver high accuracy, but 
a multiplicity of approaches is essential in order that potential
systematics can be diagnosed -- or the possible need for
new physics revealed.
Great scope in future cosmology also exists for ELT
studies of the intergalactic medium and space-based studies
of the CMB and gravitational waves; here the
synergy is less direct, but these areas will remain
of the highest mutual interest to the agencies.
All these recommended facilities will
produce vast datasets of general applicability, which will have a
tremendous impact on broad areas of astronomy.

\newpage
\pagestyle{plain}

\vglue -1.5em
\vbox{

\section*{Background}

Following an agreement to cooperate on science planning issues, the
executives of the European Southern Observatory (ESO) and the European
Space Agency (ESA) Science Programme and representatives of their
science advisory structures have met to share information and to
identify potential synergies within their future projects. The
agreement arose from their joint founding membership of EIROforum
(\url{www.eiroforum.org}) and a recognition that, as pan-European
organisations, they serve essentially the same scientific community.

At a meeting at ESO in Garching during September 2003, it was agreed
to establish a number of working groups that would be tasked to
explore these synergies in important areas of mutual interest and to
make recommendations to both organisations. The chair and co-chair of
each group were to be chosen by the executives but thereafter, the
groups would be free to select their membership and to act
independently of the sponsoring organisations. During the second of
these bilateral meetings, in Paris during February 2005, it was
decided to commission a group to address the current state of
knowledge and future prospects for progress in fundamental cosmology,
especially the nature of `dark matter' and `dark energy'. By summer
2005, the following membership and terms of reference 
for the group were agreed:

\addcontentsline{toc}{section}{Membership}
\section*{Membership}

John Peacock (Chair) jap@roe.ac.uk\\
Peter Schneider (Co-Chair) peter@astro.uni-bonn.de \\
George Efstathiou gpe@ast.cam.ac.uk\\
Jonathan R. Ellis johne@mail.cern.ch\\
Bruno Leibundgut bleibund@eso.org\\
Simon Lilly simon.lilly@phys.ethz.ch\\
Yannick Mellier mellier@iap.fr\\
\\
{\bf Additional major contributors:} 
Pierre Astier, 
Anthony Banday,
Hans B\"{o}hringer,
Anne Ealet,
Martin Haehnelt,
G\"{u}nther Hasinger,
Paolo Molaro,
Jean-Loup Puget,
Bernard Schutz,
Uros Seljak,
Jean-Philippe Uzan.\\
\\
{\bf ST-ECF Support:} Bob Fosbury, Wolfram Freudling\\
\\
Thanks are also due to many colleagues who provided further useful comments on draft
versions of the report at various stages.

}

\addcontentsline{toc}{section}{Terms of Reference}

\section*{Terms of Reference}

\begin{enumerate}
\item
To outline the current state of knowledge of the field (this is not
intended as a free-standing review but more as an introduction to set the
scene);

\item
To review the observational and experimental methods used or envisaged
for the characterisation and identification of the nature of Dark Matter
and Dark Energy;

\item
To perform a worldwide survey of the programmes and associated
instruments that are operational, planned or proposed, both on the ground
and in space;

\item
For each of these, to summarise the scope and specific goals of the
observation/experiment; also to point out the limitations and possible
extensions;

\item
Within the context of this global effort, examine the role of ESO and
ESA facilities. Analyse their expected scientific returns; identify areas
of potential overlap and thus assess the extent to which the facilities
complement or compete; identify open areas that merit attention by one or
both organisations and suggest ways in which they could be addressed;

\item
Make an independent assessment of the scientific cases for large
facilities planned or proposed.

\item
The working group membership will be established by the chair and
co-chair. The views represented and the recommendations made in the final
report will be the responsibility of the group alone.

\end{enumerate}

\vskip 8em

\noindent{\bf Catherine Cesarsky (ESO) \hfill Alvaro Gim\'enez (ESA)\break}
\vskip 3em
\centerline{\bf September 2006}

\newpage

\tableofcontents

\newpage

\pagenumbering{arabic}
\setcounter{page}{1}

\japsec{Executive summary}\label{sc:Summ}

This report is written for ESA and ESO jointly, in order to summarize
current understanding of the fundamental properties of
the universe, and to identify the key areas in
which Europe should invest in order to advance this understanding.
There is an increasingly tight connection between
cosmology and fundamental physics, and we have concentrated on this area.
We thus exclude direct consideration of the exciting recent progress in
astrophysical
cosmology, such as the formation and evolution of galaxies, the
processes of reionization and the first stars etc.
However, many of our recommended actions will
produce vast datasets of general applicability, which
will also have a tremendous impact on these broader areas of astronomy.

This is an appropriate time to take stock. The past
10-15 years have seen huge advances in our cosmological
understanding, to the point where there is a
well-defined standard model that accounts in detail for (nearly) all
cosmologically relevant observations.
Very substantial observational resources have already been invested,
so the next generation of experiments is likely to be expensive.
Indeed, the scale of future cosmological projects will approach that
of particle physics, both in financial and in human terms.
We therefore need to identify the problems that are both the most
fundamental, and which offer the best prospects for solution. In doing
this, it is hard to look too far ahead, as our views on
priorities will doubtless evolve; but planning and executing large new
experiments will take time. We intend this report to cover the period
up to about 2020.

The standard model consists of a universe described by Einstein's
theory of general relativity, with a critical energy density
dominated today by a component that is neither matter nor radiation,
but a new entity termed `dark energy', which corresponds
to endowing the vacuum with energy. The remaining energy
consists of collisionless `cold dark matter' (about 22\%) and
ordinary `baryonic' material (about 4\%), plus trace amounts
of radiation and light neutrinos. The universe is accurately
homogeneous on the largest scales, but displays a spectrum of
inhomogeneities whose gravitationally-driven growth is
presumed to account for the formation of galaxies and
large-scale structure. The simplest consistent theory for
the origin of these features is that the universe underwent
an early phase of `inflation', at which time the density in
dark energy was very much higher than at present.
Given this background, there follows a natural
set of key questions:

\japitem{(1)} What generated the baryon asymmetry? Why is there negligible
antimatter, and what set the ratio of baryons to photons?

\japitem{(2)} What is the dark matter? Is it a relic massive supersymmetric
particle, or something (even) more exotic?

\japitem{(3)} What is the dark energy? Is it Einstein's cosmological
constant, or is it a dynamical phenomenon with an observable degree
of evolution?

\japitem{(4)} Did inflation happen? Can we detect
relics of an early phase of vacuum-dominated expansion?

\japitem{(5)} Is standard cosmology based on the correct physical principles?
Are features such as dark energy artefacts of a different law of
gravity, perhaps associated with extra dimensions?
Could fundamental constants actually vary?

Whereas we have not attempted to rank these prime science
questions in importance, we took into consideration the likelihood
that substantial progress can be made with astronomical
techniques. Additional information may in some cases be
provided by particle-physics experiments. For example,
in the case of the baryon asymmetry (1), we may hope
for major progress from particle-physics experiments
studying CP violation at the LHC, or at a neutrino factory in the
longer term. The nature of dark matter (2) will be investigated
by accelerators such as the LHC or underground dark matter
experiments, while astronomical observations will constrain
its possible properties;
similarly, tests of the law of gravity (5) will also be conducted
in the laboratory as well as on cosmological scales.
Empirical studies of the properties of dark energy (3)
and the physics of inflation (4) are possible only with the largest possible
laboratory available, namely the universe as a whole. On the other
hand, there may be some synergy with searches for the Higgs boson
at the LHC; these could provide a
prototype for other scalar fields,
which can be of cosmological importance.
Given their fundamental nature, studies of dark
energy and inflation are of the utmost
interest to the science community well beyond astrophysics.

Of all these cosmological issues, probably
the discovery of a non-vanishing dark energy density poses
the greatest challenge for physics.
There is no plausible or `natural' model for its nature, and
we must adopt empirical probes of its properties. For
example, undoubtedly one of the most important questions is whether
the dark energy is simply the cosmological constant introduced by
Einstein, or whether it has an equation of state that differs from
$w=-1$, where $w$ is the ratio of the pressure to the energy density.
Several highly promising methods for
studying the value of $w$ have been identified that can be actively
pursued within the next decade and which will lead to qualitatively
improved insights.  However, new ingredients such as $w$ will often be almost
degenerate in their effect with changes in standard parameters;
numbers such as the exact value of the dark matter density
must be determined accurately as
part of the road to $w$.

Progress in answering the foregoing questions will thus require a set of
high-accuracy observations, probing subtle features of cosmology that
have been largely negligible in past generations of experiment.
Our proposed approach is to pursue multiple independent techniques,
searching for consistency between the resulting estimates of
cosmological parameters, suitably generalised to allow for the
possible new ingredients that currently seem most plausible. If these
estimates disagree, this could indicate some systematic limitation of
a particular technique, which can be exposed by internal checks and
by having more than one external check. In addition, there is also the
exciting possibility of something unexpected.

The first step in these improvements will be statistical in nature:
because the universe is inhomogeneous on small scales,
`cosmic variance' forces us to study ever larger volumes
in order to reduce statistical errors in measuring the global
properties of the universe. Thus, survey astronomy inevitably
looms large in our recommendations. Remarkably, the recent
progress in cosmology has been so rapid that the next
generation of experiments must aspire to studying a large
fraction of the visible universe -- mapping a
major fraction of the whole sky in a range of wavebands,
out to substantial redshifts.

The key wavebands for these cosmological studies are the
cosmic microwave background (CMB) around 1~mm, for the
study of primordial structure at the highest redshifts
possible; optical and infrared wavebands for the provision of
spectroscopy and photometric redshift estimates,
plus data on gravitational-lensing image distortions; the
X-ray regime for the emission signatures of intergalactic gas,
particularly in galaxy clusters.
In principle the radio waveband can also be of
importance, and the ability to survey the universe via
21-cm emission as foreseen by the Square Kilometre Array
will make this a wonderfully powerful cosmological tool.
However, according to current estimates, the SKA will become
available close to 2020. We believe that great progress in
cosmology is however possible significantly sooner than
this, by exploiting the opportunities at shorter wavelengths.

The principal techniques for probing inflation and the properties of
dark matter and dark energy involve the combination of the CMB with at
at least  one other technique: gravitational lensing; baryon acoustic
oscillations; the supernova Hubble diagram; and studies of the
intergalactic medium.  The CMB alone is the richest source of direct
information on the nature of the initial fluctuations, such as whether
there exist primordial gravitational waves or entropy perturbations. But the
additional datasets allow us to study the cosmological model in two
further independent ways: geometrical standard rulers and the growth
rate of cosmological density fluctuations.  The majority of these
techniques have common requirements: large-area optical and near-IR
imaging with good image quality, leading to photometric redshifts.
This leads to the strongest of our recommendations, which we now list:

\begin{itemize}

\item ESA and ESO should collaborate in executing an imaging
survey across a major fraction of the sky by
constructing a space-borne high-resolution wide-field optical imager and
providing the essential multi-colour component from the ground,
plus also a near-IR component from space. The
VST KIDS project will be a pathfinder for this sort of data, but
substantial increases in grasp and improvements in image
quality will be needed in order to match or exceed global efforts
in this area.

\item Near-IR photometry is essential in order to extend photometric
redshifts beyond redshift unity. VISTA will be able to perform this
role to some extent with regard to KIDS. However, imaging in space
offers huge advantages in the near-IR via the low background, and
this is the only feasible route to quasi all-sky surveys in this band
that match the depth of optical surveys.
We therefore recommend that ESA give the highest priority
to exploring means of obtaining such near-IR data, most simply
by adding a capability for near-IR
photometry to the above satellite for high-resolution optical imaging.

\item In parallel,
ESO should give high priority to expanding its wide-field optical imaging
capabilities to provide the complementary ground-based
photometric data at wavelengths $\ls 700$~nm.
The overall optical/IR dataset (essentially 2MASS with a 7 magnitude increase in depth
plus an SDSS imaging survey 4 magnitudes deeper and with $\sim 3$
times larger area) would also be a profound resource for astronomy in
general, a legacy comparable in value to the Palomar surveys some 50 years ago.

\item Photometric redshift data from multi-colour imaging of this
sort enable two of the principal tests of dark energy:
3D gravitational lensing, and baryon oscillations in projection
in redshift shells.
Photometric redshifts are also essential in order to
catalogue clusters at high redshift, in conjunction with an X-ray
survey mission such as eROSITA. The
same is true for identifying the clusters to be detected by Planck
using the Sunyaev-Zeldovich effect.

\item Calibration of photometric redshifts is key to the success of this plan, thus
ESO should plan to conduct large spectroscopic surveys spread
sparsely over $\sim 10,000$ deg$^2$, involving $>100,000$ redshifts.
This will require the initiation of a large key programme with
the VLT, integrated with the imaging survey. Ideally, a new facility for
wide-field spectroscopy would be developed, which would improve the
calibration work, and also allow the baryon oscillations to be
studied directly and undiluted by projection effects.

\item A powerful multi-colour imaging capability can also carry out a
supernova survey extending existing samples of high-redshift SNe by an
order of magnitude, although an imager of 4m class is required if 
this work is to be pursued from the ground.
In order to exploit the supernova technique fully, an improved local
sample is also required. The VST could provide this, provided that time is
not required for other cosmological surveys, in particular lensing.

\item Supernova surveys need to be backed up with spectroscopy to assure
the classification for at least a significant subsample and to check
for evolutionary effects. The spectroscopy requires access to
the largest possible telescopes, and a European
Extremely Large Telescope (ELT) will be essential for the
study of distant supernovae with redshifts $z>1$.

\item A European ELT will also be important in fundamental
cosmology via the study of the intergalactic medium. Detailed
quasar spectroscopy can limit the nature of dark matter by
searching for a small-scale coherence length in the mass
distribution. These studies can also measure directly the
acceleration of the universe, by looking at the time dependence
of the cosmological redshift.

\item ELT quasar spectroscopy also offers the possibility of better
constraints on any time variation of dimensionless
atomic parameters such as the fine-structure constant $\alpha$ and the
proton-to-electron mass ratio. There presently exist controversial
claims of evidence for variations in $\alpha$, which potentially
relate to the dynamics of dark energy. It is essential to validate
these claims with a wider range of targets and atomic tracers.

\item In the domain of CMB research, Europe is well positioned with the
imminent arrival of Planck. The next steps are (1) to deal with the
effects of foreground gravitational lensing of the CMB and (2) to measure
the `B-mode' polarization signal, which is the prime indicator of
primordial gravitational waves from inflation.  The former effect is
aided by the optical lensing experiments discussed earlier. The latter
effect is potentially detectable by Planck, since simple inflation
models combined with data from the WMAP CMB satellite predict a
tensor-to-scalar ratio of $r\simeq 0.15$.  A next-generation
polarization experiment would offer the chance to probe this signature
in detail, providing a direct test of the physics of inflation and
thus of the fundamental physical laws at energies $\sim 10^{12}$ times
higher than achievable in Earth-bound accelerators.  For reasons of
stability, such studies are best done from space; we thus recommend
such a CMB satellite as a strong future priority for ESA.

\item An alternative means of probing the earliest phases
of cosmology is to look for primordial gravitational waves
at much shorter wavelengths. LISA has the potential to
detect this signature by direct observation of a background
in some models, and even upper limits would be of extreme
importance, given the vast lever arm in scales between
direct studies and the information from the CMB.
We thus endorse space-borne gravity-wave studies as an essential
current and future priority for ESA.

\end{itemize}

\japsec{Introduction}\label{sc:Intro}

The current human generation has the good fortune to be
the first to understand a reasonable fraction of the large-scale
properties of the universe. A single human lifespan ago, we
were in an utterly primitive state where the nature of galaxies was unknown,
and their recessional velocities undreamed of. Today, we know empirically
how the current universe emerged from a hot and dense early state,
and have an accurate idea of how this process was driven dynamically
by the various contributions to the energy content.

Proceeding initially on the assumption that general relativity is
valid, cosmology has evolved a standard model in which all of
astronomy is in principle calculable from six parameters. This
success is impressive, but it is bought at the price of introducing
several radical ingredients, which require a deeper explanation:

\blob An asymmetry between normal matter and antimatter.

\blob A collisionless component of `dark matter', which has been
inferred purely from its gravitational effects.

\blob A homogeneous negative-pressure component of `dark energy',
which has been inferred only from its tendency to accelerate the
expansion of the universe.

\blob A set of density fluctuations with a power-law spectrum, which are
acausal in the sense of having contributions with wavelengths that
exceed $ct$ at early times.

It is assumed that an understanding of these ingredients relates
to the initial conditions for the expanding universe. Since about
1980, the standard assumption has been that the key feature of
the initial conditions is a phase of `inflation'. This represents
a departure from the older singular `big bang' expansion history
at high energies (perhaps at the GUT scale of $\sim 10^{15}$~GeV).
With this background, we can attempt a list of some of the big
open questions in cosmology, which any complete theory must attempt
to address:

\japitem{(1)} What generated the baryon asymmetry?

\japitem{(2)} What is the dark matter?

\japitem{(3)} What is the dark energy?

\japitem{(4)} Did inflation happen?

\japitem{(5)} Are there extra dimensions?

\japitem{(6)} Do fundamental constants vary?

There are many further features of the universe that one would
wish to understand -- notably the processes that connect initial
conditions to complex small-scale structures: galaxies,
stars, planets and life. Nevertheless, the remit of the current
Working Group is restricted to what we have
termed {\it Fundamental Cosmology\/}, on the assumption that the
complex nonlinear aspects of small-scale structure formation
involve no unknown physical ingredients. We are therefore primarily
concerned with what astronomy can tell us about basic laws of
physics, beyond what can be probed in the laboratory.

Not all the key questions listed above are amenable to attack by astronomy
alone. For example, an important source of progress in the study of the
baryon asymmetry  will probably be via pure particle-physics experiments that
measure aspects of CP violation. Given a mechanism for CP violation,
it is relatively straightforward in principle to calculate the relic baryon
asymmetry -- the problem being that the standard model yields far too
low a value (see e.g. Riotto \& Trodden 1999).
Experiments that measure CP violation and are sensitive
to non-standard effects are thus automatically of cosmological
interest. This applies to experiments looking at the unitarity triangle
within the quark sector (BaBar, BELLE, LHCb), and to a future neutrino factory
that could measure CP violation in the neutrino sector -- which would be
studying physics beyond the standard model by definition. Results from
the LHC are expected to be relevant also to many other fundamental
cosmological problems. For example, the discovery of a Higgs boson
might provide some insight into the nature of dark energy or the
driving force for inflation. Likewise, the discovery of supersymmetry or
extra dimensions at the LHC might provide direct evidence for a dark
matter candidate whose properties could be determined and compared
with astrophysical and cosmological constraints. Underground direct
dark matter searches may detect the constituent of the dark matter
even before the LHC has a chance to look for supersymmetric signatures.
Current direct detection limits for Weakly Interacting
Massive Particles (WIMPs) that form the dark matter constrain a
combination of the particle mass and the spin-independent WIMP-nucleon
scattering cross-section. The best sensitivity is currently
achieved at masses around 100~GeV, where the cross-section must be
below $10^{-46}$~m$^2$. The simplest supersymmetric extensions of the
standard model predict cross-sections somewhere in the four orders
of magnitude below this limit, so success in direct WIMP searches is
certainly plausible, if hard to guarantee.
A future linear electron-positron collider would also be of cosmological
relevance, since it could study in more detail the spectrum of particles
accompanying such dark matter candidates. However, it is
beyond our remit to consider the capabilities and relative priorities of
different particle accelerators.

Within the compass of astronomy, then, one can consider the following
observables and techniques, together with their associated
experiments. This really focuses on a small list of observable
signatures (tensor modes and the `tilt' of the density power
spectrum, plus possible non-Gaussian and
isocurvature contributions in order to probe inflation; the equation of state parameter
$w(z)$ as a route to learn about dark energy):

\blobb Cosmic Microwave Background (CMB) anisotropies

\blobb Large-Scale Structure (LSS) from large galaxy redshift surveys

\blobb Evolution of the mass distribution from cluster surveys

\blobb Large Scale Structure from gravitational lensing surveys

\blobb The Hubble diagram for Supernovae of Type Ia

\blobb Studies of the intergalactic medium (IGM)

\blobb Gravitational waves

Subsequent sections in this report are constructed around asking what
each of these techniques can contribute to the study of fundamental
cosmology. Before proceeding, we now lay out in a little more detail
some of the relevant pieces of cosmological background that will be
common themes in what follows.

\japsec{The cosmological context}\label{sc:cosmo}

\ssec{Overview}\label{sc:cosmo.1}

The expansion of the universe is one of the most fundamental
observables in cosmology. On the one hand, according to the
Cosmological Principle, the universe is essentially homogeneous and
isotropic. In this limit, the history of the scale size describes the
universe completely, and its evolution in time is the primary source
of information about the history of the universe. On the other hand,
the rate of expansion of the universe is controlled by the density of
energy it contains, and hence is sensitive to all types of matter and
their interactions.  The initial discovery of the expansion of the
universe was a great surprise in itself. Together with the laws of
gravity, described by General Relativity,
the current expansion implies that the universe
evolved from a state of tremendous density and temperature, known as
the Big Bang. The predictions of Big Bang theory include the abundance
of the lightest chemical species generated a few minutes after the Big
Bang, predominantly helium and deuterium, as well as the existence of
a thermal radiation as a leftover from the early stages of cosmic
evolution. Both of these predictions have been verified with an
impressive accuracy.

The present acceleration of the cosmic expansion has also come as a
surprise. The past history of the universe may also have featured
periods of anomalous expansion, for example during an early
inflationary epoch. This is thought to have left traces in the
fluctuations in the microwave background radiation and to have seeded
the formation of structures in the universe, whose growth has been
sensitive to the subsequent expansion of the universe. {\it For these
reasons, the detailed measurement of the history of the expansion of
the universe throughout its visible epoch since the decoupling of the
microwave background is one of the critical frontiers in cosmology}.

There are thought to be at least five major components in the energy
density of the universe during this visible epoch, each of which has its
own characteristic signature and importance. The most obvious is
{\it conventional baryonic matter}, whose density is in principle
well constrained by the concordance between the values inferred
from astrophysical observations of light-element abundances and
the cosmic microwave background, though there may still be
discrepancies that need to be resolved. A second component is the
{\it cosmic microwave background radiation} itself. The contributions of
these components to the cosmological energy density have
well-understood time evolutions, but the same cannot be said for all
the other important contributions.

A third component
in the energy density is the analogous {\it cosmic neutrino background},
whose evolution with time depends on the masses of the neutrinos.
The most stringent upper limits on their masses are currently
provided by cosmology, with oscillation experiments providing
lower limits that are considerably smaller. {\it A key objective of
future cosmological measurements will be to bring the cosmological
limits into contact with the oscillation limits, and thereby
determine the overall neutrino mass scale}.

A fourth component in the cosmological energy density is {\it cold
dark matter}, whose density is thought to dominate the previous three
throughout the visible epoch. Measurements of the cosmic microwave
background, large-scale structures, the abundance of galaxy clusters
and high-redshift supernovae all contribute to the present constraints
on the cold dark matter density, which is currently known with an
accuracy of a few percent. {\it One of the key objectives of future
cosmological observations will be to refine this estimate of the cold
dark matter density, and thereby sharpen the confrontation with
particle theories of dark matter, such as supersymmetry}. Studies have
shown that, within the frameworks of specific theories, measurements
of particle properties at future colliders such as the Large Hadron
Collider (LHC) and International Linear Collider (ILC) may enable {\it
ab initio} calculations of the cold dark matter density with an
accuracy approaching the percent level. Future cosmological
measurements should strive to meet this challenge.

The fifth, largest, most recently identified and most surprising
component in the cosmological energy density is the {\it dark
energy}. The concordance of data on the cosmic microwave background,
structure formation and high-redshift supernovae points unambiguously
to the current accelerated expansion of the universe, implying the
existence of some distributed component of the energy density that is
not associated with concentrations of matter and which exerts negative
pressure. This dark energy density in the cosmic vacuum may be a
constant, in which case it can be identified with the `cosmological
constant' that was first postulated by Einstein. He subsequently
regarded it as his greatest mistake, but it should perhaps rather be
regarded as one of his deepest insights. However, the dark energy
density might not be constant; indeed, particle theories suggest that
it has varied by many orders of magnitude during the history of the
universe. {\it The big issue is whether the dark energy density has
still been varying during the visible epoch, and specifically whether
it is fixed today}.

In a theoretical sense, the greatest puzzle may not be that dark
energy exists, since we know of no fundamental reason why it should
vanish, but rather why its present value is so small. The generally
accepted theory of the strong interactions, QCD, makes a contribution
to the vacuum energy that is over 50 orders of magnitude larger than
the observed value. The standard electroweak theory makes a
contribution via the Higgs field that is a dozen orders of magnitude
larger still. The energy density during inflation would have been more
than 100 orders of magnitude larger than at present. Indeed, in our
present state of understanding, we would not know how
to exclude, with a plausible physical model, a density of vacuum
energy 120 orders of magnitude larger than the present value. One way
to confront this dilemma is to postulate that the vacuum energy has
been relaxing towards a very small value, in which case it may be
possible to observe its evolution during the visible epoch and even
today.

It is probable that a full understanding of dark energy will
require a quantum theory of gravity. The creation of such a theory
is the most profound problem in fundamental physics today; in
its absence, we can only speculate on the
origin and nature of dark energy. 
The most prominent candidate for a quantum theory of gravity is string
theory, but the guidance currently offered by string theory is
difficult to interpret. It has only recently been realized that the
myriads of potential string vacua are far more numerous even than had
been considered previously (Susskind 2003).
The overwhelming majority of the vacua in
this `string landscape' possess non-zero, constant dark
energy. However, in view of the past history of theoretical ideas on
this problem, it would surely be premature to conclude that no further
insights remain to be gained.  Under these circumstances, the most
appropriate observational approach is pragmatic, seeking direct
measurements of the possible evolution of the dark energy density. The
only place where such an empirical approach is feasible is the largest
laboratory available: our universe.

\ssec{The global contents of the universe}\label{sc:cosmo.2}

The expansion of the universe is described by the cosmic
scale factor, $R(t)$, which governs the separation of particles
in a uniform model.
The matter content of the universe is included in the Friedmann
equation which describes the scale factor as a function of time,
independent of the equation of state
\be
\dot R^2 - {8\pi G\over 3}\, \rho\, R^2 = -k c^2,
\ee
where $k$ is a constant describing the `curvature' of the universe
The density $\rho$ is
conveniently written in terms of density parameters
\be
{\Omega\equiv{\rho\over\rho_c} = {8\pi G\rho\over 3H^2}, }
\ee
where the `Hubble parameter' is $H=\dot R/R$.
Thus, a flat $k=0$ universe requires $\sum \Omega_i =1$ at all
times, whatever the form of the contributions to the density.
Empirically, it appears that the total curvature is
very small (Spergel et al. 2006 suggest $|\Omega_{\rm total}-1| \ls 0.02$),
and it is therefore often assumed that $\Omega_{\rm total}=1$ is exact.
The curvature does however influence conclusions about other
cosmological parameters, and conclusions that depend on $k=0$ 
can weaken considerably if this assumption is relaxed.

If the vacuum energy is a
cosmological constant, then
\be
{8\pi G\rho\over 3}=H_0^2\left(\Omega_{\rm v} + \Omega_{\rm m} a^{-3} +\Omega_{\rm r} a^{-4} 
\right)
\ee
(introducing the normalized scale factor $a=R/R_0$), where here and in
the following, the density parameters $\Omega$ are taken at the
current epoch. 
The Friedmann equation then becomes an expression for
the time dependence of the Hubble parameter:
\be
{ H^2(a)=H_0^2\left[ \Omega_{\rm v}
+\Omega_{\rm m} a^{-3} + \Omega_r a^{-4}
    -(\Omega_{\rm total}-1)a^{-2}\right].
}
\ee

More generally, we will be interested in the vacuum equation of state
\be
{ w\equiv P/\rho\, c^2}
\ee
If this is constant, adiabatic expansion of the vacuum
gives
\be
{8\pi G\rho_{\rm v}\over 3H_0^2} = \Omega_{\rm v} a^{-3(w+1)}.
\ee
If $w$ is negative at all, this leads to models that become
progressively more vacuum-dominated as time goes by. When this
process is complete, the scale factor should vary as a power of
time. In the limiting case of $w= -1$, i.e. a cosmological constant,
the universe approaches de Sitter space, in which
the scale factor attains an exponential behaviour.
The case $w<-1$, sometimes known as \key{phantom dark energy},
is interesting, in particular if $w$ is constant with time. Here the
vacuum energy density will eventually diverge, which has two
consequences: this singularity happens in a finite time, rather than
asymptotically; as it does so, vacuum repulsion will overcome the
normal electromagnetic binding force of matter, so that all objects
will be torn apart in the `big rip'.

\def\d{{\rm d}}

The comoving distance-redshift relation
is one of the chief diagnostics of the matter content. The general definition is
\be
D(z) = \int_0^z {c\over H(z')} \,\d z'.
\ee
Perturbing this about a fiducial $\Omega_{\rm m}=0.25$, $w=-1$ model shows a
`sensitivity multiplier' of about 5 -- i.e. a measurement of $w$ to 10\% requires an
accuracy of distance measures to 2\%. Also, there is a near-perfect
degeneracy with $\Omega_{\rm m}$, as can be seen in Fig.\ \ref{fig:w_effect}.

The other potential way in which the matter content
can be diagnosed is via the growth of structure.
The equation that governs the gravitational
amplification of fractional density perturbations, $\delta$, is
\be
\ddot\delta + 2 {\dot a\over a}\dot\delta =
  \delta \left( 4\pi G\rho_{\rm m} - c_s^2 k^2/a^2\right),
\ee
where $k$ is the comoving wavenumber (or $2\pi$ times reciprocal wavelength) of a perturbation.
Pressure appears through the speed of sound, $c_s$,
and the gravitational effects appear twice in this growth equation:
explicitly via $\rho_{\rm m}$ and implicitly via the
factor of $H = \dot a/a$ in the damping term.
This growth equation in general has two solutions, one of which is
decreasing in time and thus of little interest. The other one
describes the growth of density fluctuations.
While the universe is matter dominated and curvature
is negligible, the growing mode is just $\delta = g(a) \propto a$, but this growth
tends to slow at later times:
\be
g(a) \propto a\; f(\Omega); \quad f(\Omega) \simeq
\frac{5}{2}\Omega_{\rm m}\left[\Omega_{\rm m}^{4/7}-\Omega_{\rm v}+
 (1+\half\Omega_{\rm m})(1+\frac{1}{70}\Omega_{\rm v})\right]^{-1},
\ee
where the approximation for the growth suppression in low-density
universes (assuming $w=-1$) is due to Carroll, Press \& Turner (1992).
For flat models with
$\Omega_{\rm m}+\Omega_{\rm v}=1$, this says that the growth suppression is less
marked than for an open universe -- approximately $\Omega_{\rm m}^{0.23}$ as
against $\Omega_{\rm m}^{0.65}$ if $\Lambda=0$. This reflects the more rapid
variation of $\Omega_{\rm v}$ with redshift; if the cosmological constant
is important dynamically, this only became so very recently, and the
universe spent more of its history in a nearly Einstein--de Sitter
state by comparison with an open universe of the same $\Omega_{\rm m}$.

However, it is important to realise that, for $w\ne-1$, these standard
forms do not apply, and the second-order differential equation has to
be integrated numerically to obtain the growing mode.
In doing this, we see from Fig.~\ref{fig:w_effect}
that the situation is the opposite of the
distance-redshift relation: the effects of changes in $w$ and
$\Omega_{\rm m}$ now have opposite signs. Note however that
the sensitivity to $w$ displays the same `rule of 5' as
for $D(z)$: $|d\ln g/d w |$ tends not to exceed 0.2.

\epsfigsimptwo{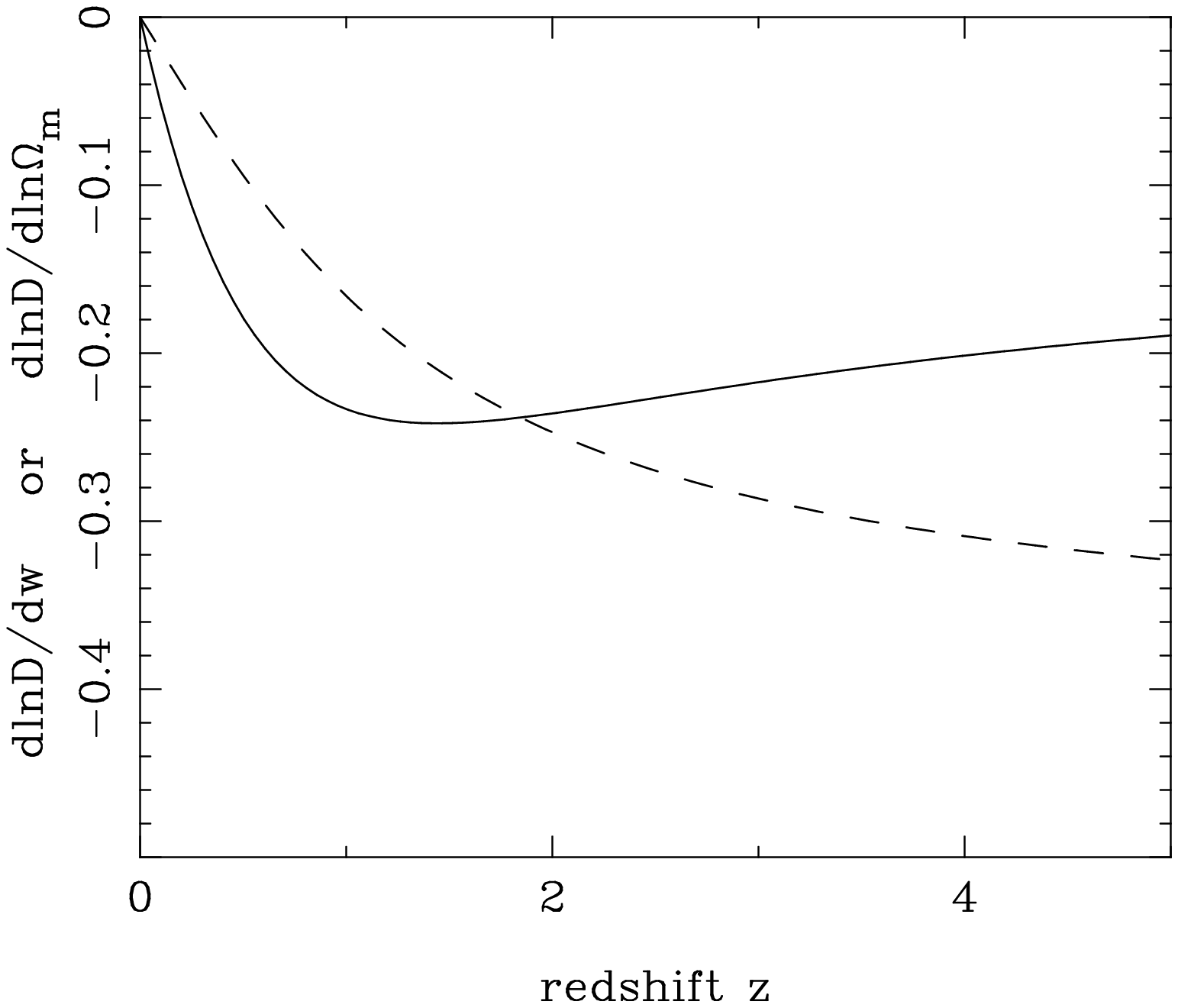}{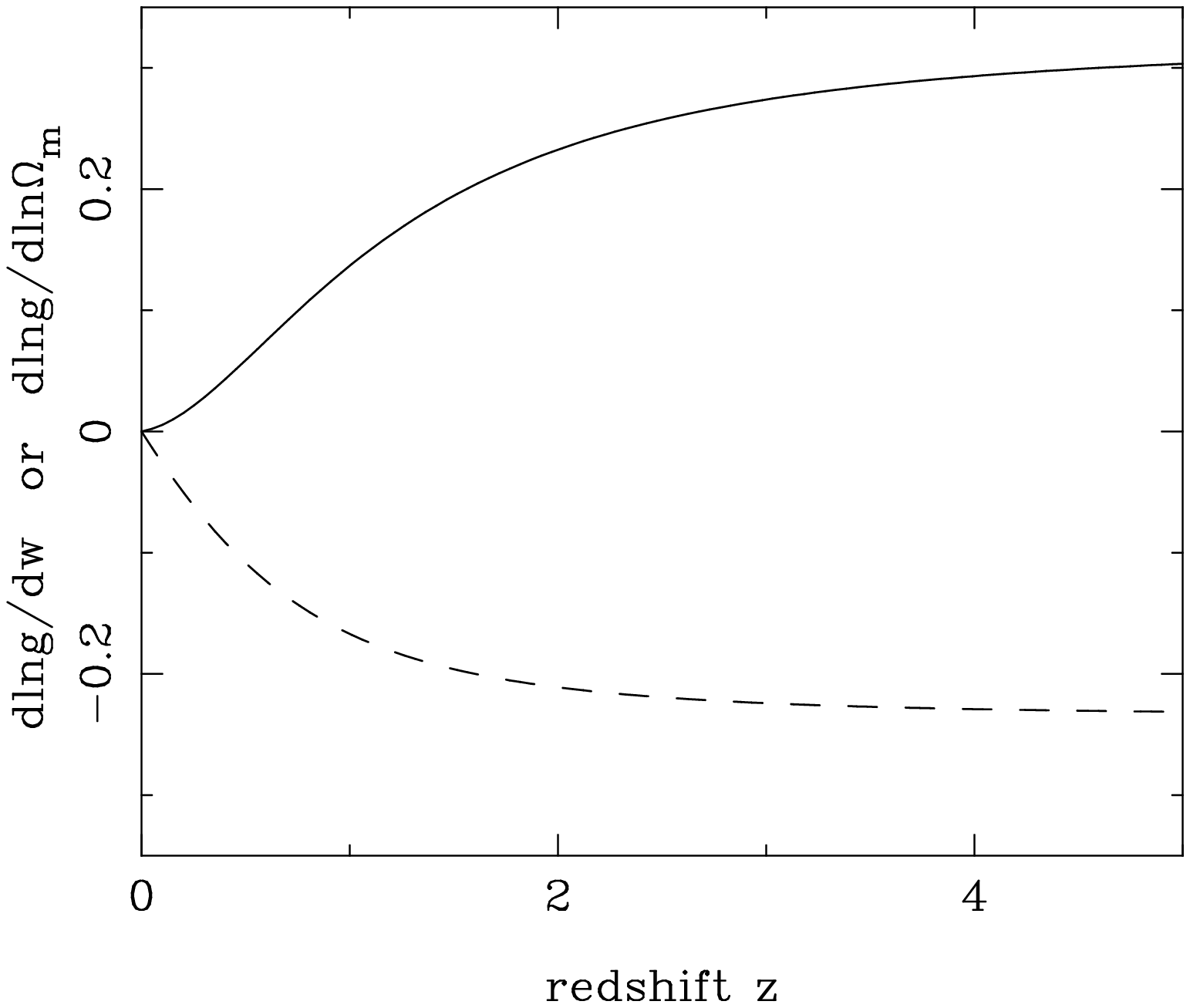}{0.52}
{Perturbation of cosmological observables around a
fiducial flat model with $\Omega_{\rm m}=0.25$ and $w=-1$ for the dark
energy of the distance-redshift
$D(z)$ and growth factor-redshift $g(z)$ relations.
The solid line shows the effect of increase in $w$; the dashed line
the effect of an increase in $\Omega_{\rm m}$.}
{fig:w_effect}

\sssec{Models for dark energy}
The existence of a negative-pressure component to the universe
was first strongly indicated from the mid-1980s via the lack of small-scale
structure in the CMB, plus indications of a low
matter density from large-scale structure
(e.g. Efstathiou, Sutherland \& Maddox 1990). Following the
more direct results from the supernova Hubble diagram in the late 1990s
(Riess et al. 1998; Perlmutter et al. 1999), it is
not seriously disputed that the universe contains something close
in nature to a cosmological constant.

However, there are no firm physical grounds for assuming that the
dark energy is indeed a simple time-independent vacuum density.
In the absence of any unique or even plausible theory, it is best to
be empirical and
allow the equation of state $w$ to be different from $-1$ or even
to vary; in this case, we should
regard $-3(w+1)$ as $d\ln\rho/d\ln a$, so that
\be
{8\pi G\rho_{\rm v}\over 3H_0^2} = \Omega_{\rm v} \exp\left(\int -3[w(a)+1]\; \d\ln
a\right).
\ee
In general, we therefore need
\be
{ H^2(a)=H_0^2\left[ \Omega_{\rm v} e^{\int -3[w(a)+1]\; \d\ln a}
+\Omega_{\rm m} a^{-3} + \Omega_r a^{-4}
    -(\Omega_{\rm total}-1)a^{-2}\right].
}
\ee
Some complete dynamical model is needed to calculate $w(a)$. Given
the lack of a unique model, the simplest non-trivial
parameterization is
\be
w(a) = w_0 + w_a(1-a).
\ee
Generally here we will stick with constant $w$; a given experiment
is mainly sensitive to $w$ at a particular redshift of order unity,
so one can treat it as constant for the purposes of comparing
raw sensitivities.

The simplest physical model for dynamical vacuum energy is
a scalar field, sometimes termed `quintessence'.
The Lagrangian density for a scalar field is as usual of the form of
a kinetic minus a potential term:
\be
{\cal L}=\half\partial_\mu\phi \, \partial^\mu\phi - V(\phi).
\ee
In familiar examples of quantum fields, the potential would be
a mass term:
\be
V(\phi) = \half\,m^2\, \phi^2,
\ee
where $m$ is the mass of the field. However, it
will be better to keep the potential function general at this stage.
Note that we use \key{natural units}
with $c=\hbar=1$ for the remainder of this section.
Gravity will be treated separately, defining the
Planck mass $m_{\japsub P}=(\hbar c/G)^{1/2}$, so that
$G=m_{\japsub P}^{-2}$ in natural units.

The Lagrangian lacks an explicit dependence on spacetime, and Noether's
theorem says that in such cases there must be a conserved
energy--momentum tensor. In the specific case of a scalar field, this is
\be
T^{\mu\nu}=\partial^\mu\phi\partial^\nu\phi-g^{\mu\nu}\displaystyle{{\cal L}}.
\ee
From this, we can read off the energy density and pressure:
\be
\eqalign{ \rho&=\half\dot\phi^2 +V(\phi)+\half(\nabla\phi)^2 \cr p
&=\half\dot\phi^2 -V(\phi)-
  {\textstyle{\scriptstyle 1\over\scriptstyle 6}}(\nabla\phi)^2. \cr
}
\ee
If the field is constant both spatially and temporally, the equation
of state is then $p=-\rho$, as required if the scalar field is to
act as a cosmological constant; note that derivatives of the field
spoil this identification.

Treating the field classically (i.e. considering the
expectation value $\langle\phi\rangle$), we get from energy--momentum
conservation ($T^{\mu\nu}_{;\nu}=0$) the equation of motion
\be
{ \ddot\phi+3H\dot\phi-\nabla^2\phi+dV/d\phi=0. }
\ee
Solving this equation can yield any equation of state,
depending on the balance between kinetic and potential
terms in the solution.
The extreme equations of state are: (i) vacuum-dominated,
with $|V| \gg  \dot\phi^2/2$, so that $p=-\rho$; (ii)
kinetic-dominated, with $|V| \ll  \dot\phi^2/2$, so that $p=\rho$. In
the first case, we know that $\rho$ does not alter as the universe
expands, so the vacuum rapidly tends to dominate over normal matter.
In the second case, the equation of state is the unusual $p=\rho$,
so we get the rapid behaviour $\rho\propto a^{-6}$. If a
quintessence-dominated universe starts off with a large kinetic term
relative to the potential, it may seem that things should always
evolve in the direction of being potential-dominated. However, this
ignores the detailed dynamics of the situation: for a suitable
choice of potential, it is possible to have a \key{tracker
field}, in which the kinetic and potential terms remain in a
constant proportion, so that we can have $\rho\propto a^{-\alpha}$,
where $\alpha$ can be anything we choose.

Putting this condition in the equation of motion shows that the
potential is required to be exponential in form. More importantly,
we can generalize to the case where the universe contains scalar
field and ordinary matter. Suppose the latter dominates, and obeys
$\rho_{\rm m}\propto a^{-\alpha}$. It is then possible to have the
scalar-field density obeying the same $\rho\propto a^{-\alpha}$ law,
provided
\be
V(\phi) \propto \exp[-\lambda\phi/M],
%V(\phi)={2M^4\over \lambda^2} (6/\alpha -1) \exp[-\lambda\phi/M],
\ee
where $M=m_{\japsub P}/\sqrt{8\pi}$.
The scalar-field density is $\rho_\phi = (\alpha/\lambda^2)\rho_{\rm
total}$ (see {\it e.g.} Liddle \& Scherrer 1999). The impressive
thing about this solution is that the quintessence density stays a
fixed fraction of the total, whatever the overall equation of state:
it automatically scales as $a^{-4}$ at early times, switching to
$a^{-3}$ after matter-radiation equality.

This is not quite what we need, but it shows how the effect of the
overall equation of state can affect the rolling field. Because of
the $3H\dot\phi$ term in the equation of motion, $\phi$ `knows'
whether or not the universe is matter dominated. This suggests that
a more complicated potential than the exponential may allow the
arrival of matter domination to trigger the desired $\Lambda$-like
behaviour. Zlatev, Wang \& Steinhardt (1999) tried to design a potential
to achieve this, but a slight fine-tuning is
still required, in that an energy scale $M\sim 1$~meV
has to be introduced by hand, so
there is still an unexplained coincidence with the energy scale of
matter-radiation equality.

\sssec{Modified gravity}
It should be emphasised that current inferences concerning dark energy
rest on the assumption of the validity of the Friedmann equation,
which is based on Einstein's gravitational field equations. This is
the simplest possibility, but something more complex could still be
permissible in general relativity: i.e. there might still be a
Robertson-Walker metric, but non-standard dynamics.

The most interesting possibilities of this sort to emerge from recent
work are modifications motivated by the predictions from
string theory of the existence of higher dimensions. The hidden
scale associated with these dimensions allows a scale dependence of
the strength of gravity, which can mimic cosmic acceleration.
For example, in the DGP model (Dvali, Gabadadze \& Porrati 2000), we have the
relation
\be
H^2(z) =H_0^2 \left( {1-\Omega_{\rm m}\over 2} +
\sqrt{\left({1-\Omega_{\rm m}\over 2}\right)^2 + \Omega_{\rm m} a^{-3}} \; \right)^2
\ee
(neglecting radiation), so that the universe tends to a de Sitter
model with constant $H$ even without an explicit vacuum energy.
This particular model seems to be a less good
fit than standard $\Lambda$CDM (Sawicki \& Carroll 2005), but it
serves to remind us that dark energy may be more complex in nature
than is suggested by the standard parameterization. One way in
which this can be addressed is to pursue multiple probes
of dark energy within the standard framework and to search
for concordance. If this fails to appear, then either the model
is more complex than assumed, or there is some unidentified systematic.

An even more radical possibility under this heading is the
attempt to dispense with dark matter altogether,
by introducing Modified Newtonian Dynamics (MOND). This
model was introduced by Milgrom in order to account for the
flat rotation curves of galaxies, and for many years
it remained on an ad hoc basis. However, Bekenstein (2004)
has proposed a covariant generalisation in
a theory whereby gravity is a mixture of tensor, vector
and scalar fields (the TeVeS hypothesis). It is not yet clear
whether this theory is well-defined from a field theory point of view
nor whether it agrees with local constraints in the Solar System,
but TeVeS certainly allows a
much richer spectrum of possible tests, ranging from
global cosmological models through large-scale structure
and gravitational lensing. The indications are that the
model still requires both dark matter and $\Lambda$ in
order to be consistent (Zhao et al. 2006; Skordis et al. 2006).
It has also received an impressive challenge from observations
of the `bullet cluster' (Clowe et al. 2006), in which an apparently
merging pair of cluster shows the X-ray emitting gas lying
between the two groups of galaxies -- each of which is inferred
to contain dark matter on the grounds of their gravitational
lensing signature. The standard interpretation is that the
baryonic material is collisional, whereas the dark matter is collisionless;
this object seems to dramatise that view most effectively.

\ssec{The perturbed universe}\label{sc:cosmo.3}

It has been clear since the 1930s that galaxies are not
distributed at random in the universe (Hubble 1934). For decades, our
understanding of this fact was limited by the lack
of a three-dimensional picture, but current studies have
amassed redshifts for over $10^6$ galaxies. Following the
detection of structure in the CMB by
NASA's COBE satellite in 1992, we are
able to follow the growth of cosmological structure over
a large fraction of the time between last scattering at
$z\simeq 1100$ and the present. In effect, such studies
provide extremely large standard rulers, which have done
much to pin down the contents of the universe.

In discussing this area, it will be convenient to adopt a notation,
already touched on above,
in which the density (of mass, light, or any property) is
expressed in terms of a dimensionless density
contrast $\delta$:
\be
1+\delta({\bf x}) \equiv \rho({\bf x}) / \langle\rho\rangle,
\ee
where $\langle\rho\rangle$ is the global mean density.
The existence of density fluctuations in the universe raises two
questions: what generated them, and how do they evolve?
A popular answer for the first question is inflation, in which quantum
fluctuations, inflated to macroscopic scales, are able to seed density
fluctuations. So far,
despite some claims, this theory is not proved, although
it certainly matches current data very well. The second question is
answered by the growth of density perturbations through gravitational
instabilities, as was discussed earlier.

The Fourier power spectrum, $P(k)$, of this fluctuation field, $\delta$,
contains a rich structure. It can be written dimensionlessly
as the logarithmic contribution to the fractional
density variance,
\be
\Delta^2(k) \equiv {k^3\over 2 \pi^2} \, P(k)   \propto k^{3+n_{\rm s}}\, T^2(k),
\ee
where $k$ is again the wavenumber,
$T(k)$ is the matter transfer function, and $n_{\rm s}$ is the primordial
scalar perturbation spectral index. This is known empirically
to lie close to the `scale-invariant' $n_{\rm s}=1$, but one of the key
aims in testing inflation is to measure `tilt' in the form of $n_{\rm s}\ne1$.
Excitingly, the 3-year WMAP data has given the first evidence in this
direction, measuring $n_{\rm s}\simeq 0.96$ (Spergel et al. 2006). 

The transfer function is sensitive
to the class of perturbation. This is
normally assumed to be adiabatic, so that both nonrelativistic matter
and photon density are perturbed together. The alternative is entropy
or isocurvature perturbations, in which only the equation of state is altered.
In the limit of very early times, these correspond to keeping the
radiation uniform and perturbing the matter density only.
The resulting behaviour for the transfer functions is very
different for these two modes, as shown in Fig.~\ref{fig:tkplot}.
The isocurvature mode also makes rather distinct predictions for
the CMB anisotropies, so that we can be confident that the initial
conditions are largely adiabatic. But an isocurvature admixture at the
10\% level cannot be ruled out, and this certainly relaxes many
parameter constraints (e.g. Bean, Dunkley \& Pierpaoli 2006).

%\epsfigsimp{tkplot.ps}{0.55}
\epsfigsimp{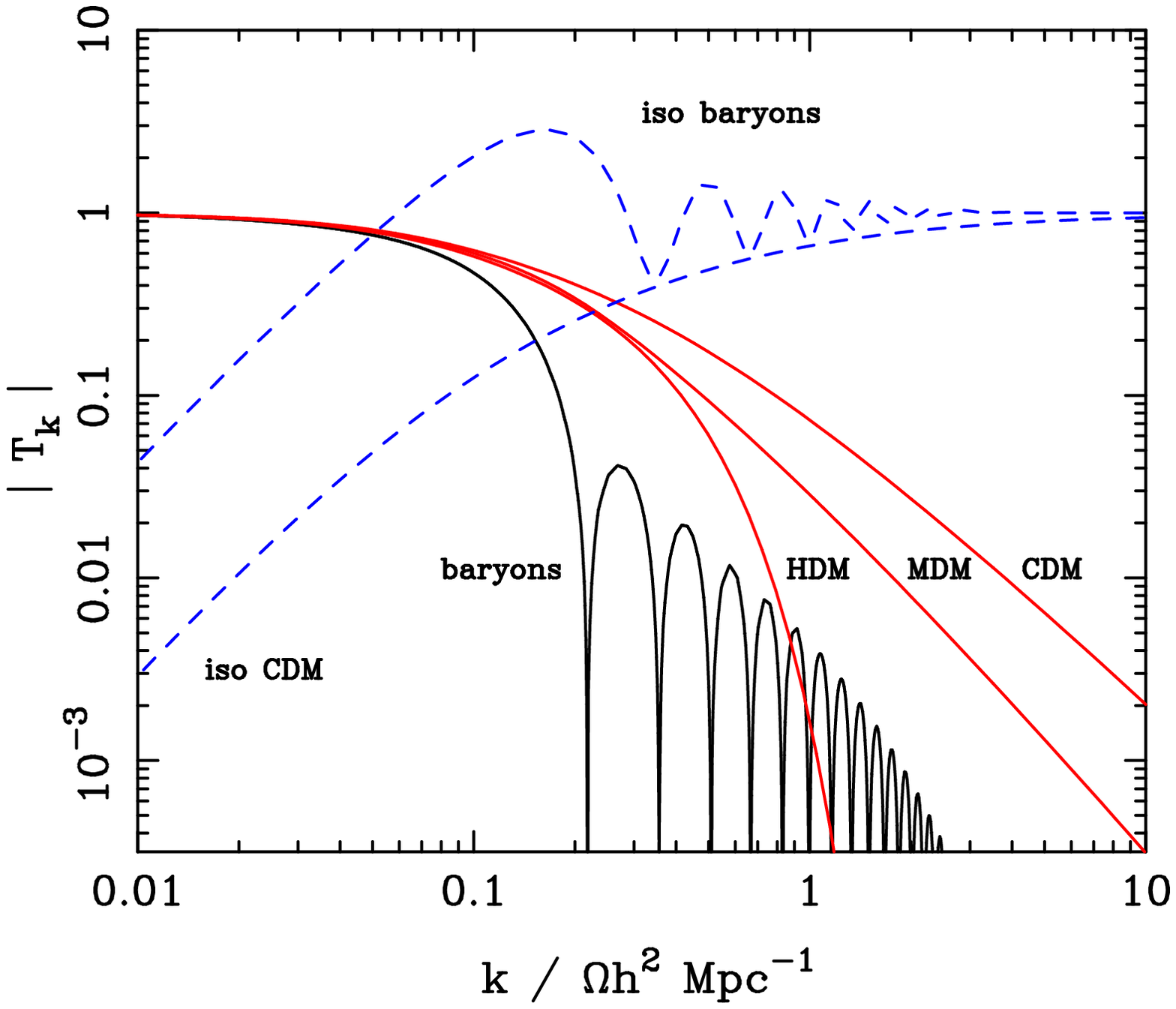}{0.80}
{A plot of cosmological transfer functions for matter perturbations.
Solid lines show adiabatic models, in which both matter and radiation are
perturbed; dashed lines are isocurvature perturbations.
A number of possible matter contents are illustrated:
pure baryons; pure CDM; pure HDM.
For dark-matter models, the characteristic wavenumber scales
proportional to $\Omega_{\rm m} h^2$, marking the break
scale corresponding to the horizon length at
matter-radiation equality. The scaling for baryonic
models does not obey this exactly; the plotted case corresponds
to $\Omega_{\rm m}=1$, $h=0.5$.
}
{fig:tkplot}

The transfer function also depends on
the density parameters, as well as on the physical properties of
the dark matter. Fluctuations of wavelength smaller than the cosmological
horizon have their growth affected, so that there should be a break
in the spectrum at around the horizon size at the redshift of matter-radiation equality:
\be
D_{\japsub EQ}^{\japsub H} \simeq 123\, (\Omega_{\rm m} h^2/0.13)^{-1}\; {\rm Mpc}.
\ee
In addition, small-scale perturbations will be erased by free-streaming
of collisionless dark-matter particles until the point at which they
become non-relativistic. This marks out the horizon size at this
non-relativistic era, which depends on particle mass:
\be
L_{\rm free-stream} = 112\, (m/{\rm eV})^{-1}\, {\rm Mpc}.
\ee
It is this large coherence length for the case of light neutrinos
that led to such hot dark matter being rejected as the dominant
constituent of the universe. We need a mass large enough that
free-streaming preserves small-scale structures such as
high-redshift galaxies and the Lyman-alpha forest: thus
$m\gs 1$~keV (unless the particle is something like the axion,
which was never in thermal equilibrium).
Even if the mass of the dominant dark-matter particle should
turn out to be too high to affect astronomy, however, we know from
neutrino-oscillation results that light neutrinos must exist
with $m\gs 0.05$~eV, and these can have an effect on
high-precision measurements of large-scale structure.
Measuring the absolute mass scale of neutrinos through
such effects is important not only for particle physics, but
because the neutrino effects are degenerate with changes
in the vacuum equation of state, $w$ 
(e.g. Goobar et al. 2006).

Finally, we should remark on the important possibility of tensor-mode
perturbations, or primordial gravitational waves. These have a negligible
effect on the matter distribution, and only manifest themselves
via perturbations in the CMB -- unless they can be detected
directly via experiments sensitive to the local strain of spacetime.

\ssec{Statistical methodology}\label{sc:cosmo.4}

The standard methodology for forecasting errors in
cosmological parameters is in terms of the Fisher Matrix,
which is the expected curvature matrix of the likelihood of a given model
in the face of data:
\be
F_{ij} = -\left\langle {\partial^2\ln L\over \partial p_i \partial p_j}
\right\rangle\;,
\ee
where the $p_i$ are a set of parameters characterising the model.
The inverse of $F_{ij}$ gives a pseudo-covariance matrix, such that
the diagonal elements are lower limits to the true variance in a
given parameter -- although they are usually taken to be a reasonable
estimate of the actual expected error. 
The Fisher matrix thus defines a multi-dimensional Gaussian ellipsoid
that gives the joint confidence region for the parameters.
It is straightforward to marginalize over unwanted
parameters (i.e. integrating the likelihood over all values of
the hidden parameters). This yields a lower-dimensional projected
likelihood, which is 2D may be plotted as a confidence ellipse,
as shown in several plots below (e.g. Figs \ref{fig:wmapinf}, \ref{fig:fish7})
The results of doing this can often be
a little counterintuitive when experiments are combined. Here,
we are inspecting the intersection of two ellipsoids in a
multidimensional space, and sometimes the parameter degeneracies
can work out such that the intersection has a very compact
projection onto a two-parameter plane, whereas this is not
true of the individual constraints (think of two planes intersecting
in a line).

Using the $w(a) = w_0 + w_a(1-a)$ model for the equation of state of
dark energy, this exercise normally will show a strong 
correlation between $w_0$ and $w_a$. This is readily
understandable: the bulk of the sensitivity comes from data at higher
redshifts, so the $z=0$ value of $w$ is an unobserved
extrapolation. It is better to assume that we are observing the value
of $w$ at some intermediate {\it pivot redshift\/}:
\be
w(a) = w_{\rm pivot} + w_a(a_{\rm pivot}-a).
\ee
The pivot redshift is defined so that $w_{\rm pivot}$ and $w_a$ are
uncorrelated -- in effect rotating the contours on the $w_0 - w_a$ plane.
If we do not want to assume the linear model for $w(a)$, a more general
approach is given by Simpson \& Bridle (2006), who express the
effective value of $w$ (treated as constant) as an average over its
redshift dependence, with some redshift-dependent weight. Both these
weights and the simple pivot redshifts depend on the choice of
some fiducial model. With reasonable justification (both from existing
data, and also because it is the fiducial model that we seek to disprove),
this is generally taken to be the cosmological constant case.

One question with the whole issue of measuring $w$
and its evolution is what our target should be. In some
areas of cosmology, such as the scalar spectral tilt, there are
classes of simple inflationary models that make clear predictions
(deviations from $n_{\rm s}=1$ of order 1\%), and a sensible
first target is to test these models. With dark energy,
we have much less idea what to expect. The initial
detections of dark energy were made on the assumption of a
cosmological constant, and this remains a common prejudice.
If this is in fact the truth, how small do the errors around
$w=-1$ have to be before we are convinced? Trotta (2006)
gives a nice analysis of this sort of situation from a Bayesian
point of view. Rather than only ever rejecting theories, he shows
how it is possible to develop statistical evidence in favour
of a simple model ($w=-1$ in this case). Errors on a constant $w$
of around 0.5\% would place us in this situation (unless
the $w=-1$ model has been rejected before this point).
This represents a 10-fold improvement on the current state
of knowledge, or roughly a 100-fold expansion in data volume.
This is challenging, but eminently feasible -- and such
advances will in any case have other astronomical motivations.
Therefore, in practice the ability to confirm or disprove $\Lambda$
is most likely within reach.

In the following sections we will discuss a number of astronomical
methods that have the capability of yielding significant constraints on
cosmological parameters via the sensitivity of their results to
the growth function $g(z)$ of density perturbations and the
redshift-distance relation $D(z)$. Both of these quantities contain the
expansion history $a(z)$ of the universe and are thus sensitive to the
function $w(z)$. Furthermore, those methods that probe the density
fluctuation field can constrain the slope $n_{\rm s}$ of the primordial
power spectrum of density fluctuations and thus test predictions from
inflationary models.

\ssec{Photometric redshifts}\label{sc:phot-z}
Many of the possible projects that we describe below
will require redshifts of
$\sim 10^6$ to $\sim 10^9$ galaxies, most of which are faint enough
to require many hours of integration on current 8m-class telescopes.
Such observations may be possible with facilities we can anticipate
for 2020, but are presently impossibly expensive.

Therefore, an alternative approach has been developed
and tested in recent years: {\it photometric redshifts\/}. This method
uses a set of photometric images in several broad-band filters and
attempts to obtain an approximate redshift by isolation of the (small)
effect that a given spectral feature will produce as it moves through
a given bandpass. The direct approach to this problem assumes that the
filter profiles are known exactly and that galaxy spectra can be
expressed as a superposition of some limited set of template
spectra. The results are thus potentially vulnerable to errors in the
assumed filter properties and/or templates.  Given a large enough
calibration sample with exact spectroscopy, such errors can be
removed; it is also then possible to employ `blind' methods such as
artificial neural nets, with an impressive degree of success. These
methods effectively fit for the empirical redshift in multi-colour
space, without needing to know about filters or spectra (see Collister
\& Lahav 2004).

Photometric redshift codes generate a probability distribution in
redshift for each object, based on the multi-colour flux values and
their estimated errors. If this probability distribution shows a
single narrow peak, the corresponding redshift is a good and, most likely,
correct estimate of the true redshift, and the redshift accuracy can
be estimated from the width of the distribution.
However, in case of very broad peaks, or multiple peaks, the
interpretation and use of these redshift probabilities is less
straightforward. Depending on the application, one can either
discard such galaxies, take the highest peak (or the peak with the
largest integrated area) as the most likely redshift, though with a
potentially large error, or make use of the full redshift probability
distribution.

The performance of photometric redshift estimates on a set of objects
can be quantified by at least three numbers. The first is the rms
deviation of the estimated redshift from the true value.
Second, one characterises the
deviation of the mean estimated redshift from the mean of the true
redshift, say within a given photometric redshift interval. This
number then yields the {\it bias\/} of the method, i.e. a systematic offset of
the photometric redshifts from the true ones -- this error is more
of a limitation for large samples than the single-object random error.
And third, there is the fraction
of estimated redshifts that are grossly wrong, which are called
catastrophic outliers. All these indicators for the performance are
improved if the number of photometric bands and their total
$\lambda$-coverage are increased. In particular, the fraction of
catastrophic outliers is substantially reduced with the inclusion of
near-IR photometry, as shown in Fig.~\ref{fig:photz}.
The performance can also be improved by including prior
information, such as assumptions regarding the
evolving type- and redshift-dependent galaxy luminosity functions.
In addition, the way in which the magnitudes
of galaxies are measured (fixed aperture photometry vs. seeing-matched
photometry) can have an important effect.
See Hildebrandt et al. (2006) for a comparative study of
some of these issues.

The performance of photometric redshifts depends
particularly strongly on the galaxy
type. Early-type galaxies are known to have very homogeneous
photometric properties, due to their old stellar populations and the
smallness of their intrinsic extinction. Their 4000{\AA} break
provides a clean spectral feature that yields reliable photometric
redshifts, provided that the wavelength coverage is sufficiently large that
this 4000{\AA} break can be distinguished from the Lyman break,
corresponding to much higher redshifts. For spiral galaxies,
quite reliable redshift estimates can also be obtained. In contrast,
irregular and star-forming galaxies tend to have spectra dominated
by a featureless blue continuum, making redshift estimation very
difficult except at the highest redshifts where UV breaks from
intervening absorption can be seen.

Currently, the best performance comes from the COMBO-17 survey,
which extended the method to include intermediate-width filters, for a
total of 17 bands. Their brightest objects have a precision $\delta
z/(1+z) \simeq 0.01$ (note the inevitable factor $1+z$, which arises
when we are able to measure observed wavelengths of some feature to a
given precision).
For surveys with fewer bands, the fractional precision in $1+z$ is
typically larger. Current photometric redshifts calibrated with deep
enough spectroscopic galaxy samples have typical errors of $\Delta
z/(1+z)=0.035$ (Ilbert et al. 2006) without showing a systematic
offset. The accuracy on the mean redshift is obtained via the huge
number of galaxies in each bin (several tens of thousands) in this
study. An rms scatter of the photometric redshifts around the true
one of 3\% to 5\% is typical for data with several optical broad-band
filters, although clearly the accuracy depends on magnitude: we expect
a precision that scales with the typical magnitude error, so that the
bulk of an imaging catalogue (the $5\sigma$ objects) is of little use
for redshift estimation. In addition, going to fainter magnitudes, the
spectroscopic redshifts also become more uncertain (and less
complete), particularly affecting the `redshift desert' between $z\simeq
1.2$ and $z\simeq 2$ where few spectroscopic signatures are located in
the optical passband.  Hence, the precision depends on redshift
itself. We are looking to track the motion of features such as the
4000{\AA} break though the set of bandpasses, and this procedure fails
when we reach the reddest band. Thus, in order to obtain reliable
photo-z's at $z\gs 1$, near-IR photometry is essential.

\epsfigsimp{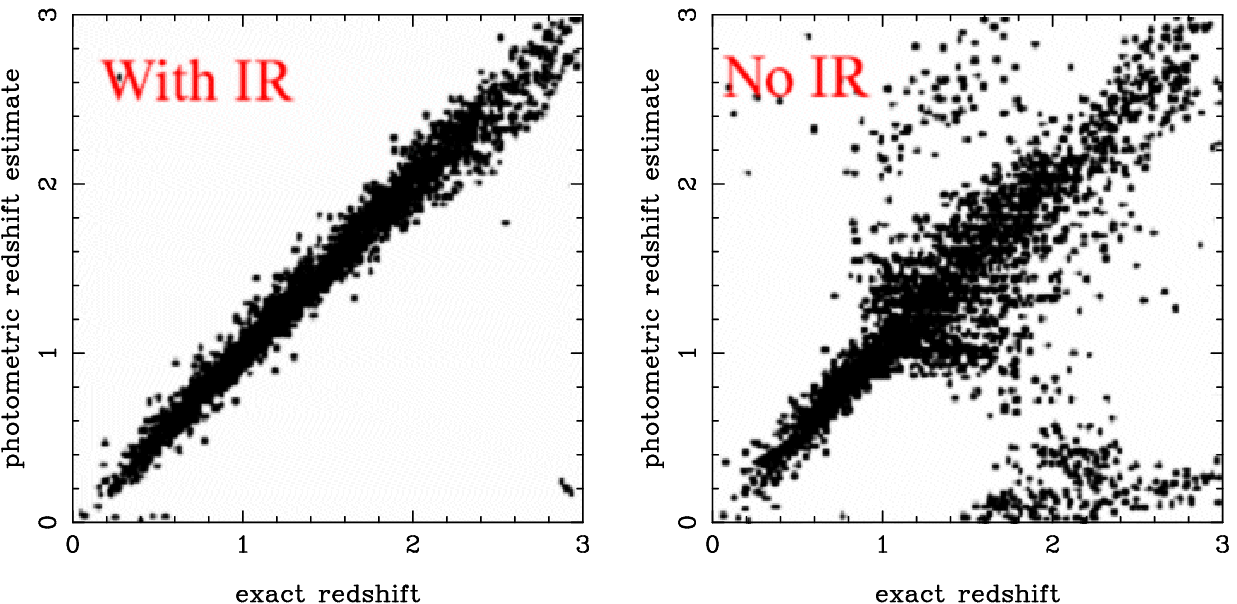}{0.85}
{An illustration of how the accuracy of photometric redshift
estimation is affected by the number of wavebands,
based on data from R. Pell\'o. On the
left, we show that a full optical/IR dataset ($ugrizJHK$) yields
well controlled redshift estimates out to $z\simeq 3$. On
the right, we see that removing the near-IR data induces
catastrophic failures in the predicted redshifts at $z\gs 1$.
}
{fig:photz}

Some applications such as weak lensing require the knowledge
of the error distribution of photometric redshift to exquisite
accuracy (see Sect.~\ref{sc:GL}). The mean redshift of a given sample
can be affected significantly by even a small fraction of catastrophic
outliers. A key question is thus the fraction of galaxies with catastrophic or
unknown photometric redshift. In principle, such objects
should be identified from the redshift probability distribution obtained from
the photo-z codes, but this capability is not yet routinely available.
Unless we can be certain of the accuracy of each photo-z, the
simple method of discarding all objects with uncertain or
possibly catastrophic redshifts will not work. It is therefore
much better to choose a filter set that eliminates these problems.
The best filter choice depends on
the redshift distribution of the underlying sample and should
therefore be defined on a case by case basis.

As surveys go deeper, the typical redshift increases, so that
(1) galaxies become so faint that there is no
spectroscopy to calibrate the photo-z's;
(2) more and more galaxies enter the redshift range $1.2<z<2.5$ --
which cannot be sampled without the use of deep near-IR data.
It is likely that the
uncertainty of the redshift distribution will continue to be a
dominant source of error in cosmological parameter estimation,
e.g. from faint weak lensing surveys, unless we can obtain
sufficiently extensive multi-colour photometry.  A major step in
improving the calibration of photometric redshift surveys are the VVDS
and the DEEP2 spectroscopic samples on fields of the CFHT Legacy
Survey fields.  For much deeper surveys like those proposed with the
LSST or for the JDEM/SNAP/DUNE missions, it is therefore urgent to
start extremely deep visible and near infrared spectroscopic surveys
that will sample galaxies at the depth of these surveys.
The VIMOS instrument on the VLT has the best
potential of current ESO instruments to produce spectroscopic calibration
data of this sort.

\japsec{The Cosmic Microwave Background}\label{sc:CMB}

\ssec{Current status}\label{sc:CMB.1}

The anisotropies in the CMB form arguably the pre-eminent tool of
modern cosmology. Almost immediately after the discovery of the
background radiation,  over forty years ago, cosmologists realised that
the CMB could provide an entirely new way of studying
the early universe. The fluctuations responsible for structures that we
see today -- galaxies, clusters and superclusters -- must have
imprinted small differences in the temperature of the CMB (so-called
`anisotropies') coming from different directions of the sky.
The CMB anisotropies were first discovered in 1992 by the COBE satellite
and since then the CMB sky has been mapped with great precision
by many ground-based and balloon-borne experiments, culminating
in the highly successful WMAP satellite.

The CMB anisotropies provide an especially powerful cosmological probe
because they are imprinted at a time when the universe was only
400,000 years old. At this early stage, all of the structure in the
universe is expected to be of small amplitude and to be well
characterised by linear perturbation theory. The likely absence of
non-linear phenomena at the time that the CMB anisotropies were
generated makes them a uniquely clean probe of cosmology.  We believe
we understand the physical processes at that epoch and can set up and
solve the corresponding equations describing anisotropies to obtain
very accurate predictions that can be compared with observations.

By studying the CMB anisotropies, we are also looking back to great
distances ($z\simeq 1100$).  The angles subtended by structures in the
CMB, together with the simple atomic physics of the recombination
process that sets physical scales, provide a direct route to the
geometry of the universe.  This geometrical sensitivity can be used to
construct an accurate inventory of the matter and energy content of
the universe. In addition, the statistical properties of the CMB
anisotropies provide a unique window on conditions in the ultra-early
universe and, in particular, on the generation of the fluctuations.

As pointed out forcefully by Guth (1981), an early period of
inflation, corresponding to near exponential expansion, offers
solutions to many fundamental problems in cosmology.  Inflation can
explain why our universe is so nearly spatially flat without recourse
to fine-tuning, since after many e-foldings of inflation spatial
curvature will be negligible on the scale of the present Hubble
radius. Furthermore, the fact that our entire observable universe
might have arisen from a single causal patch offers an explanation of
the so-called horizon problem (e.g., why is the temperature of the CMB
on opposite sides of the sky so accurately the same if these regions
were never in causal contact in the framework of standard Friedmann
expansion?).  But perhaps more importantly, inflation offers an
explanation for the origin of fluctuations.

In the simplest models, inflation is driven by a single scalar field $\phi$
with a potential $V(\phi)$. As well as the characteristic energy
density of inflation, $V$, inflation can be characterised by two
`slow-roll'  parameters, $\epsilon$ and $\eta$, which are given by
the first and second derivatives of $V$ with respect to $\phi$:
\be
\epsilon = { m^2_{\japsub P} \over 16 \pi} \left ( V^\prime \over V \right)^2,
\quad  \eta =  { m^2_{\japsub P} \over 8 \pi} \left ( V^{\prime\prime}
\over V \right)  ,
\ee
where $m_{\japsub P}$ denotes the Planck mass.  Generally, a successful
model of inflation requires $\epsilon$ and $\eta$ to be less than
unity.  In terms of these parameters, the inflationary predictions for
the scalar perturbation spectral index is
\be
n_{\rm s} = 1-6\epsilon +2\eta.
\ee
In simple single field models, $\epsilon$ and $\eta$ are small during
the early stages of inflation but necessarily must become of order
unity for inflation to end. Thus, generically, some `tilt' (deviation
from a scale invariant spectrum, $n_{\rm s} = 1$) is expected ({\it e.g.}
$n_{\rm s} \simeq 0.97$ for a quadratic potential $V(\phi) \propto
\phi^2$). Furthermore, some deviation from a pure power law (a `run'
in the spectral index) should be seen to second order in the slow roll
parameters.

\epsfigsimp{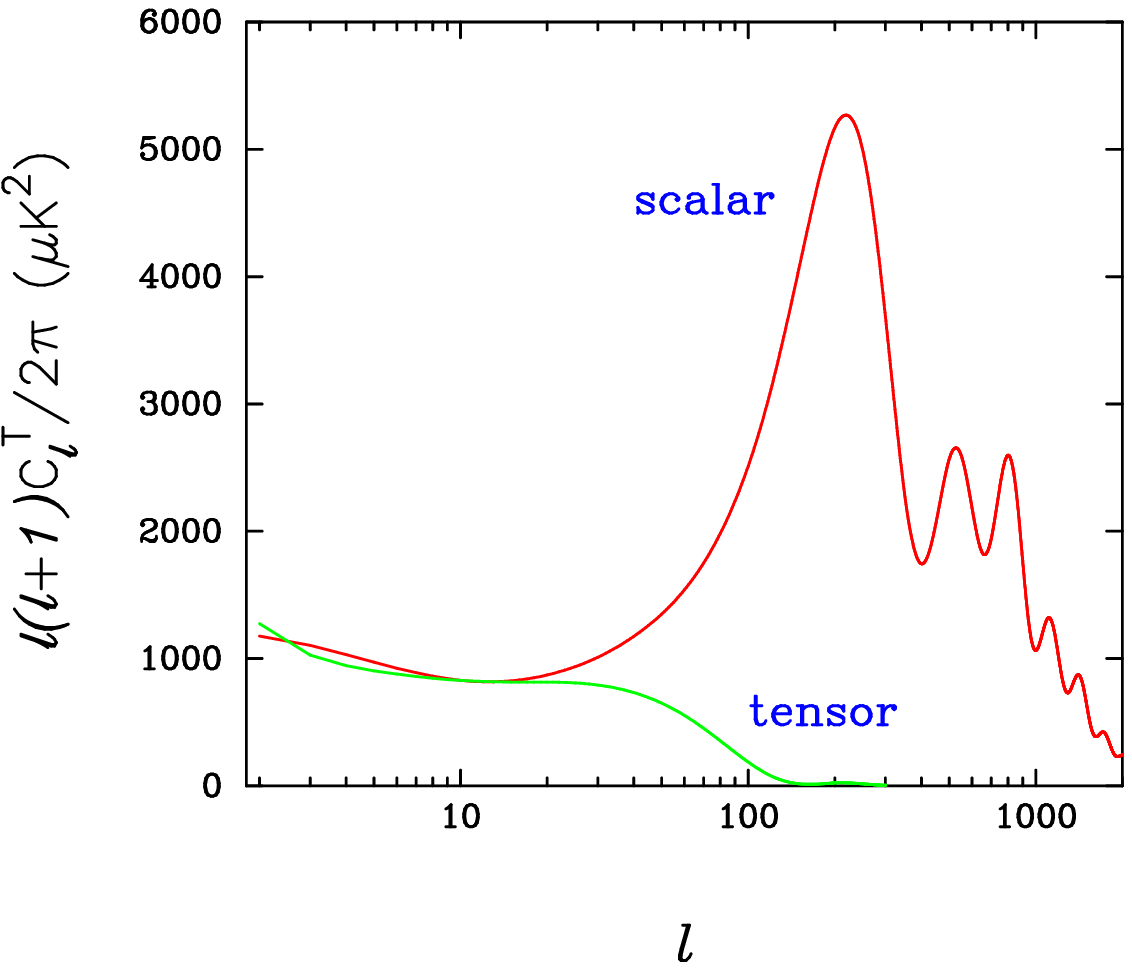}{0.65}
{The contributions to the power spectrum of CMB anisotropies
from scalar and tensor (gravitational wave) perturbations generated during
inflation. The concordance  $\Lambda$CDM model has been assumed with exactly
scale-invariant initial fluctuations.
The amplitudes of the tensor and scalar power
spectra have been chosen arbitrarily to be equal at $\ell = 10$.}
{fig:cmbfig1}

In addition to scalar modes, inflation generates a spectrum of tensor
perturbations (gravitational wave fluctuations), as first described by
Starobinsky (1985). The characteristic CMB power spectra of these two
modes are shown in Fig.~\ref{fig:cmbfig1} for the concordance
$\Lambda$-dominated cosmology favoured by WMAP. The relative amplitude
of tensor and scalar contributions, $r$, is given to first order by
the inflationary parameter $\epsilon$ alone:
\be
r \equiv {C_\ell^{\japsub T}\over C_\ell^{\japsub S}} \simeq 12\epsilon.
\ee
Determination of the scalar spectral index $n_{\rm s}$ and the tensor-scalar
ratio $r$ thus provides important information on the slow roll parameters
and hence on the shape of the inflationary potential $V(\phi)$.

The amplitude of the tensor component unambiguously fixes the energy scale of inflation,
\begin{equation}
V^{1/4} \simeq 3.3 \times 10^{16} r^{1/4} \; {\rm GeV}. \label{theory1}
\end{equation}
The detection of a tensor mode, via its signature in the polarization
of the CMB, is an important target for future experiments. Such a
detection would confirm that inflation really took place. It
would determine the energy scale of inflation -- a
key discriminant between physical models of inflation -- and would
constrain the dynamics of the inflationary phase.

Observations of the CMB, and in particular from the WMAP satellite,
have revolutionised our knowledge of cosmology. These observations
favour a `concordance' cosmology in which the universe is spatially
flat and currently dominated by dark energy. However, even after three
years of WMAP data, we still have only crude constraints on the
dynamics of inflation. The data are consistent with an adiabatic
spectrum of fluctuations generated during an inflationary phase, but
the tensor mode has not yet been detected and it is not known whether
there are other contributions to the fluctuations arising, for
example, from cosmic strings or isocurvature perturbations (entropy
perturbations, in which the photon-to-matter ratio varies). The third
year WMAP data give tentative indications of a slight tilt in the
scalar spectral index ($n_{\rm s} \simeq 0.96$) and at face value marginally
exclude quartic potentials $V(\phi) \propto \phi^4$. Current
constraints on $r$ and $n_{\rm s}$ from WMAP and other data are summarized
in Fig.~\ref{fig:wmapinf}.

%\epsfigsimp{wmap_newfig14.ps}{0.85} {The marginalized $68$ and $95$
\epsfigsimp{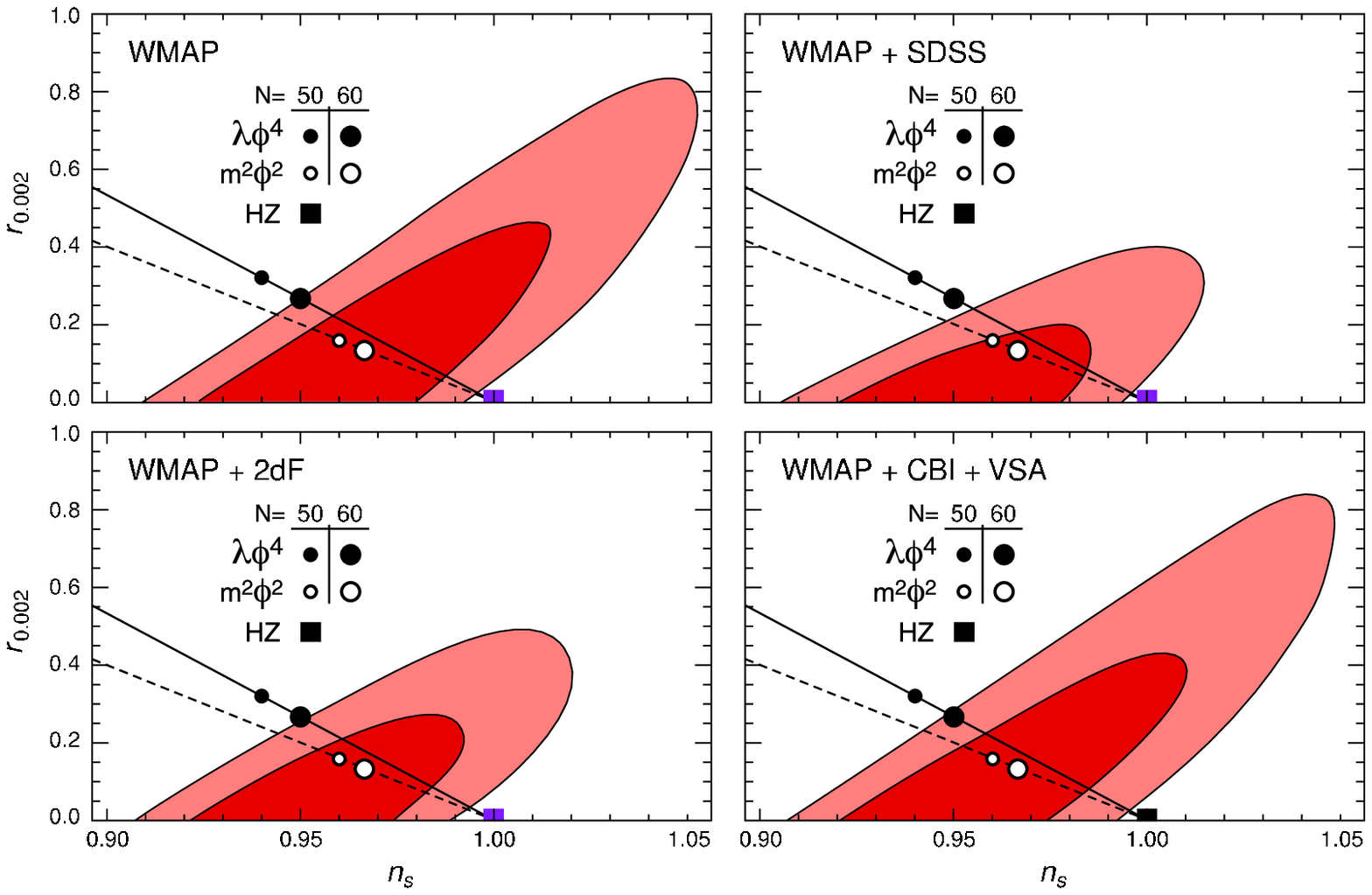}{0.95} {The marginalized $68$ and $95$
percentile confidence contours on the inflationary $r-n_{\rm s}$ plane for
WMAP3 data combined with other data sets (Spergel et al. 2006).
Upper left panel shows WMAP
data alone. Lower left and upper right panels show WMAP data combined
with 2dFGRS and SDSS redshift survey data on galaxy clustering. Lower right panel
shows WMAP combined with CMB measurements of smaller scale anisotropies.
The lines show predictions for power-law potentials with $N=60$ and
$N=50$ e-folds of inflation from the time that fluctuations on our
present horizon scale were frozen until the end of inflation. Filled
and open circles show the predictions for quartic and quadratic potentials
respectively.}
{fig:wmapinf}

\ssec{Future prospects}\label{sc:CMB.2}

It is expected that Planck will effectively complete the mapping of
the primordial temperature pattern in the CMB. For $\ell\ls 2000$ the
temperature power spectrum from Planck will be limited by cosmic
variance.  At higher multipoles, beam calibration uncertainties,
unresolved point-sources and various other factors will need to be
understood in order to disentangle the cosmological CMB signal. At these small
angular scales, non-linear anisotropies from {\it e.g.} the
Sunyaev--Zeldovich and Ostriker--Vishniac effects will become important,
as will gravitational lensing of the CMB. Planck will have
polarization sensitivity over the frequency range $30$--$353$ GHz and should
provide an accurate estimate of the polarization power spectrum
up to $\ell \sim 1000$.

High sensitivity polarization measurements will be an especially
important goal for future CMB experiments.  The polarization arises
from the Thomson scattering of an anisotropic photon distribution,
especially its quadrupolar anisotropy.  The resulting linear
polarization pattern on the sky can be decomposed into scalar E-modes
and pseudo-scalar B-modes.  These patterns are illustrated in
Fig.~\ref{fig:bmode}, and are similar to the phenomenon whereby a 3D
vector field can be decomposed into the gradient of a scalar
potential, plus the curl of a vector potential. Here, polarization is
represented by a $2\times 2$ symmetric matrix, whose components
$\gamma_{ij}$ can be constructed from symmetric and antisymmetric
combinations of a double derivative $\partial_i\partial_j$ acting on
potentials.  This decomposition was initially introduced in the
context of gravitational-lens shear, which has the same mathematical
properties as polarization, by Kaiser (1992) and Stebbins (1996).  Scalar primordial
perturbations generate only an E-mode polarization signal, while
tensor perturbations generate E- and B-modes of roughly comparable
amplitudes (e.g. Zaldarriaga \& Seljak 1997;
Kamionkowski, Kosowsky \& Stebbins 1997). 
The detection of a primordial B-mode polarization
pattern in the CMB would therefore provide unambiguous evidence for a
stochastic background of gravitational waves generated during
inflation. A B-mode detection would thus prove that inflation
happened, and would determine the energy scale of inflation.

\epsfigsimp{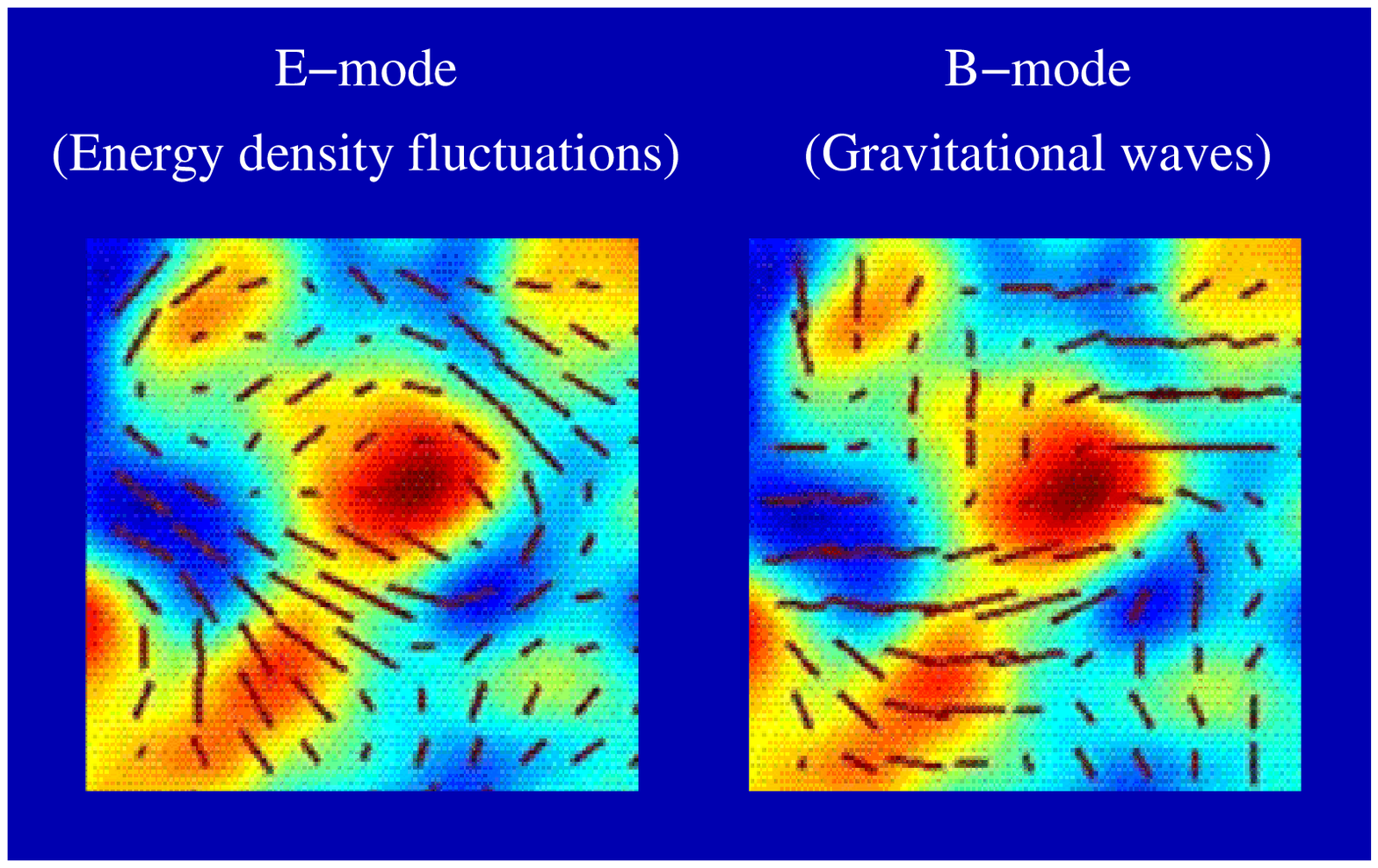}{0.7}
{An illustration of the two modes of CMB polarization, and how
they might be expected to correlate with total intensity. The
B-mode pattern resembles the E-mode pattern but with all polars
(indicated by the sticks)
rotated by $45^\circ$. Scalar perturbation modes generate E-mode
polarization only, whereas tensor perturbations generate both
modes. The B-mode signal is thus a unique signature of the
presence of primordial gravitational waves. An analogous decomposition
exists for the image shear field induced in gravitational lensing.
Adapted from Seljak \& Zaldarriaga (1998).}
{fig:bmode}

However, the detection of primordial B-modes from inflation poses a
challenging problem for experimentalists. The {\it indirect\/} limits
of $r \ls 0.4$ set by WMAP3 (Fig.~\ref{fig:wmapinf}) translate to an
rms B-mode signal of less than $0.35 \mu{\rm K}$ on the sky. This is
to be compared with an rms of $160 \mu{\rm K}$ for the temperature
anisotropies and $8 \mu{\rm K}$ for the E-mode signal. How well will
Planck do in measuring these small polarization signals?  Already the
Planck HFI bolometer noise figures are all below the photon noise for
all CMB channels, with the exception of the polarized 353-GHz
channel. The readout electronics shows white noise at the level
6~nV/Hz$^{1/2}$ from 16~mHz to 200~Hz (between the sampling period and
the 1 minute spin period), comparable to or below the photon noise.
Thus Planck HFI is fundamentally photon noise limited. Yet even so,
the detection of B-modes will pose a formidable challenge to Planck.

\begin{figure}[h]
\begin{center}
\scalebox{0.4}[0.4]{\includegraphics[angle=-90]{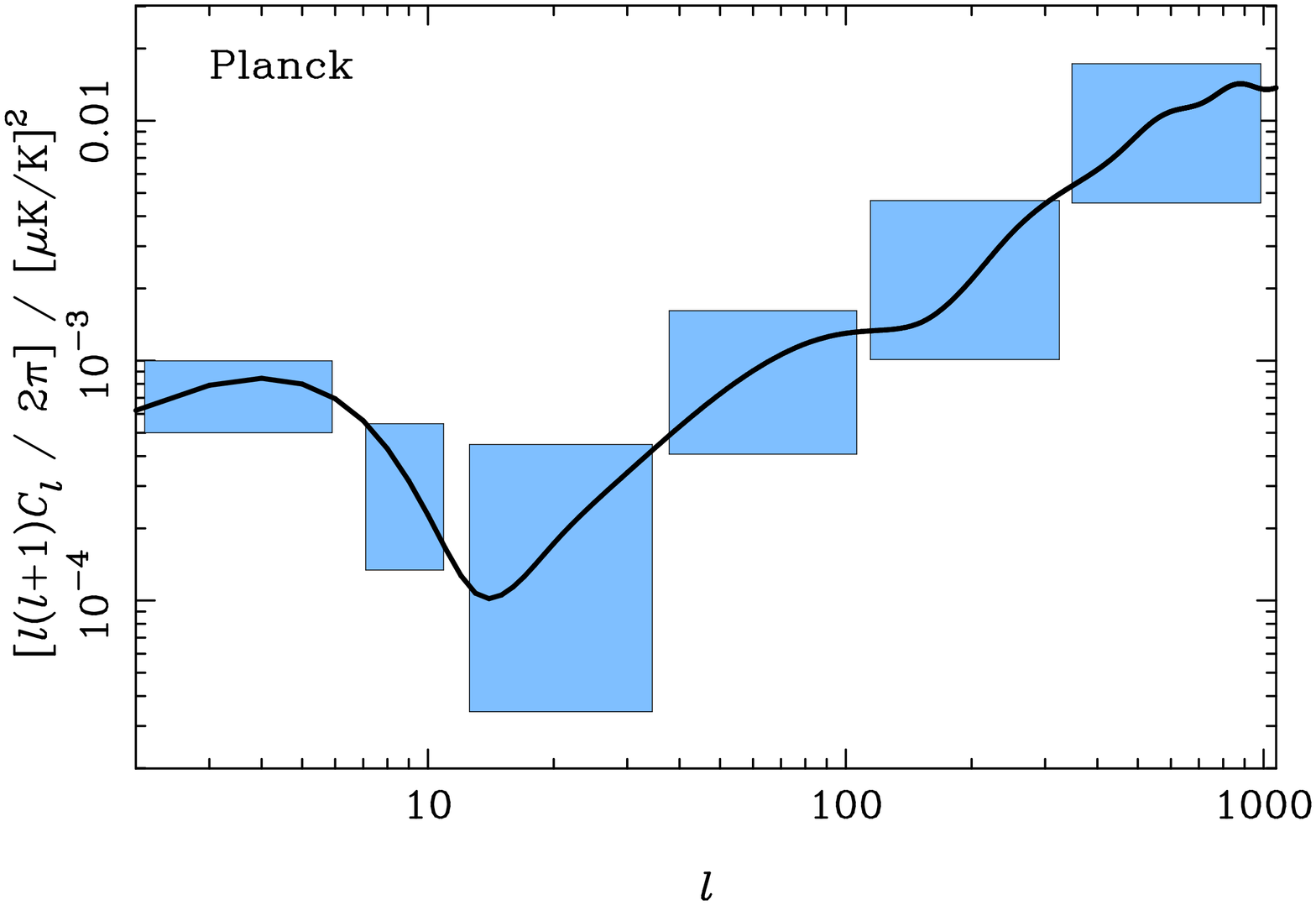}}
\end{center}
\caption{Forecasts for the $\pm 1\sigma$ errors on the magnetic polarization
power spectrum $C_\ell^B$ from Planck, assuming $r=0.1$.
Above $\ell \sim 150$ the primary spectrum
is swamped by weak gravitational lensing of the E-modes.}
\label{fig:cmbfig2}
\end{figure}

This is illustrated by Fig.~\ref{fig:cmbfig2}, which shows the
errors on $C_\ell^B$ expected from Planck for a model with $r$
arbitrarily set to $0.1$. The figure suggests that for $r=0.1$ Planck
can characterise the primordial power spectrum in around four
multipole bands.  The B-mode polarization signal generated by weak
gravitational lensing (which is, of course, independent of $r$)
dominates the primary signal above $\ell \simeq 150$. At the very
least, this lensing-induced signal should be detectable by
Planck. However, even if systematic errors and polarised Galactic
emission can be kept under control, Planck will at best only be able
to detect tensor modes from inflation if the tensor-scalar ratio is
greater than about 10 percent.

The key to achieving even higher sensitivities than Planck is to
build experiments with large {\it arrays} of polarization sensitive
detectors. This is being done for ground-based experiments (for
example,  Clover will use large bolometer arrays, while QUIET will
use large arrays of coherent detectors) and in future balloon
experiments ({\it e.g.} SPIDER, which plans to map about half the sky
at low angular resolution using large bolometer arrays). In addition
to these experiments, a number of studies are in progress in Europe
and the USA for a B-mode optimised satellite designed to map the
polarization pattern at high sensitivity over the whole sky.

What does theory predict for the energy scale of inflation and the
amplitude of the tensor component? Many theoretical models of
inflation can be grouped into one of two classes, `chaotic' inflation in
which the inflaton potential has a simple polynomial type form and
`hybrid' inflation in which the potential is relatively flat but
inflation ends abruptly (as happens in string-inspired brane inflation
models or because the inflationary dynamics is controlled by more than
one scalar field). In chaotic inflationary models, the predicted
tensor-scalar ratio is expected to be observably high, with $r \gs
10^{-2}$, while in hybrid-type models the value of $r$ could be as little
as $10^{-10}$,  or even lower. There is therefore no guarantee that the
amplitude of the tensor component will be high enough to permit detection.

Nevertheless, CMB experiments designed to probe tensor-scalar ratios
as low as $r \sim 10^{-2}$ are feasible and well motivated. Chaotic
inflation models have played an important role in cosmology and, if
they are correct, a tensor mode should be detectable. Inflation would
then be fact, rather than conjecture. Also, all such models involve
high field values in excess of the Planck scale.  The physics
underlying such models is therefore exotic and probes new territory
beyond the realms of conventional field theory. If, on the other hand,
tensor modes are not detected at this level, then this would rule out
chaotic inflation and point towards an abrupt end to inflation. If
this is the case, it may be more profitable to design high angular
resolution experiments to search for signatures of an abrupt end to
inflation ({\it e.g.}  from cosmic strings) rather than focusing
single-mindedly on searching for primordial tensor modes of even lower
amplitude.

A detection of the intrinsic B-modes will require extremely good
foreground removal, which will require a broad frequency coverage.
The nature of the polarized  foregrounds is not known in great detail, although
WMAP has provided important new information on the Galactic polarized
synchrotron contribution. The WMAP data suggest that the optimal
frequency for measuring primordial CMB polarization at low multipoles,
$\ell \ls 30$,  is about 70~GHz, though the optimal value may be
higher (perhaps 100~GHz) at higher multipoles. Realistically,  to
achieve a precision of $r \sim 10^{-2}$ will require sufficient
sensitivity and frequency coverage to remove foregrounds using the
internal data of the experiment itself, rather than relying on
detailed modelling and/or extrapolation from data at different
frequencies and angular resolutions. Ground-based experiments, even if
they fail to detect a primordial tensor mode, will provide valuable
information of the Galactic polarized synchrotron component and
anomalous dust-correlated emission below 50~GHz. Ultimately, the best
control of foregrounds, particularly over large areas of the sky,
would come from a B-mode optimised satellite unconstrained by
atmospheric absorption and emission covering as wide a frequency range
as possible (say $70$ -- $500$~GHz for a satellite flying bolometer
arrays).

In addition to the primary goal of detecting gravitational waves from
inflation,  a B-mode optimised satellite experiment could also tackle
the following problems:

\smallskip

\noindent
$\bullet$ Measuring the weak lensing effects in CMB polarization, which would
result in improved constraints on neutrino properties and the physics of
dark energy.

\smallskip

\noindent
$\bullet$ Constrain the history of reionizaton and the astrophysical
processes that ended the dark ages with a cosmic variance-limited
measurement of the large-scale E-mode polarization signal.

\smallskip

\noindent
$\bullet$ Explore whether anomalous features, such as multipole alignments,
seen in the WMAP temperature maps are also present in polarization.

\smallskip

\noindent
$\bullet$ Search for the B-mode signal of cosmic strings predicted by
 models of brane inflation.

\smallskip

\noindent
$\bullet$ Search for non-Gaussian signatures expected in some inflation models
(e.g. Bartolo et al. 2004). This is one of the main ways to distinguish
different models for the generation of perturbations (single-field
inflation vs multiple-field vs curvaton etc.).

Although most of this discussion has been focused on the detection of
B-mode anisotropies, there is still much science to be extracted from
precision measurements of temperature anisotropies on smaller angular
scales than those accessible to Planck.  Several large projects are
underway with sub-arcminute resolution designed to detect large
samples of galaxy clusters via the Sunyaev-Zeldovich (SZ) effect (for
example, ACT and SPT in the USA, AMI and APEX in Europe). These
surveys will be especially useful in constraining the amplitude of the
primordial fluctuations and probing the evolution of fluctuations. In
combination with other data, large SZ surveys will provide strong
constraints on the nature of dark energy.  To fully utilise the power
of such surveys, the CMB SZ surveys must be supplemented by optical
and infra-red imaging surveys to provide photometric redshifts for the
clusters, and to provide structural parameters. Associated X-ray
observations would provide valuable information on the physical
properties of the hot intra-cluster gas. The high amplitude in the
temperature power spectrum at $\ell \gs 2000$ observed by the CBI
experiment is still not understood. It is important to continue high
resolution interferometric observations of the CMB over a range of
frequencies to resolve this discrepancy.\footnote{Possible
explanations for the CBI excess include: inaccurate subtraction of
point sources; a high amplitude SZ signal; modified structure growth
rates arising from a dynamic `quintessence' field.} Finally, since
weak gravitational lensing generates B-modes with an effective
amplitude of $r \simeq 10^{-2}$, low resolution ground based
experiments will be limited to a statistical detection of the
primordial B-mode polarization signal and will not be able to
improve on this limit no matter what their sensitivity. An all-sky
B-mode satellite experiment could, in principle, achieve a tighter
limit of $r \sim 10^{-3}$ by measuring the `reionization' B-mode
bump at $\ell \ls 10$ (see Fig.~\ref{fig:cmbfig2}), but of course, this is
conditional on understanding large-scale polarized foregrounds at this
level.  To achieve even lower values of $r$ than this via CMB
polarization will require reconstruction of the weak lensing
deflection field. But to do this, a fundamentally different strategy
is required -- since an angular resolution of about an arcminute is
required to map the weak lensing potential. Thus probing below $r \sim
10^{-2}-10^{-3}$ will be formidably difficult, even if polarized
foregrounds can be controlled at these levels.  A high sensitivity,
high angular resolution polarization experiment would be required
covering a large part of the sky. It is premature to contemplate such
an experiment until (at least) polarized foregrounds are better
understood and experiments have already ruled out inflationary models
with $r \gs 10^{-2}$.  There is, however, a strong case for measuring
the weak gravitational potential by mapping restricted patches of the
CMB sky at high sensitivity and angular resolution. This type of
measurement would be complementary to conventional weak lensing
surveys based on the distortion of galaxy images.

\ssec{Conclusions on the CMB}\label{sc:CMB.3}

In summary, there is a compelling case to continue studies of the CMB
beyond Planck. In particular, a high priority should be given to
sensitive polarization measurements
of the CMB, from balloons and the ground, culminating in a B-mode optimised
satellite mission.
We make the following specific proposals:

\smallskip

\noindent
$\bullet$ Continue to develop polarization-optimised ground-based and
 sub-orbital CMB experiments.

\smallskip

\noindent
$\bullet$ Invest funds in developing large arrays of bolometers and coherent
detectors.

\smallskip

\noindent
$\bullet$ Investigate fully the scientific case and design of a low
resolution ($ \sim 30^{\prime\prime}$) B-mode optimised satellite. This
will require: (i) identifying a suitable detector technology and
associated cryogenics; (ii) optimising the frequency range to subtract
polarized foregrounds to very low levels; (iii) determining a suitable
scanning strategy and method of modulating the polarization pattern seen
by the detectors; (iv) assessment of stray-light, polarized point sources
and other potentially limiting systematic effects.

\smallskip

\noindent
$\bullet$ Support specialised ground-based measurements designed to
provide information on polarised foregrounds as, for example,
measurements of the polarization of radio and sub-millimetre sources.

\smallskip

\noindent
$\bullet$ Continue support for high-resolution experiments to measure
the Sunyaev--Zeldovich clusters, using both targeted observations and
surveys of `blank' fields of sky.

\smallskip

\noindent
$\bullet$ Investigate the development of high-sensitivity
interferometers to measure accurately the non-linear contributions to
the CMB power spectrum at multipoles $\ell \gs 2000$ and to study weak
lensing of the CMB.

\japsec{Large-scale structure}\label{sc:LSS}

\ssec{Principles}\label{sc:LSS.1}

The CMB contains information about the early seeds of cosmological
structure formation. Its natural counterpart  in the local universe
is the distribution of
galaxies that arises as a result of
gravitational amplification of these early small density
fluctuations. The pattern of present-day cosmological density
inhomogeneities contains certain preferred length-scales, which permit
large-scale structure to be used in a geometrical manner to diagnose
the global makeup of the universe. These characteristic scales are
related to the horizon lengths at certain key times -- i.e. to the
distance over which causal influences can propagate. There are two
main length-scales of interest: the horizon at matter-radiation equality
($D_{\japsub EQ}$) and the acoustic horizon at last scattering
($D_{\japsub LS}$). The former governs the general break scale in
the matter power spectrum and is contained in the transfer function
$T(k)$, and the latter determines the {\it Baryon
Acoustic Oscillations\/} (BAO) by which the power is modulated through sound
waves prior to recombination:
\be
\eqalign{
D_{\japsub EQ}^{\japsub H} &\simeq 123\, (\Omega_{\rm m} h^2/0.13)^{-1}\; {\rm Mpc} \cr
D_{\japsub LS}^{\japsub H} &\simeq 147\, (\Omega_{\rm m} h^2/0.13)^{-0.25}
(\Omega_{\rm b} h^2/0.023)^{-0.08}\; {\rm Mpc}, \cr
}
\ee
Distances deduced from redshift surveys automatically include
an inverse power of the Hubble parameter, being measured in
units of $\mpcoh$. Thus the main length deduced from LSS
scales as $(\Omega_{\rm m} h)^{-1}$, and so $\Omega_{\rm m} h$ is the main
parameter that can be measured, plus the baryon fraction $f_{\rm b}\equiv
\Omega_{\rm b}/\Omega_{\rm m}$ if the baryon oscillations are detected.

These scales can be seen projected on the sky via the CMB power
spectrum. Following the superb recent 3-year WMAP results (Spergel
et al. 2006), the detailed shape of the CMB power spectrum breaks
many of the degeneracies implicit in the above scaling formulae.
Thus, the combination of observations of cosmological perturbations
at $z\simeq 1100$ and $z\simeq 0$ makes an appealingly
self-consistent package within which most of the cosmological
parameters can be determined very accurately.

\sssec{The power spectrum}
The practical tool for measuring the length-scales encoded in
large-scale structure is the density power spectrum. This is
factorized into a product of the primordial spectrum and the
transfer function, which controls the different growth rates of
perturbations of different wavelength, according to the matter
content of the universe. The primordial spectrum is generally taken
to be a power law, $P(k) \propto k^{n_s}$, although the improving
accuracy of the data is such that it is increasingly common to
consider the {\it running\/} of the spectral index -- i.e. a
curvature of $P(k)$ in log-log space. The power spectrum is best
presented in dimensionless form, as the contribution to the
fractional variance in density per unit $\ln k$. In 3D, this is
$\Delta^2(k)\propto k^3 P(k)$.

\epsfigsimp{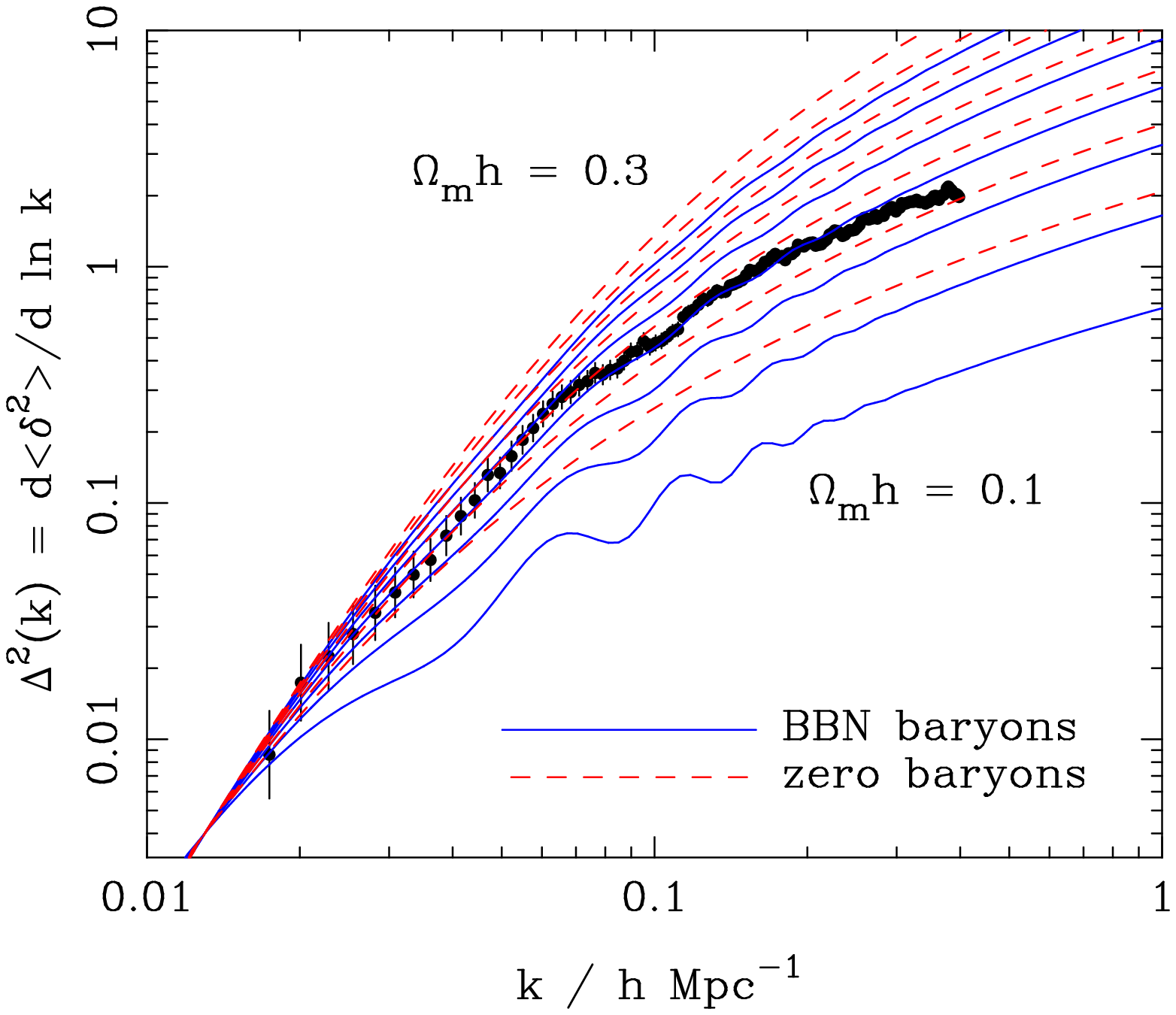}{0.55}
{The 2dFGRS redshift-space dimensionless power spectrum, $\Delta^2(k)$.
The solid and dashed lines show various CDM models, all assuming
$n=1$. For the case with non-negligible baryon content,
a big-bang nucleosynthesis value of $\Omega_{\rm b} h^2=0.02$ is
assumed, together with $h=0.7$. A good fit is clearly obtained
for $\Omega_{\rm m} h \simeq 0.2$. Note the baryon acoustic oscillations
that develop as the baryon fraction is raised, and how their
location depends on the overall density.
}
{p_2df}

We can probe large-scale density fluctuations by Fourier transforming the
galaxy number density field. The result is limited by two forms of
statistical noise: {\it cosmic variance\/}, reflecting the finite
volume sampled, and {\it shot noise\/}, reflecting the small-scale
nature of the galaxy distribution as a discrete point process. Both
these effects can be seen in the standard expression for the
fractional error in the galaxy power spectrum $P_{\rm g}(k)$. For a
survey of volume $V$ and galaxy number density $n$, measuring power in
a wavenumber range $\Delta k$, this is given by
\be
\sigma_{\ln P} = {2\pi \over (V k^2 \Delta k)^{1/2}}\,\left(
{1+nP_{\rm g}\over nP_{\rm g}}\right)
\ee
(Feldman, Kaiser \& Peacock 1994). Provided $nP_{\rm g}$ is greater
than of order unity, this represents largely a cosmic-variance limited
measurement. Assuming that there are more targets than our capability
for simultaneous spectroscopy, the volume covered in a fixed time
scales as $1/n$, so that the overall power error is minimised at
$nP_{\rm g}=1$ (Kaiser 1986).

\sssec{Bias and other nonlinearities}
The use of the galaxy power spectrum in cosmology faces a number of
practical obstacles compared to the ideal in which the linear power
spectrum at $z=0$ could be measured.  Most fundamentally, small-scale
information in the spectrum is destroyed by cosmic evolution up to the
present because the modes involved have become nonlinear. The point at
which $\Delta^2(k)$ reaches unity is around $0.25 \hompc$, and the
power at smaller scales reflects mainly the internal structure of dark
matter haloes, rather than any relic from early times (this is the
view of the `halo model': see e.g. Cooray \& Sheth 2002).  Even for
somewhat larger scales, the power spectrum is significantly altered in
shape, and probably we can only measure the shape of the linear
spectrum cleanly for $k<0.1 \hompc$. In order to use the information
at smaller scales to $k\simeq 0.2 \hompc$, an approximate nonlinear
correction has to be made; alternatively, measurements have to be
made at higher redshifts where non-linear evolution has had less effect
on these smaller scales, a fact employed by the Ly$\alpha$ forest technique
(see Sect.\ \ref{sc:Lya}).

Nonlinear modification of the density power spectrum is not the only issue.
One of the main results of nonlinear evolution is that the CDM
fragments into a population of dark-matter haloes and subhaloes,
and the latter are the hosts of galaxies. Haloes of different masses
and in different environments will form stars with varying
degrees of efficiency, so the galaxy distribution has a complex
relation to the overall density field. This changes not only the
small-scale shape of the galaxy power spectrum, but also the
overall normalization even in the long-wavelength limit.
Thus we have absolute and relative {\it biased clustering\/}.
The scale-dependent relative bias means there is a danger of
deducing a systematically incorrect value of $\Omega_{\rm m} h$ from
the shape of the galaxy spectrum. This was one of the main
topics addressed in the final 2dFGRS power spectrum paper
(Cole et al. 2005), where it was concluded that an empirical
correction could be made so that the shape of the power
spectrum was unaffected within the random measuring errors.
For $k<0.15\hompc$, the correction is at most 10\% in power,
so it is not a large effect -- but this aspect will need more
detailed treatment in future surveys of greater size and precision.

The absolute level of bias is a separate problem, and a frustrating
one: a direct measurement of the normalization of the
power spectrum (conventionally given via $\sigma_8$, which is the
fractional linear density rms when smoothed with a sphere of radius
$8\mpcoh$) would allow
us to measure the amount of perturbation growth since last scattering,
which is sensitive to the properties of the dark energy. Within the
framework of galaxy redshift surveys, two methods
have been used to try to overcome this barrier. Higher-order
clustering can distinguish between a two-point amplitude that is high
because $\sigma_8$ is high and one where $\sigma_8$ is low but the
degree of bias is high. In essence the distinguishing factor is the
filamentary character of the galaxy distribution that arises with
increased dynamical evolution. This can be diagnosed by a three-point
statistic such as the bispectrum; in this way, Verde et al. (2002)
were able to show that 2dFGRS galaxies were approximately unbiased. An
alternative route is by exploiting the fact that 3D clustering
measurements are made in redshift space, where the radial coordinate
is affected by peculiar velocities associated with structure
growth. This introduces a characteristic anisotropy in the clustering
pattern, which depends on the combination $\beta\equiv
\Omega_{\rm m}^{0.6}/b$, where $b$ is the linear bias parameter. This has
been measured via detection of the anisotropy signal, but so far
only at a level of precision of around 15\% (Peacock et al. 2001).
To be useful for dark energy studies, this accuracy will need to be very
substantially improved, to better than 1\% (a measurement of $\sigma_8$
to this precision would yield $w$ to 5\%, as shown earlier).
A novel method to determine the bias factor directly consists in
measuring the cross-correlation between galaxies and the shear from
weak gravitational lensing, as will be discussed in Sect.\ \ref{sc:GL}

\ssec{Current status}
\label{sc:LSS.2}

\sssec{Local clustering}
The current state of the art in studying LSS in the local universe
(defined arbitrarily as corresponding to $z<0.2$) is given
by the 2dFGRS, SDSS and 6dFGS surveys. Their properties are
summarized in Table~\ref{tab:lss_loz}, and they collectively
establish the 3D positions of around 1,000,000 galaxies.

\begin{table}
\caption{Local galaxy redshift surveys. The numbers in brackets
represent the size of samples not yet analyzed.}
\begin{center}
\begin{tabular}{|l|l|l|l|}
\hline
\topstrut\botstrut Survey & Depth & $N_z$ & Ref \\
\hline
\topstrut 2dFGRS & $b_{\japsub J} < 19.45$ & 221,414 & Cole et al. (2005) \\
SDSS & $r<17.77$ & 205,443 & Tegmark et al. (2004) \\
&& (674,749) & \\
6dFGS & $K<14$ & 83,014 & Jones et al. (2005) \\
\botstrut && (167,133) & \\
\hline
\end{tabular}
\end{center}
\label{tab:lss_loz}
\end{table}

As discussed earlier, measurements of the galaxy power spectrum
principally yield an estimate of $\Omega_{\rm m} h$ (in a 
manner that is degenerate
with the assumed spectral index $n_{\rm s}$) and the baryon fraction.
These constraints have the greatest value in combination with
CMB data, and the most accurate parameter estimates in the 3-year
WMAP study of Spergel et al. (2006) come in combination with
the 2dFGRS data.
Earlier CMB datasets lacked the power to break degeneracies such as
that between $\Omega_{\rm m}$ and $h$ very strongly, but the latest WMAP
data allow this do be done, so that the individual parameters and thus
the key horizon lengths can be estimated. Datasets at lower
redshifts probe the parameters in a different way, which also depends
on the assumed value of $w$: either directly from $D(z)$ as in SNe, or
via the horizon scales as in LSS. Consistency is only obtained for a
range of values of $w$, and the full CMB+LSS+SNe combination already
yields impressive accuracy:
\be
w= -0.926^{+0.051}_{-0.075}
\ee
(assuming $w$ to be constant and a spatially flat model); this number
is the marginalized limit from the full multi-dimensional
parameter constraints, as in Fig.~\ref{fig:wmap_w}, and is derived
assuming the SNe datasets to be independent.
Any future experiment must aim for a
substantial improvement on this baseline figure.

\epsfigsimp{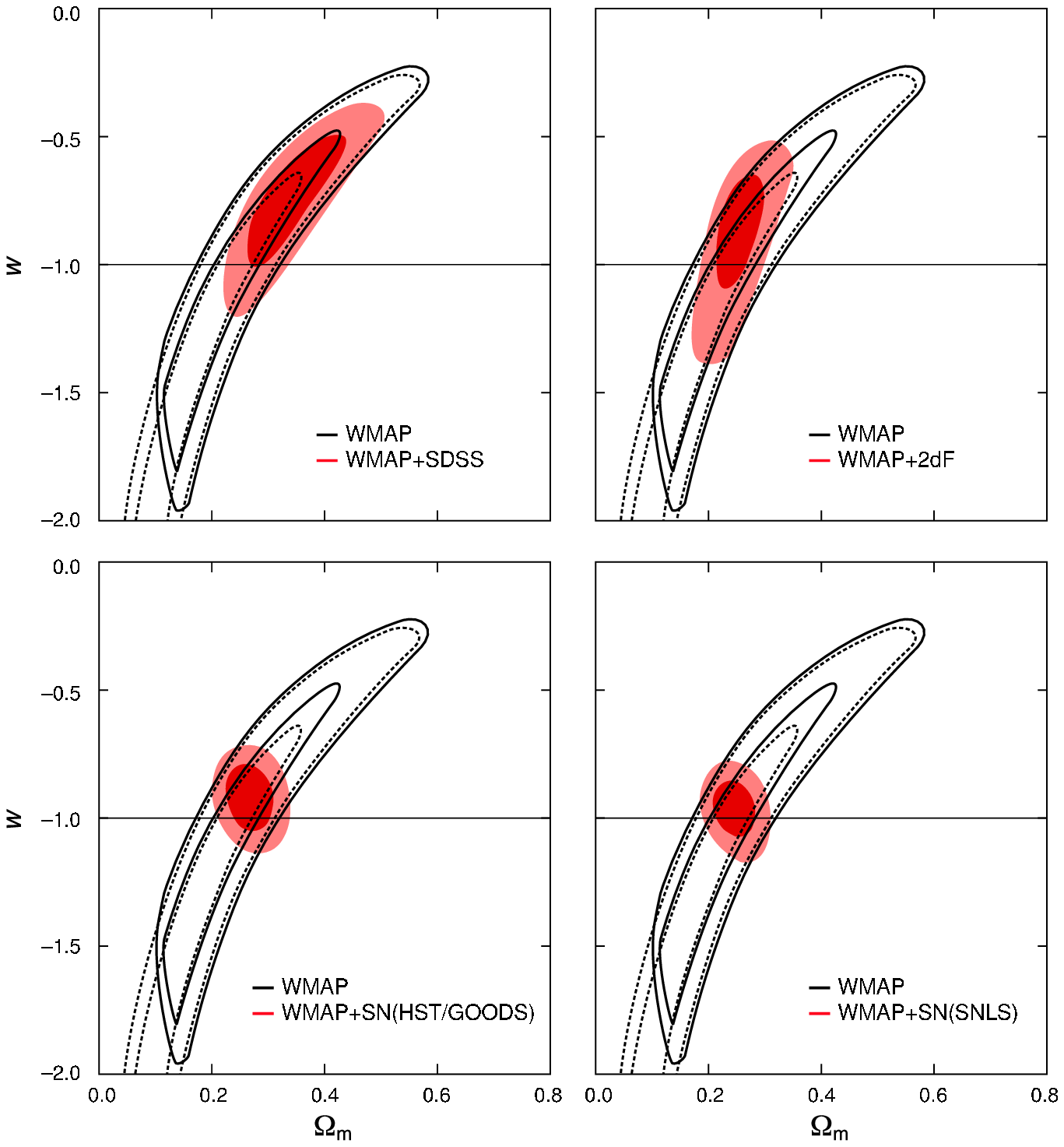}{0.80} {Confidence contours on the plane of
$w$ vs $\Omega_{\rm m}$ from WMAP year-3 results, allowing for dark-energy
perturbations with $c$ as the speed of sound (Spergel et al. 2006).
It is assumed that $w$ does not evolve, and also that the universe
is spatially flat, although this latter assumption is not so
critical.} {fig:wmap_w}

\sssec{High-redshift clustering}
Current activity in redshift surveys
is summarized in Table~\ref{tab:lss_hiz}; it tends to concentrate
at higher redshifts, under two distinct headings. Over the
intermediate range $0.2 \ls z \ls 0.7$, Luminous Red Galaxies
have been selected from the SDSS over areas of several thousand
square degrees. At the other extreme, we have deeper pencil-beam
surveys covering a handful of square degrees, but extending
to $z\simeq 3$ or beyond.

\begin{table}
\caption{High-redshift galaxy surveys. We include
samples based on photometric redshifts as well as
full spectroscopy (denoted by `(p)'.)}
\begin{center}
\begin{tabular}{|l|l|l|l|}
\hline
\topstrut\botstrut Survey & Depth & $N_z$ & Ref \\
\hline
\topstrut COMBO-17 & $R<24$ & 25,000(p) & Bell et al. (2004) \\
VVDS & $I_{\japsub AB}<24$ & 11,564 & Le F\`evre et al. (2004) \\
DEEP2 & $R_{\japsub AB}<24.1$ & 30,000 & Coil et al. (2006) \\
CFHTLS-VVDS & $I_{\japsub AB}<24$ & 550,000 & Ilbert et al. (2006) \\
COSMOS & $I_{\japsub AB}<25.5$ & 300,000(p) & Scoville et al. (2006) \\
SDSS LRG & & 46,748 & Eisenstein et al. (2005) \\
\botstrut SDSS LRG & & 600,000(p) & Padmanabhan et al. (2006) \\
\hline
\end{tabular}
\end{center}
\label{tab:lss_hiz}
\end{table}

\ssec{Future LSS experiments}\label{sc:LSS.3}

It seems likely that future LSS experiments in fundamental cosmology
will focus on the BAO signature. This is because it is a
sharp feature in the power spectrum, and thus defines a length-scale
that is relatively immune to the slow tilting of the spectrum
introduced by nonlinearities, scale-dependent
bias etc. Furthermore, the BAO signature has a direct and
clean relation to the corresponding
oscillations in the CMB power spectrum
(Fig.~\ref{fig:wiggles_cmblss}).

%\epsfigsimp{wiggles_cmblss.ps}{0.6}
\epsfigsimp{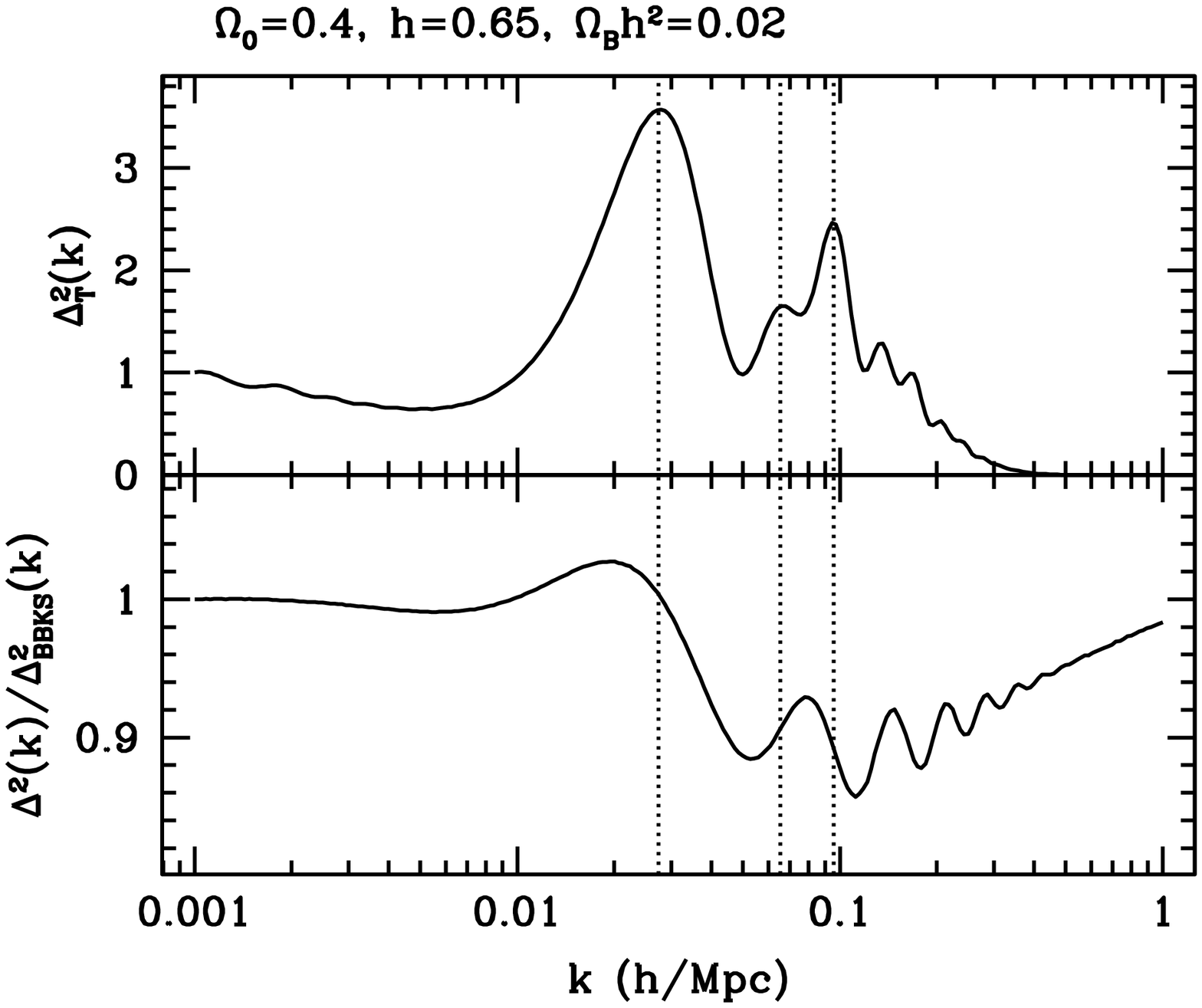}{0.80}
{Acoustic oscillations in the
radiation-baryon fluid imprint a pattern of harmonics in the Fourier
spectrum of both CMB and density fluctuations
(e.g. Meiksin, White \& Peacock 1999).
In the latter case for which the ratio of the power
spectrum to that of a model with zero baryon content is plotted in the
lower panel, the effect is much smaller, because the dominant dark
matter has no intrinsic oscillations. Nevertheless, features
corresponding to the same physical effect can be picked out at low and
high redshift, opening the way to a relatively clean geometrical tool
in cosmology.}  {fig:wiggles_cmblss}

Assuming that the survey of interest can be made to operate
somewhere near the optimal number density with $nP_{\rm g}=1$,
the following rule of thumb gives the
accuracy with which $D(z)$ can be measured by picking out the
acoustic scale:
\be
\% \ {\rm error\ in\ } D(z) = (V/5\, h^{-3}\, {\rm Gpc})^{-1/2}\;
(k_{\rm max} / 0.2 \hompc)^{-1/2}.
\ee
From the earlier discussion of nonlinearities, one would not want
to push much beyond $k=0.2 \hompc$. It is sometimes claimed that the
high-$k$ limit in BAO analyses should be increased at high redshift
because the density field is less evolved at this point.  The problem
with this argument is that, precisely because of this evolution, the
galaxies that are found at these distances are rarer and more strongly
biased.  Fig.~\ref{fig:ms_galp} contrasts the mass and galaxy power
spectra at various epochs.  To some extent, this is encouraging for
$z=3$: the nonlinear mass spectrum clearly shows a third peak at
$k\simeq 0.2\hompc$, which is not really apparent at either $z=0$ or
$z=1$. On the other hand, the shapes of the galaxy and mass power
spectra start to diverge at smaller $k$ at $z=3$ than at $z=1$,
reflecting the larger degree of bias for galaxies at that redshift --
and indeed the galaxy spectrum shows no more evidence for a third peak
at $z=3$ than it does at $z=1$.  This indicates that one should be
extremely cautious about expecting to use galaxy data beyond the
second BAO peak at any redshift.

\epsfigsimp{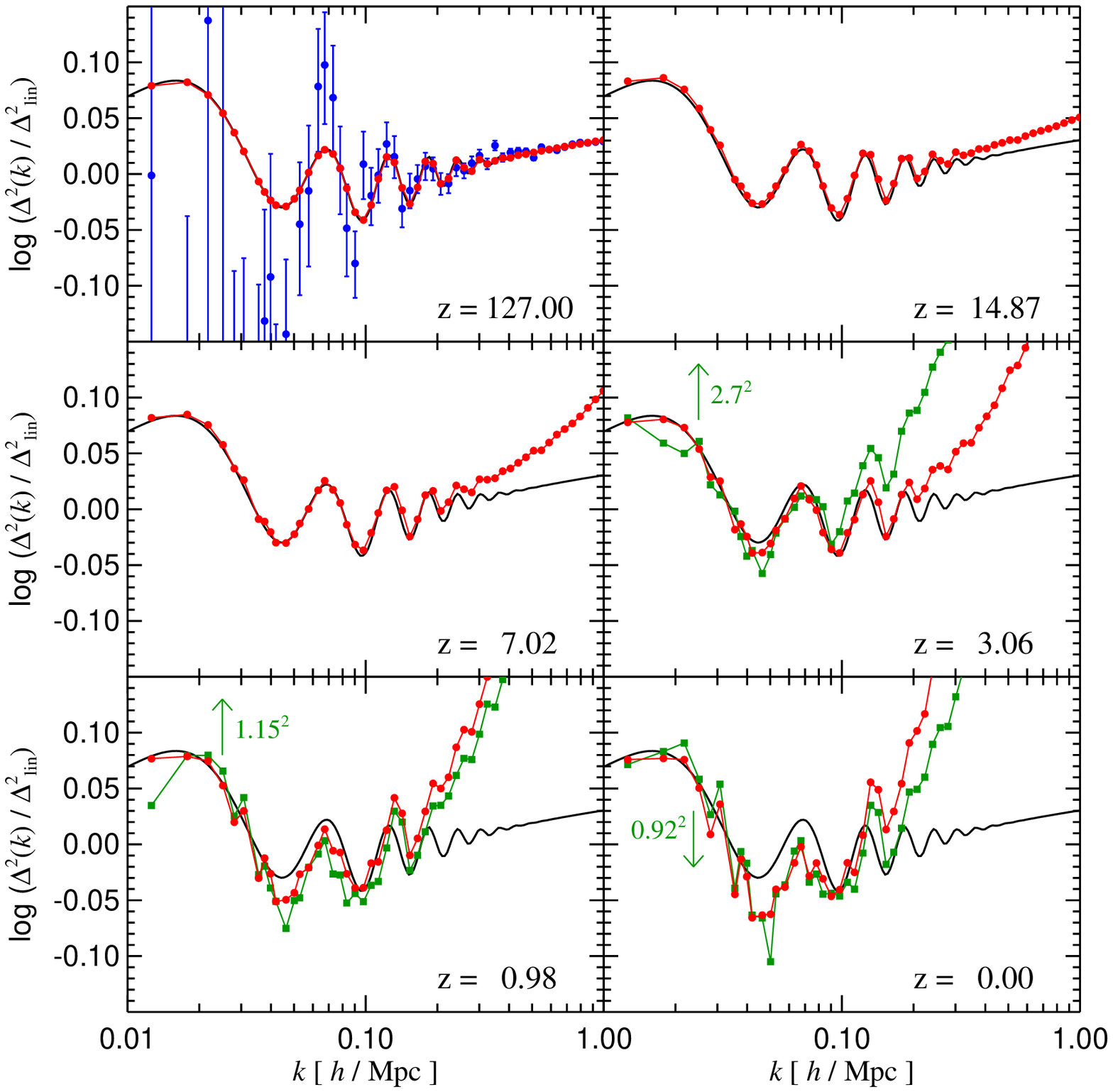}{1.0}
{The power spectrum from the Millennium Simulation (Springel et al. 2005).
The plot shows the linear power spectrum ratioed to a smooth model
(black), the nonlinear mass power spectrum (red), and the galaxies (green),
scaled to allow for bias.}
{fig:ms_galp}

This discussion suggests that the minimum interesting volume
for a BAO survey is $5\, h^{-3}\, {\rm Gpc}^3$ (1\% error
in distance, thus 5\% in $w$).
How many galaxies are implied by such a survey? We need to choose
a typical power level to set the $nP_{\rm g}=1$ sampling.
At the wavenumber of the main
acoustic feature ($k=0.065\hompc$), the observed galaxy power
spectrum for a variety of high-redshift tracers displays rather
little evolution, and a canonical value of $P_{\rm g}\simeq 2500(\mpcoh)^3$
is a reasonable choice, suggesting a number density of
$n=4\times 10^{-4}\, h^3\,{\rm Mpc}^{-3}$.
Over a broad redshift band such as can be selected by photometric
means (e.g. the DEEP2 $0.7 < z < 1.3$), the corresponding
surface densities are of order 1000 deg$^{-2}$.
A minimal BAO survey
thus requires around 2,000,000 galaxies, and pushing the errors
on $w$ towards 1\% is clearly demanding in terms of direct spectroscopy.
The alternative is to carry out BAO studies using photo-z's,
but this has two related drawbacks. The large radial smearing
associated even with good  photo-z's, i.e, $\delta z/(1+z)=0.03$,
means that the signal level is greatly reduced. To compensate for this,
a photo-z sample needs to be roughly ten times larger than one
with full spectroscopy (e.g. Blake \& Bridle 2005). Also, more of the signal now comes
from angular variations in galaxy density, placing more
stringent requirements on photometric uniformity.

\epsfigsimp{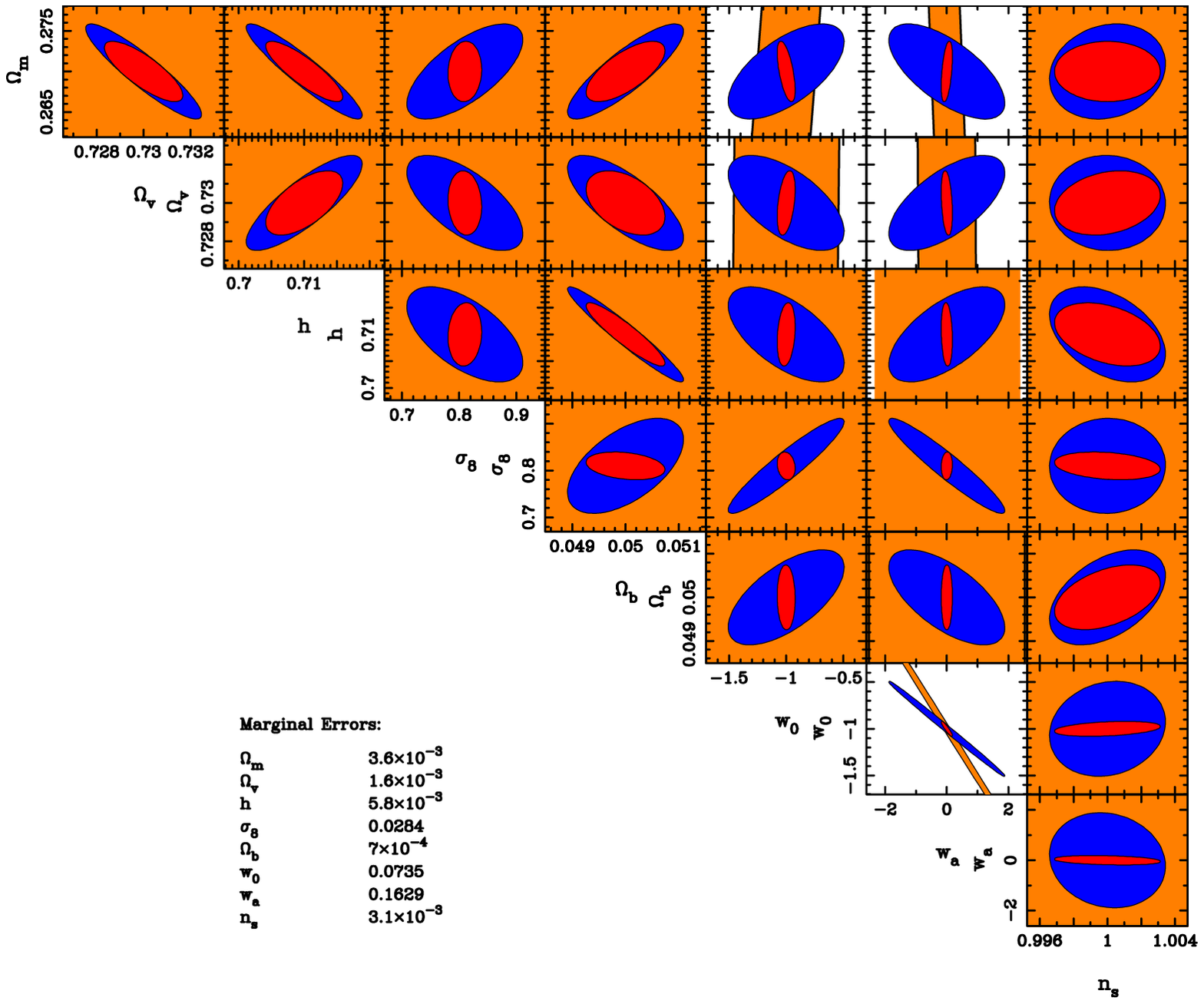}{1.0} {Fisher matrix confidence contours for the
main cosmological parameters, assuming a single $z=1$ BAO experiment
of $2\times 10^6$ galaxies and CMB data.  This shows the expected
one-parameter 1-$\sigma$ contours on a Gaussian approximation to the
likelihood distribution (i.e. $\Delta\ln L=1$). The blue ellipse
shows the expected result from Planck alone; orange is BAO alone; red
is the combined constraint.
Plot courtesy of T. Kitching.}  {fig:fish7}

\sssec{Ground-based BAO studies}
The most powerful existing facility for BAO work is AAOmega: the
upgraded 2dF on the Anglo-Australian 4m telescope.  This delivers 400
spectra over a 2-degree diameter field.  AAOmega will carry out the
`wigglez' project: a survey of 600,000 emission-line galaxies at
$z\simeq 1$, selected using a mixture of SDSS and GALEX data. When
complete (in approximately 2010), this will probably measure $w$ (if
constant) to around 10\% accuracy, so is more in the nature of a proof of
concept rather than being competitive with current constraints.  In
the longer term, it is proposed to construct WFMOS, which would offer
2000--4000 fibres over a 1.5-degree field on the Japanese Subaru
8m. For details of the project and the science, see WFMOS study (2005)
in the reference list.
This facility would be capable of carrying out BAO surveys of
several million galaxies, pushing down the limits on constant $w$ and
yielding in parallel constraints on evolution.  However, construction
is not yet approved, and operation would not begin until at least
2012. In the meantime, a pilot project at the 500,000-redshift level
will probably proceed using FMOS, which offers 400 near-IR fibres over
the existing 0.5-degree SuprimeCam field.

A number of other projects exist that have the potential to
contribute to this area, although these are generally at a
less advanced stage of planning. HETDEX intends to 
us the 10m Hobby-Eberly Telescope to obtain
spectra of around $10^6$ galaxies at $1.8 < z < 3.7$ using their
Lyman-alpha emission, and this would be a powerful
and competitive BAO probe. Although not yet funded for construction,
the claim is that the instrument could be ready within 3 years,
and that the experiment would take a further 2.5 years.
Other possibilities include planned twin Korean 6.5m
telescopes, to be sited at San Pedro Mart\'\i r, Mexico.
One of these telescopes is planned to have a multi-fibre 1.5-degree field,
and the intention is to begin operation around 2012. It is not yet clear
if there are plans to focus on BAO, but this could well be a competitive
facility. Finally, one might mention Hectospec, which offers
spectroscopy over a 1-degree field on the 6.5m MMT. This presently
offers only 300 fibres, but would be a powerful facility if
upgraded. Given the rapid growth in the number of telescopes
worldwide in the 6m--8m class, we predict with some
confidence that one of these will produce a spectroscopic
facility of impressive grasp within the next 5--10 years.
So far, there is no European proposal under this heading.

A number of projects will also attempt to measure BAO signals using
photo-z's.  The ESO/VST KIDS project (1400 deg$^2$ imaging) will
probably be the first to achieve this, although the area as originally
proposed is sufficiently small that the measurement of $w$ will only
be possible at the 15\% level. This survey will profit from the
parallel near-IR surveys UKIDSS and VIKING, the latter
using the new VISTA telescope.  The USA's Dark Energy Survey aims to
cover 5000 deg$^2$ using a new camera on the CTIO 4m
(see \url{www.darkenergysurvey.org}), and will do a little better, with projected error
on $w$ of 10\%, but for this survey no near-IR data are yet planned,
with potentially significant consequences for photometric redshift
estimates.  However, this photometric technique will only come into
its own with imagers of larger grasp, that are able to survey most of
the sky to deep limits in multiple wavebands. The Japanese
HyperSuprimeCam, recently approved for construction on Subaru, has
this capability, although the telescope will not be dedicated to
surveys. In the longer term, LSST (\url{www.lsst.org}) should be able to measure $w$ to
about 2\% using the photo-z BAO method, although again the lack of
near-IR information may be an issue.

\sssec{Space-based BAO studies}
As part of the NASA call for Dark Energy missions under the JDEM
heading, the ADEPT mission has been proposed, which would carry out
slitless spectroscopy of $\sim 10^7$ galaxies at $z\gs 1$. Although
there is a tendency to think of spectroscopy as being something
best suited to ground-based studies, the low background in
space means that there is an impressive speed advantage.

\ssec{ESO's capability for LSS studies}\label{sc:LSS.4}

How well are ESO facilities capable of meeting the challenge outlined
above, of measuring tens of millions of redshifts over an area of many
thousands of square degrees?  The widest-field spectroscopic
facilities at the VLT are VIMOS, with a field of view of 224
arcmin$^2$, and FLAMES, with a field of view of 490 arcmin$^2$.  Based
on experience with the VVDS and zCOSMOS surveys, VIMOS is able to
deliver $\sim 1000$ redshifts per night to limits of order
$I_{\japsub AB}\ls 23$, and it is therefore clear that the maximum
size of survey that is practical with this instrument is a few times
$10^5$ redshifts. This size of database is potentially significant
from the point of view of photo-z calibration, since one of the key
issues will be to verify that photo-z estimates are uniform and
consistent over the whole sky. In the presence of a photo-z error of
at least $\delta z = 0.05$, a single VIMOS pointing would be able to
establish that the mean redshift was unbiased at the level of a few
parts in 1000 -- as required if we are to measure $w$ to order 1\%. A
sample of $10^5$ redshifts would thus yield approximately 1000
calibration points, spread across several thousand square
degrees. This is the best that can be done with existing instrumentation,
and it would be an important contribution towards validation of the
photo-z technique.

Beyond such calibration work, ESO presently lacks a facility that
could undertake the large-scale spectroscopic studies needed for
spectroscopic BAO work.  VIMOS is optimized to multi-object
spectroscopy at source densities of $\sim 10^4\,{\rm deg}^2$, about a
factor of $\sim 10$ above the optimal source density of $n P_{\rm
g}=1$. This is a principal drawback of multi-slit spectrographs,
compared to fibre instruments. The rather small field of view covered
by current VLT instruments therefore prevents
significant BAO work.  This seems a pity,
given that there will be many other science drivers for such massive
spectroscopy (e.g. galaxy evolution, star formation history,
photo-z's, clusters).  We also see a need for optical imaging with a
greater grasp than can be provided by the VST, in order to provide
photo-z data at significantly higher rate, to be able to compete with
world-wide efforts. Both these needs would be most efficiently
satisfied by the provision of a wider field on one of the VLTs, which
would involve the construction of a prime focus, entailing substantial
modification of the telescope structure, or the refurbishment of an
existing 4-m telescope, to allow a very wide-field camera of several
square degree. In addition, ESO might consider seeking to
become involved in one of the existing projects of this type elsewhere.

\japsec{Clusters of galaxies}\label{sc:clus}

Going beyond the primordial relics of the initial conditions on the
largest scales, it is also possible to test cosmological models and to
assess the origin, geometry, and dynamics of our universe using galaxy
clusters as nonlinear tracers of the large-scale structure.  These
systems mark the scale of transition between large-scale linearity and
the complicated physics of galaxy formation, thus occupying a special
place in the hierarchy of cosmic structures. They are the
largest virialized dark
matter aggregates (dark matter halos) in our universe, and have
formed relatively recently. During their formation, they collect galaxies
and diffuse gas from the surrounding cosmological environment. While
their overall dynamics is still dominated by Dark Matter,
the astrophysical processes taking place at the galactic
scale have sizable effects on observable properties of the gas
component trapped in their gravitational potential. Galaxy
clusters are more easily described theoretically and modelled in
simulations than other tracer objects of the large-scale matter
distribution. In this sense, clusters represent the place where
astrophysics and cosmology meet.

Cluster properties such as
their number density and its redshift evolution, as well as their
distribution on large scales, are very sensitive to the cosmological
model, thus providing strong constraints on cosmological parameters
(e.g. Vikhlinin et al. 2003; Schuecker et al. 2003a,b).
Detailed XMM-Newton and Chandra studies revealed that clusters follow
closely a self-similar structure of basic properties such as
intracluster medium density and temperature profile (e.g.
Vikhlinin et al. 2005), hence allowing more precise cluster
modelling than expected from the mere inspection of high resolution
X-ray images returned by Chandra and XMM. Among the success of cluster
cosmology, the constraints on the density fluctuation amplitude
$\sigma_8$ should be mentioned. Cluster X-ray data suggested low values of 0.7 -- 0.8
(e.g. Schuecker et al. 2003a, Vikhlinin et al. 2003), which have recently been
confirmed by the 3-year WMAP results (Spergel et al. 2006; see also
Table\ \ref{tab:lensdonesur} for estimates of $\sigma_8$ from weak lensing).
This has strongly revived the interest in cosmological cluster surveys
in X-rays (e.g. Haiman et al. 2005), SZ (e.g. Holder, Haiman \& Mohr 2001), and the
optical (e.g. Annis et al. 2005).

The galaxy cluster population provides information on the cosmological
parameters in several
complementary ways:
\begin{enumerate}
\item
The cluster mass function in the local universe mainly depends on the
matter density $\Omega_{\rm m}$ and the amplitude $\sigma_8$ of the density
fluctuation power spectrum.
\item
The evolution of the mass function $n(M,z)$ is governed by the growth
of structure in the universe and thus is sensitive to the density
parameters and the equation of state of dark energy.
\item
The amplitude and shape of the power spectrum $P_{\rm cl}(k)$ of the
cluster distribution, and its growth with time, depends on the
underlying matter power spectrum, so that the cluster distribution
can be used to determine cosmological parameters.
In particular, wiggles due to the BAO at the time of
recombination are imprinted on the large-scale distribution of
clusters and thus can be employed in a similar way to the
study of BAO features in the galaxy distribution, as described in
Sect.\ \ref{sc:LSS.1}.
\end{enumerate}

The constraints provided by the different cosmological tests with
clusters are complementary; parameter degeneracies in any of the tests
can be broken by combining them. The simultaneous constraint of
$\Omega_{\rm m}$ and $\sigma_8$ by combining methods 1 and 3 above is one
such example (Schuecker et al. 2003a). In addition, the combination of
several tests provides important consistency checks as explained
below.

In addition to the above applications, galaxy clusters have been used as
cosmological standard candles to probe absolute distances, analogous to the
cosmological tests with supernovae type Ia:
\begin{itemize}
\item
The assumption that the cluster baryon fraction is constant with time,
combined with observations of this quantity, provides constraints on
cosmological parameters (e.g. Allen et al. 2004).
\item
In a very similar way, combined X-ray and SZ-measurements provide a means for
absolute distance measurements and constraints on the geometry of the universe
(e.g. Molnar et al. 2004).
\end{itemize}

\ssec{Cosmological galaxy cluster surveys}\label{sc:clus.1}
Large, well-defined and statistically complete samples of galaxy
clusters (which are dynamically well evolved and for which masses are
approximately known) are obvious prerequisites for such
studies. Substantial progress in the field requires samples of tens to
hundreds of thousands of clusters. Surveys in several wavelength
regions are currently used or planned to achieve this goal:

Galaxy clusters are detected in X-ray surveys by the radiation of the
hot intracluster medium ($T\simeq 3\,{\rm keV}$) which provides a good
tracer of the gravitational potential of the cluster and a measure of
its mass.  In the radio and microwave sky below 1.4\, mm, galaxy
clusters appear as shadows on the CMB sky due to the
Sunyaev--Zeldovich effect, the Comptonization of CMB photons by the
same intracluster plasma that also gives rise to the X-ray
radiation. In the optical, galaxy concentrations with high velocity
dispersions ($\sigma_{\rm v} \simeq 500 - 1500\kms$) mark the appearance of
galaxy clusters. While the optical characterisation of galaxy clusters
is affected by confusion of the cluster galaxy distribution with the
fore- and background (projection effects), multi-colour imaging and
highly multiplexed spectroscopy can be employed to construct effective
cluster samples. Finally, clusters of galaxies can also be
detected due to their gravitational lensing effect.

\epsfigsimp{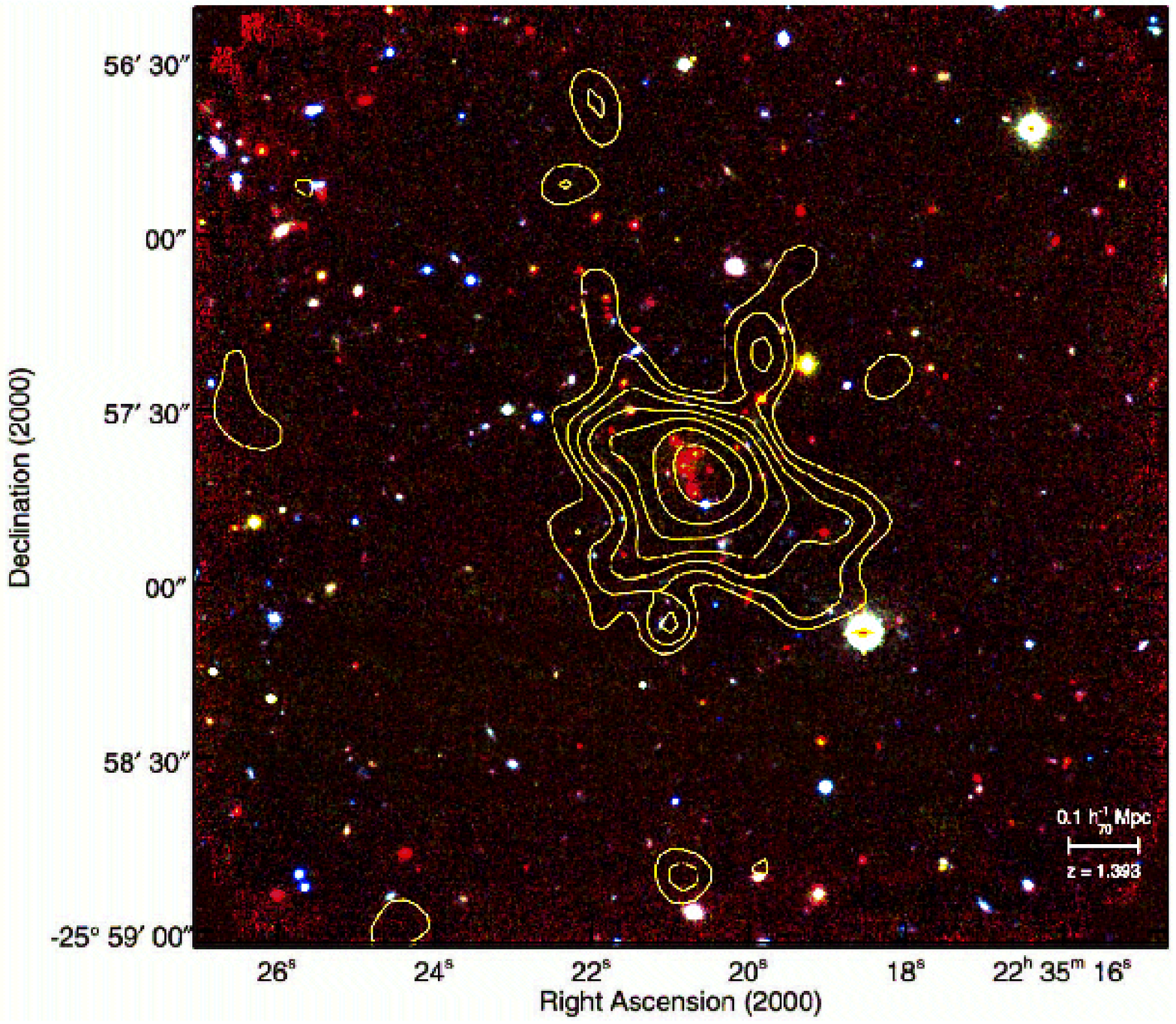}{0.75}
{A newly discovered massive evolved X-ray luminous
galaxy cluster at $z=1.39$ (Mullis et al. 2005) discovered
in a serendipitous XMM-Newton archive cluster survey. This shows that
X-ray luminous clusters exist at high redshifts. We need the redshift
leverage out to these redshifts for cluster evolution studies as a means to
probe the properties of dark energy.}
{fig:xmmm_hizclus}

Apart from providing clean and complete cluster detections,
these surveys also need to provide an estimate of the cluster masses.
X-ray observations are presently still the most efficient means of
providing cluster samples with these qualities, since the X-ray
luminosity is tightly correlated to the gravitational mass (Reiprich
\& B\"ohringer 2002), and because the X-ray emission is highly peaked,
thus minimizing projection effects. Furthermore, one might expect that
bright X-ray emission is only observed when the cluster is well
evolved, showing a very deep gravitational potential well.  As a
specific illustration of these advantages,
Fig.~\ref{fig:xmmm_hizclus} shows a recent detection of a cluster at
$z=1.39$ using XMM.  Based on simulations,
Kravtsov, Vikhlinin \& Nagai (2006)
claim that the total mass of such clusters can be estimated
using data on the amount and temperature of baryons alone, with an
impressively small uncertainty in mass of only 8\% (see
Fig.~\ref{fig:mx_proxy}).

\epsfigsimp{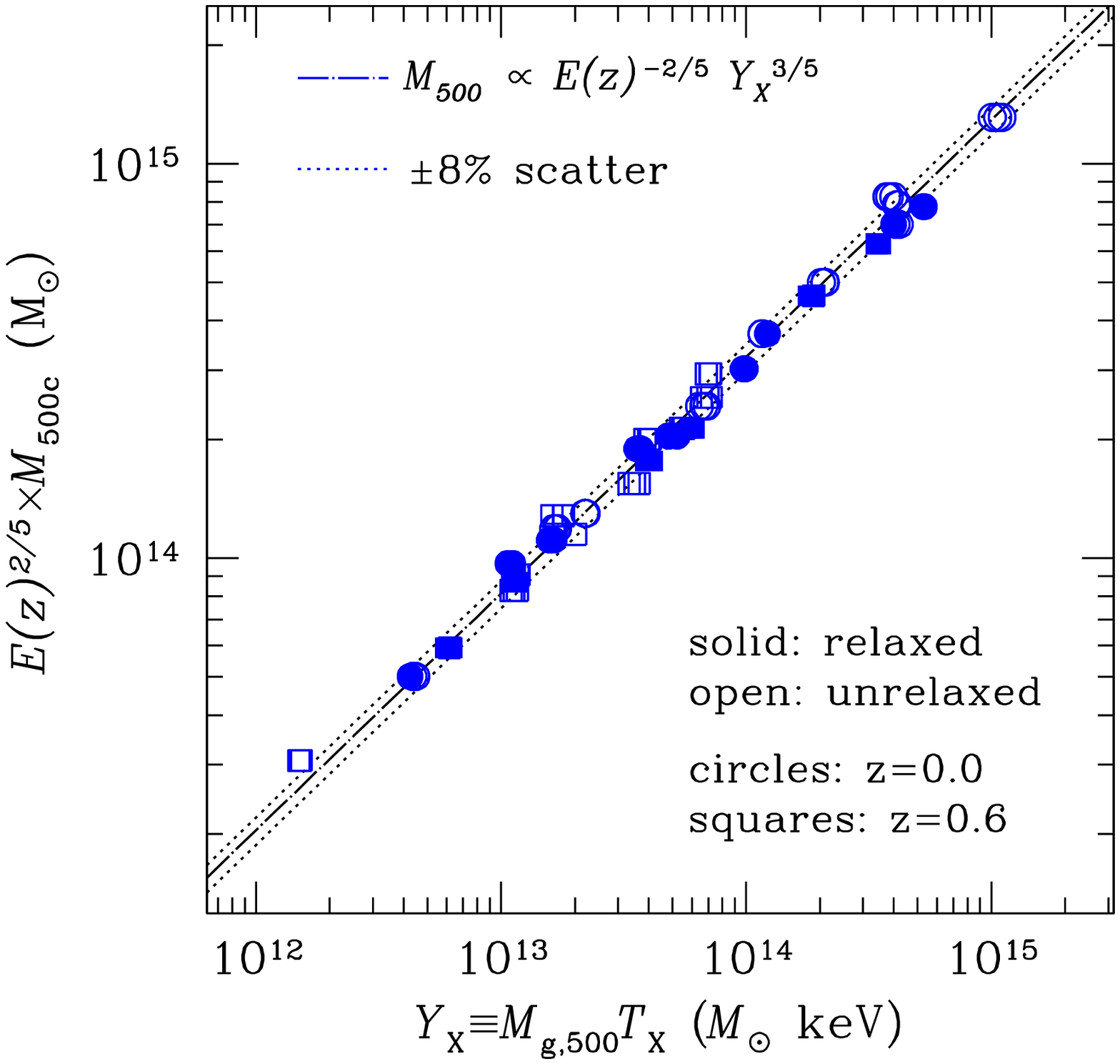}{0.5}
{A plot of total cluster mass versus a proxy based on the total
baryon mass and temperature, both of which can be inferred from
X-ray observations (Kravtsov, Vikhlinin \& Nagai 2006). This plot
is based on simulated X-ray data, and shows that total cluster
masses can be estimated to an rms precision of only 8\%.
Including the factor $E(z)\equiv [\Omega_{\rm m}(1+z)^3+\Omega_{\rm v}]^{1/2}$
allows a single redshift-independent relation to be used.
}
{fig:mx_proxy}

Thus most cosmological studies involving galaxy clusters are
based on X-ray surveys. In optical cluster surveys, great progress has been
made in cluster detection and characterisation from the large
galaxy redshift surveys such as the SDSS
(e.g. Popesso et al. 2005).
The detection of clusters in blind SZ surveys is a prospect for the near future,
and several very promising projects are on their way,
including the APEX telescope.  Similarly, gravitational lensing
surveys offer interesting possibilities for cluster detection. Many clusters have
already been found from weak lensing surveys (Wittman et al.\ 2006; Schirmer et al.\ 2006).
However, these lensing-selected `mass peaks' will not provide an
entirely clean sample of galaxy clusters (Hennawi \& Spergel 2005), and they
are more logically discussed in Sect.\ \ref{sc:GL} below (see also
Fig.~\ref{fig:Clus-princ}).

Galaxy clusters are also widely
used to study cosmic evolution of the visible matter: the formation
and evolution of galaxies as a function of their environment, the
feedback and chemical yields of supernovae, the thermal evolution of
the intergalactic medium and its enrichment by heavy elements. In this
way cosmological galaxy cluster studies also provide a route towards
a better understanding and use of galaxies and the intergalactic
medium as cosmological probes.

\ssec{Systematic uncertainties}\label{sc:clus.2}

\epsfigsimp{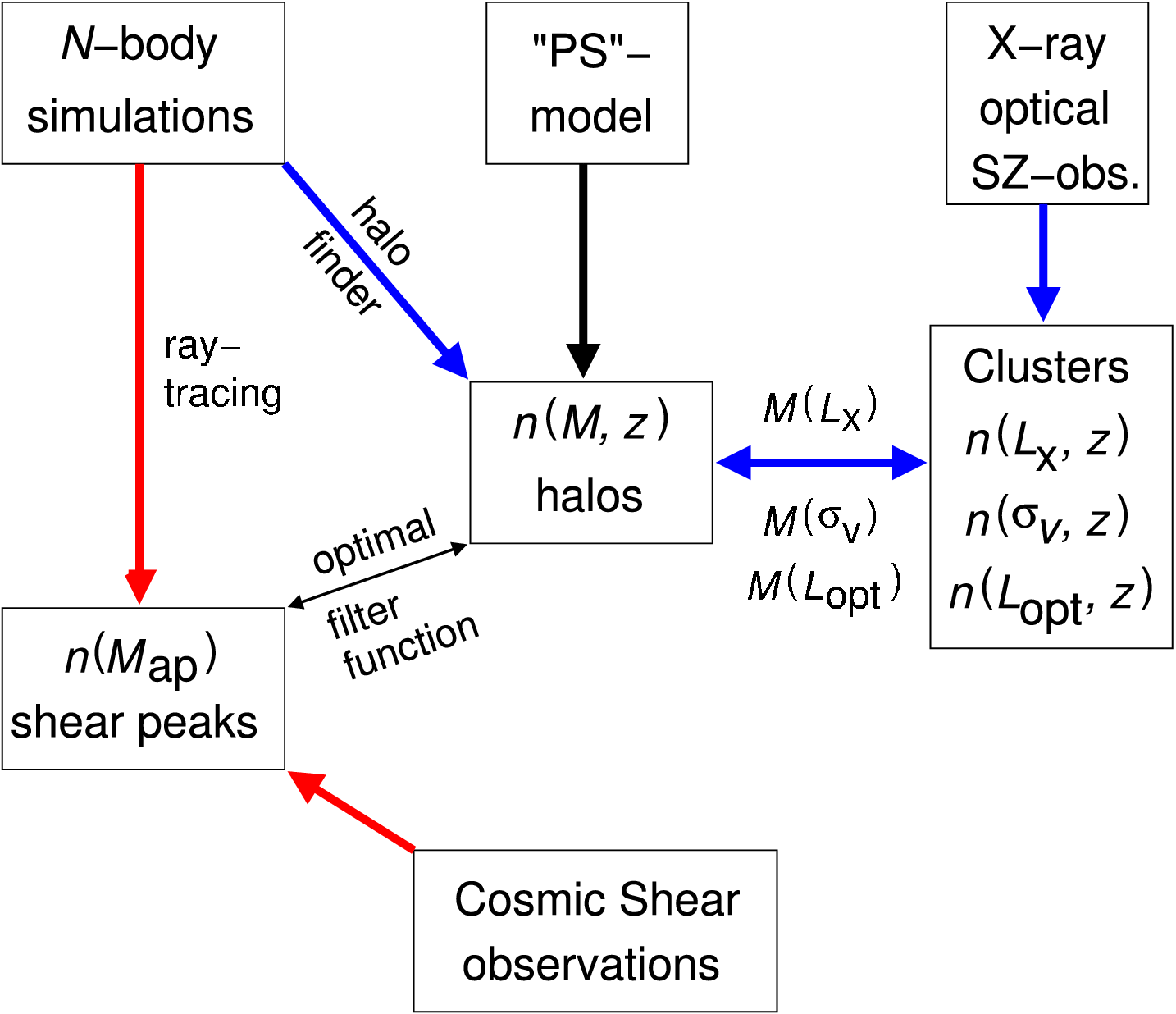}{0.6}
{The principle of using the cluster abundance as a cosmological tool.
The expected abundance of dark matter haloes $n(M,z)$ is obtained as a
function of cosmological parameters either from a model or fitting
function, or determined directly from cosmological simulations. In
order to relate them to observed cluster samples, obtained from X-ray,
optical or SZ-surveys, the redshift of the latter need to be
determined. Furthermore, an observable quantity (like X-ray or optical
luminosity, X-ray temperature, or velocity dispersion) needs to be
related to the halo mass of the cluster. The mass-observable relation
is  the critical aspect of this method. Also illustrated is the use of
mass peaks in weak lensing studies. Here, the number density $n(M_{\rm ap})$
of such peaks can be predicted directly from N-body simulations
through ray-tracing. Note that weak lensing mass peaks do not provide
exact cluster samples, owing to projection effects; but this
does not weaken their sensitivity to cosmology. By relating
the weak lensing properties to cluster masses, the latter can be
calibrated.}
{fig:Clus-princ}

The largest current limitation in using galaxy cluster data to
constrain cosmological parameters is the uncertainty in the mass
estimates that are derived from
observed properties, such as X-ray or optical luminosities
and temperatures. Cosmologists can predict the
abundance and spatial distribution of dark matter haloes as a function
of their mass, whereas the prediction of observable properties is
substantially more difficult, due to the involved physical processes
(see Fig.~\ref{fig:Clus-princ}). It is therefore essential that the mass
calibration of cluster samples can be performed with a minimum of bias.

Cluster masses are usually determined from their X-ray properties by
mapping the surface brightness and temperature profiles, and assuming
hydrostatic equilibrium of the intracluster gas. The accuracy of the
mass estimates depends on the S/N of the X-ray data, the spatial
resolution with which the temperature can be measured, and the spatial
extent out to which the X-ray gas can be studied.
The temperature profile will be measurable only for the
brightest and largest clusters; for the bulk of clusters,
only an average temperature will be measurable. In this case one has to
rely on scaling relations between temperature and X-ray luminosity with
mass.

For the envisioned precision cosmological tests, the accuracy of the
mass calibration bias has to improve from currently worse than 10\%
(for a representative selection of cluster morphologies) down to about
one percent. In addition, the scatter in the relation has to be well
quantified. This can be achieved with three different approaches:

\begin{itemize}
\item
Detailed observations with XMM-Newton and Chandra in conjunction with
comparisons to numerical simulations are currently improving our
understanding of cluster structure and reducing the bias uncertainty
to less than 10\%.  Complementary measurements of cluster structure
properties with the Sunyaev--Zeldovich effect, gravitational lensing
and comprehensive spectroscopic observations will help to
improve on this.

\item
The rapidly improving precision of numerical N-body + hydrodynamical
simulations is expected to allow reasonably accurate predictions of
several cluster observables in addition to cluster masses within a few
years. Thus the comparison between observations and theory
will involve simultaneously many parameters such as X-ray luminosity,
shapes etc., further reducing the uncertainty in the model testing.

\item
Last and not least, the consistency checks between several cosmological
tests with galaxy clusters described above will provide a calibration
check of the mass-observable relations. One of the most direct ways to
see how this works is to consider how the
amplitude of $P_{\rm cl}(k)$ depends on
the cluster mass limit, since this mass-dependent bias is what
naturally emerges from theory. In this way, we can use
clustering to achieve an independent determination of the true
cluster masses in a given subsample.
In a recent breakthrough, it has been shown that the
information in deep surveys is rich enough to solve for the unknown
mass-observable parameters with only modest degradation of constraints
on the nature of dark energy (e.g. Majumdar \& Mohr 2003, 2004;
Lima \& Hu 2004). Hence, self-calibration of a
large, clean cluster sample over large, contiguous regions of the sky
can be expected to overcome the cluster mass uncertainties.
\end{itemize}

However, it is difficult to predict the accuracy to which these
relations and their redshift dependence can be calibrated. Whereas
theoretical models suggest that a very accurate calibration can be
achieved (see Fig.~\ref{fig:mx_proxy}), we have also learned that
clusters are more complex than previously thought. The lack of massive
cooling flows is currently attributed to heating of the intracluster
gas by a recurrently active galactic nucleus in their centre.
This implies that the central regions of clusters are more
difficult to model, although the global cluster characterisation
is probably less affected.

A further point that merits more detailed study is the handling of
non-relaxed clusters, and cluster mergers. Whereas for those clusters
that have a well-resolved brightness (and temperature) profile,
signatures of merging can be readily detected (and those cluster can
then be excluded from further statistical consideration), for the more
typical cluster in large future surveys the data will be insufficient
for such an identification. Thus, proper theoretical modelling of such
merging and unrelaxed clusters will be required.

Whereas we have confined our discussion on systematics to X-ray
clusters, many of the foregoing remarks apply in a similar way also to
optical and SZ-cluster samples. However, since the X-ray band seems to
be most appropriate to use clusters as a cosmological probe, we have
not considered systematics of cluster surveys obtained by different
means.

\ssec{Prospects with a 100k cluster survey}\label{sc:clus.3}
The prospects of a large cosmological cluster survey are best
illustrated by taking the example of the well-studied
eROSITA project, which has a
projected launch date on the Russian Spectrum-R\"ontgen-Gamma-Mission
in 2010. The eROSITA mission will perform the first imaging all-sky
survey in the medium energy X-ray range up to 10 keV with an
unprecedented spectral and angular resolution. The flux limit of the
survey in the 0.5 to 2 keV band will be about $5 \times 10^{-14}\;\rm
erg\, s^{-1}\, cm^{-2}$ over most of the sky and about ten times
deeper near the poles of the survey scan pattern. The wider energy
band and the better angular resolution will make the survey about 30
times more sensitive than the ROSAT All-Sky Survey. At this depth the
X-ray sky is dominated by clusters and AGN, which can be separated
with an angular resolution of 20 arcsec. The number-flux relationship
is well known to the expected depth, and predicts that the proposed
survey will identify around 100,000 clusters (see Table~\ref{tab:erosita}).
The cluster population will essentially cover the
redshift range $z = 0 - 1.5$ and will reveal all evolved galaxy
clusters with masses above $3.5 \times 10^{14}\; h^{-1} M_\odot$ up to
redshifts of 2. Above this mass threshold the tight correlations
between X-ray observables and mass allow direct interpretation of the
data.  This sample size is necessary for the following reasons:
\begin{itemize}
\item
to characterise the cluster mass function and power spectrum 
accurately in at
least ten redshift bins, to follow the growth of structure with time,
\item
to study in detail the biasing of the cluster power spectrum as a
function of the cluster mass, in order to obtain a better understanding
of the cluster mass calibration. The biasing
describes the ratio of the amplitude of the fluctuations in
the cluster number density versus the fluctuations in the mass density. 
This parameter can
be determined theoretically as a function of mass and the comparison
with observations will serve as an important calibration check.
\item
A sample of 50,000 to 100,000 clusters is necessary to reveal
the baryonic oscillations in the cluster distribution power spectrum
(Angulo et al. 2005).
\end{itemize}
Multi-band optical (and near-IR) surveys will be needed to obtain
photometric redshifts for the clusters. The issues here are discussed
in Sect.~\ref{sc:phot-z}. However, the required
redshift accuracies ($\Delta z/(1+z) \simeq 0.02$) to $z<1.5$ for
$>100,000$ clusters are less demanding than for the applications in
galaxy surveys (Sect.~\ref{sc:LSS}) and weak lensing (Sect.~\ref{sc:GL}),
for two reasons. First, the cluster redshift estimates are obtained
from several of its member galaxies, so that random errors in the
galaxy redshifts average out to some degree. Second, the cluster
population tends to be dominated by early-type galaxies, for which
more reliable photometric redshifts can be obtained than for other
galaxy types. Nevertheless, an accurate calibration of the photo-z's is
required here as well in order to avoid a bias. The photometric
surveys that are useful for the optical follow-up of X-ray clusters
are essentially the same as can be used for weak lensing and LSS
studies; those that are currently underway or planned are summarized
in Table~\ref{tab:lenssurv}.

The expected constraints on dark energy parameters from a mission of
the scope of eROSITA have been modelled in detail by Haiman et
al. (2005), taking into account self-calibration. The cosmological
sensitivity is extracted from $\d N/\d z$, the cumulative counts of
clusters above a given X-ray flux, and their distribution in redshift
(in $\Delta z = 0.05$ wide bins), combined with measurements of
$P_{\rm cl}(k)$ in wider ($\Delta z = 0.2$) bins. Note that $\d N/\d
z$ represents a unique, exponential sensitivity to dark energy through
a combination of the comoving volume element, and through the growth
function $g(z)$. The power spectrum contains cosmological information
from the intrinsic shape of the transfer function and also from baryon
acoustic oscillation features (Blake \& Glazebrook 2003; Hu \& Haiman
2003; Angulo et al., 2005). 
These wiggles will be detectable at $\simeq 3.5\sigma$
significance in 5 separate redshift bins, varying in width between
$\Delta z = 0.2 - 0.5$, each containing about 20,000 clusters. Their
use as standard rods account for roughly half of the $P(k)$
constraints on the dark energy (Hu \& Haiman 2003). The depth of the
survey also allows a measurement of the redshift evolution of the
$P(k)$ normalization, which is an independent, direct assessment of
fluctuation growth.

The results of the Haiman et al. analysis on the $1\sigma$
uncertainties on the dark energy density and its equation of
state parameters are given in Table~\ref{tab:Rosita-constr}. The
analysis allows for 4 additional free cosmological parameters and
optionally 3 parameters describing the expected dependence of the
X-ray flux $f_X(M,z)$ on the cluster mass $M$, and redshift
$z$. Leaving the latter 3 parameters free allows for the calibration
of the mass-X-ray luminosity relation self- consistently within the
survey data, making use of the complementarity of the cluster
cosmological tests (self-calibration). The upper part of the table
refers to these results, while for the lower set of results an
external calibration of this relation to 1\% was assumed.  Due to the
complementarity of the Planck CMB observations, a combination of the
eROSITA and Planck data yields a further reduction in
the allowed parameter
space, as illustrated in the Table. The parameter degeneracies arising
from cluster constraints are also highly complementary to those from
Type Ia SNe (Wang \& Steinhardt 1998; Holder, Haiman \& Mohr 2001;
Levine, Schulz \& White 2002).

\begin{table}
\caption{Parameter uncertainties from a 100,000--cluster sample (with
an area of 20,000 deg$^2$ and limit flux of $f_X=2.3\times
10^{14}\rm\, erg\, cm^{-2}\, s^{-1}$), for the density parameter in
dark energy (DE), and the two parameters $w_0$ and $w_a$
parameterizing the equation of state of dark energy.  The results are
based on the assumption of a spatially flat universe and have been
marginalized over $\Omega_{\rm b} h^2$, $\Omega_{\rm m} h^2$, $\sigma_8$, and
$n_{\rm s}$ (from Haiman et al. 2005; based on Wang et al. 2004).  The
models with superscript (a) assume a constant $w$ ($w_a = 0$). Models
(b) correspond to using a 7-parameter cosmology--only Fisher matrix
that assumes that the $f_X$--mass relation has been externally
calibrated to 1\% accuracy (i.e., it effectively assumes the
self-calibration parameters are known to 1\% precision).}
\label{tab:Rosita-constr}
\begin{center}
\begin{tabular}{|l|c|c|c|}
\hline
\topstrut\botstrut {Self-Calibrated Experiment(s)} & $\sigma(w_0)$ & $\sigma(w_a)$ & $\sigma({\rm DE})$ \\
\hline
\topstrut X-ray & 0.093 & 0.490 & 0.0067 \\
X-ray+Planck & 0.054 & 0.170 & 0.0052 \\
\botstrut X-ray+Planck$^a$ & 0.016 & --- & 0.0045 \\
\hline
\topstrut\botstrut {Ideal Experiment}$^b$ & & & \\
\hline
\topstrut X-ray & 0.021 & 0.120 & 0.0030 \\
X-ray+Planck & 0.013 & 0.066 & 0.0027 \\
\botstrut X-ray+Planck$^a$ & 0.0087 & --- & 0.0019 \\
\hline
\end{tabular}
\end{center}
\end{table}

The modelling by Haiman et al. (2005) makes use of only the part of the cluster data
that can be included in an overall cosmological test and it mostly relies on the
application of self-calibration. This limitation could be overcome
if a powerful X-ray observatory of the type envisaged for XEUS were available
to complement the survey projects. Some of main applications
of such a facility within the context of this report are
the detailed characterisation of high redshift galaxy cluster
properties to assist the cosmological X-ray, optical, and SZ cluster
surveys and to confirm
the nature of massive black hole mergers detected with LISA.
Therefore a comprehensive use of the future
observational data that can be envisaged from the X-ray waveband
should offer a robust and
competitive way to learn more about the properties of our
universe and those of Dark Energy in particular.

\begin{table}
\caption{Sensitivity and yields for the cluster surveys planned with
eROSITA.}
\begin{center}
\begin{tabular}{|l|c|c|c|}
\hline
\topstrut\botstrut Survey & All-Sky Survey & Wide Survey & Deep Survey \\
\hline
\topstrut Solid Angle & 42000 & 20000 & 200 \\
Exposure time & 1 yr & 2.5 yrs & 0.5 yrs \\
Clusters: 0.5--5 keV $S_{\rm min}$ & $1.6\times 10^{-13}$ & $3.3 \times 10^{-14}$
& $8\times 10^{-15}$ \\
\botstrut Clusters: numbers & 32000 & 72000 & 6500 \\
\hline
\end{tabular}
\end{center}
\label{tab:erosita}
\end{table}

\japsec{Gravitational Lensing}\label{sc:GL}

\ssec{Techniques}\label{sc:GL.1}

Gravitational lensing arises when cosmic density inhomogeneities
deflect the light from distant objects, causing their images to be
distorted.  Galaxies subject to lensing can change their apparent
magnitude, but the main observable is an image distortion, principally
in the form of a shear.  The coherent pattern of image distortions
from lensing gives a direct probe of the location and distribution of
mass concentrations in the universe. Because the lensing effect is
insensitive to the dynamical state and the physical nature of the mass
constituents of the mass distribution causing the gravitational field,
lensing has been widely used over the past 15 years to analyse the
distribution of dark matter in complex systems like clusters of
galaxies. Lensing and X-ray mass estimates for clusters show an
excellent degree of agreement between both techniques (e.g. Allen,
Ettori \& Fabian 2001), and this is a fundamental result: it verifies
Einstein's factor of 2 in light deflection over cosmological scales,
which is an important piece of knowledge when considering alternative
theories of gravity.

As datasets have expanded, lensing has increasingly concentrated on
the `weak' regime, in which image ellipticities are altered by only a
few per cent. By averaging the distortions of many background
galaxies, it has been possible to map structures in the dark-matter
distribution statistically, measuring correlations in the shear field
-- the so-called `cosmic shear', as envisaged by the beginning of the
past decade (Blandford et al. 1991; Miralda-Escud\'e 1991; Kaiser
1992). The first conclusive detections of cosmic shear were made in
2000. Remarkably, all teams found similar signals, with shape and
amplitude in good agreement with the gravitational instability
paradigm in a CDM-dominated universe. The impressive consistency of
these results stimulated a rapidly growing field, involving an
increasing number of teams and techniques.

Cosmological weak lensing is also among the most recent dark energy
probes. The sensitivity to dark energy arises because lensing
is sensitive to a ratio of distances from us to
the lens and the lens to the source, and to the amplitude of
the projected mass density contrast along the line of sight.
The convergence
$\kappa(\vec\theta)$ governs the strength of lensing,
\be
\kappa(\vec\theta) = {4 \pi G\over c^2} \, {D_{\japsub L} D_{\japsub LS}\over
D_{\japsub S}} \; \Sigma(D_{\japsub L}\vec\theta),
\ee
where $\Sigma$ is the projected surface mass density of the lensing
structure, $D_{\japsub LS}$ is the angular-diameter distance between
lens and source, $D_{\japsub L}$ is the distance between the observer and 
the lens and $D_{\japsub S}$ is the distance between the observer and the source.

The image shear, $\gamma$, is related to
derivatives of $\kappa$.  The two signatures of dark energy arise from
its presence in the distance factors (a purely geometrical effect),
and because the effective mass of lensing structures reflects the
power spectrum and growth rate of large-scale density perturbations.
Both of these effects, geometry and growth, can be probed by taking
two-point functions (e.g. shear power spectra, or correlation
functions) of distant galaxy images as a function of redshift
(e.g. Heavens 2003).  Alternatively, the geometry of the universe can
be isolated from weak lensing by measuring the ratio of shear behind
dark matter haloes as a function of redshift (Jain \& Taylor 2003;
Zhang et al. 2005; hereafter the geometry-power spectrum decoupling
technique).  These methods are challenging, since the characteristic
lens distance ratio scales out some of the $w$ sensitivity, so that
the precision multiplier $|\partial w/\partial\, \ln {\rm shear}|$ exceeds
10.  Nevertheless, lensing is unique in being able to measure both
signatures of a dynamical vacuum.  The key to making this technique
work is distance information, but the typical galaxies involved are
too faint and too numerous for direct spectroscopy to be
feasible. Therefore, one has to rely on photometric redshift estimates
in order to perform 3D lensing analyses (see Sect.~\ref{sc:phot-z}).

Lensing thus provides an attractive and physically based probe of dark
energy. In contrast to the more empirical supernova method
(Sect.~\ref{sc:SN}), the lensing signal depends both on the geometry of
the universe, via the angular distances, and on the dark matter power
spectrum and its evolution with redshift. The direct connection with
the gravity field generated by dark matter means that lensing is
simultaneously a tool to explore modified gravity theories
(Uzan \& Bernardeau 2001; White \& Kochanek 2001).
Weak gravitational lensing experiments also provide
byproducts: the mass properties of galaxy halos via the so-called
galaxy-galaxy lensing (see e.g. Hoekstra et al.\ 2003); a sample of
shear-selected clusters of galaxies (e.g. Schirmer et al.\ 2006); a
sample of arcs and Einstein rings in clusters of galaxies or around
galaxies; the direct measurement of the relations between light
and mass through weak lensing-galaxy cross correlation to determine
the galaxy bias parameter (e.g. Hoekstra et al.\ 2002b; Seljak et al.\ 2005a); 
and the relation of lensing signals from galaxies to
lensing effects on the CMB, which must be understood in order to
interpret a B-mode signal in the CMB polarization.
These data are important probes of the hierarchical model of
structure formation and the history of galaxy formation. Several of
these byproducts also depend on the dark energy content of the
universe and provide ways of cross-checking the constraints.

\ssec{Current status}\label{sc:GL.2}

Past and ongoing surveys are still focused on the primary outcome
expected from lensing: constraints on the dark matter density
$\Omega_{\rm m}$ and in particular on the amplitude of density
fluctuations, $\sigma_8$. The inferred normalization depends on the
matter density roughly as $\Omega_{\rm m}^{-0.6}$, so one really probes the
product $\sigma_8 \Omega_{\rm m}^{0.6}$.  Table~\ref{tab:lensdonesur}
summarizes the present status of cosmological weak lensing surveys
(from Hetterscheidt et al. 2006). The constraints derived so far on
$\sigma_8(\Omega_{\rm m})$ and on $w$ are also given in that Table. All
these results are based solely on two-point statistics. Surveys with
constraints on dark energy are still few and very recent (Jarvis et
al. 2006; Semboloni et al. 2006a; Hoekstra et al. 2006). They mostly
demonstrate the capability of the techniques but the results are not
yet as impressive as supernovae or BAO.  Overall, $\sigma_8(\Omega_{\rm m})$
is now derived with a 5\%-10\% accuracy, whilst $w$ is derived with a
50\% accuracy, assuming it to be constant. Nevertheless, lensing
provides interesting constraints on $w$, when used jointly with CMB,
BAO or SNe Ia surveys (Contaldi, Hoekstra \& Lewis 2003; Schmid et al. 2006).

At the forefront of observations and data analysis, the most recent
survey is the CFHTLS that now explores angular scales up to 2 degrees
to a depth of $I_{\rm AB}=24$ (Fu et al. 2006). Several teams
have also gone beyond two-point ellipticity correlations and explored
territory that will be important in next generation surveys. Pen et
al. (2003a) derived an upper limit on $\Omega_{\rm m}$ by computing the
skewness on the convergence field in the Virmos-Descart survey. They
then tentatively broke the degeneracy between $\Omega_{\rm m}$ and
$\sigma_8$ contained in the two-point statistics. Analyses of
three-point statistics from real data were also carried out by
Bernardeau, Mellier \& van Waerbeke (2002) and Jarvis et al. (2004) who found
cosmological signatures.  But the signal is noisy and its
contamination by systematics is as yet poorly understood.  So, despite
its potential for the exploration of non-Gaussian signatures in
cosmological structures (Bernardeau, van Waerbeke \& Mellier 1997; Jain \& Seljak 1997;
Takada \& Jain 2002; Kilbinger \& Schneider 2005), cosmology from
higher-order shear statistics can hardly be fully exploited from
present surveys.  Pen et al. (2003b) performed the first 3D dark matter power
spectrum reconstruction using weak lensing and compared the results
with the WMAP1 power spectrum. Both datasets are in remarkable
agreement and show a continuous and monotonic shape that is consistent
with a $\Lambda$CDM power spectrum. The power of employing redshift
information on individual galaxies has been recognised and already
applied to a weak lensing survey (Bacon et al. 2005). It was also used
in its most simplistic way to derive constraints on $w$ by a joint
analysis of the deep and the wide CFHTLS surveys (Semboloni et al. 2006a;
Hoekstra et al. 2006).  A reliable and conclusive use of the
redshift distribution in weak lensing surveys has been carried out by
Heymans et al. (2005) using the COMBO-17 photometric redshifts. Their
derived value for $\sigma_8$ is in excellent agreement with
WMAP3. This demonstrates that redshift information will be a
prerequisite for all new weak lensing surveys.

Overall, most weak lensing statistics that will be used in next
generation surveys have thus already been tested and validated on
present-day surveys. The results to date demonstrate that the field
has the statistical tools needed to use lensing to study dark energy.
An especially impressive development has been the rapid adoption of a
standard test for systematics based on the E-mode/B-mode decomposition
of the shear field. As with the CMB, this corresponds to separating
the part of the shear field that can be generated by symmetric
combinations of derivatives acting on a potential (the E mode) from
the B-mode shear, which effectively corresponds to rotating all E-mode
shear directions through $45^\circ$, and which is not caused by
lensing (in leading order). Requiring a negligible level of B-mode
contamination has been an important part of demonstrating the
consistency of current results on cosmic shear.  Nevertheless, as the
required precision increases, the challenge will grow of assuring that
results are not affected by small systematics that presently lurk
beneath the random noise.  We make a detailed assessment of these
effects below.

%%%%%%%%%%%%%%%%%%%%%%%%%%%%%%%%%%%%%
% Summary table
%%%%%%%%%%%%%%%%%%%%%%%%%%%%%%%%%%%%%
\begin{table}
\caption{Summary of properties of present lensing surveys, with
constraints on cosmological parameters -- assuming a flat universe
with $n_{\rm s}=1$.}
\begin{center}
{\tiny
%\relscale{1.20}
\begin{tabular}{|l|c|c|c|c|c|c|l|}
\hline
\topstrut Survey & Telescope & Sky & n & Mag & $\sigma_8$ & $w_0$ & Ref.  \\
\botstrut & & coverage & $\rm gal\; arcmin^{-2}$ & & ($\Omega_{\rm m}=0.3$) & & \\
\hline
\topstrut VLT-Descart & VLT & 0.65 deg$^2$ & 21 & I$_{\rm AB}=24.5$ & 1.05$\pm 0.05$ & &
Maoli et al. 2001 \\
 & & & & & & & \\
Groth Strip & HST/WFPC2 & 0.05 deg$^2$ & 23 & I=26 & 0.90$^{+0.25}_{-0.30}$ & &
Rhodes et al. 2001 \\
 & & & & & & & \\
MDS & HST/WFPC2 & 0.36 deg$^2$ & 23 & I=27 & 0.94$\pm 0.17$ & & R\'efr\'egier
et al. 2002 \\
 & & & & & & & \\
RCS & CFHT & 16.4 deg$^2$+ & 9 & R=24 & 0.81$^{+0.14}_{-0.19}$ & & Hoekstra et
al. 2002a \\
 & CTIO & 7.6 deg$^2$ & & & & & \\
 & & & & & & & \\
Virmos-Descart & CFHT & 8.5 deg$^2$ & 15 & I$_{\rm AB}$=24.5 & 0.98 $\pm 0.06$ & -
& van Waerbeke et al. 2002\\
 & & & & & & & \\
RCS & CFHT & 45.4 deg$^2$+ & 9 & R=24 & 0.87$^{+0.09}_{-0.12}$ & & Hoekstra et
al. 2002b \\
 & CTIO & 7.6 deg$^2$ & & & & & \\
 & & & & & & & \\
COMBO-17 & 2.2m & 1.25 deg$^2$ & 32 & R=24.0 & 0.72 $\pm 0.09$ && Brown et al.
2003\\
 & & & & & & & \\
Keck + & Keck & 0.6 deg$^2$ & 27.5 & R=25.8 & 0.93 $\pm 0.13$ & & Bacon et al.
2003 \\
 WHT& WHT & 1.0 deg$^2$ & 15 & R=23.5 & & & \\
 & & & & & & & \\
CTIO & CTIO & 75 deg$^2$ & 7.5 & R=23 & 0.71$^{+0.06}_{-0.08}$ & & Jarvis et al.
2003 \\
 & & & & & & & \\
SUBARU & SUBARU & 2.1 deg$^2$ & 32 & R=25.2 & 0.78$^{+0.55}_{-0.25}$ & & Hamana
et al. 2003 \\
 & & & & & & & \\
COMBO-17 & 2.2m & 1.25 deg$^2$ & R & R=24.0 & 0.67 $\pm 0.10$ && Heymans et al.
2004\\
 & & & & & & & \\
FIRST & VLA & 10000 deg$^2$ & 0.01 & 1 mJy & 1.0 $\pm0.2$ & & Chang et al. 2004
\\
 & & & & & & & \\
GEMS & HST/ ACS & 0.22 deg$^2$ & 60 & I=27.1 & 0.68 $\pm 0.13 $ & & Heymans et
al. 2005 \\
 & & & & & & & \\
WHT + & WHT & 4.0 deg$^2$ + & 15 & R$_{\rm AB}$=25.8 & & & Massey et al. 2005 \\
COMBO-17 & 2.2m& 1.25 deg$^2$ & 32 & R=24.0 & 1.02 $\pm 0.15$ && \\
 & & & & & & & \\
Virmos-Descart & CFHT & 8.5 deg$^2$ & 12.5 & I$_{\rm AB}$=24.5 & 0.83 $\pm 0.07$ &
- & van Waerbeke et al. 2005\\
 & & & & & & & \\
CTIO & CTIO & 75 deg$^2$ & 7.5 & R=23 & 0.71$^{+0.06}_{-0.08}$ &
$-0.89^{+0.16}_{-0.21}$& Jarvis et al. 2006 \\
 & & & & & & & \\
CFHTLS Deep+ & CFHT & 2.1 deg$^2$ + & 22 & i$_{\rm AB}$=25.5 & 0.89 $\pm 0.06$& $\le -0.80 $& Semboloni et al.
2006a \\
Wide & & 22 deg$^2$ & 13 & i$_{\rm AB}$=24.5 & 0.86 $\pm 0.05$& &
Hoekstra et al. 2006 \\
 & & & & & & & \\
GaBoDS & 2.2m & 15 deg$^2$ & 12.5 & R=24.5 & 0.80 $\pm 0.10$ & - &
Hetterscheidt et al. 2006\\
 & & & & & & & \\
ACS parallel + & HST/STIS & 0.018 deg$^2$ & 63 & R=27.0 ? &
0.52$^{+0.13}_{-0.17}$ & & Schrabback et al. 2006 \\
\botstrut GEMS+GOODS & HST/ACS & 0.027 deg$^2$ & 96 & V=27.0 & & & \\
\hline
\end{tabular}
}
\end{center}
\label{tab:lensdonesur}
\end{table}
%%%%%%%%%%%%%%%%%%%%%%%%%%%%%%%%%%%%
% Figure
\begin{figure}
\begin{center}
\includegraphics[width=0.49\hsize,height=0.49\hsize]{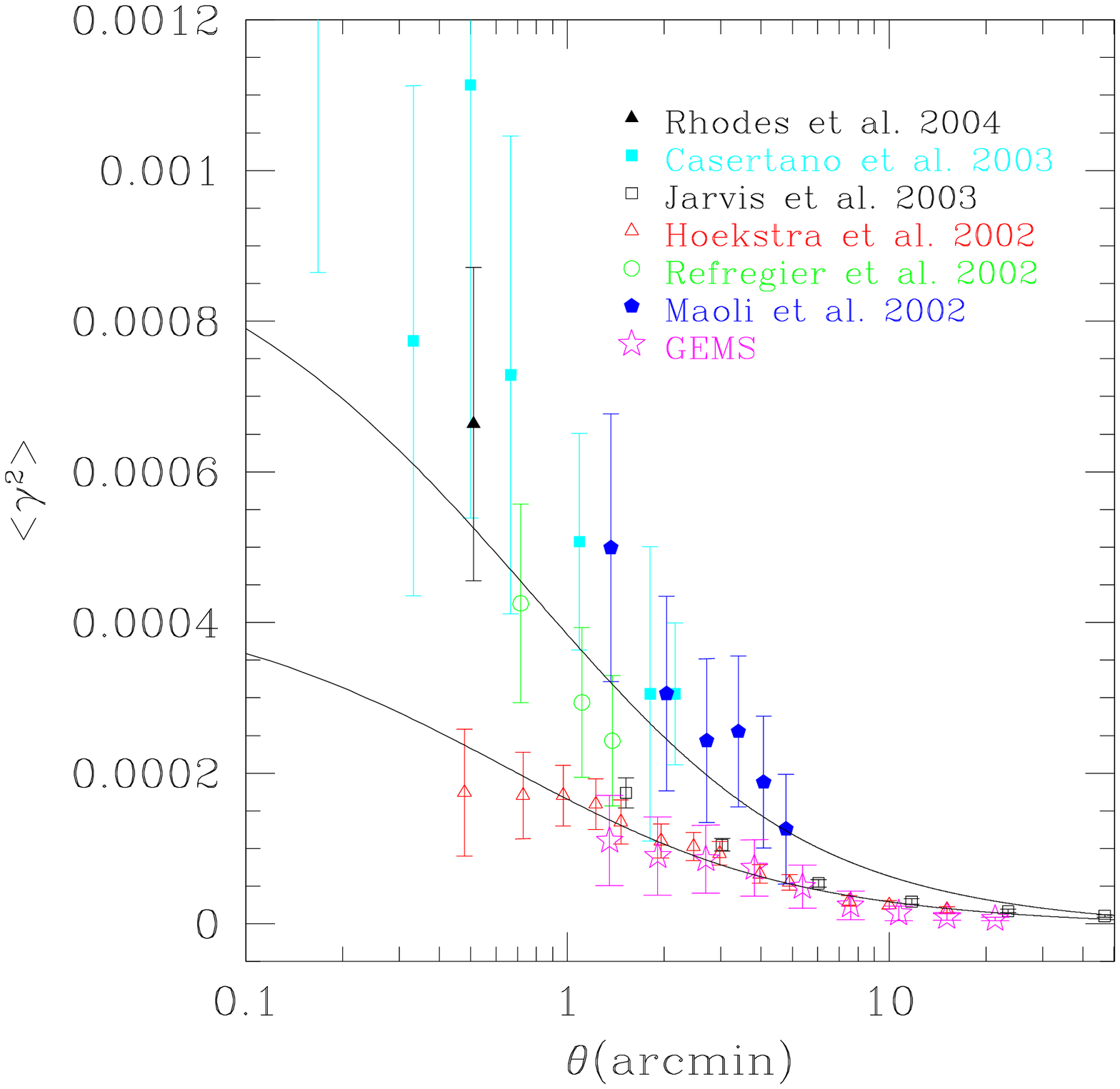}
\includegraphics[width=0.49\hsize,height=0.49\hsize]{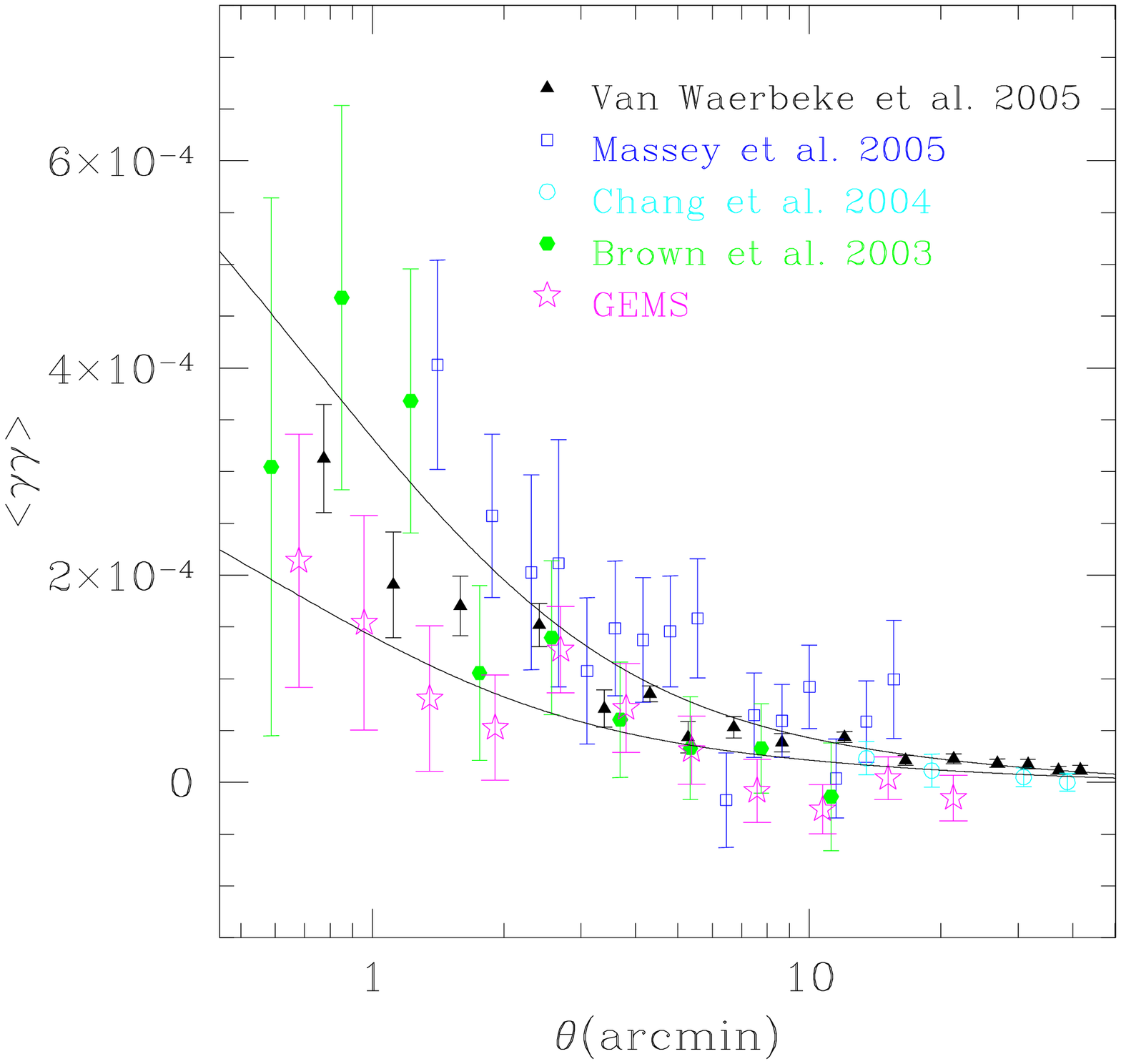}
\caption{Left: top-hat shear variance $E[\gamma^2]_\theta$ as
measured from recent space or ground based surveys. Only the E-modes
are shown. The lines show theoretical $\Lambda$CDM predictions for a
mean source redshift $z_{\rm m} = 1$ with $\sigma_8 = 0.7$ (lower) and
$\sigma_8 = 1.0$ (upper).  Right: the total shear correlation
function $E[\gamma \gamma]_\theta$ derived from few recent ground
based or space surveys. The lines shows the same models as on the
left panel. From Heymans et al. (2005). }
\end{center}
\label{fig:hey}
\end{figure}
%%%%%%%%%%%%%
\begin{figure}
\begin{center}
\includegraphics[width=0.49\hsize,height=0.49\hsize]{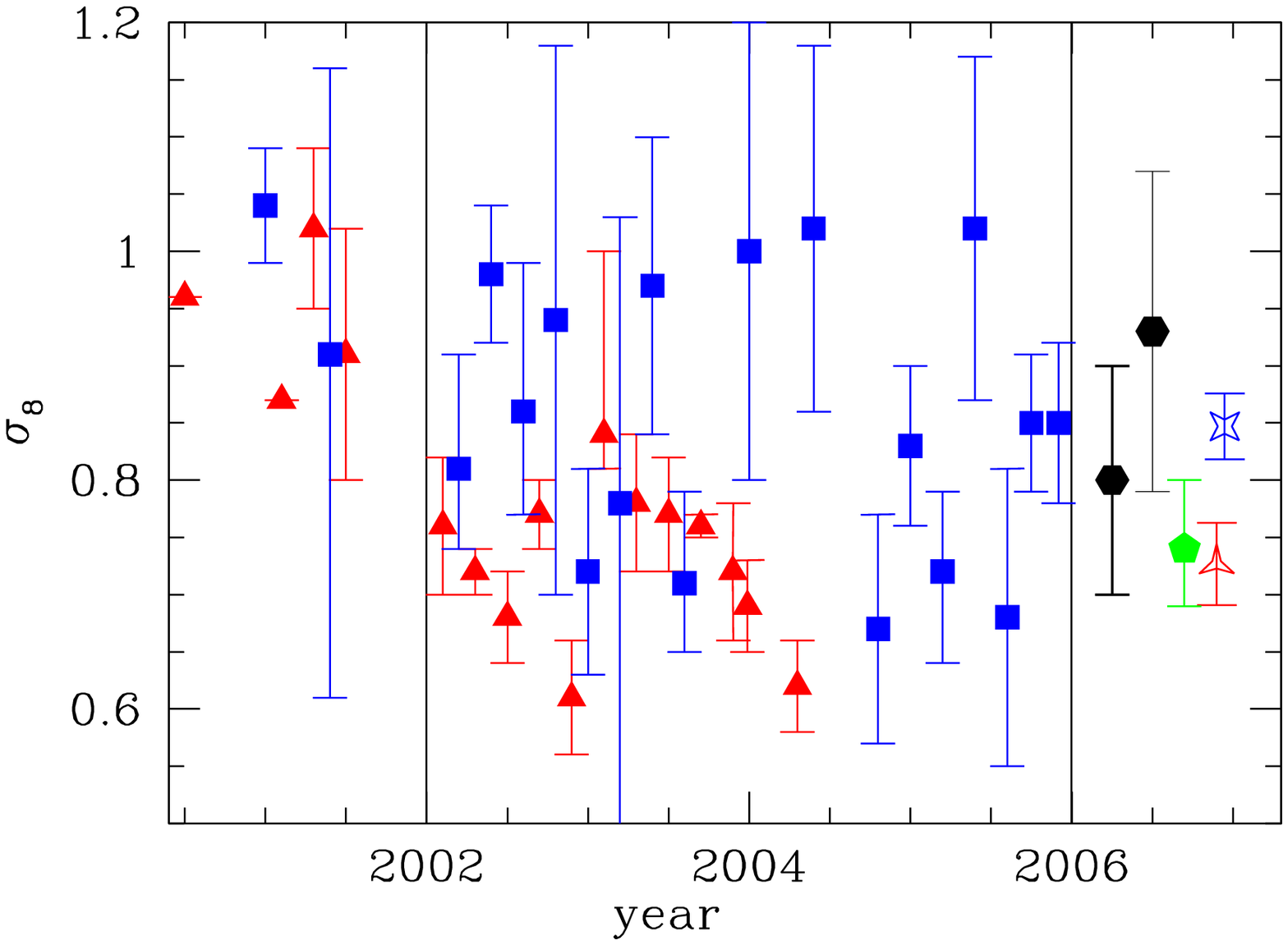}
\includegraphics[width=0.49\hsize,height=0.49\hsize]{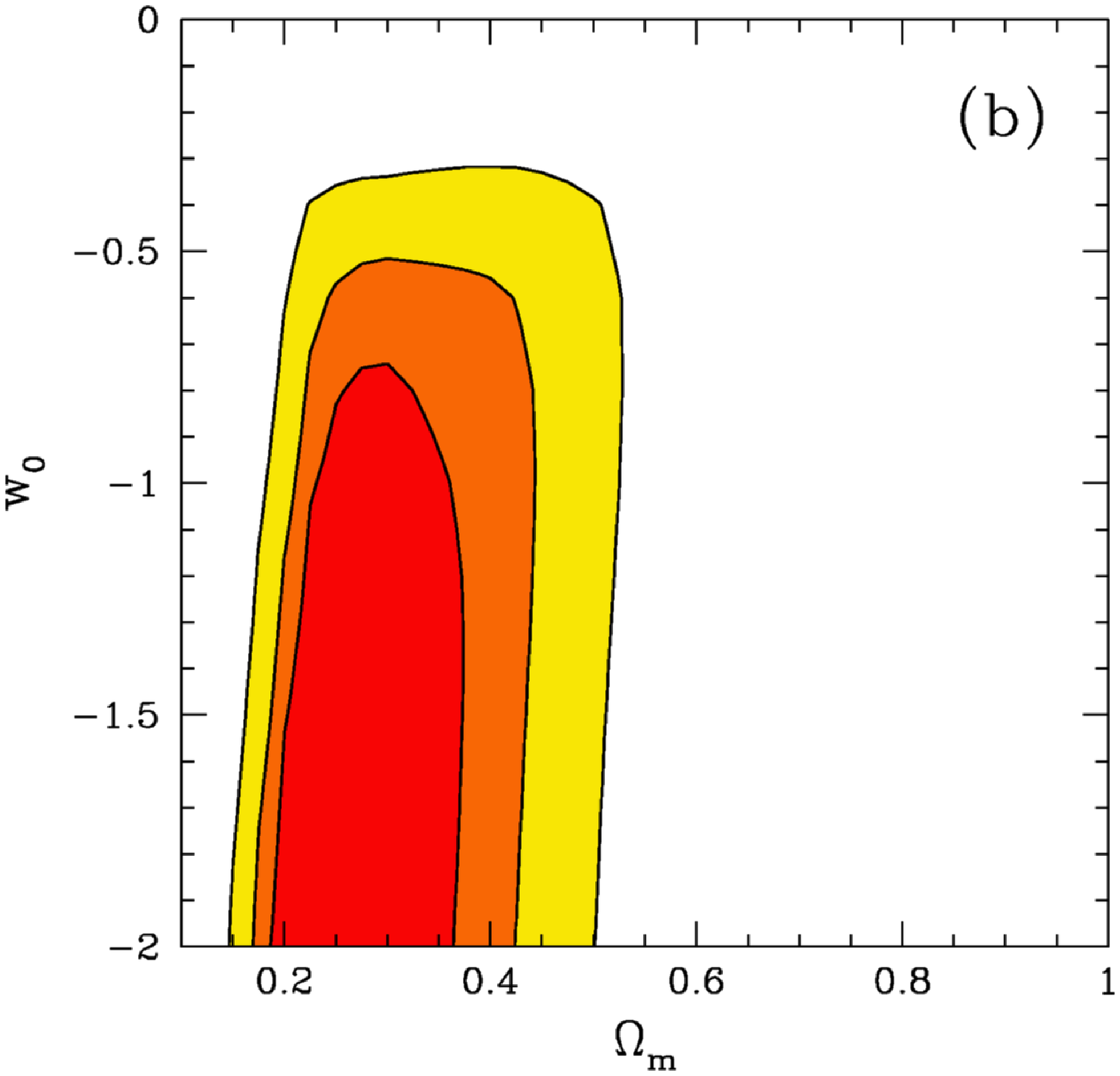}
\caption{Left: a compilation of the most recent determination
of $\sigma_8$ from the analysis of clusters of galaxies (red) and
cosmological weak lensing. {}From Hetterscheidt et al. (2006). 
Right: constraints on dark energy derived from the CFHTLS Deep and
Wide weak lensing survey, assuming a constant $w$ and a flat
universe. From Hoekstra et al. (2006).}
\end{center}
\label{fig:hethoe}
\end{figure}
%%%%%%%%%%%%

\ssec{Systematic uncertainties and errors}\label{sc:GL.3}

There are three major types of bias that may contaminate the lensing
signal: (1) PSF correction, (2) biased selection of the galaxy sample
(e.g. the redshift distribution of galaxies used for weak lensing may
differ from the galaxy selection used to measure the redshift
distribution of the survey) and (3) intrinsic (astrophysical)
distortion signal. Most systematics that are discussed below turn out
to be negligible or well under control for present day surveys. However,
they may start to become severe for the short and mid-term weak lensing
projects and could limit progress unless we improve present
day corrections by one order of magnitude.
The most critical issues are
\begin{itemize}
\item
the interpretation of the distortion
signal on angular scales below 10 arcmin where non-linear evolution of the
dark matter power spectrum necessitates numerical modelling for
accurate predictions; furthermore, on small angular scales galaxy
shapes may be
intrinsically correlated;
\item
the
measurements at very large angular scales for which the PSF
corrections need to be very accurate
to measure very weak gravitational distortion correlations;
\item
and the
redshift distribution of the galaxy sample, or the redshifts of
individual galaxies that will be needed in precision application of
weak lensing.
\end{itemize}

\begin{figure}
\begin{center}
\includegraphics[width=0.5\hsize,angle=270]{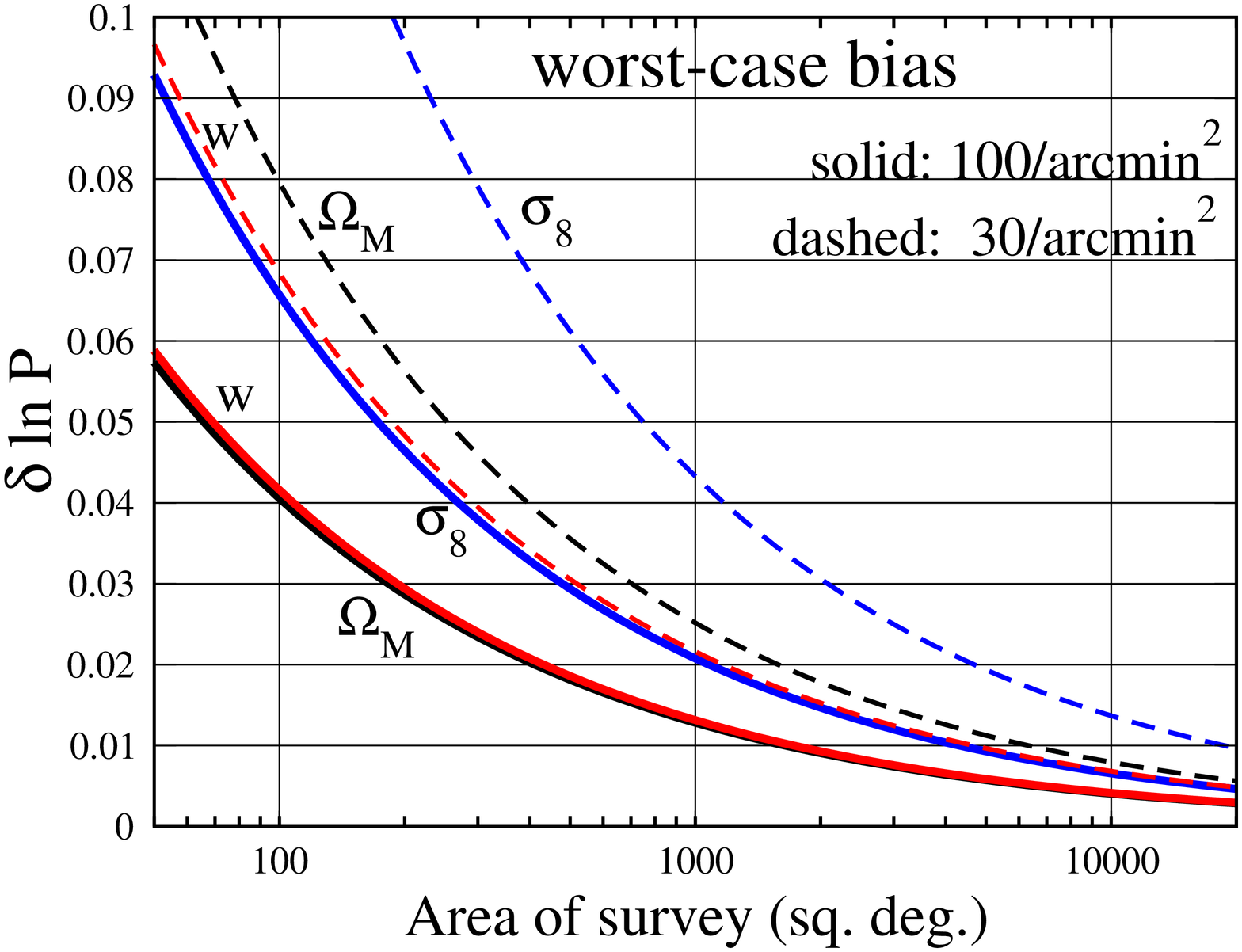}
\caption{Required accuracy of cosmic shear power spectrum
  predictions. The curves show the fractional uncertainty of the
  predicted power spectrum that can be afforded such that the bias in
  the estimated cosmological parameters is not larger than the
  statistical uncertainties of the measured cosmic shear signal. The
  latter depends on the survey area, as well as on the survey depth
  and thus number density of source galaxies. In particular this plot
  demonstrates that, for large future surveys, the predictions for the
  lensing power spectrum need to be known with an accuracy of $\sim
  1\%$ in order not to degrade the accuracy of cosmological parameter
  estimates (from Huterer \& Takada 2005). }
\label{fig:hut05}
\end{center}
\end{figure}

For some of these issues, high-resolution numerical simulations are
necessary in order to understand and calibrate systematics (nonlinear
evolution, as well as the influence of the baryonic matter on the dark
matter power spectrum; the validity of the Born and Limber
approximations; intrinsic correlations; cosmic and sampling variance
as a function of the survey design). Overall, none of these obstacles
appears insuperable, but some systematics need further investigation
in order to assess how they may impact on the survey designs. It is
currently unclear what is the most efficient combination of sky
coverage, survey geometry, and survey depth.  The survey size seems,
however, the primary goal. Fig.\ref{fig:hut05} shows that any survey
covering more than 10,000 deg$^2$ with a reasonable depth can recover
the dark matter power spectrum with an accuracy of $\sim 1\%$ (Huterer
\& Takada 2005).

\sssec{PSF correction}
The PSF correction is the most challenging and debated technical issue
in weak lensing. Both the isotropic smearing and the PSF anisotropy
corrections bias the amplitude of the lensing signal. The result is an
additive bias produced by any error in the PSF anisotropy correction,
and a multiplicative bias produced by any error in the calibration of
the PSF smearing effect.  So far, the multiplicative bias is the
dominant source of error. The amplitude of the bias may also depend on
the apparent size, the surface brightness and the intrinsic
ellipticity of each galaxy (and therefore it also depends on the mean
redshift of the galaxy sample). A dozen or so techniques have been
proposed to handle these corrections. They provide reasonable
corrections that satisfy the requirements of present-day surveys.

The status of the PSF corrections for weak lensing is discussed in
great detail in Heymans et al. (2006a), who describe the thorough joint
Shear TEsting Program (STEP). Most teams involved in weak lensing
agreed to use blindly the same simulated data sets to assess the
capability of their techniques for both space- and ground-based
images. The good news is that a few techniques demonstrated that they
can already measure weak lensing shear signals correctly down to 1\%
-- but most current techniques are limited by calibration errors of
order $\simeq 5$\%.  Using this calibration, one can measure a shear
signal up to angular scales of 2-3 degrees without showing measurable
B-modes or being contaminated by systematics, but we can hardly go
beyond this scale until calibration errors are decreased to less than
$\simeq 2$\%. The bad news is that none of the techniques currently
under evaluation surpasses others and prevails for the next generation
surveys. Even the best ones still show bias residuals or turn out to
be only valid in a given range of distortion amplitude. This shows that the
PSF correction problems are not fully understood and under control and
thus have not yet reached the goal for the next generation surveys,
where the calibration error should be less than $0.1$\%. But the
results of the STEP project (which is continuing) are very encouraging
and demonstrate there is scope for improvements, as well as the will
of the community to collaborate towards developing more accurate
methods.

\sssec{Clustering}
Source/lens clustering has two effects on weak lensing.  First, it
produces an apparent B-mode signal, plus additional E-modes that would
not exist otherwise. This is a subtle effect, which arises because the
angular correlation of galaxies depends on their redshift, and the
angular distance between galaxies that scales the lensing amplitude
also depends on the apparent angular separation between galaxy
pairs. It results in a coupling between the shear two-point
correlation function and the clustering terms. The amplitude is always
below 2\% on scales larger than one arcminute (Schneider, van Waerbeke
\& Mellier 2002). Though it is negligible in present-day surveys, it could
however bias the E- and B-modes on small angular scales in next generation
surveys. Since it only affects very small scale, even if there were no
way to correct the lensing signal from this clustering effect, a safe
and simple solution is to discard angular scales below one arcminute.

Second, source clustering changes the amplitude of higher-order
statistics whilst keeping the second-order statistics
unaffected. Bernardeau (1998) pointed out that both the skewness and the
kurtosis are affected. It is likely that the three-point shear correlation
functions are also affected by a similar amount. Hamana et al. (2002)
showed that on scales smaller than $100'$ and for a $\Lambda$CDM
model, the source/lens clustering can modify the amplitude of the
skewness by 5 to 40\% as compared to an uncorrelated galaxy
distribution. It is worth noticing that the bias is important and is
almost flat from $100'$ to $1'$. However, its amplitude can be considerably
reduced by using narrow redshift distributions. The correction drops
below 1\% for a mean redshift of $\sim 1$ and by using a width of the
redshift distribution below 0.15. Photometric redshifts should
therefore have sufficient accuracy to permit redshift bins of this size.
However, the correction is still about 10\% for galaxies
at redshift 0.5. The use of higher-order correlations thus definitely
strengthens the need for accurate photometric redshifts.

%%%%%%%%%%%%%%%%%%%%%%%%%%%%%%%%%%%%
% Figure
%%%%%%%%%%%%%
%%%%%%%%%%%%%%%%%%%%%%%%%%%%%%%%%%%%

\sssec{Contamination by overlapping galaxies}
Very close galaxy pairs have overlapping isophotes that contaminate
both the first and second surface brightness weighted moments of
galaxies that are used to derive the galaxy centroid and the galaxy
ellipticity. This yields incorrect shape estimates and potentially
correlations in ellipticity at small angular scales. The contamination
may be worst in the deepest weak lensing surveys, with a very high
galaxy number density; galaxy clustering also increases the number of
galaxies that may be affected. Van Waerbeke et al. (2000) found that
pairs separated by less than 10 arcsec in ground-based surveys 
may be affected in this
way. Therefore, surveys with more than 100 galaxies per arcmin$^{2}$
may be seriously contaminated by isophote overlaps.

\sssec{Intrinsic alignment}
An intrinsic correlation of galaxy ellipticities may result from
tidal interactions of close physical pairs. Compression of dark haloes
or transfer of angular momentum during the interaction will modify the
shapes of galaxies (e.g. Catelan, Porciani \& Kamionkowski 2000).
Close multiplets are
then correlated, just as for the weak lensing effect. Since only
close physical pairs are affected, this intrinsic signal is stronger
at small angular scales. It also prevails on shallow surveys, where
the fraction of true physical pairs as compared to non-physical
(projected) is larger. Numerical simulations show that the intrinsic
alignment is one or even two orders of magnitude below the lensing
signal for a mean galaxy redshift of $\langle z\rangle\simeq
1$. However, it dominates for shallow surveys like the SDSS.

The intrinsic alignment can be corrected by a judicious weighting of
galaxy pairs as function of their redshift difference. Close physical
pairs can therefore be downweighted while the relevant lensing signal
between distant pairs is strengthened. Provided the redshift of each
galaxy is known with only a rough accuracy ($\Delta z \simeq 0.1$) ,
the intrinsic alignment can be almost perfectly corrected (King \&
Schneider 2003; Heymans \& Heavens 2003).

\sssec{Intrinsic foreground-background correlation}
The intrinsic foreground-background correlation results from the
coupling between the tidal field responsible for the intrinsic
alignment of close physical pairs and the gravitational distortion
this same field may produce on distant galaxies (Hirata \& Seljak
2004): when the tidal field is strong, its lensing effect is also
strong and both effects produce a distortion of galaxy isophotes. In
contrast with pure intrinsic alignment, the coupling produces a
negative lensing signal and seems to depend on the morphology of
foreground galaxies (Heymans et al. 2006b). It also increases with
redshift, but not in the same way as weak lensing. The intrinsic
foreground-background correlation could contribute up to 10\% of the
lensing signal for $\langle z\rangle\simeq 1$ sources.

When not corrected, the shear-ellipticity coupling may result in an
underestimate of $\sigma_8$ that could be as large as 20\% for some
survey parameters. However, as with intrinsic alignment, one can
handle this coupling using redshift information, provided the accuracy
of redshift estimates is sufficiently good (King 2005). The noise
residual of this correction has however not yet been addressed.

\sssec{Non-linear variance}
Kilbinger \& Schneider (2005) and Semboloni et al. (2006b) have
pointed out that non-Gaussian corrections to cosmic variance in weak
lensing surveys are not negligible. On scales where the nonlinear
regime dominates, below 10~arcmin, the error budget may increase by a
factor 2 to 3.  While non-Gaussian effects do not produce extra
systematics, they significantly affect the estimated accuracies on
dark energy constraints obtained from small angular scales.

\sssec{Non-linear matter distribution}
For most weak lensing surveys that explore cosmological parameters,
the non-linear evolution of the dark matter power spectrum is a
serious limitation to a cosmological interpretation of the lensing
signal on scales below 20 arcmin. Past and current surveys use either
the Peacock--Dodds approximation (Peacock \& Dodds 1996) or the
halo-fit model (Smith et al. 2003). Unfortunately, these analytic
equations are not sufficiently accurate for the analysis of future
cosmic shear surveys with their expected small statistical
errors. Furthermore, no accurate analytic predictions exist for
higher-order correlation functions of the mass distribution, and thus
of the corresponding shear correlations. Numerical ray-tracing
simulations, employing cosmological matter distributions, are
therefore essential for providing accurate predictions; without them,
cosmic shear surveys must either disregard the distortion signal below
$\sim 20$ arcmin or use more complex corrections, like the nulling
tomography techniques (Huterer \& White 2005), at least to derive
constraints on dark energy. Such simulations will require a
substantial effort by modellers and must be an integrated part of any
future large lensing survey. Pure N-body simulations may not suffice
to accurately predict the lensing signal on scales below $\sim 1'$;
here, cooling of the baryons start to play a significant role
(Jing et al. 2006).

% Figure
\begin{figure}
\begin{center}
\includegraphics[width=0.5\hsize,angle=270]{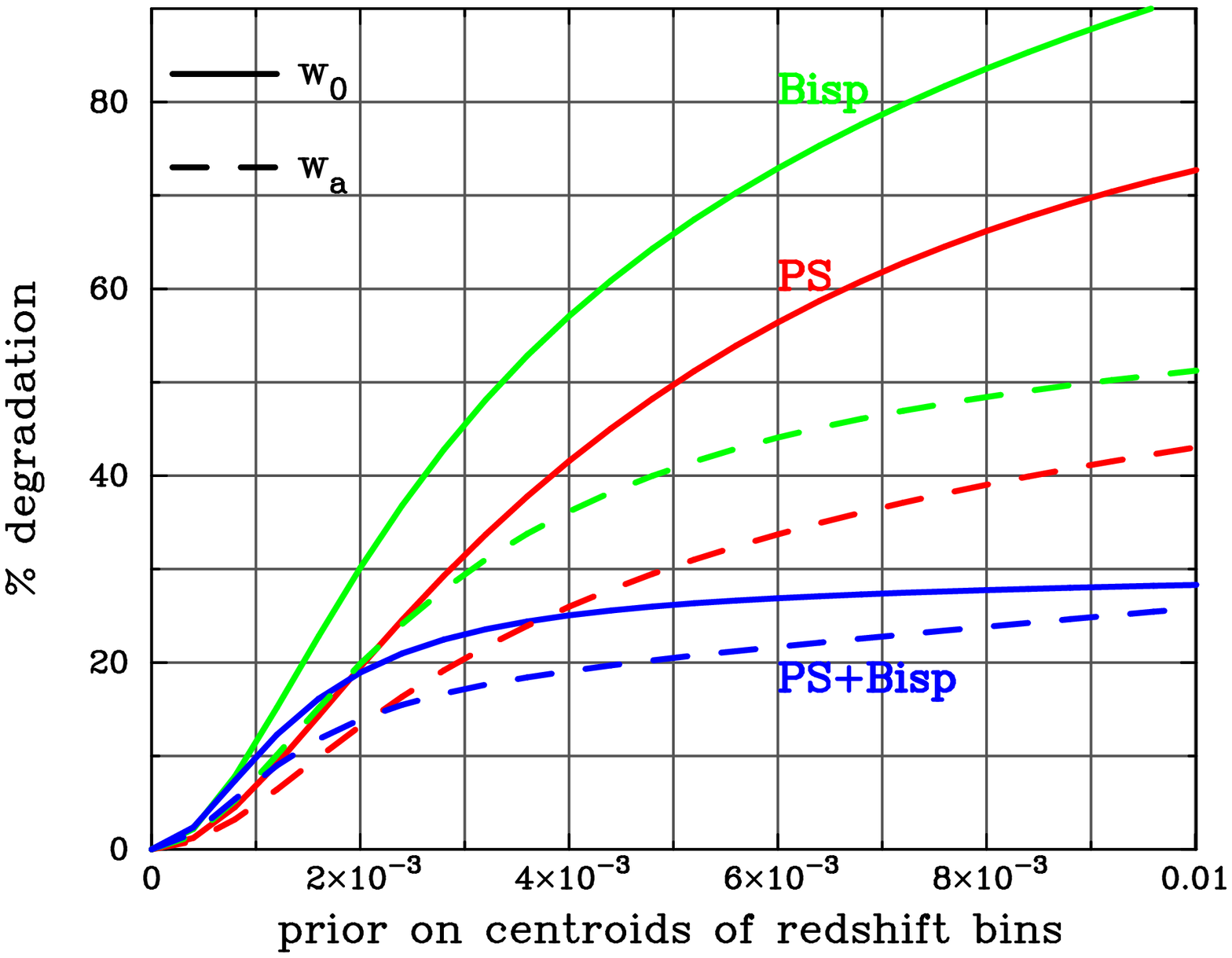}
\caption{Effects of the accuracy of the mean redshift of source
  galaxies in photometric redshift bin. The curves show the degradation
  of the accuracies of dark energy parameters from a bias of the mean
  redshifts of galaxies in $\Delta z_{\rm phot}=0.3$-broad
  bins. `PS' and `Bisp' stand for
  power spectrum and bispectrum, respectively. They refer to information
  derived from the 2-point and the 3-point shear correlation functions.
  The plots show that photometric redshift estimates must have a very small
  bias of below $\delta z\sim 3\times 10^{-3}$ in order not to degrade
  the parameter accuracies by more than $\sim 50\%$ compared to the
  statistical uncertainties of a large-scale survey. Also noticeable
  is the fact that the combination of second- and third-order
  statistics not only increases the accuracies, but can also provide a
  self-calibration of photometric redshift uncertainties, seen by the
  saturation of the combined curve (from Huterer et al.\ 2006). }
\label{fig:hut06}
\end{center}
\end{figure}

\sssec{Redshift distribution}
Without redshift information for the lensed sources, weak lensing
cannot provide reliable information on cosmological parameters, even
for present-day surveys. Tomography, i.e. the variation of the lensing
signal as function of source redshift, and geometry-power spectrum
decoupling techniques will be even more demanding on redshift
accuracy, in particular for galaxies beyond the most distant lens
planes. Furthermore, redshift information on individual galaxies is
required to control systematic effects from intrinsic shape
alignments, shear-shape correlations, and effects of source
clustering.  Therefore, full exploitation of the statistical
power of future weak lensing surveys (particularly with regard to
dark energy parameters) requires that the redshift
properties of the galaxy samples
be very accurately known, with the uncertainties of the mean
and dispersion of galaxy redshifts in redshift slices limited to
$\Delta z/(1+z) \sim 3\times 10^{-3}$ (see Fig.\ref{fig:hut06}).
Since future weak lensing surveys will analyse the shapes of $\gs
10^8$ galaxies, it is impossible to obtain spectra for all galaxies;
instead, photometric redshifts techniques need to be employed (see
Sect.~\ref{sc:phot-z}) as was already done by Heymans et al. (2005).

%%%%%%%%%%%%%%%%%%%%%%%%%%%%%%%%%%%%%%%%%%%%%%%%%%%%%%%%%%%%%%%%%%%%
% Table updated

%
%\begin{center}
\begin{table}
\caption{Wide-field surveys that prioritize weak lensing as their top scientific goal.
We distinguish three classes of project:
on-going; next (funded or probably funded); future (still under discussion).
VIKING complements the KIDS survey with near-infrared bands. The WL and DE acronyms in the last
column stand for weak lensing and dark energy, respectively. 
}
\tiny
%\relscale{1.05}
\begin{center}
%\small
\begin{tabular}{|l|c|c|c|c|c|l|}
 \hline
\topstrut Survey & Telescope/ & Sky & Filters & Depth & Period & Main goals\\
\botstrut  & Instrument& coverage & & & &  \\
 \hline
\multicolumn{7}{c}{\bigstrut{\bf Current surveys}}\\
%\noalign {\phantom{a}}
%\noalign {\bf Current surveys}
%\noalign {\phantom{a}}
\hline
\topstrut Deep Lens& Mayall+ & 7$\times$4 deg$^2$ & BVRz' & R=25. & 2001-2005 & WL
\\
Survey & Blanco& & & & & DE, Clusters\\
 & & & & & & High-z Univ.\\
%ESSENCE& Blanco & 32$\times$0.36 deg$^2$ & V/R/I & R=23.5 & 2002-2005
%&$0.2<z<0.8<$ SNIa  \\
% & 8Kmosaic& & & & & DE\\
CFHTLS Deep & CFHT/ & 4$\times$1 deg$^2$ & ugriz &i$_{AB}$=27 & 2003-2008 &
$0.3<z<1.$ SNIa\\
 & Megacam& & & & & DE\\
 & & & & & & Clusters, P(k)\\
 & & & & & & WL ($z<2.0$)\\
 & & & & & & High-z Univ.\\
 CFHTLS Wide & CFHT/ & 3$\times$50 deg$^2$  & ugriz &i$_{AB}$=24.5 & 2003-
2008 & WL ($z<1$), \\
 &Megacam & & & & & DE, P(k),\\
 & & & & & & Bias\\
SDSS-II  & APO& 250 deg$^2$&  ugriz& r'=22.& 2005-2008&
$0.1<z<0.3<$ SNIa\\
SN Survey  & & & & & & DE\\
 & & & & & & \\
 SUPRIME-33 & SUBARU/ & 33 deg$^2$ &  R & R=26 & 2003-? & WL ($z<1.$), \\
 & Suprime& & & & & DE, P(k), Bias\\
 & & & & & & High-z Univ.\\
 RCS2 & CFHT/ & 1000 deg$^2$  & grz & i$_{AB}\simeq$22.5 & 2003-? & WL
($z<0.6$), \\
 & Megacam& & & & & DE, P(k),\\
 & & & & & & Clusters, Bias\\
 CTIO-LS & CTIO & 12$\times$ 2.5 \ deg$^2$ & R & R=23 & 2002-2006 & WL ($z<0.6$) \\
  & & & & & & \\
 COSMOS & HST/ACS & 1$\times$2 deg$^2$  & I & I$_{AB}$=25.5& 2003-? & WL

($z<1$), \\
 & & & & & & DE, P(k),\\
\botstrut  & & & & & & Clusters, Bias\\
\hline
\multicolumn{7}{c}{\bigstrut{\bf Funded  surveys}}\\
%\noalign {\phantom{a}}
%\noalign {\bf Funded surveys}
%\noalign {\phantom{a}}
\hline
\topstrut  KIDS-Wide & VST/ & 1500 deg$^2$ &  ugriz & i$_{AB}$=22.9& 2006-2009 & WL ($z<0.6$), \\
 & Omegacam& & & & & DE, P(k), Bias\\
 & & & & & & High-z Univ.\\
 UKIDSS-Large & UKIRT/ & 4000 deg$^2$  & YJHK & K=18.4& 2006-2012 & Clusters
\\
 & WFCam& & & & & $z>7$ Univ.\\
 UKIDSS-Deep & UKIRT & 3$\times$10 deg$^2$ & JK & K=21& 2006-2012 & Clusters \\
 & WFCam& & & & & High-z Univ.\\
 UKIDSS-Ultra & UKIRT & 0.77 deg$^2$  & JHK & K=25& 2006-2012 & Gal. Formation
\\
 Deep& WFCam& & & & & \\
WIRCam Deep   & CFHT/ & 4$\times$0.75 deg$^2$ & J/H/K &K$_{AB}$=23.6 & 2005-2008
& High-z Univ.\\
Survey(CFHTLS) & WIRCam& & & & & Clusters, P(k)\\
 VISTA-Wide & VISTA &  5000 deg$^2$  & JHK & K=20.5& 2006-2018 &  \\
 VISTA-Deep & VISTA &  250 deg$^2$  & JHK & K=21.5& 2006-2018 &  \\
 VISTA-VeryDeep & VISTA &  25 deg$^2$  & JHK & K=22.5& 2006-2018 &  \\
% Deep& & & & & & \\
 PanSTARRS & MaunaKea & $\sim$30000 deg$^2$ & giz & i$_{AB}$=24. & 2008-
2012? & WL ($z<0.7$), \\
\botstrut  & TBD& & & & & DE, P(k), Bias\\
\hline
\multicolumn{7}{c}{\bigstrut{\bf Planned surveys}}\\
%\noalign {\phantom{a}}
%\noalign {\bf Planned surveys }
%\noalign {\phantom{a}}
\hline
\topstrut  VIKING & VISTA/ & 1500 deg$^2$ &  zYJHK & i$_{AB}$=22.9& 2007-2010 & WL ($z<0.6$), \\
 & & & & & & DE, P(k), Bias\\
 & & & & & & High-z Univ.\\
 Dark Energy & CTIO & 5000 deg$^2$  & griz &i$_{AB}$=24.5 & 2009-2014 & WL ($z<0.8$), \\
 Survey& DECam& & & & & DE, P(k),\\
 DarkCam & VISTA & $\sim$10,000 deg$^2$ &  ugriz &i$_{AB}$=24. & 2010-2014 &
WL ($z<0.7$), \\
 & & & & & & DE, P(k),\\
 HyperCam & SUBARU/ & $\sim$3500 deg$^2$ &  Vis. & ?& $>$2012? & WL
($z<2$), \\
 & Suprime& & & & & DE, P(k),\\
 SNAP/JDEM  & Space & 100/1000/ deg$^2$  & Vis.+NIR & -& $>$2013 & WL
($z<1.5$), \\
 & & 5000 deg$^2$& & & & DE, P(k),\\
 & & & & & & SNIa, Bias\\
 DUNE & Space & $\sim$20000 deg$^2$ &  ugriz+NIR? & i=25.5 & $\sim$2015? & WL ($z<1$), \\
 & & & & & & SNIa, DE, P(k),\\
 LSST & Ground & 20000 deg$^2$ &  ugrizy & i$_{AB}$=26.5& $>$2014 & WL ($z<2.$), \\
 & TBD& & & & & DE, P(k)\\
\botstrut  Dome-C & SouthPole & ? deg$^2$ &  ? & ?& $\sim$2012? & SNIa, DE \\
 \hline
\end{tabular}
\end{center}
\label{tab:lenssurv}
\end{table}
%\end{center}

\ssec{Future prospects for weak lensing surveys}\label{sc:GL.4}

The key ingredients for next generation weak lensing surveys are
\begin{itemize}
\item
the field-of-view, which determines the statistical accuracy of the
measured shear signal; for precision measurements, many thousands of
square degrees will be required;
\item
the depth of the survey, which determines the redshift up to which the
mass distribution can be probed, but also the number density of
objects;
\item
the number of filter bands, which determines the accuracy of the
redshift information on individual galaxies;
\item
the
accuracy of shear measurements, which is governed by the accuracy of
the PSF correction made to raw data.
\end{itemize}
These parameters are internal to the survey design and to the data
analysis technique. They can be set in advance to design the optimal
survey as a function of intrinsic limitations and the survey goals. In
addition, one needs external ingredients from numerical
simulations. The most important is the non-linear evolution of the dark
matter power spectrum, as well as predictions for higher-order shear
signals. These precision simulations will be a challenge in their own right.

The present-day surveys have focused on rather simple but robust
cosmological analyses based on two-point ellipticity correlation
functions of lensed galaxies. Next generation surveys will go further,
with the goal of deriving $w_0$ and $w_a$ with a 1-5\% and 5-10\%
accuracy, respectively. The goal is then to measure gravitational
distortion as weak as 0.1\% and to use novel techniques to derive
cosmological signatures from different and independent methods,
breaking the intrinsic degeneracies of the simple two-point shear
correlation function. The most promising methods are the
reconstruction of the three-dimensional dark matter power spectrum,
lensing tomography, 3-point and higher-order statistics or the
geometry-power spectrum decoupling analysis, which is much less
dependent on accurate modelling of the large-scale structure.

The ongoing and next generation surveys are summarized in
Table~\ref{tab:lenssurv}.  The CFHTLS Cosmic Shear Survey is an
example of a second generation weak lensing survey, which should
eventually provide $\sigma_8(\Omega_{\rm m})$ with $\simeq 5\%$ accuracy and
$w$ with $\simeq 20-50\%$ accuracy. In addition to its depth, field of
view and image quality, the CFHTLS benefits from its joint DEEP and
WIDE components, plus its photometric redshifts that can be calibrated
using the VVDS and the DEEP2 spectroscopic surveys. When used jointly
with SNLS and WMAP3, one can reasonably expect a gain of accuracy by a
factor 2 to 3 by 2008-2009.

\epsfigsimp{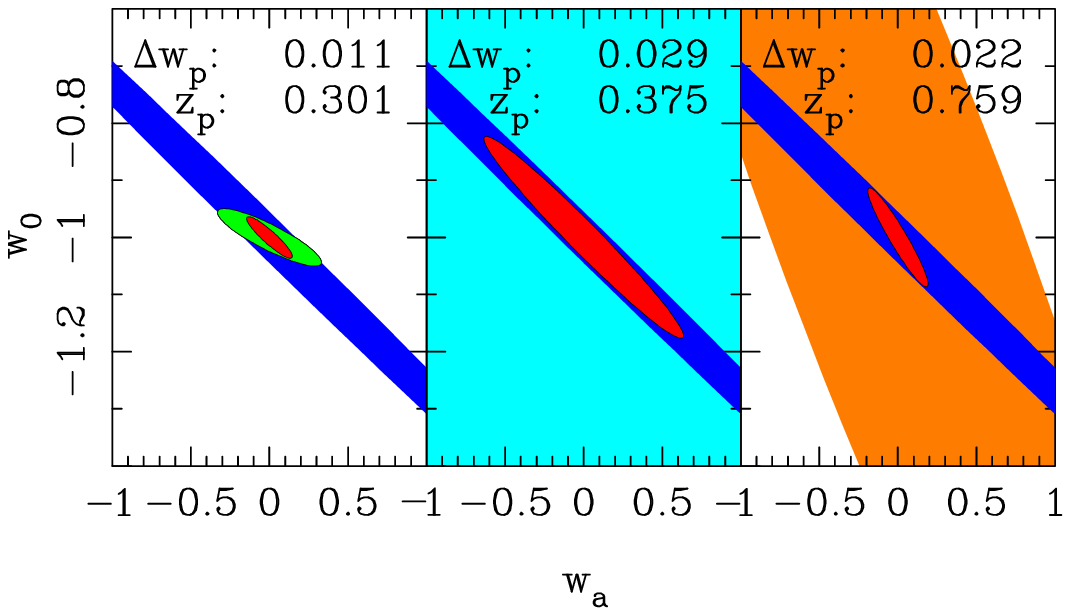}{0.9}
{Comparison of Fisher-matrix uncertainties in
the dark-energy parameters from weak lensing, SNe Ia and
baryon oscillations (left to right). The 3D lensing survey
assumes a space-borne 5-band survey covering 20,000 deg$^2$;
the SNe experiment is as specified in the SNAP proposal;
the baryon oscillation experiment assumes a survey of 2000 deg$^2$,
yielding 2,000,000 galaxies at $z\simeq 1$. The current
parameter $w_0$ and the evolution parameter are shown, and
errors are quoted on $w_p$, which is $w$ at the pivot redshift;
this error is effectively the width of the confidence ellipse
in the narrow direction. In all cases, the plots show
projected errors from
Planck alone, from the selected technique alone, and
the two combined, with marginalization over hidden parameters.
Adapted from Heavens, Kitching \& Taylor (2006).
}
{fig:fishcompare}

Beyond 2008, the ESO joint KIDS+VIKING optical/near-IR imaging survey
looks promising. It will cover at least 1500 deg$^2$ with 4 optical
and 5 near-infrared filters, about two magnitudes deeper than SDSS and
one magnitude shallower than CFHTLS. If the survey could be extended
by a factor 2-3 from its original goal of 1500 deg$^2$, KIDS+VIKING
would be a unique bridge between CFHTLS and generation 3 surveys. By
2011, a joint CFHTLS+KIDS+VIKING weak lensing analysis could easily
explore the equivalent of 3 to 5 redshift bins with good sampling of
each bin. As compared to CFHTLS alone, one can expect a gain in
accuracy by a factor 1.5 (conservative) to 3 (optimistic), depending
on the genuine dependence on $w$ with time (and the yet unknown image
quality of the VST/OmegaCAM). The gain could be qualitatively much
more important if the evolution of $w$ with time could be measured. It
is worth noticing that by 2011, one also expect the first results from
Planck and therefore more stringent constraints from a join analysis
than with WMAP.

PanSTARRS and DES are expected to be
third generation weak lensing surveys. They exceed the
grasp of VST by an order of magnitude, and have much more ambitious
goals in terms of sky coverage and depth. Thus, unless
ESO takes steps now to plan a more powerful facility,
i.e. a DarkCam equivalent, European
facilities will not be competitive for work on weak lensing
surveys from about 2011.

From 2015 and beyond, LSST, DUNE or JDEM/SNAP are designed in order to
make a major step forward in the field. We can envisage nearly
all-sky imaging with resolution below 0.5$''$, and this will push
the measurement of $w$ to 1\% or below (see Fig.~\ref{fig:fishcompare}).
The main technical challenge
of these projects is the image quality. Space missions therefore seem
a logical strategy, in particular with respect to long-term stability
of the instrument, whereas LSST follows an alternative strategy by
using short exposure times on each field. For satisfying the
requirements on photometric redshift estimates, multi-colour surveys
including near-IR photometry are mandatory, with the latter being only
available to the necessary depth from space-based photometry.

The DUNE mission is indeed a joint ground-based and space concept
where the critical shear data would be obtained from space,
exploiting the image quality and stability unique to that environment.
The other optical data needed for photometric
redshifts as well as their spectroscopic calibrations would be gathered
from the ground. This kind of joint mission provides a very attractive
opportunity for synergy between the activities of ESA and ESO. 
This attraction would be strengthened still further if it were possible
to add near-IR photometric channels to DUNE, since the background
in space is greatly reduced at these wavelengths. This would
greatly improve the photometric redshifts in terms of extending to
$z>1$ and improving robustness and precision at $z<1$.
It should be emphasised that the required quantity of data is well beyond
the capabilities of VST and VISTA: what is required is a survey at
least one magnitude deeper than KIDS+VIKING and covering a field of
view $\sim 10$ times larger. The ideal capability would be a combination
of DUNE (one optical filter for shape measurement, two or three
near-infrared bands for photometric redshifts) and a DarkCam equivalent (4
optical filters for photometric redshift), together with a campaign of
deep spectroscopic surveys for photo-z calibration. The enormous
data rate from such a project will be challenging, and close
cooperation with high-energy particle physicists may be very
useful. In addition, the time-scale of this project allows a smooth
transition of expertise from the Planck community to DUNE.

\japsec{Supernovae}\label{sc:SN}

\ssec{Current status}\label{sc:SN.1}

Type Ia Supernovae have been among the most successful cosmological
probes. As shown in Fig.~\ref{fig:sn_hub_near},
%Fig.~\ref{fig:snls1},
their peak luminosities can be calibrated empirically to yield individual
%based on the timescale of the explosion, yielding
relative distances accurate to about 7\% (Phillips et al. 1999;
Goldhaber et al. 2001; Tonry et al. 2003; Riess et al. 2004).
Their use as distance indicators provides a
geometric probe of the expansion of the universe, and provides
the most direct evidence for a recent period of accelerated
expansion, as well as the most accurate direct measurement
of the Hubble constant (see Fig.~\ref{fig:sn_hub_near} and Leibundgut
2001 for a review of SNe~Ia as cosmological distance indicators).
%Fig.~\ref{fig:snls2}).
SNe Ia are thus an important pillar of the concordance model (cf.
Fig.~\ref{fig:wmap_w}) and
are currently the only method that dynamically establishes the
acceleration. They remain one of the most promising tools for further
study of the properties of dark energy.

Table~\ref{tab:sn_surveys} lists the major current and planned
supernova survey projects.  The introduction of rolling searches,
i.e. observations of the same fields nearly continuously (only
interrupted by the bright phases of the moon and the seasonal
observability of the fields) has solved the problem of incomplete
light and colour curves. The success of the CFHT Supernova Legacy
Survey (and to some extent the ESSENCE project) shows that with
four-colour continuous observations, exquisite light curves can be
obtained for large numbers of SNe (Astier et al. 2006).
From Table~\ref{tab:sn_surveys} it
can be noted that the largest ground-based telescopes are used for the
supernova spectroscopy. The current searches are still limited by the
spectroscopic observing time, despite large time allocations at these
facilities (e.g. Lidman et al. 2005; Matheson et al. 2005; Hook et al.
2005; Howell et al. 2005).  These current experiments aim to determine the constant
equation of state parameter $w$ to better than 10\%. This is achieved
by accurately measuring distances to several hundred supernovae at
$z>0.3$ in projects of five years (or more) duration.  At the same
time the local sample has to be increased to provide the local
expansion field and hence the comparison to the more distant
objects. They provide the `anchor' in the supernova Hubble diagram
(cf. Fig.~\ref{fig:sn_hub}).  These samples are increasing steadily
and we can expect to have several hundred supernovae within the local
Hubble flow in the next few years (cf. Table~\ref{tab:sn_surveys}).

%\epsfigsimptwo{snls_04D3fk.eps}{snls_resvss.eps}{0.6}
%{The basic multi-colour standard-candle approach to supernova
%cosmology, illustrated by SNLS 04D3fk. The top panel shows
%the multi-colour light-curve data, as fit by a single scaled model.
%The bottom panel shows the main element of the standard-candle
%adjustment, which is the magnitude residual versus the
%`stretch factor' in timescale of outburst. Open circles
%show SNe at $z>0.8$, as an example of the sort of test
%for consistency that is necessary to establish this as a
%cosmological probe.}
%{fig:snls1}

\epsfigsimp{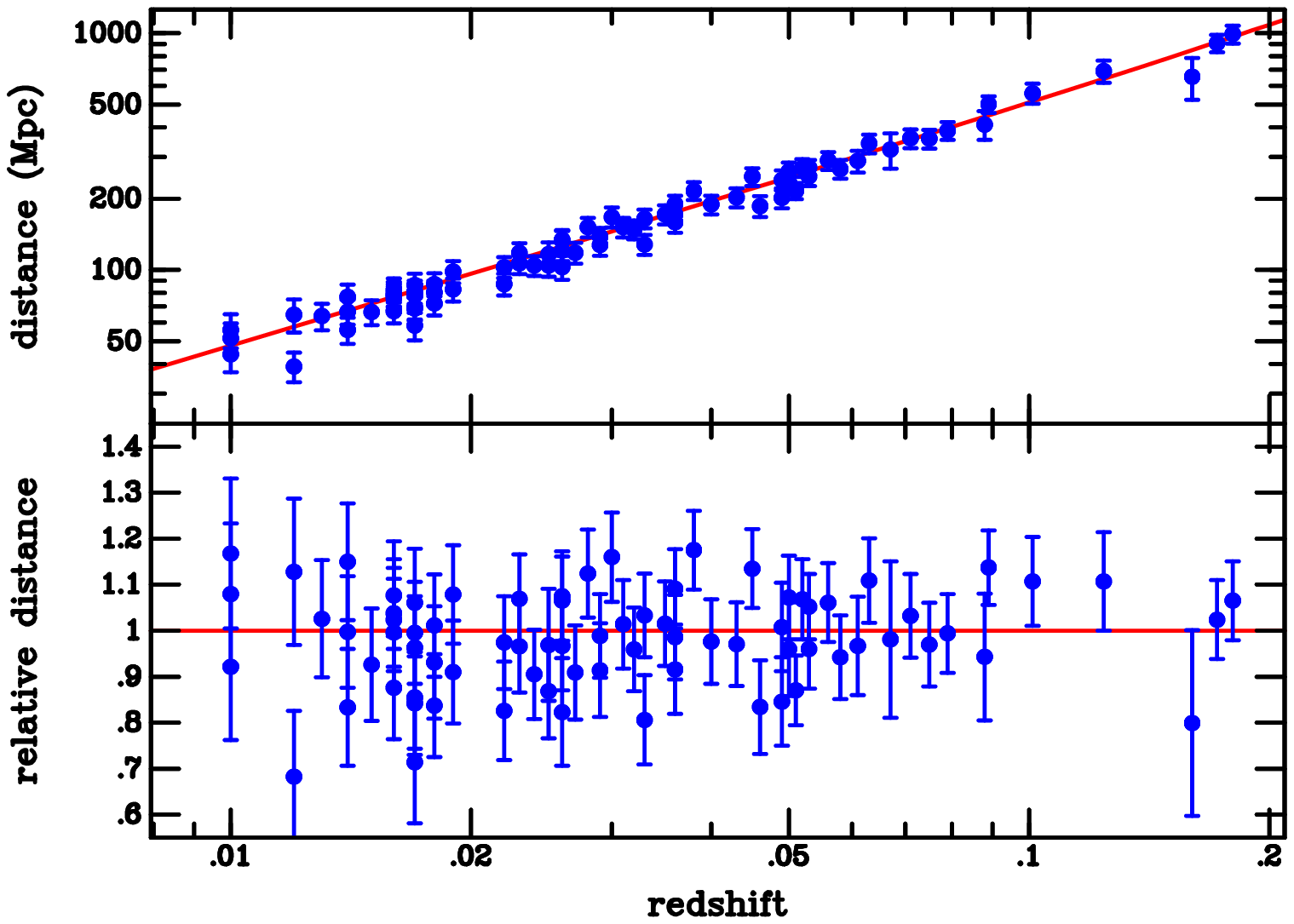}{0.8}
{Hubble diagram for nearby Type Ia supernovae. The small scatter indicates the
exquisite quality of these objects as relative distance indicators. With an
absolute calibration, e.g. Cepheid distances, they also provide the most
accurate value of the Hubble constant so far. Data from Riess et al.
(2004).}
{fig:sn_hub_near}

\epsfigsimp{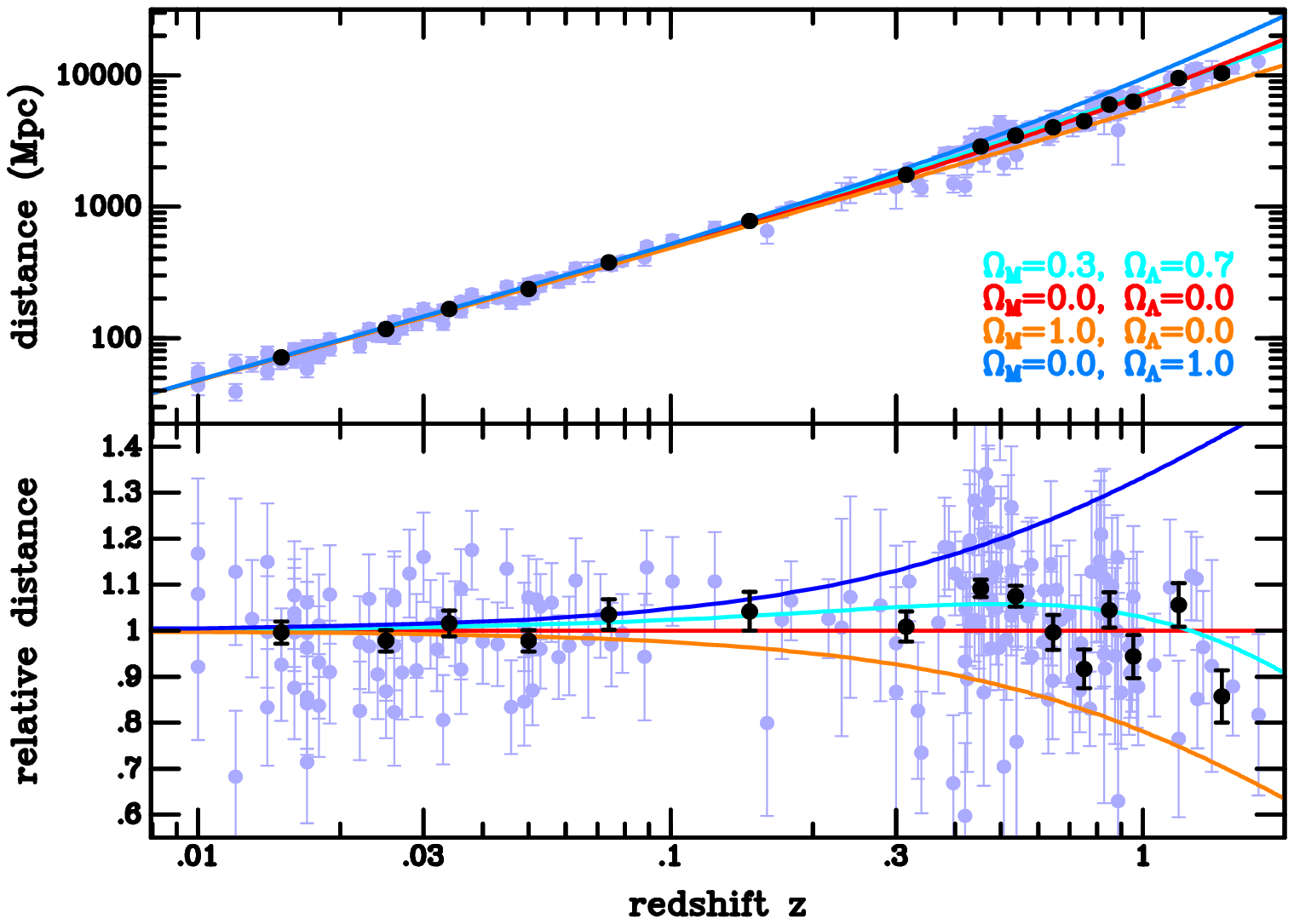}{0.8}
{Type Ia supernovae Hubble diagram. The relative faintness of the
distant supernovae relative to their nearby counterparts is apparent.
A comparison with various cosmological models is made. Data from Riess
et al. (2004).}
{fig:sn_hub}

%\epsfigsimp{snls_hubble.eps}{0.6}
%{The SNLS first-year supernova Hubble diagram, from Astier et al. (2006).
%The lower panel shows the impressively small residuals with
%respect to a concordance model.}
%{fig:snls2}

Thermonuclear supernova explosions are, however, complex events and
the explosion physics as well as the radiation transport is not fully
understood (e.g. Hillebrandt \& Niemeyer 2000).  Rather, supernova distances are based on an empirical
relation that connects the light curve shape to the peak luminosity
of the event (cf.\ Fig.~\ref{fig:sn_normalization}).  Several methods
have been proposed and all give roughly the same results (Phillips et
al. 1999; Goldhaber et al. 2001; Riess et al. 1996; Jha 2002; Guy et al.
2005).

\epsfigsimp{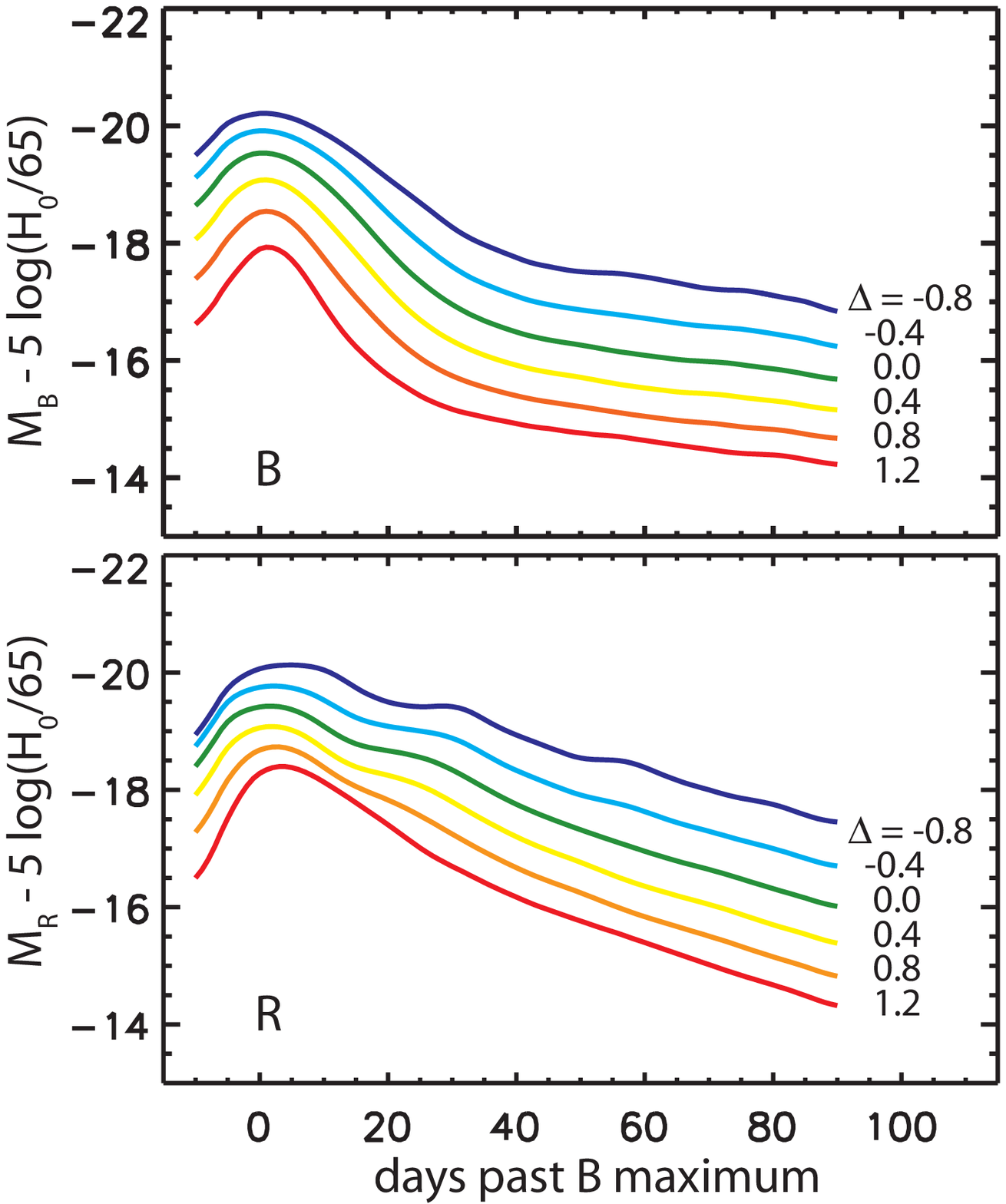}{0.75}
{The basic multi-colour approach to supernova
cosmology. The relation of brighter supernovae with slower light curve
evolution in several filter bands is demonstrated. This correlation is
used to normalize the maximum luminosity of Type Ia supernovae. Figure
adapted from Jha (2002).}
{fig:sn_normalization}

An additional difficulty is the correction for host extinction
suffered by the supernova light. This correction is mostly based on
the observed colour, but there are variations in the intrinsic colours
of supernovae. There are now clear indications that the standard
extinction law for the solar neighbourhood is not applicable in
external galaxies, where the supernovae are observed (e.g. Krisciunas et
al. 2000; Elias-Rosa et al. 2006; Astier et al. 2006).  A possible
solution to both the above problems may be the near-infrared.  Recent studies
have shown that nearby Type Ia supernovae are indeed nearly standard
candles in the rest-frame JHK bands (Krisciunas et al. 2004). Together with the reduced
extinction at these wavelengths, the infrared is a promising route for
improved supernova distances. Current projects are trying to sidestep
the problem by either concentrating on supernovae in elliptical
galaxies, i.e. galaxies with limited extinction, and to improve the
distances to local supernovae through IR observations of Cepheid
stars. This should yield a further improved value of the Hubble
constant. An IR-optimised space telescope would provide a unique
chance to sharpen the cosmological distances from supernovae by limiting uncertainties
in extinction correction in late-type host galaxies and permitting the
use of rest-frame bands for galaxies at higher redshift than present-day samples --
although clearly we will not be able to probe beyond 1$\,\mu$m in the rest frame.

The main systematic uncertainties at the moment are the unknown cause
of the variety of Type Ia supernovae, possible evolution
of the supernova luminosity with redshift (or age of the progenitor
star), the extinction in the host galaxies (there are indications
that the average extinction of the supernovae is different from the
local extinction law in the solar neighbourhood), any intergalactic
extinction, K-corrections and even the definition of photometric
systems. Another uncertainty at the moment is the accuracy with which
the normalization of the peak luminosity can be achieved. The light
curve shape and colour correction remain mysterious, but ongoing
intense modelling efforts may soon lead to substantial progress in our
understanding (e.g. Blinnikov et al. 2006). Increasing the statistics for SNe will be useful to

\blobb decrease the statistical error and reduce the intrinsic error,

\blobb give information on the statistical properties and dispersion
of the SN,

\blobb and to investigate evolution effects. With the current
error estimates in present day and next generation
survey strategies, SN surveys will probably be
limited by systematics, with 50 SNe per redshift bin (statistical error
lower than  $\simeq 2\%$). This is achievable in a few years with currently
planned projects (see Table \ref{tab:sn_surveys}). Correlated errors
across redshift bins will further bias the measured value of $w_a$
(see below).  Supernova observations at high redshifts,
$z\gs 1.5$, are best suited to yielding an
order of magnitude improvement is the accuracy
with which $w_a$ can be measured.

\begin{table}
\caption{Current and planned cosmological Supernova surveys.}
\begin{center}
\tiny
%\relscale{1.07}
\begin{tabular}{|l|c|c|c|c|c|c|l|}
\hline
\topstrut Survey & Telescope/ & Sky & Filters & \# SNe & Spectroscopy & Period & Main goals\\
\botstrut  & Instrument& coverage & & & & &  \\
\hline
\multicolumn{8}{c}{\bigstrut{\bf Low-z searches; $\bf z<0.1$}}\\
%\noalign {\phantom{a}}
%\noalign {\bf Low-z searches; $\bf z<0.1$}
%\noalign {\phantom{a}}
\hline
\topstrut LOTOSS & KAIT (70cm) & Northern & BVRI & $>$200 & Lick/Keck &
1992- & discover and follow \\
 & & Hemisphere & & & & & nearby SNe \\
European Supernova & 2m and 4m & -- & UBVRIZJHK &
$\sim$20 & various 4m & 2002-2006 & SN Ia physics, \\
Collaboration & & & & & & & early epochs \\
Supernova Factory & NEAT (1.2m) & Northern & BVRI & 300 &
SNIFS & 2002- & establish local \\
 & & Hemisphere & & & & & Hubble diagram, \\
\botstrut  & & & & & & & study systematics \\
\hline
\multicolumn{8}{c}{\bigstrut{\bf Intermediate-z searches; $\bf 0.1<z<0.5$}}\\
%\noalign {\phantom{a}}
%\noalign {\bf Intermediate-z searches; $\bf 0.1<z<0.5$}
%\noalign {\phantom{a}}
\hline
\topstrut Carnegie Supernova & 1m,2.5m,6.5m & -- & UBVRIYJH & $\sim$250 & Dupont/ & 2004--2009 & all SN types, \\
Project & & & & & Magellan & & I Hubble diagram \\
SDSS II & Sloan 2.4m & 250 deg$^2$ & u'g'r'i'z' & $>$500 & various 4m & 2005--2008 &
fill in redshift gap \\
\botstrut & & & & & & & $0.1<z<0.3$ \\
\hline
\multicolumn{8}{c}{\bigstrut{\bf High-z searches; $\bf z>0.5$}}\\
%\noalign {\phantom{a}}
%\noalign {\bf High-z searches; $\bf z>0.5$}
%\noalign {\phantom{a}}
\hline
\topstrut Supernova & CTIO 4m & - & RI & $\sim$100 & Keck/Gemini/
& 1990--2000 & established \\
Cosmology Project & & & & & VLT & & acceleration \\
High-z Supernova & CTIO 4m & - & RI & $\sim$100 &
Keck/Gemini/ & 1995--2001 & established \\
Search Team & & & & & VLT & & acceleration \\
Higher-z Supernova & HST & GOODS Fields & RIz & 17 & Keck/Gemini/
& 2002--2004 & $z>1$ SNe \\
Search & & & & & VLT/Magellan & & \\
ESSENCE & CTIO 4m & 36$\times$0.36 deg$^2$ & RI & $\sim$200 & Keck/Gemini/
& 2001--2007 & $w$ to 10\% \\
 & & & & & VLT/Magellan & & \\
SNLS (within the CFHT & CFHT & 4$\times$1 deg$^2$ & ugriz & $\sim$700 & Keck/Gemini/ & 2003-2008 & $w$ to 7\% \\
Legacy Survey)  & & & & & VLT/Magellan & & \\
Accelerating and & HST & - & griz & $\sim$20 & Keck/Gemini
& 2005-- & distant SNe in \\
dustfree & & & & & Gemini & & elliptical galaxies \\
PAENS/SHOES & HST & -- & JHK & $\sim$15 & ??? & 2006-- & distant SNe \\
PanSTARRS-4 & 4$\times$1.5m & $\sim$10,000 deg$^2$ & BVrIz & thousands & ???? & 2011? & \\
Dark Energy Survey & CTIO 4m & 10,000 deg$^2$ & griz & thousands & ???? &
2010--2015 & \\
%ALPACA & & & & & & & \\
LSST & 7.5m & $>$20000 deg$^2$ & ugriz & thousands & -- & $>$2014 & \\
JDEM/SNAP/DUNE & space & $>$10,000 deg$^2$ & optical/IR & $\sim$2000 & onboard & $>$2015 & space missions\\
\botstrut JEDI/DESTINY & & & & & & & \\
\hline
\end{tabular}
\end{center}
\label{tab:sn_surveys}
\end{table}

Due to the degenerate dependence on
cosmological parameters of the luminosity distance over limited redshift
ranges, the supernova results need to be compared with independent constraints
on the cosmological parameters. Constraints on the
dark matter density $\Omega_{\rm m}$ from LSS are nearly orthogonal to the
likelihood contours from supernovae in the $\Omega_{\rm v}$ vs.
$\Omega_{\rm m}$ plane.
This is true for the determination of a constant $w$
as well as for a measurement of $w_a$, and an accurate determination
of the matter density is crucial for meaningful limits on $w$.

For a time-variable equation of state parameter the redshift range has
to be extended beyond $z>1$ to provide the sufficient leverage (e.g.
Linder \& Huterer 2003). Currently, the
sampling in redshift and the accuracy with which the most distant
objects can be observed are not sufficient to constraint $w_a$ and new
surveys are being proposed. At the same time ways to sharpen the
supernovae as distance indicators and to reduced the intrinsic scatter
will need to be found. In addition, systematic effects will need to be
controlled even more tightly than in the current projects.

\ssec{Systematic uncertainties}\label{sc:SN.2}

\sssec{Contamination}
All current surveys make extensive use of the largest existing
telescopes for spectroscopy. It is critical to exclude any
non-Type Ia events. In particular, some Type Ib/c supernovae have
light curves and colours similar to the Type Ia supernovae. The
spectroscopic identification and classification has been very
successful so far, but only single-epoch spectroscopy has typically
been obtained. With thousands of supernovae from the future surveys,
spectroscopic classification may become impractical. It will have to
be demonstrated that tightly sampled light and colour curves can
distinguish between the different supernova types (e.g. Sullivan et al. 2006).

\sssec{Photometry}
Current supernova photometry is good to about 2 to 5\%. The main
limitations are the colour and atmospheric extinction corrections. The
colour corrections from standard stars to the object in question, if applied,
inherently assume a black-body spectral energy distribution. The
non-thermal spectrum of Type Ia supernovae leads to small errors in
the photometry. Spectroscopy can typically help (Stritzinger et al.
2002), but is not feasible
for very large samples. Additional problems are the variations in
mirror reflectivity, filter transmission curves and detector
sensitivity, which need to be monitored closely. Light curve fitting
typically alleviates the problem somewhat as the peak brightness can
be determined based on several individual light curve points and hence
is accurate to 2\% and better, so systematic residuals may be
more critical than random errors.

\sssec{Malmquist bias}
Brightness limited samples suffer from boundary effects.
Intrinsically more luminous objects can be observed to larger
distances. This means that the most distant sample bins are dominated
by the most luminous events. A small intrinsic scatter limits the
effects of Malmquist bias. The small scatter in the normalized peak
luminosity is not sufficient to eliminate the bias, but there is hope
that the normalization works even for limited samples.  In principle
the Malmquist bias can be corrected by using the
brightness distribution of galaxies that are within
the redshift range where the survey is complete.

\sssec{Luminosity normalization} The methods with which this is
achieved are quite varied. The light curve shape is the most commonly
used technique, but there are other proposals including colour evolution and
spectral line strengths.  The successful normalization has been
demonstrated and appears reliable (cf.
Fig.~\ref{fig:sn_hub_near}). The peak luminosity is related to the
amount of $^{56}$Ni synthesised in the explosions (Arnett 1982,
Stritzinger \& Leibundgut 2005), but the physics
that drives the isotope composition remains unsolved.
The corrections are typically derived for local samples and then
applied to the distant objects. This intrinsically assumes no
evolution in the normalization, something very difficult to check
(but comparing low- and high-redshift samples may test
whether there is any evolution of the normalization). Alternatively, the
comparison with infrared peak luminosities, which appear to be much more
uniform, could test the different normalization methods.

\sssec{Local expansion field}
If the local expansion field is not uniform, but distorted through
some large-scale flow, the local
supernovae will not fairly represent the Hubble flow. The influence of
the Virgo cluster is typically avoided by choosing objects with
redshifts $z>0.01$.
There have been claims for a deviation at larger
recession velocities (a `Hubble bubble' to $z\simeq 0.027$; Zehavi et
al. 1998). Even very
small deviations could result in a mis-interpretation of the
cosmological result.
Future supernova samples will allow us to examine the local expansion
field in great detail, determine the Hubble constant in velocity
shells and resolve this problem. Infrared supernova observations would
be particularly useful as the distance determinations are more
accurate.

\sssec{K-corrections}
K-corrections for the distant supernovae are essential and critical for an accurate
determination of the rest-frame luminosity of the objects (e.g. Nugent
et al. 2002). Since the spectral
appearance of supernovae changes dramatically during their evolution,
the K-corrections are time-dependent and small errors in phase can
introduce additional scatter. Of course, the K-corrections are also
correlated with the (intrinsic) supernova colour, and incorrect
determination of the extinction towards the supernova can introduce
additional errors.
Fundamentally, the K-correction also requires precise photometric calibrations
because the spectral energy distribution of
the comparison stars must be known with high accuracy.  Current projects are
now re-measuring the SEDs of Sirius and Vega to make sure that the
K-corrections do not introduce additional uncertainties.

\sssec{Extinction}
Galactic extinction is fairly well understood and can be corrected
with reasonable accuracy (most projects use the Schlegel et al. 1998
maps). Small variations in
different directions could introduce an additional scatter.
Much more serious, however, is the lack of knowledge of the extinction
law in other galaxies. Currently, the absorption in the host galaxy is
determined by the reddening of the supernova. The reddening law is
known to have different dependencies on the size of the dominant dust
particles. From infrared observations of supernovae it has been
deduced that the reddening law in the Galaxy is not applicable to
other galaxies. This remains one of the major error sources in
supernova cosmology. If there is a systematic change with redshift,
this would further introduce a systematic error.

\sssec{Evolution}
Evolution is probably the most difficult systematic uncertainty to
tackle. The lack of our understanding of the progenitor evolution
leading to a Type Ia supernova (e.g. Livio 2000) and the missing pieces in the explosion
physics make this uncertainty the most difficult to constrain. Sample
evolution must be considered here as well. If supernovae can come from
different progenitor channels and one of them has a longer gestation
time then the samples at high redshifts will be dominated by a
different population than the one observed in the local (and older)
universe. Only careful observations and comparison of objects at all
redshifts can lead to a better understanding.  Normalization of
the luminosity distribution is likely to be different for each population,
which may also produce a bias.

Comparison of supernova properties in individual redshift bins will
become an important tool for large samples. Light curve shapes, line
velocities and spectroscopy will be essential tools for these
comparisons and the use of space-based telescopes (for low-resolution
spectroscopy) and extremely large ground-based telescopes will be
necessary. On the other hand, environment-induced evolution does
not require space spectroscopy from space, and can
better be addressed by splitting the samples according to host galaxy
types rather than in redshift bins.

Improved progenitor and explosion models will also provide clues on
possible evolutionary effects. The progress in our understanding of
the progenitor evolution (e.g. Hamuy et al. 2003; Stritzinger et al.
2006), the explosion physics (R\" opke et al. 2006) and the radiation
transport (Blinnikov et al. 2006) have been considerable recently and a tighter connection
with the observations is providing first indications
of what evolutionary effects could become important
(e.g. metallicity of progenitor star). The modelling
effort will have to be maintained.

\ssec{Future applications of supernovae to dark energy}\label{sc:SN.3}
The previous section lists many systematics that will have
to be overcome in future cosmological applications of SNe.
But this does not mean that the technique compares poorly
with alternative methods; rather, the length of the list indicates
the maturity of the field, and is the result of more than a decade
of careful study. There is thus every reason to expect that
supernovae are capable of measuring the properties of dark
energy and its evolution. From our present understanding of
structure formation, it is likely that dark energy became dominant after
redshift $\sim 1$. The critical redshift range where the role of
dark energy is increasingly important is therefore
$0\ls z\ls 2$.
Extending SNe samples to redshifts $z>1$ will therefore expand our
knowledge of the critical transition between the matter-dominated and
dark energy-dominated periods and will provide tighter constraints
on dark-energy evolution with lookback time. This requires observations
from space, as demonstrated by the successful SNe searches with HST
(Strolger et al. 2004, Riess et al. 2004). The sky
brightness from the ground prevents the current telescopes to reach
sensitivities required for supernovae at such high redshifts.
Only an extremely large ground-based telescope will be able to
obtain spectroscopy of these distant supernovae in an efficient manner.
Space-based telescopes will be able to obtain low-resolution
spectroscopy. The real issue here are the wide-field searches, which
appear to be best done from space.

In general, results are referred to a fiducial $\Lambda$CDM model with
$w_0= -1$ and $w_a = 0$. A prior on $\Omega_{\rm m}$ is applied with an
error of 0.01 for the calculations below. This is already a very small
uncertainty on $\Omega_{\rm m}$. To discriminate among various theoretical
models, a stringent precision at the level of 0.02 magnitudes at large
$z$ is needed. This translates in the fit to a tiny variation on the
evolution of the parameter $w$. Covering a large redshift range is not
only mandatory to discriminate between theoretical interpretations but
also to control various systematic effects.

In the following, we will quantify the expected statistical errors for
a large number of SNe and emphasise the level of control of
systematic errors required to match the residual intrinsic limitation.
It is important to stress that, even if systematics can be controlled,
all next-generation surveys must in addition be backed up
with a better calibrated sample of nearby SNe.

\sssec{The intrinsic dispersion of supernovae}
Supernovae have an intrinsic dispersion of 0.12-0.15 in magnitude at
maximum (e.g. Jha 2002), which constitutes the dominant error today. It can be treated
as a statistical error, thus increasing statistics will reduce it to a
negligible level. Indeed, if $\sigma (m) = 0.15/N^{1/2}$, using a
typical sampling of 50 SNe
per redshift bin would reduce the magnitude scatter to 0.02. Therefore, for
large future surveys systematic errors of $>$2\% would dominate the result.

\sssec{Statistical and systematic limitation of SN surveys}
The uncertainty in the determination of the model parameters is
dominated today by the intrinsic uncertainty in the luminosity at
maximum. The current projects are designed to match
the systematic uncertainty with the expected intrinsic variation in the
peak luminosity of the supernovae. Increasing the statistics will not
improve on the current results, if the systematics cannot be improved.
As it is difficult to characterise these types of effects, we have
studied what will be the impact of various types of errors on the
observed brightness and estimated the control needed to let its impact
decrease well below the statistical error. In doing so, we also assume
that the systematic uncertainties do not introduce an intrinsic offset,
but can be treated statistically. This is a somewhat naive approach, but
it allows us to make an estimate on the required improvements in the
experiments. Each systematic uncertainty needs to be examined
individually for possible offsets as well.

The uncertainty in the luminosity distance introduced in the
cosmological fit can be written for a redshift bin
in the form $\sigma (m) = (0.15^2/N + \Sigma \delta
m_i^2)^{1/2}$ where N is the number of SNe per redshift bin, 0.15
is the currently measured intrinsic dispersion, and the various
$\delta m_i$ are systematic errors that add quadratically with the
statistical error. The error $\delta m_i$ on the magnitude can be
correlated between redshift bins.

\sssec{Statistical limitation and the effect of systematics}
At the required statistical level, any unexpected bias in the magnitude
measurement of more than $\delta m=0.02$, which evolves with redshift, will
mimic a cosmological evolution. Note that the relative error with
redshift is important. An absolute error in all bins will only modify
the normalization, but will not introduce a cosmological bias.

The sources of systematics are reviewed in Table~\ref{tab:sn_sys}; it is
difficult today to estimate the future limitation of each. The table
summarises the factors presently identified and the level needed to control
the measurement at the required 2\% level. The current values are
estimates or upper limits from the literature. The `needed' value is
the current SNAP estimate. The statistical values indicate the level
accepted per supernova if this error is corrected in a redshift bin.

\begin{table}
\caption{
Magnitude errors estimation: `Current' is the estimation or limit
in the literature, the `Needed' accuracy is the level of control, if we
assume a large statistic ($>$50~SNe/bin) extracted from a future
experiment (e.g. SNAP). When the
error can be corrected, the requirement on one SN can be relaxed but a
correction in the bin should be applied. The error on the correction is
quoted as systematic. Note that cross-filter calibration and the
K-correction are mixed together, although they are different sources of systematics.
Gravitational lensing refers to flux amplification, which increases the variance
on SN light curves.  }
\begin{center}
\begin{tabular}{|l|c|c|l|l|}
\hline
\topstrut \botstrut Error source & Current & Needed & Stat & Correlated/bin \\
\hline
\topstrut {\bf Experimental} & & & & \\
Data reduction & 2-3\% & $<0.5$\% & -- & Yes \\
(cross filter calibration, Kcorr) & & & & \\
Malmquist Bias & 4\% & $<0.5$\% & -- & \\
\botstrut Non-SN Ia Contamination & 5\% & $\sim$0\% & -- & \\
\hline
\topstrut {\bf Astrophysical} & & & & \\
Galactic extinction & 4\% & $\sim0.5$\% & 10-20\% & Yes \\
{\small (Galactic extinction model)} & & & & \\
Host galaxy extinction & 10\% & $\sim1$\% & 10--20\% & Yes \\
{\small (reddening dust evolution)} & & & & \\
SN Type Ia Evolution & 10\% & $\sim1$\% & 10--20\% & Yes ?? \\
{\small progenitor mass,} & & & & \\
{\small metallicity (C-O)} & & & & \\
{\small change of explosion} & & & & \\
{\small (Amount of nickel} & & & & \\
{\small synthesised)} & & & & \\
{\small radiation transport} & & & & \\
{\small Host galaxy properties} & & & & \\
Gravitational Lensing & 6\% & $\sim0.5$\% & $\sim10$\% & Yes \\
\botstrut Absolute Calibration & & 1\% & & No \\
\hline
\topstrut \botstrut TOTAL & 17\% & 2\% & & \\
\hline
\end{tabular}
\label{tab:sn_sys}
\end{center}
\end{table}

\japsec{The intergalactic medium}\label{sc:Lya}

The principal means of learning about the initial conditions for
cosmological structure formation has been the study of large-scale
fluctuations using the CMB, galaxy clustering and gravitational
lensing. But all these methods lack the ability to probe the very
smallest fluctuations, those responsible for the generation of
galaxies and the first stars.  The best way of overcoming this limit
is to use the Ly$\alpha$ forest measured from the absorption of light
in quasar spectra. These absorptions are caused by Ly$\alpha$
transitions caused by neutral hydrogen along the line of sight and are
thus measuring its one-dimensional (1D) distribution. Through
theoretical modelling this can be related to the distribution of dark
matter along the line of sight.

This is done in two steps. First, neutral hydrogen is generated
through electron-proton recombination and destroyed by ionizing
photons that fill the universe. This allows one to relate the neutral
hydrogen density to the gas density. Second, gas and dark matter trace
each other on large scales since they have the same equations of
motion, while on small scales gas pressure counters gravity and
prevents gas from clustering, in contrast to the dark matter. These
processes are well understood and allow one to relate the information
in the Ly$\alpha$ forest to the dark matter distribution. The process is
mildly nonlinear and requires hydrodynamic simulations for proper
calibration.

In this way, the Ly$\alpha$ forest can measure the distribution of
dark matter at smaller scales and higher redshifts than other tracers,
which are strongly affected by nonlinear evolution. The critical
scales are around $1-40\hompc$, where we have little other information
on the matter power spectrum and the Ly$\alpha$ forest provides
powerful constraints on the nature and composition of dark matter.
Because of the long lever arm between the CMB measurements at the
largest scales and the Ly$\alpha$ forest measurements at the smallest
scales, the Ly$\alpha$ forest measurements are also particularly
valuable in constraining the overall spectral shape of primordial
density fluctuations. The \lya\ measurements help discriminating
between different flavours of inflationary models of the early
universe and allows us to measure the effect of non-zero neutrino
masses on the matter power spectrum.

Another example where the Ly$\alpha$ forest can play an important role
are the dark matter models that erase structure on small scales, such as
warm dark matter (WDM). These models are tightly limited
since no evidence of any suppression of power is seen even on the
smallest scales observed by Ly$\alpha$ forest.

\ssec{Method and systematic uncertainties}\label{sc:Lya.1}

\sssec{Method}
The \lya\ forest blueward of the \lya\ emission line in QSO spectra is
produced by the inhomogeneous distribution of the warm ($\sim 10^4$ K)
and photoionized intergalactic medium (IGM) along the line-of-sight.
The opacity fluctuations in the spectra trace the gravitational
clustering of the matter distribution in the quasi-linear regime on
scales larger than the Jeans length of the photoionized IGM.

The relevant physical processes can be readily modelled in
hydro-dynamical simulations.  The physics of a photoionized IGM that
traces the dark matter distribution is, however, sufficiently simple
that considerable insight can be gained from analytical modelling of
the IGM opacity, based on the so-called fluctuating Gunn--Peterson
approximation, which neglects the effect of peculiar velocities and
thermal broadening.  In this approximation and with the assumption of
a power-law relation between density and temperature, the optical
depth for \lya\ scattering is related to the overdensity of baryons
$\delta_{\rm b}(= \Delta \rho_{\rm b}/\rho_{\rm b})$ as

\begin{eqnarray}
\tau(z) &\propto & [1+\delta_{\rm b}(z)]^2 \, T^{-0.7}(z)
~=~ \displaystyle{{\cal A}}(z) \, [1+\delta_{\rm b}(z)]^\beta~,
\label{eqtau} \\
{\rm with}\qquad \displaystyle{{\cal A}}(z) & \simeq & 0.43
\left(\frac{1+z}{3.5}\right)^6
\left(\frac{\Omega_{\rm b} h^2}{0.02}\right)^2
\left(\frac{T}{6000\;{\rm K}}\right)^{-0.7}  \nonumber \\
\times & & \left(\frac{h}{0.65}\right)^{-1}
\left(\frac{H(z)/H_0}{3.68}\right)^{-1}
\left(\frac{\Gamma_{\rm HI}}
{1.5\times 10^{-12}\;{\rm s}^{-1}}\right)^{-1}\; ,\nonumber
\end{eqnarray}
where $T$ is an effective average temperature, $\beta \equiv 2 - 0.7\,
(\gamma-1)$ is in the range $1.6-1.8$, and $\Gamma_{\rm HI}$ is the HI
photoionization rate.  For a quantitative
analysis, however, full hydro-dynamical simulations are needed that
properly take into account the non-linear evolution of the IGM and its
thermal state.

Equation (\ref{eqtau}) shows how the observed flux $F=\exp{(-\tau)}$
depends on the underlying local gas density $\rho_{\rm b}$, which in
turn is simply related to the dark matter density, at least on large
scales where the baryonic pressure can be neglected. Statistical
properties of the flux distribution, such as the flux power spectrum,
are thus closely related to the statistical properties of the
underlying matter density field. Note, however, that inferring the
amplitude of matter fluctuations from fluctuations of the flux level
requires knowledge of the mean flux level, which has also to be
determined empirically from \lya\ absorption spectra.

\sssec{QSO absorption spectroscopy}
The \lya\ forest absorption in QSO spectra is superposed on the
intrinsic emission spectrum of the QSO, which normally is itself a
superposition of a broad continuum and numerous strong and weak
emission lines. The absorption by the \lya\ forest is due to the warm
photoionized intergalactic medium, and the width of typical absorption
features ranges from about $10-100 \kms$. Fully resolving these
features requires high resolution spectroscopy to be performed with
Echelle-type spectrographs like HIRES@KECK and UVES@VLT.  At the
smallest scales the information on the spatial clustering of the
matter distribution is masked by the superposed metal lines and the
thermal broadening in real and redshift space.  The spatial scales
utilised in studies of the matter power spectrum with \lya\ forest
data correspond to velocity scales ranging from about $50\kms$ to
about $2000\kms$.

\sssec{Systematic Uncertainties}
As discussed in detail by Viel, Haehnelt \& Springel (2004) and
McDonald et al. (2005) there is a wide range of
systematic uncertainties in estimates of the matter power spectrum
from the \lya\ flux power spectrum.  These uncertainties fall broadly
into five categories:
\begin{itemize}
\item{necessary corrections due to finite S/N, continuum
normalization, associated metal absorption and damped absorption systems;}
\item{residual uncertainty in estimates of the mean flux level;}
\item{uncertainty of the thermal state of the IGM;}
\item{limited ability to make accurate   predictions of the flux power
spectrum for a large parameter space;}
\item{limited ability to accurately model other physical processes that
potentially affect the flux power spectrum, such as galactic winds and
temperature fluctuations.}
\end{itemize}

%------------------------------------------------------------------------------
% Table 7
%------------------------------------------------------------------
\begin{table}
\caption{Error budget for the determination of the
rms fluctuation amplitude of the matter density field reproduced from
Viel, Haehnelt \& Springel (2004).}
\begin{center}
\label{tab7}
\smallskip
\begin{tabular}{|lc|l|}
\hline
%\noalign{\smallskip}
\topstrut statistical error&\ \ \ \ \ &4\%\\
systematic errors&\ \ \ \ \ &\\
\ \ \ $\tau_{\rm eff}(z=2.125) = 0.17\pm 0.02$& &8\%\\
\ \ \ $\tau_{\rm eff}(z=2.72) = 0.305\pm 0.030$& &7\%\\
\ \ \ $\gamma=1.3 \pm 0.3$&&4\%\\
\ \ \ $T_{0} = 15,000\,{\rm K} \pm 10,000\,{\rm K} $&&3\%\\
\ \ \ method&&5\%\\
\ \ \ numerical simulations&&8\% (?)\\
\botstrut \ \ \ further systematic errors&&5\% (?)\\
%\noalign{\smallskip}
\hline
\end{tabular}
\end{center}
\end{table}

Table~\ref{tab7} reproduces the estimates of the contribution of most
of these uncertainties to the total error budget given by
Viel, Haehnelt \& Springel (2004). On some of these, progress has been made the
estimates may be slightly pessimistic.

\ssec{Major cosmological results from the IGM}\label{sc:Lya.2}

\sssec{Measurements of the \lya\ flux power spectrum}
The pioneering measurements of the \lya\ flux power spectrum by Croft
et al. (2002) made use of a mixture of 23 high-resolution Echelle
(HIRES) and 30 lower-resolution spectra (LRIS).  Further measurements
using high-resolution spectra were obtained by McDonald et al. (2000,
8 HIRES spectra) and Kim et al. (2004a, 2004b, 23 UVES spectra).  In 2004,
McDonald et al. published the flux power spectrum obtained from 3035
SDSS QSO spectra with $z>2.2$, nearly two orders of magnitude larger
than previous samples.  This large data set allows one to determine
the amplitude of the flux power spectrum to better than 1\%.  The
spectra of the SDSS sample have much lower resolution and lower S/N
than the high-resolution spectra.  However, due to the large number
and wide redshift coverage the flux power spectrum could be measured
for a wider redshift range, and the statistical errors are smaller than
those of the previous measurements.  Note, however, that significant
corrections have been applied because of the rather low S/N and
resolution.  There is good agreement between the different
measurements of the \lya\ flux power spectrum to within the quoted
errors.

\sssec{Measurements of the matter power spectrum}
On the scales probed by the \lya\ forest the matter density field is
mildly non-linear. This together with the fact that the relation
between density and flux is non-linear makes the use of realistically
simulated mock spectra mandatory for a quantitative analysis of the
inferred underlying matter power spectrum. The first serious quantitative
attempts were made by Croft et al. (2002) who developed an `effective
bias' method that uses mock spectra from numerical simulations to
define the scale-dependent relation between linear matter and flux
power spectrum:
\begin{equation}
P_{\rm flux}(k) = b^2(k) P_{\rm mat}^{\rm lin}(k).
\end{equation}
The matter power spectrum is then inferred by assuming that this
relation is indeed linear (i.e.~that the dependence of $b(k)$ on
$P_{\rm mat}(k)$ can be neglected).  As mentioned above, in practice
the flux power spectrum is not only sensitive to the underlying matter
power spectrum but amongst other things also to the assumed mean flux
level.  Initially there was some controversy what values to assume for
the mean flux level but recent studies have reached a consensus in
this question. This is the main reason that there is now not only good
agreement between the observed flux power spectra but also between the
matter power spectrum inferred by different authors.

Theoretical analysis of this flux power spectrum shows that at the
pivot point close to $k=1\hompc$ in comoving coordinates for standard
cosmological parameters, the power spectrum amplitude is determined to
about 15\% and the slope to about 0.05.  This is an accuracy
comparable to that achieved by WMAP.  More importantly, it is at a
much smaller scale, so combining the two leads to a significant
improvement in the constraints on primordial power spectrum shape over
what can be achieved from each data set individually.

The results for the shape of the power spectrum are consistent with
the concordance $\Lambda$CDM model.  A joint analysis with the first
year data of WMAP gave a rather large amplitude $\sigma_8 \sim 0.9$
(Viel, Haehnelt \& Springel 2004; McDonald et al. 2005).  The third
column of Table~\ref{tab:lya} shows the inferred fluctuation amplitude
in terms of $\sigma_{\rm 8} $ for a range of recent studies.  Not all
of these results are completely independent but there is a wide range
of data sets and analysis methods involved. The excellent agreement is
encouraging.  Note, however, that the CMB data has significant weight
in the joint analysis, and with the new WMAP three year results the
amplitude of a joint analysis has become somewhat lower $\sigma_8 \sim
0.8$ (Viel, Haehnelt \& Lewis 2006a; see Table~\ref{tab:lya} and
Fig.~\ref{fig:wmap_lya} for details).

\epsfigsimptwo{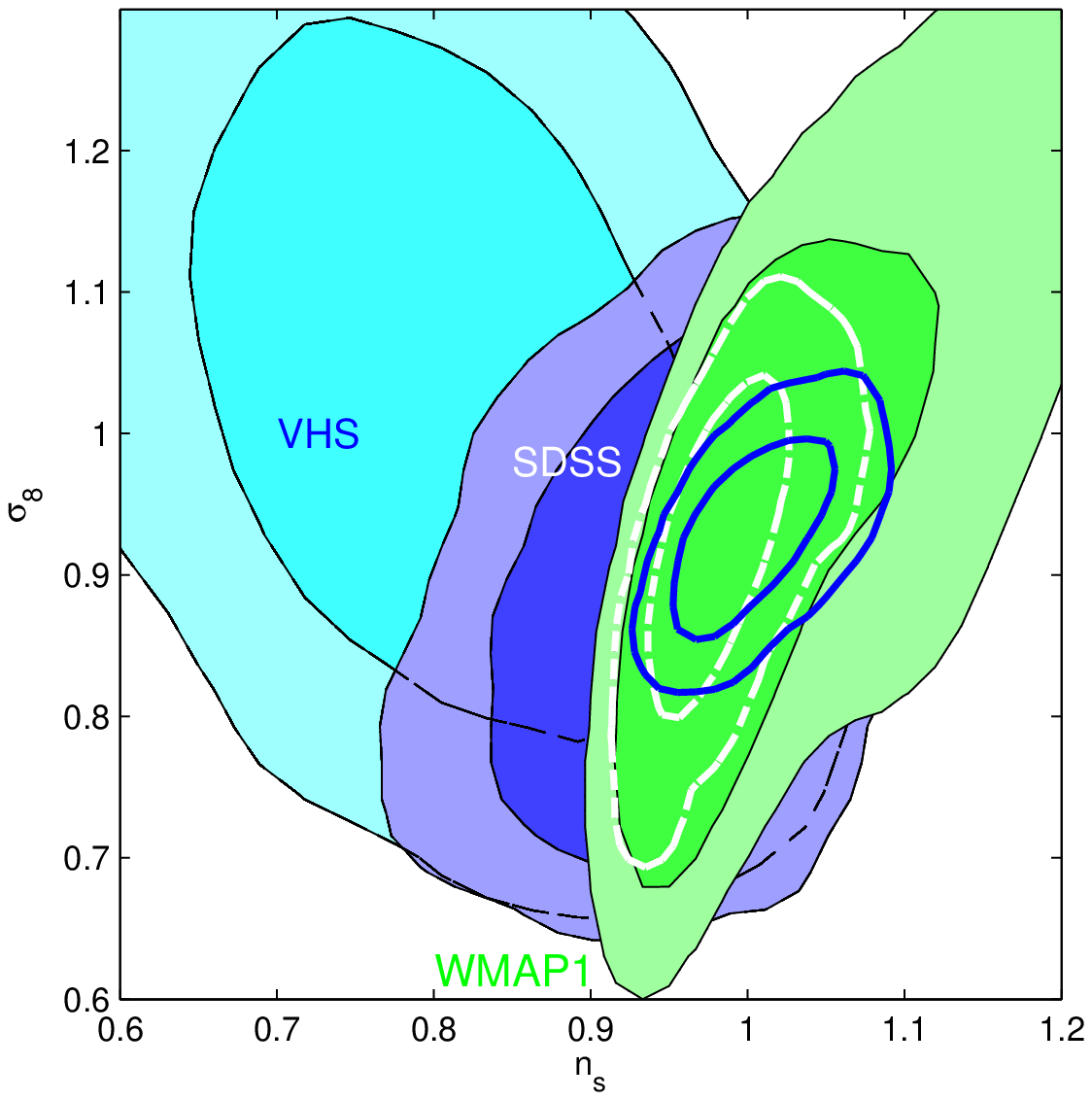}{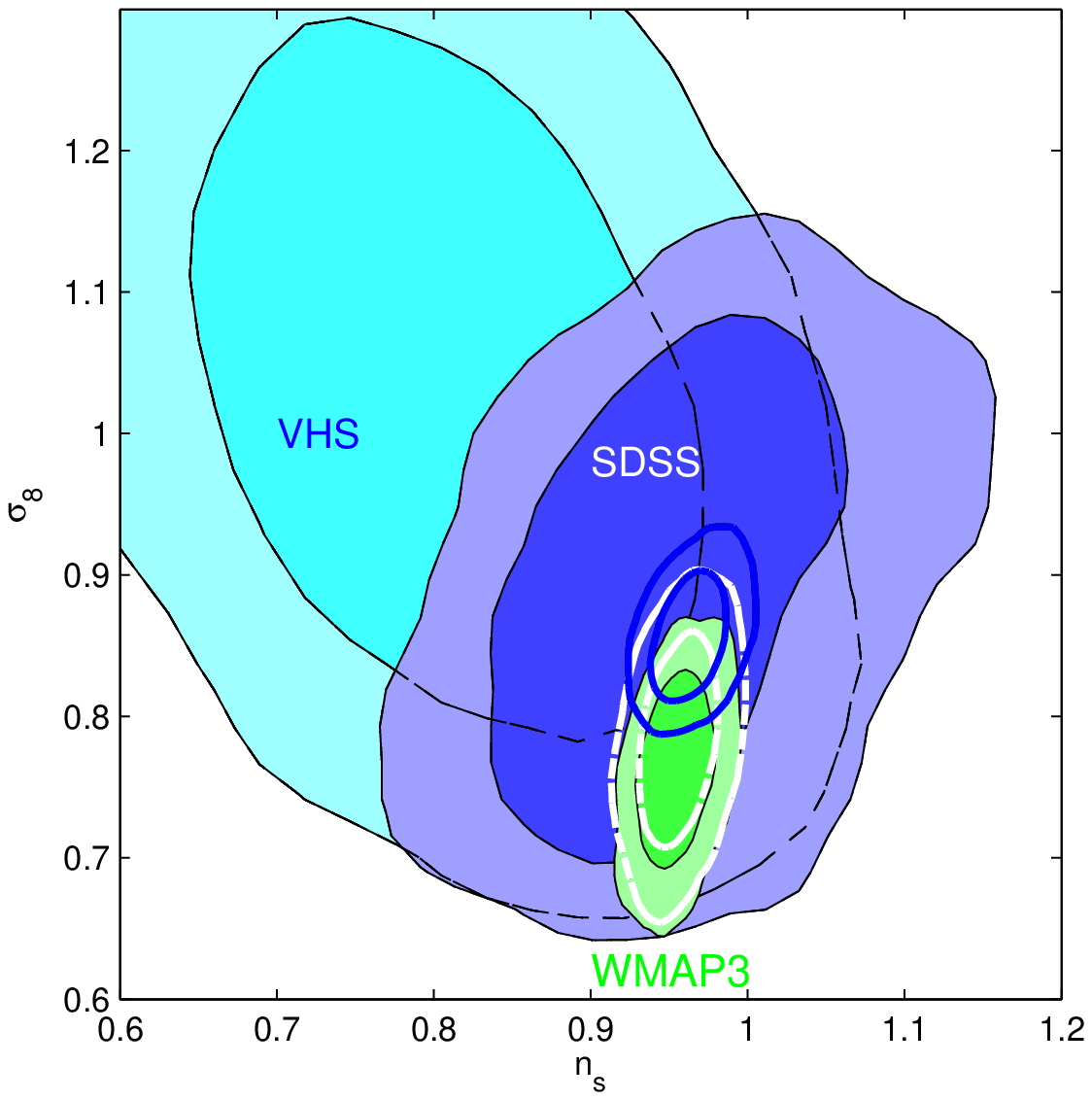}{0.55} {$1$ and $2\sigma$
likelihoods for $\sigma_8$ and $n_s$ marginalized over all other
parameters.  {\it Upper panel:} Constraints are for WMAP1 only
(green), the high-resolution \lya\ forest data analysed by Viel,
Haehnelt \& Springel (2004a; cyan) and the SDSS \lya\ forest data of
McDonald et al. (2005; blue). The thick dashed white contours refer
to WMAP1 + VHS, while the solid blue contours are for WMAP1 +
SDSS. {\it Lower panel:} As in the upper panel, but for the WMAP3 data
set. Figure reproduced from Viel, Haehnelt \& Lewis (2006a).}
{fig:wmap_lya}

%------------------------------------------------------------------------------
% Table 1
%------------------------------------------------------------------
\begin{table}
\caption{Parameter uncertainties including \lya\ data.}
\label{tab:lya}
\begin{center}
\begin{tabular}{|l|l|l|l|}
\hline
%\noalign{\smallskip}
\topstrut \botstrut &&${\rm\sigma_8}$ & $ n_{\rm s}$ \\
%\noalign{\smallskip}
\hline
%\noalign{\smallskip}
\topstrut Spergel et al. (2003)& WMAP1 only &$0.9\pm 0.1$& $0.99\pm 0.04$\\
Spergel et al. (2003)& WMAP1+\lya &--& $0.96\pm 0.02$\\
Viel et al. (2004a)& COBE+\lya &$0.93\pm 0.1$& $1.01\pm 0.06$\\
Viel et al. (2004b)& WMAP1+\lya&$0.94\pm 0.08$& $0.99\pm 0.03$\\
Desjaques \& Nusser (2005)& \lya+priors &$0.90\pm 0.05$&--\\
Tytler et al. (2004)& \lya+priors & $0.90$&--\\
McDonald et al. (2005)& \lya+priors&$0.85\pm 0.06$& $0.94\pm 0.05$\\
Seljak et al. (2005b)& WMAP1+\lya &$0.89\pm 0.03$& $0.99\pm 0.03$\\
&+other data&&\\
Viel \& Haehnelt (2006)&\lya+priors&$0.91\pm 0.07$&$0.95\pm 0.04$\\
\botstrut Zaroubi et al. (2006)& 2dF+HST+\lya& $0.92\pm 0.04$&--\\
%\noalign{\smallskip}
\hline
%\noalign{\smallskip}
\topstrut Viel et al. (2006a)& WMAP3+\lya & $0.80\pm 0.04$&$0.96\pm0.01$\\
Seljak et al. (2006)& WMAP3+\lya&   $0.85\pm
0.02$&$0.965\pm 0.012$\\
\botstrut &+other data&&\\
\hline
%\noalign{\smallskip}
\end{tabular}
\end{center}
\end{table}
%---------------------------------------------------------------------------

\sssec{Constraints on inflationary parameters}
The real strength of the \lya\ forest data comes into play when it is
combined with data probing matter distribution on larger scales.  Due
to the long lever arm provided by a joint measurements of the power
spectrum from CMB and \lya\ forest data it is possible to constrain
the overall shape of the power spectrum and in particular the spectral
index of the primordial density fluctuations $n_s$. The WMAP team used
the results of Croft et al. (2002) to perform such an analysis and
claimed evidence for a moderate tilt $n_s = 0.96\pm 0.02$ and/or a
running of the spectral index also at the $2\sigma$ level. Such an
analysis has obviously very interesting implications for constraining
the wide range of inflationary models.  Viel, Haehnelt \& Springel (2004) were able
to confirm the suggestion of Seljak, McDonald \& Makarov (2003) that the claim of
the WMAP team was due to a too low mean flux level assumed by Croft et
al.  As demonstrated by the fourth column in Table~\ref{tab:lya} there
is consensus that a combined analysis of the WMAP first year results
and the \lya\ forest data is consistent with a Harrison--Zeldovich
spectrum ($n_s=1$) and no running of the spectral index. This situation has
changed with the release of the three year results of WMAP. The lower
Thomson optical depth has lead to a significant reduction of the
fluctuation amplitude $\sigma_8$, and there is now a very significant
detection of a spectral tilt (Fig.~\ref{fig:wmap_lya}, Table~\ref{tab:lya}).

As a further example of the power of these small-scale constraints,
adding Ly$\alpha$ forest information reduces the errors on the running
of the spectral index, one of the most important tests of inflation,
by a factor of 3 relative to the case without it.

\epsfigsimptwo{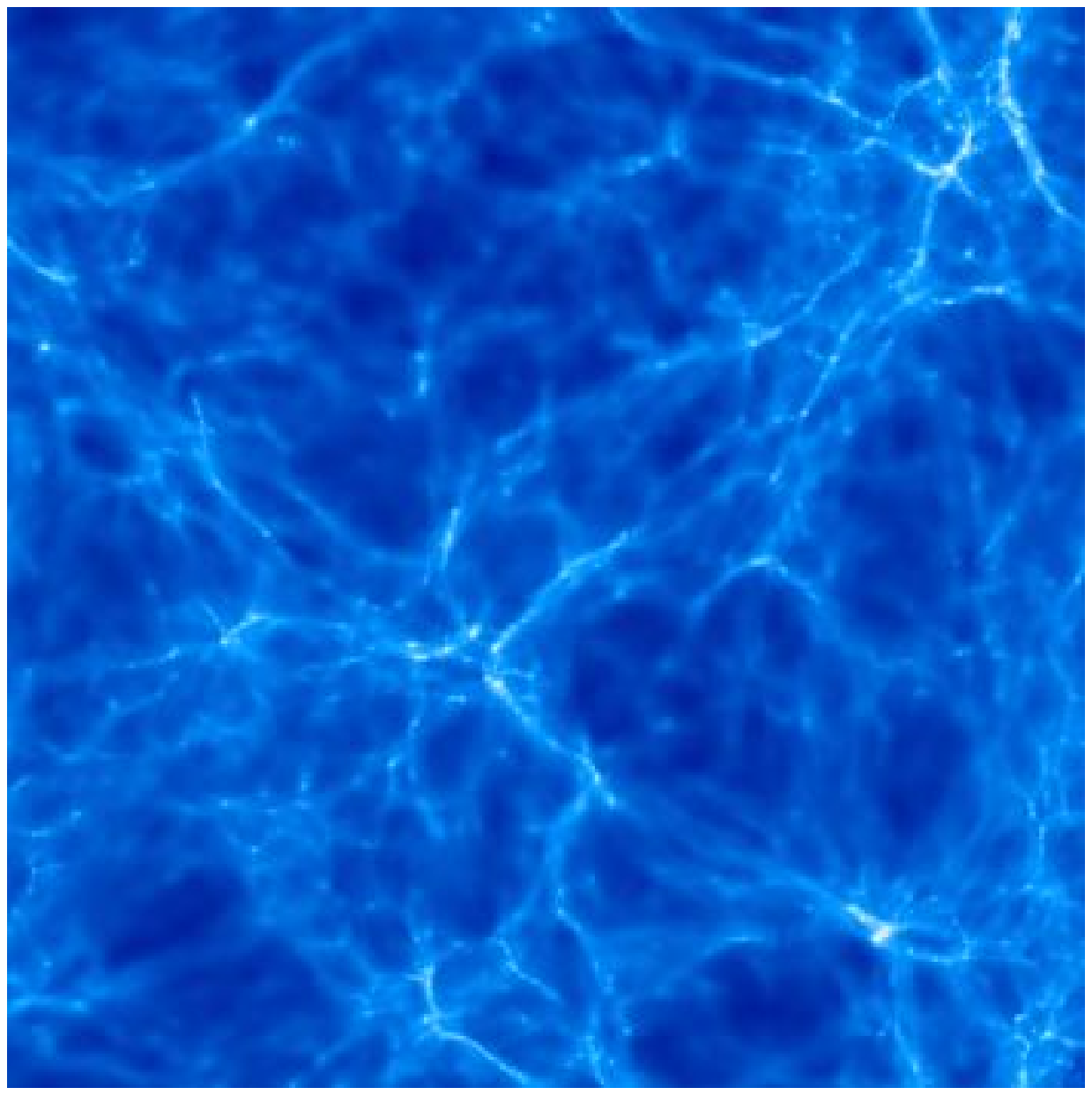}{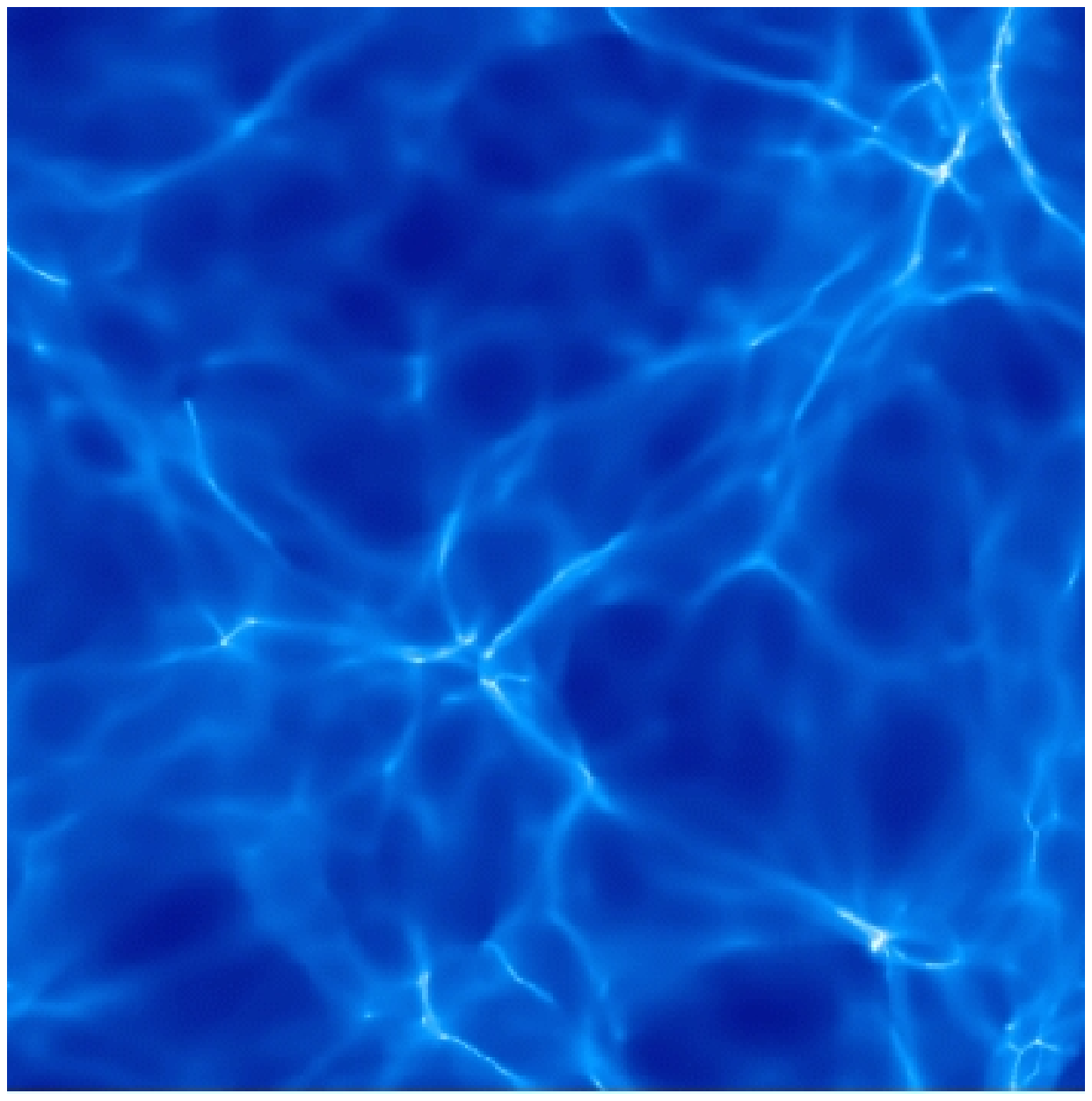}{0.4}
{Numerical simulations of the matter distribution
of a cold/warm (left/right) dark matter model. The
reduced small scale structure due to the free-streaming of the warm
dark matter ($m_{\rm WDM}$ = 0.5\,keV) is clearly seen and well probed
by the Ly$\alpha$ forest. Simulations were run with the numerical
hydrodynamical  code Gadget-II on the COSMOS computer in Cambridge
(box-size 30$h^{-1}$ Mpc).}
{fig:lyasim}

\sssec{Constraints on warm dark matter}
As demonstrated in Fig.~\ref{fig:lyasim}, the matter distribution is
sensitive to effects of free-streaming of dark matter particles on
scales probed by the \lya\ forest data.  Interesting constraints can
be obtained for a range of putative warm dark matter particles.  The
lack of the signature of a cut-off in the matter power spectrum due to
free-streaming constrains the mass of warm dark matter particle to be
$> 2$ keV for early decoupled thermal relics and $>10$ keV for sterile
neutrinos (Viel et al. 2005, 2006b; Seljak et al. 2006).  It also
limits the mass of gravitinos in models for supersymmetric gauge
mediation to be $<16$~eV. (Viel et al. 2005; all limits are
$2\sigma$). Together with the upper limits from the X-ray background,
sterile neutrinos can be ruled out as a source of (warm) dark matter.

\sssec{Constraints on neutrino masses}
The \lya\ forest data in combination with other data currently gives
also the tightest upper limit on neutrino masses.  In a combined
analysis of the WMAP 1st year data, the SDSS galaxy power spectrum and
the SDSS \lya\ forest data, Seljak et al.  (2005b) obtained $\sum
{m_{\nu}} < 0.42$ eV ($2\sigma$) for the case of three neutrino
families (see Elgaroy \& Lahav 2005 for a review of astrophysical
measurements of neutrino masses).  Seljak et al. (2006) performed a
similar analysis for the WMAP three year data and further tightened
the upper limit for the sum of the neutrino masses to $\sum {m_{\nu}}
< 0.17$ eV. The (moderate) discrepancy of the amplitude of the matter
power spectrum inferred from the SDSS \lya\ forest and the WMAP year
three data mean this result merits further scrutiny.

\ssec{Future Prospects} \label{sc:Lya.3}

\sssec{The \lya\ flux power spectrum}
Current estimates of the \lya\ flux power spectrum utilised data sets
taken for different purposes. The smallest scale resolved by the
high-resolution spectra is about a factor 20 smaller than can be used
to infer the dark matter power spectrum.  This means that many more
photons have been collected than necessary for this purpose even
though the additional information on smaller scales has been
invaluable for an understanding of systematic uncertainties. The SDSS
spectra mark the opposite regime. Substantial corrections due to
insufficient resolution and S/N are required.  A future spectroscopic
survey with intermediate resolution of $R\sim 5000$ and sufficient S/N
tailored to a determination of the \lya\ flux power spectrum could
reduce the statistical errors of the flux power spectrum significantly
(probably by about a factor three with a moderate effort). There is
also room for a more moderate improvement for samples of high- and
low-resolution spectra of the kind currently studied.  It does not appear
worthwhile to take more high resolution spectra for this purpose but
there is a significant number of spectra in observatory archives that
have not yet been used.  SDSS already has a data sample 3-4 times
larger than previous analysis waiting to be analyzed. This could lead
to a reduction of errors if increase in statistical power reduces the
degeneracies between the cosmological parameters and astrophysical
parameters.

Looking further into the future, a qualitative jump in the statistical
power of the data will be achieved once the surface density of the
quasars is sufficiently high that most of the information will come
from the cross-correlation between the spectra rather than
auto-correlation itself.  This would significantly increase the
statistical power by using 3D information instead of current 1D
information.  There are plans to have surveys with two orders of
magnitude higher surface density of measured quasar spectra than
currently available, which would be ideal for this purpose.

\sssec{Matter power spectrum and systematic uncertainties}
The error budget of measurements of the matter power spectrum from
\lya\ forest data is currently limited by the lack of knowledge of
several astrophysical processes.  There are a number of these that
affect the spectrum of Ly$\alpha$ forest fluctuations, such as the UV
background (responsible for maintaining the ionizing balance between
ionizations and recombinations), temperature-density relation for the
gas, and gas filtering length determined by time averaged gas
pressure. In addition, there may be possible additional physical
effects affecting the fluctuation spectrum, such as the galactic winds
blowing out of the galaxy and the effects of fluctuations in UV
background. The current analyses account for many of these effects by
parameterizing them within a model with free parameters that are then
marginalized over. In the current SDSS analysis this leads to a 5-fold
increase in the error on primordial amplitude and slope of the power
spectrum.

The prospects for a better understanding of some of the systematic
uncertainties are good. Improved determinations of the temperature of
the IGM should reduce the uncertainty due to the thermal state of the
IGM.  Another major uncertainty is due to the numerical modelling,
where it is computationally demanding to demonstrate convergence on
small and large scales simultaneously; the constant increase in
computing power will obviously help here.  Systematic comparisons
between simulations with different methods and codes are under way
that should also reduce systematic uncertainties due to the numerical
simulations.  Furthermore a wide redshift coverage of the data appears
to be able to break the degeneracy between assumed mean flux level and
amplitude of the matter power spectrum. This merits further
investigation and may remove another important systematic
uncertainty. Significant improvements in the measurements of the
matter power spectrum can therefore be expected.

The remaining concern for the current and future prospect of
Ly$\alpha$ forest is whether there are additional physical effects
that have not yet been considered and that may affect the results at
the level larger than the current or future statistical errors. To
some extent the same concern can be raised for all of the tracers,
which is why cross-checks between the different data sets and other
types of analysis are needed to make conclusions robust. It is often
argued that Ly$\alpha$ forest is more problematic because of many
astrophysical effects that could in principle influence it, contrary
to cleaner probes such as weak lensing or CMB. However, it should be
remembered that not that long ago CMB was not viewed as clean at all
and that only through detailed investigations of many possible effects
were we able to conclude that the contaminants were subdominant
relative to the signal.

\sssec{Inflationary parameters}
Improved measurements of the matter power spectrum from \lya\ forest
data should reduce the errors for the spectral index $n_s$ and its
running. This should further shrink the parameter space of viable
inflationary models.

\sssec{Warm dark matter}
The main obstacle to improving the  lower limits on the mass of dark
matter particles  are the thermal cut-off of the flux power spectrum
and the increasing somewhat uncertain contribution from
metal lines to the flux power spectrum at small scales.

\sssec{Neutrino masses}
Further improved measurements of the matter power spectrum will
certainly also lead to tighter  upper limits on neutrino masses.
Together with the  improved constraints from CMB measurements, which
should break some of the  degeneracies (mainly with the
spectral index $n_s$, and the Thomson optical depth $\tau$), it may
be possible to close the gap to the lower limit for the neutrino masses
from neutrino oscillations. An actual measurement of neutrino
masses appears possible.

\sssec{Beyond power spectra}
Just as with other large-scale structure data (Sections \ref{sc:LSS}
and \ref{sc:GL}), there is much to be gained by going beyond the
power spectrum analysis. This is particularly important for the
Ly$\alpha$ forest, since there is a strong degeneracy between the UV
  background intensity (affecting the mean level of absorption) and the
amplitude of the fluctuation spectrum.  This degeneracy can be broken
by adding non-Gaussian information, such as the bispectrum or the
one-point distribution function. Within the existing data the expected
improvement should be a factor of 3-4, and if this remains true even
with the larger data set already available, but not yet analyzed, then
one can expect an order of magnitude better determination of the
amplitude of the power spectrum on megaparsec scales in the near
future.  This could lead to a comparable improvement in some of the
cosmological parameters such as running of the spectral index and the
mass of warm dark matter candidate.

A further advantage of a more general approach lies in demonstrating
robustness of the conclusions.  The same physical model with a small
number of parameters must be able to explain not only the observed
power spectrum as a function of scale and redshift, but also all the
higher-order correlations, cross-correlations between the close lines
of sight, cross-correlations between galaxies and Ly$\alpha$ forest
etc.  By applying these tests we should be able to test the
reliability of the models, refine them and converge on the correct
cosmological model.

\japsec{Variability of fundamental constants}

\ssec{Background}

Fundamental constants play an important role in our understanding of
nature. Any variation of these constants would, for example,
question the validity of the General Relativity, since it
explicitly assumes laws of physics to be independent of time.
Astronomical observations provide a unique way to probe
any such variability. In particular, cosmological variations in the
fine-structure constant $\alpha$ and proton-electron mass ratio
$\mu=m_p/m_e$ can be probed through precise velocity measurements of
metallic and molecular absorption lines from intervening gas clouds
seen in spectra of distant quasars.  The fine-structure constant
characterises the strength of the interaction between electrons and
photons and is defined as $\alpha = e^2/4\pi \epsilon_0 \hbar c$.  It
is a dimensionless quantity and its value is $\alpha
=1/137.03599911(46)$. The proton-electron mass ratio $\mu$ is related
to the quantum chromodynamic and quantum electrodynamic scale
variations (Flambaum et al. 2004).

In the 1930s, Milne and Dirac first suggested the time variation of
fundamental physical constants and in particular of Newton's
gravitational constant. In the subsequent decades the possible
variability of fundamental physical constants has occupied quite a
prominent place in theoretical physics and many modern theories
predict variations of various fundamental constants.  String theory
predicts seven as yet undiscovered dimensions of space.  In such
higher-dimensional theories, the constants' values are defined in the
full higher-dimensional space, and the values we observe in the
four-dimensional world need not be constant. The effective values of
the constants in our 4D spacetime depend on the structure and sizes of the
extra dimensions. Any evolution of these sizes, either in time or
space, would lead to dynamical constants in the 4D spacetime. Thus, any
change in the scale of extra dimensions would be revealed by a change
in the constants of our 4D spacetime.

There is a class of theoretical models in which a cosmological scalar
field, which may be closely related to the cosmological acceleration,
manifests itself through a time-dependent fine structure constant
(see e.g. the review by Uzan 2003). Varying $\alpha$ is obtained by
arbitrarily coupling photon to scalar fields. If the scalar fields are
extremely light they could produce variations of constants only on
cosmological time scales. Scalar fields provide negative pressure and
may drive the cosmological acceleration (Wetterich 2003;
Bento, Bertolami \& Santos 2004;
Fujii 2005).  The functional dependence of the gauge-coupling constants on
cosmological time is not known and even oscillations might be possible
during the course of the cosmological evolution (Marciano 1984; Fujii
2005). In this regard, astronomical observations are the only way to
test such predictions at different space-time coordinates.

In 2001, observations of spectral lines in distant quasars brought the
first hints of a possible change in the fine-structure constant over
time, with a variation in $\alpha$ of 6 parts per million (Murphy et
al.  2004). More recently, also $\mu$ has been claimed to vary
(Ivanchik et al. 2005; Reinhold et al. 2006).  If true, such a
variation would have profound implications, possibly providing a
window into the extra spatial dimensions required by unified theories
such as string/M-theory.  However, recent results from VLT/UVES
suggest no variation in $\alpha$.  The debate is still open and makes
strong demands for a high-resolution spectrograph at a large telescope
for significant progress. A precision increase in $\delta
\alpha/\alpha$ of 2-to-3 orders of magnitude will resolve the present
controversy and probe the variability in a presently unexplored
regime.

These astronomical observations will also rival in precision the local
measures that will be provided by the forthcoming satellite-borne atomic
clock experiment ACES. The ACES (Atomic Clock Ensemble in Space)
project, foreseen to fly on the International Space Station in 2007,
will operate several cold atomic clocks in microgravity to test
general relativity and search for a possible drift of the
fine-structure constant. Neutral atomic clocks in microgravity have
the potential to surpass ground-based clocks both in the microwave and
optical domains.  Rubidium clocks should enter the 10$^{-17}$
stability range with a gain of two orders of magnitude with respect to
present laboratory constraints.

\ssec{Constraints on variations in the fine-structure constant}

\sssec{General limits}
The mere existence of nucleons, atoms and stars constrains $\delta
\alpha/\alpha \leq 10^{-2}$ where we define $\delta \alpha/\alpha
=(\alpha_z - \alpha_0)/\alpha_0$, with $\alpha_z$ and $\alpha_0$ the
values of $\alpha$ at an epoch $z$ and in the laboratory,
respectively. A change of 4\% in $\alpha$ shifts the key resonance
level energies in the carbon and oxygen nuclei that are needed for C
and O synthesis from He nuclei.  Stable matter, and therefore life and
intelligent beings, would probably then not exist. Similar limits are
imposed by Big Bang nucleosynthesis (BBN) at $z\simeq 10^{10}$ and the
CMB power spectrum at $z\simeq 1100$.

Constraints on the present-day variation of $\alpha$ in laboratory
experiments are based on the comparison of atomic clocks using
different types of transitions in different atoms, such as $^{87}$Rb
and $^{133}$Cs. The time-dependence of $\alpha$ is restricted to the
level of ($\dot{\alpha}/\alpha)_{t_0} \ls 10^{-15}$
year$^{-1}$. This limit transforms into $|\delta \alpha/\alpha| <
10^{-5}$, at a cosmological time-scale of $t \sim 10$~Gyr,
corresponding to $z > 1$, assuming $\alpha_z$ is a linear function of
t.  Meteoritic data on the radioactive $\beta-$decay of $^{187}$Re
place a bound around $\delta \alpha/\alpha \leq 10^{-7}$, but this is
somewhat model dependent. An intriguing geophysical constraint comes
from the Oklo uranium mine in Gabon, Africa, where a fission reaction
took place about 1.8 Gyrs ago ($z \simeq 0.14$), naturally
self-sustained and moderated for about 200,000 years. The isotopic
abundances in the rock surrounding the mine provide information about
the nuclear rates and therefore about the value of $\alpha$ at that
time. A key quantity is the ratio of two light isotopes of Samarium
that are not fission products. This ratio is 0.9 in normal Sm but
about 0.02 in Oklo samples due to the transformation of Sm after
neutron capture while the reactor was active. Recent analysis of the
isotopic abundances in the Oklo samples provides a hint for a
variation at a very low level: $\delta \alpha/\alpha \geq 4.5 \times
10^{-8}$ (Lamoreaux \& Torgerson, 2004), but this result still
needs further confirmation.

\sssec{Constraints from QSO spectroscopy}
The astronomical measurements of the fine-structure splittings of
emission lines in galaxies provide a sensitivity of
$\delta \alpha/\alpha \simeq 10^{-4}$ at
relatively low redshift $0.4 < z < 0.8$. Early high-resolution
spectroscopy of distant absorption systems lying along the
lines-of-sight to background QSOs focused on the alkali-doublets (AD)
such as C{\sc iv}, Si{\sc ii}, Si{\sc iv}, Mg{\sc ii} and Al{\sc iii},
since the comparison between AD separations in absorption systems with
those measured in the laboratory provides a simple probe of the
variation of $\alpha$. The best current constraints come from the
analysis of Si{\sc iv} absorption systems in $R \simeq 45,000$ spectra:
$\delta \alpha/\alpha = (0.15 \pm 0.43) \times 10^{-5}$ (Chand et
al. 2004, 15 systems: $1.6 < z_{\rm abs} < 3$).

The many-multiplet (MM) method utilises many transitions from
different multiplets and different ions associated with each QSO
absorption system (Dzuba, Flambaum \& Webb 1999; Webb et al. 1999).
This is because the energy of each transition depends differently on
changes in $\alpha$. The relativistic correction to the frequency of each
transition is expressed by the coefficient $q$ (Dzuba, Flambaum \& Webb 1999;
Dzuba et al. 2002).
The MM approach compares the line shifts of the species particularly
sensitive to a change in $\alpha$ to those with a comparatively minor
sensitivity, which are referred to as anchor-lines. Mg, Si and Al act as
anchors against which the larger expected shifts in Cr, Fe, Ni and Zn
transition wavelengths can be measured. The method provides an effective
order-of-magnitude precision gain with respect to the AD method due to
the large differences in sensitivity of light and heavy ions to a
varying $\alpha$.

Applied to HIRES-Keck QSO absorption spectra the MM method has yielded
tentative evidence for a varying $\alpha$ (Webb et al. 1999), which
has become stronger with successively larger samples.  The most recent
value $\delta \alpha/\alpha = (-0.57 \pm 0.11) \times 10^{-5}$ comes
from the analysis of 143 absorption systems over the range $0.2 <
z_{\rm abs} < 4.2$ (Murphy et al. 2004). The deduced variation of
$\alpha$ at about 5$\sigma$ significance, if proved correct, would
have extraordinary implications. However, the result has not been
confirmed by two other different groups. Chand et al. (2004) have
analysed 23 Mg/Fe absorption systems in high signal-to-noise ratio
(S/N) spectra from a different telescope and spectrograph, the
UVES-VLT, claiming a null result over the range $0.4 < z_{\rm abs} <
2.3$ of $\delta \alpha/\alpha = (-0.06 \pm 0.06) \times 10^{-5}$.

A second group adopted a slightly different methodology. Levshakov and
collaborators make use only of pairs of Fe{\sc ii} lines observed in
individual high-resolution exposures. This approach avoids the
influence of possible spectral shifts due to ionization
inhomogeneities in the absorbers or non-zero offsets between different
exposures (Levshakov 2004; Levshakov et al. 2005, 2006; Quast, Reimers
\& Levshakov 2004).  Applied to the Fe{\sc ii} lines of the metal
absorption line system at $z_{\rm abs} = 1.839$ in the spectrum of
Q1101$-$264, and to the $z_{\rm abs} = 1.15$ system in the spectrum of
HE0515$-$4414, this methodology provides $\delta \alpha/\alpha = (0.4
\pm 1.5_{\rm stat}) \times 10^{-6}$ and $(-0.07 \pm 0.84_{\rm stat})
\times 10^{-6}$, respectively. These values are shifted with respect
to the HIRES-Keck mean at the 95\% confidence level. This discrepancy
between UVES-VLT and HIRES-Keck results is yet to be resolved.

Problems are likely to exist in both datasets and any significant
improvement in the future will require higher precision, as we explain
below. The validity of both results are still under intense scrutiny in
the literature, and the final conclusion from QSO absorption lines is
still far from clear.

\epsfigsimp{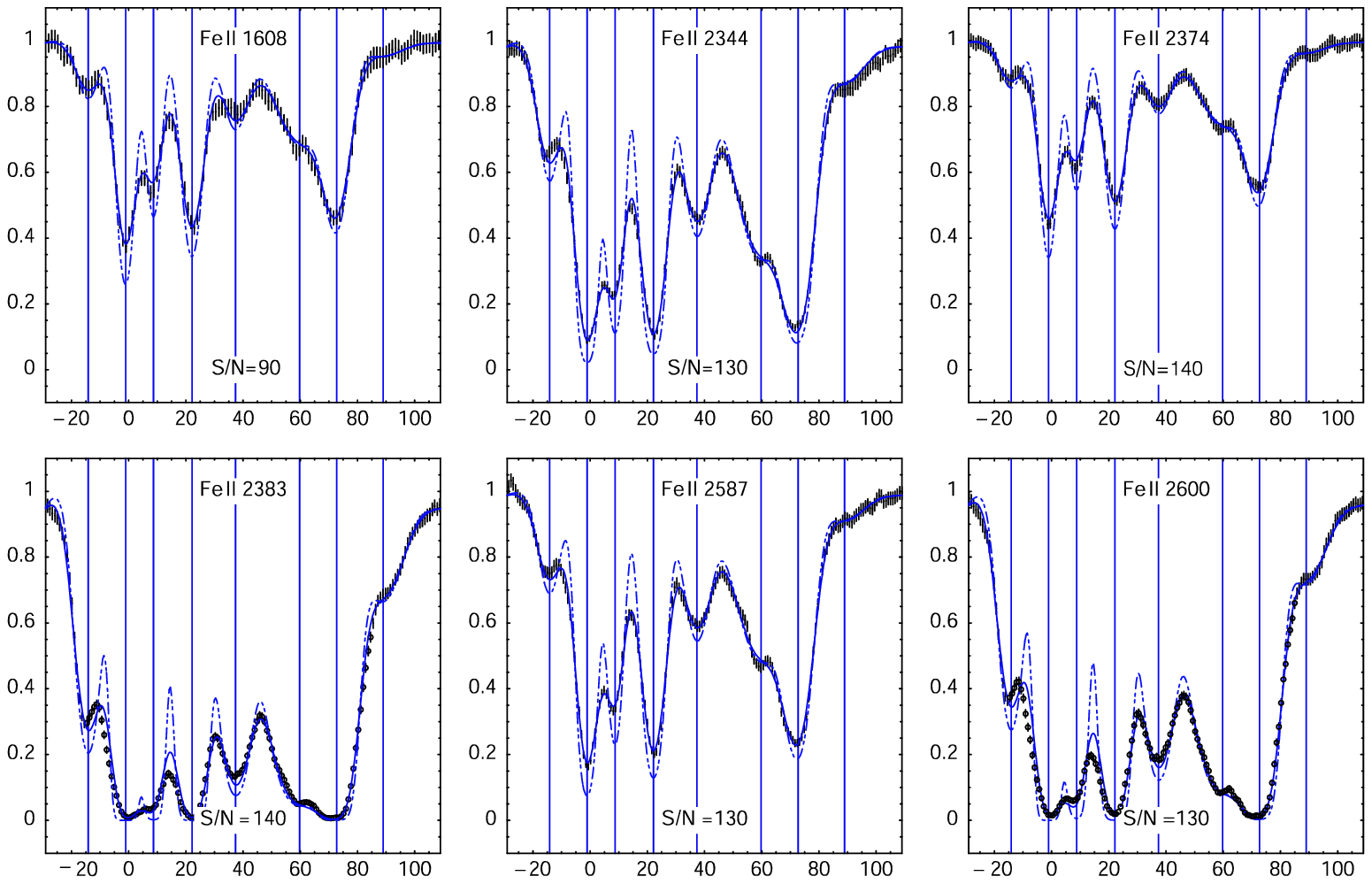}{0.9}
{The Fe{\sc ii} and Mg{\sc ii} spectroscopic
multiplets used to detect possible shifts in $\alpha$ (Quast, Reimers \& Levshakov 2004).}
{fig:var_alp_spec}

\epsfigsimp{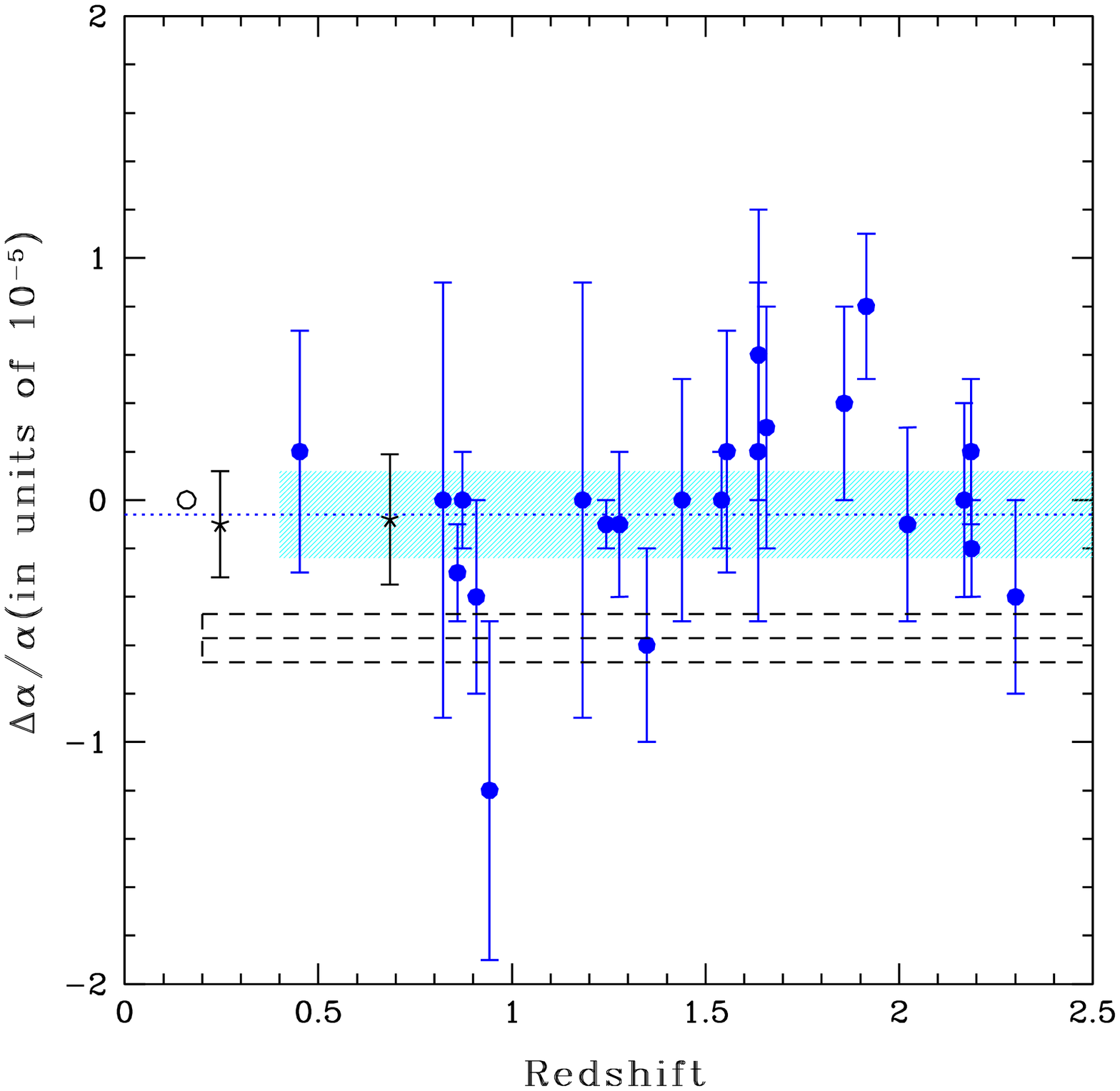}{0.6}
{The VLT/UVES data of Chand et al. (2004) on
variations in $\alpha$, compared with the previous Murphy results
which are indicated by the dashed lines.}
{fig:var_alp}

\sssec{Possible systematics: the isotopic composition in the absorbers}
The MM measures, which use Mg as an anchor, rely on terrestrial
relative composition of the Mg isotopes. This is because the frequency
shifts of $\delta \alpha/\alpha$ are of the same order of magnitude of
the typical isotope shifts; thus a departure from these values can
produce different results for the measures. Sub-solar
$^{25,26}$Mg/$^{24}$Mg ratios would make a variation of $\alpha$ even
more significant. A suggestion that this may be the case comes
indirectly from a recent upper limit in the $^{13}$C abundance
(${}^{12}\rm C/{}^{13} C > 200$, 1$\sigma$, in the system at $z= 1.15$
in the spectrum of HE0515$-$4414; Levshakov et al. 2006).  Since both
$^{25,26}$Mg and $^{13}$C are produced in the Hot Bottom Burning stage
of AGBs, a low $^{13}$C possibly implies a low $^{25,26}$Mg.  In the case
of the Chand et al. data set the relaxation of this assumption would
have implied $\delta \alpha/\alpha = (-3.6 \pm 0.6_{\rm stat})
\times 10^{-6}$. This well illustrates that the case for variability
requires a better understanding of the isotopic evolution of the
absorption clouds, a problem that can be addressed only with a
spectrograph of very high resolution able to separate the isotopic
lines.

\ssec{Constraints on variations in the proton-electron mass ratio}

The proton-electron mass ratio $\mu$ is another fundamental
constant that can be probed by astronomical observations (Cowie and
Songaila 1995). The observation of roto-vibrational transitions of $H_2$
in damped Lyman-$\alpha$ systems provides constraints on variations in the
proton-to-electron mass ratio, a method first proposed by Varshalovich
and Levshakov (1993).

In the context of Grand Unified Theories a possible variation in
$\alpha$ may be related to time-variation in other gauge couplings and
in particular to variations in the QCD scale. This is because the proton
mass is proportional, at first order, to this scale. Several authors
have argued that the quantum chromodynamic scale should vary faster than
that of the quantum electrodynamic scale producing a variation in $\mu$
even larger than expected in $\alpha$, although this is rather model
dependent (Flambaum et al. 2004; Dine et al. 2003). Theoretically the
connections between $\alpha$ and $\mu$ are quite complex, but it seems
that a varying $\alpha$ entails a varying $\mu$.

At present the $\mu$ ratio has been measured with high accuracy: 
\be
\mu = 1836.15267261(85)
\ee 
(Mohr and Taylor 2005).  By using VLT/UVES
high-resolution spectra of the quasar Q0347$-$3819 and unblended
electronic-vibrational-rotational lines of the $H_2$ molecule
identified at $z=3.025$, Levshakov et al. (2002) placed a limit on
cosmological variability of $\delta \mu/\mu < 5.7 \times
10^{-5}$. This measurement has been improved by Ubachs \& Reinhold
(2004) by using new laboratory wavelength of $H_2$ at the level of
$\delta \mu/\mu = (-0.5 \pm 1.8) \times 10^{-5} (1\sigma)$. More
recently a new measure of this system together with a new one towards
Q0405$-$443 by Reinhold et al. (2006) and Ivanchik et al. (2005)
provided $\delta \mu/\mu = (2.0 \pm 0.6) \times 10^{-5}
(1\sigma)$. This would indicate a 3.5$\sigma$ confidence level that
$\mu$ could have decreased in the past 12 Gyr.

\ssec{Outlook}

A measurement of $\delta \alpha/\alpha$ or $\delta \mu/\mu$ is
essentially a measurement of the wavelength for a pair or more
lines. Therefore the accuracy of a variability measurement is
ultimately determined by the precision with which a line position can
be determined in the spectrum.

With current spectrographs with $R\equiv \lambda/\delta\lambda \simeq
4\times 10^{4}$, the observed line positions can be known with an
accuracy of about $\sigma_{\lambda} \simeq 1$~m\AA\ (or $\Delta v =
60\,\rm m\, s^{-1}$ at 5000\,\AA).  Thus for $\delta \alpha/\alpha$
the accuracy is about $10^{-5}$ for a typical pair of lines with
typical sensitivity coefficients. This value is normally further
improved to reach one part per million when more transitions and/or
more systems are available. Any improvement with respect to this
figure is related to the possibility to measure line positions more
accurately.  This can be achieved with an increase in the resolving
power of the spectrograph up to the point in which the narrowest lines
formed in intervening physical clouds are resolved, and with an
increase of the signal-to-noise ratio in the spectrum (Bohlin et
al. 1983). The Bohlin formula gives a relatively simple analytical
expression that has also been tested by means of Monte Carlo
analysis:
\begin{equation}
\sigma_{\lambda} = \Delta \lambda_{\rm pix}(\Delta
\lambda_{\rm pix}/W_{\rm obs})(1/\sqrt{N_e})(M\sqrt{M}/\sqrt{12}),
\end{equation}
where $\Delta \lambda_{\rm pix}$ is the pixel size or the wavelength
interval between pixels, $W_{\rm obs}$ is the observed equivalent
width, $N_e$ is the mean number of photoelectrons per pixel at the
continuum level, and $M$ is the number of pixels covering the line
profile. The metal lines that are observed in the QSO absorption
systems have intrinsic widths of typically a few $\kms$ and rarely of
less than $1\kms$. One can therefore expect significant improvements
in the near future using higher spectral resolution.  The limitation
may then be in the statistics and calibration and it would be useful to have more than two
QSOs with overlapping spectra to cross-calibrate the line positions.

\japsec{Gravity-wave cosmology with LISA and its \mbox{successors}}

\ssec{LISA overview}

LISA is a joint ESA-NASA space-based gravitational wave detector,
currently expected to be launched in the middle of the next
decade. ESA will launch a technology mission called LISA Pathfinder in
2009, but this will not make gravitational wave observations. Concept
studies exist for a LISA successor, called the Big Bang Observer
(BBO), which would be dedicated to observing cosmological radiation
from the Big Bang.

LISA will observe gravitational radiation in the broad frequency band
from about 0.1~mHz up to 1~Hz, with best sensitivity between 1 mHz and
10 mHz. Sources that radiate in this band include massive and
supermassive black holes (SMBH) in the range $10^4$--$10^7M_{\odot}$,
and compact binary systems with periods smaller than an hour or
so. Unlike ground-based gravitational wave detectors such as LIGO, LISA will have
great sensitivity at these low frequencies, as shown in Fig.~\ref{fig:lisaligo}.
LISA will detect coalescences of binary SMBHs at redshifts of
order 1 with signal-to-noise ratios exceeding 1000; this means that
binary coalescences in LISA's band will be easily visible to it even
at redshifts of 10 or more. Because of the redshift, LISA is sensitive
at $z\sim 10$ to black holes in the shifted mass range
$10^3$--$10^6M_{\odot}$.

LISA consists of three spacecraft in a triangular array, with
separations of $5\times10^6$~km, in a stable configuration orbiting
the Sun at 1~AU -- which means that it will return three independent
gravitational wave signals. It can therefore measure the polarization
of the signals and its antenna pattern allows it to locate strong
sources on the sky to accuracies of tens of arcminutes.  Below 10~mHz
the wavelength of the gravitational waves is longer than the LISA
arms, and in this regime the three signals are linearly dependent on
one another, which allows a linear combination that does not contain
any gravitational wave signal. This provides an important check on
instrumental noise and will assist in the detection of random
cosmological backgrounds, as described below. As another check on
LISA's operation, several known galactic binary systems must be
detected in the first few weeks of operation.

\ssec{Science goals for LISA}

LISA will make the first observations of gravitational waves in this
frequency band, so it is impossible to make definite predictions about
what it will see.  However, at least four kinds of observations that
have cosmological implications have been discussed. These are

\begin{itemize}

\item Study the SMBH binary population to high redshifts.
\item Measure the Hubble expansion and the dark energy at high redshifts.
\item Find or limit a cosmological gravitational wave background.
\item Search for compact components of the dark matter.

\end{itemize}

\epsfigsimp{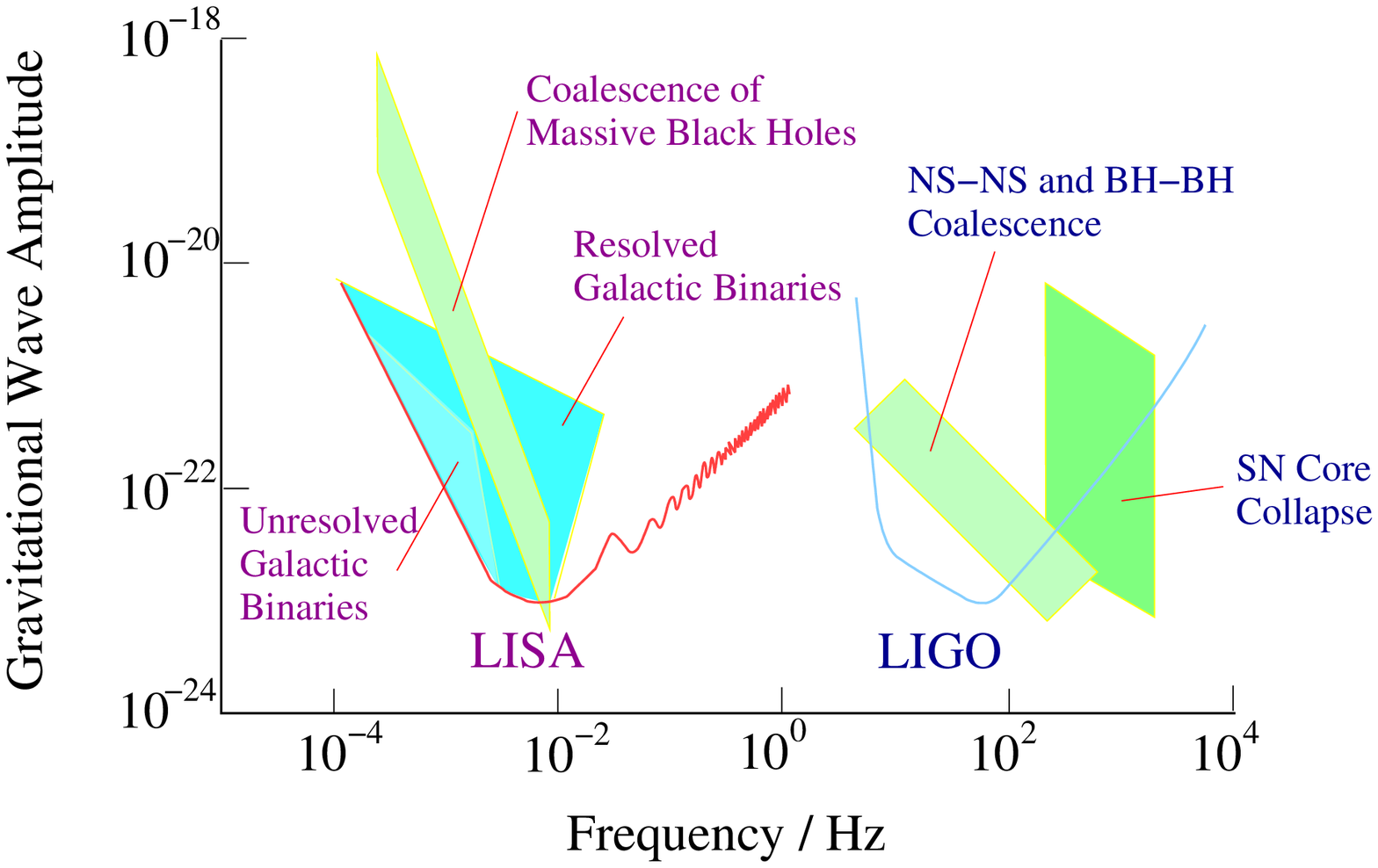}{0.8}
{A comparison of the sensitivity to gravitational strain expected
from LISA as compared to LIGO, together with illustrations of the
expected level of astronomical signals. Although the raw strain numbers
are similar, ground-based experiments cannot access the especially
interesting low-frequency regime.}
{fig:lisaligo}

\sssec{SMBH binary population}
LISA will see all coalescences of SMBH binaries in its frequency band,
no matter how far away. Black holes of masses $10^6M_{\odot}$ and
higher seem almost ubiquitous in galaxies. Black holes at the bottom
of LISA's band ($10^3$--$10^4M_{\odot}$) are less certain, but in some
models they are even more abundant. The event rate is uncertain
because it depends not only on the number of such black holes but on
the processes that lead to binaries compact enough to evolve to
coalescence in a Hubble time. Estimates of LISA's event rate range
from one event every five years to hundreds of events per year. Recent
research has tended to push up the upper bound.

Observations of such systems would tell us much about those processes
that contribute most to the event-rate uncertainty. These include
whether galaxies form from smaller fragments that contain smaller
black holes; whether black holes in merging star systems are brought
close together rapidly or slowly; whether the ubiquitous
$10^6M_{\odot}$ SMBHs are formed at that mass, or have grown either by
gas accretion or by merging with smaller black holes; how old the
oldest black holes are and whether they formed in their galaxies or
helped seed the actual formation of their galaxies. The answers to
these questions clearly have a direct bearing on theories of galaxy
formation, but they also have the potential to affect theories of
early heavy-element nucleosynthesis, considerations of the IR
background energy budget, Population III star formation and evolution,
and many other early-universe issues.

\sssec{Hubble expansion and dark energy}
LISA observations of in-spiralling SMBH binaries measure directly the
cosmological distance to the source (luminosity distance). The
accuracy of the distance measurement is limited in principle by the
signal-to-noise ratio, but in practice random gravitational lensing
introduces the major distortion. In order to use these distances to
gain cosmological information, LISA's sources would have to be
identified and their redshifts measured. It is by no means certain
that identifications will be possible, since the merger event is not
likely to emit any electromagnetic radiation. But within LISA's
observational error box, galaxies hosting merging SMBHs may be
identifiable because (1) they exhibit a distinctive disturbed
morphology, or (2) they show evidence that earlier quasar activity has
been cut off, or (3, and most interestingly) because a year or so
after the merger quasar-like activity suddenly turns on when the
accretion disc restores itself after the tidal disruption caused by
the inspiralling holes. Much more research is needed on these
questions, but it is clear that near-simultaneous observations with
suitable X-ray instruments would be useful. Recent estimates suggest
that LISA may see 100 or so merger events out to $z=2$, and in this
case it would be possible to determine the cosmological parameters
accurately even if each of the error boxes contains a dozen or more
potential host galaxies, provided the actual host is normally among
the sample.

The cosmological significance of such observations would be
enormous. Recent studies, taking into account the gravitational
lensing limits, indicate that LISA could measure the dark-energy
parameter $w$ to accuracies around 4\%. It would place strong
constraints on the time-evolution of the dark energy. This is
competitive with the expected accuracy of some proposed dedicated
dark-energy projects, but it does depend on the as-yet undemonstrated
host galaxy identifications. It is worth noting that LISA has the
sensitivity to go well beyond $z=2$ in its examination of the dark
energy, albeit with decreasing accuracy.

\sssec{Cosmological gravitational waves}
LISA can make essentially a bolometric measurement of random
backgrounds: if the noise power in the background is larger than
instrumental noise, LISA can identify it. This translates into a
sensitivity to a cosmological background with a normalized energy
density of $\Omega_{\rm gw}\sim 10^{-10}$ at 3~mHz. Note that LISA's
ability to measure its own instrumental noise at low frequencies
(mentioned earlier) allows it to make a gravitational wave background
measurement with confidence.

The sensitivity of LISA is not good enough to see standard predictions
from slow-roll inflation, which are around or below $10^{-15}$. But
there are many more exotic scenarios that produce stronger radiation,
including interesting ones based on brane models. It is interesting to
note that radiation in the LISA band today would have had a wavelength
comparable to the cosmological horizon size when the universe was
passing through the electroweak phase transition.  LISA therefore has
the potential to study not only cosmology but to make discoveries
about the electroweak interactions. The BBO mission (see below) has
been suggested in order to go beyond LISA's sensitivity down to the
predictions of inflation.

\sssec{Dark matter components}
Cold dark matter is likely to consist mainly of weakly interacting
uncharged particles, and dark matter searches are placing interesting
constraints on the nature of these particles. But there may also be
minor constituents that have such a small effect on standard
cosmological indicators of dark matter --- galaxy formation scenarios,
gravitational lensing --- that they are not predicted. These include
cosmic strings, which have been eliminated as a candidate for the
dominant dark matter, but which may nevertheless be a significant
minor component. Other minor constituents could include black-hole
systems expelled from their host star clusters during galaxy
formation, or even more exotic boson stars, composed of interacting
boson fields too massive to have been seen in accelerator experiments
so far. All these systems have predictable gravitational waveforms,
and LISA will make searches for them.

Discovering minor components of the dark matter would clearly have
far-reaching implications for early-universe physics and for unified
theories of the fundamental interactions. It is worth pointing out
that LISA has such good sensitivity that it has a chance of
discovering compact massive components even if their waveforms are not
predicted beforehand. Given our ignorance of early-universe physics,
and the number of surprises the universe has already given us, this
may well be the area where LISA will turn out to do its most important
work.

\ssec{The future of gravity-wave astronomy}

\sssec{BBO}
The Big Bang Observer is a concept developed at the request of NASA to
get an idea of what technology might be required to detect
gravitational waves from the Big Bang if the standard inflation
scenarios are correct. Two issues affect this solution: what is the
appropriate frequency window, and what technology will be available on
a 20-year time-scale.

There is little point making LISA itself more sensitive in an attempt
to see an inflation-level background. In LISA's frequency band,
astrophysically generated random backgrounds of gravitational waves
are expected to lie just under LISA's instrumental noise above 1~mHz
(and to exceed instrumental noise below this frequency). Studies
suggest that there is an accessible ``window'' where a cosmological
background is likely to exceed local-source backgrounds around
1~Hz. This lies just between the best sensitivities of LISA and the
current ground-based detectors, and so it would require a dedicated
space mission.

Moving up in frequency from the LISA band to the 1~Hz band has a
disadvantage: since the energy density per unit logarithmic frequency
of gravitational waves from cosmology is expected to be relatively
flat, the rms amplitude of the expected waves falls off as
$f^{-3/2}$. The result is that, in order to gain an energy sensitivity
of $10^6$ over LISA at a frequency $10^3$ times higher, BBO would have
to improve on the displacement sensitivity of LISA by something like 7
orders of magnitude if it were to rely simply on a LISA-like
bolometric measurement. Instead, BBO proposes two co-located LISA-like
arrays of three spacecraft each, whose output can be cross-correlated
to look for a residual correlated component of (cosmological) noise
below the (uncorrelated) instrumental noise. Even with this, BBO has
another problem: severe interference from isolated gravitational wave
sources in this window. These are mainly compact neutron-star binaries
(like the Hulse-Taylor pulsar system) on their way to
coalescence. Their signals must be measured accurately enough to be
removed before the cross-correlation is done. BBO proposes to do this
with two further LISA-like arrays placed at equal spacings in the same
1~AU orbit around the Sun; these systems can triangulate the binaries
and remove them wherever they occur, even at redshifts of 10 or more.

To do this still requires much more sensitivity in each array than
LISA will offer. It requires lasers hundreds of times more powerful and
mirrors ten times larger. It is not clear at present how realistic
such advances are, even twenty years from now. But no other design has
been proposed that is capable of seeing the cosmological gravitational
wave background. And that goal is so fundamental that the BBO is bound
to continue to be studied and to inspire near-term technology
development in this field.

\japsec{Conclusions}

\ssec{The next decade in cosmology}

The past ten years have revolutionized our knowledge about the
universe. The technical developments in observational astronomy,
together with increasingly sophisticated modelling and
simulations, have led to a vast deepening in understanding
of the processes that have shaped the cosmos we inhabit.
This understanding is best demonstrated by the great achievements
of `precision cosmology', such as the wonderful match of
WMAP data and theory -- surely one of the best examples of
a successful theoretical prediction in all of physics. Where
does the road take us from this point? Now that we have a well-established
standard model for cosmology, it is likely that a good deal of interest
in the field will seek to exploit the `phenomenology' of the
model by pursuing observations of
the early universe.  Studies of the evolution of cosmic structure, of
the formation of clusters and galaxies together with their
supermassive black holes, and of the history of the reionization of
the universe will increasingly become the focus of astrophysical
cosmology. These latter aspects have been largely neglected in this
report, owing to the specific terms of reference given to the Working
Group.

But even if we neglect purely astrophysical aspects, observational
astronomy will continue to have huge fundamental importance.
It is only through astronomy that we know that the universe consists
mainly of dark energy and dark matter -- although
both these ingredients raise key questions for particle
physicists. Inflation now counts as another effect that has become
part of the standard cosmological model, not the least due to the
spectacular results from WMAP. Again, the physics of inflation is a
great puzzle for fundamental physics. Astrophysics, and cosmology in
particular, can be regarded as the key driver for challenges and
developments in fundamental physics, and the next decade will see an
increasingly intense interaction and collaboration between these two
communities. Each working with their own tools -- accelerators and
telescopes -- the interpretation of their results will most likely be
possible only with a joint effort.

From a European perspective, the past decade has seen the opening of
the VLT, greatly increasing our capabilities for (mainly)
spectroscopic studies; a vast advance in infrared astronomy through
the ISO satellite and VLT instruments; the true power of the Hubble
Space Telescope; as well as two high-energy cornerstone observatories,
XMM-Newton and Integral. Overall, there is no doubt that the
relative impact of European cosmology on the world stage is far higher
than before these initiatives, and we should aim to maintain
this level of achievement.
Looking ahead a further decade,
the second-generation VLT
instruments will increase the capability of these telescopes; Planck
and Herschel are now close to being launched, the former expected to yield
the most precise information on cosmological parameters yet;
the VST and VISTA will provide unprecedented imaging
capabilities; GAIA will study the detailed mass structure of the Milky
Way and bring `near-field cosmology' to fruition. We will see ALMA in
action to study very high redshift objects, and we are tremendously
curious to enter the era of gravitational wave astronomy with LISA,
detecting merging supermassive black holes throughout the visible
universe, thus viewing the hierarchical formation of galaxies. Near the
end of this period, we will see JWST taking magnificent images of
unimaginable depth of the infrared sky, and taking spectra of galaxies
too faint and/or too distant to be seen in even the Hubble Deep
Fields. Finally, the construction of giant ground-based
optical/near-IR telescopes may have begun.

This suite of new tools will give astronomers plenty to do, and
is guaranteed to lead to many tremendous gains of insight.
But this great perspective falls short
on one aspect that we consider to be a key element for fundamental
cosmology: {\it we need to survey a major fraction of the sky down to
depths corresponding to a mean redshift of about unity}, because
this is the region in our visible universe where dark energy reveals its
presence. Precision measurements require us to minimize statistical
errors, implying immediately a wide sky coverage, essentially
independent of the method of investigation. The past decade has
demonstrated the great power of surveys, e.g. with the 2dFGRS and SDSS
surveys -- whose results complement WMAP data to yield a
substantial increase in cosmological accuracy.
Imaging survey work is known to be of key interest, as is
recognised in several wavebands: microwave (Planck), UV (GALEX), and
X-ray (eROSITA); furthermore, GAIA will perform an all-sky
astrometric survey. All-sky surveys in the optical and near-IR
do exist (ESO/UKST; POSS; DENIS; 2MASS), but are restricted to
shallow levels by current standards, and thus tell us about only the
local universe. We believe that the coming decade will see a revival of
sky surveys in these wavebands, and that these surveys will form a key
ingredient in our attempts to learn about the dark side of the
universe. This belief is supported by looking outside the confines of
Europe.

\ssec{The international perspective}

The outlook for cosmology has naturally been much debated
throughout the global community in the subject, and it is worth
summarising the results of some other important studies.
Chief among these are the USA's DoE/NASA/NSF interagency task forces
on Dark Energy (Kolb et al. 2006) and CMB research
(Bock et al. 2006).

\sssec{Dark Energy}
The Dark Energy Task Force takes the standard approach of parameterising
the dark energy equation of state as $w(a)=w_0 + w_a(1-a)$ and
advocates a figure of merit for any given experiment that is
the reciprocal area of the error ellipse in the $w_0 - w_a$ plane. This
is not an unreasonable choice, but it does presume that nontrivial
DE dynamics will be detected in due course, whereas the greatest
immediate advance one could imagine in the field is to rule
out the cosmological constant model. Until or unless this is achieved,
it arguably makes sense to optimise the error on a constant $w$, independent
of evolution. This difference in emphasis does not matter hugely in practice.
Kolb et al. define a number of stages for DE probes:
\begin{itemize}
\item Stage I: Current knowledge.
\item Stage II: Ongoing projects.
\item Stage III: Near-term, medium-cost projects, which should deliver a factor
of 3 improvement in figure of merit over stage II.
\item Stage IV: Long-term, high-cost projects, which should deliver a factor
of 10 improvement in figure of merit over stage II. These are taken to be LSST, SKA,
plus one space mission expected to emerge from the NASA/DoE JDEM process.
\end{itemize}

The Dark Energy Task Force report argues that all four principal
techniques (baryon oscillations; cluster surveys; supernovae; lensing)
should be pursued, since none in isolation is capable of delivering
the required accuracy. We agree with this multi-pronged strategy,
while noting that there are significant differences in the ideal
accuracies promised by the various methods. In particular, large-scale
weak lensing surveys with photometric redshifts has the best formal
accuracy. However, all techniques may be subject to unanticipated
systematic limits, so there is certainly a strong rationale for
pursuing several independent techniques. It is also worth noting
that there is scope for disagreement about timing: the DETF tend
to see space-borne imaging as the ultimate long-term stage IV approach,
whereas we believe that ESA's Cosmic Vision process offers the
opportunity of achieving these gains on a relatively accelerated
timescale.

\sssec{The CMB}
The Bock et al. report emphasises three main points: (1) there
is a huge potential science gain from detecting intrinsic
large-scale `B-mode' polarization; (2) small-scale CMB
anisotropies contain important information that indirectly
constrains the fundamental-physics aspects of the CMB; (3)
progress in both these areas will be limited without
an improved understanding of Galactic foreground emission.

The main Bock et al. recommendation under the heading of large-scale
CMB polarization is a satellite mission, termed CMBPOL, which aims
to detect intrinsic B modes if present at a level of about $r=0.01$.
This has a notional development period starting in 2011, with a
launch in 2018. In the interim, they recommend development
work on large-area polarized detector arrays for the frequency
range 30 -- 300~GHz, to be validated via balloon flights.
Their assumption is that the preferred technology for these
frequencies will continue to be bolometric.

On the small-scale CMB front, the Bock et al. outlook is dominated by
two experiments currently under construction: the South Pole Telescope
(SPT) and the Atacama Cosmology Telescope (although the European APEX
sub-mm dish at the Atacama site also receives some mention). These
experiments have apertures between 6m and 12m, and will study the CMB
with resolution of order 1 arcmin at wavelengths of a few mm or below.
This allows the use of the Sunyaev--Zeldovich effect for the detection
and characterisation of clusters of galaxies, over areas of sky of
order 1000~deg$^2$. These studies of nonlinear signatures in the CMB
will remove uncertainties in the power-spectrum normalization, and
will also impact on dark energy -- both via the use of the CMB power
spectrum and via the use of SZ-detected cluster evolution. No major
new experiments in this area are proposed by Bock et al.

Similarly, no specific experiment is proposed for the study of
Galactic foregrounds alone. Rather, it is in the main expected that
knowledge of the foregrounds will emerge through a consistent
integration of the CMB data in various frequency channels. The only
exception to this strategy is the acknowledged desirability of
obtaining improved low-frequency maps around 10~GHz, which can
constrain not only the synchrotron foreground but also the so-called
anomalous foreground, which is sometimes hypothesized to arise from
spinning dust grains.

\ssec{Recommendations}

After discussion of the issues documented in this report,
the ESA-ESO Working Group arrived at the following
set of recommendations. These are based on a
number of considerations, given here in approximately decreasing order
of relative weight:
\begin{itemize}
\item
{\it What are the essential questions in fundamental cosmology?}\\[0.3ex]
Among the many issues, we identified five key
questions that lie at the heart of our understanding of the fundamentals
of the evolution of the universe: (1) baryogenesis; (2) the nature of
dark matter; (3) the nature of dark energy; (4) the physics of inflation;
and (5) tests of fundamental physics.
\item
{\it Which of these questions can be tackled, perhaps exclusively,
with astronomical techniques?}\\[0.3ex]
It seems unlikely that
astronomical observations can currently contribute any
insight into baryogenesis. Furthermore, the nature of dark matter may
well be best clarified by experiments at particle accelerators, in
particular the Large Hadron Collider, or by direct dark matter
searches in deep underground laboratories. This particle astrophysics
approach may also tell us much about other fundamental issues,
such as the law of gravity, extra dimensions etc.
However, astronomical tools will also make essential contributions
to these problems. They can constrain the dark matter constituents via their
spatial clustering properties and/or their possible annihilation signals.
Astronomy is also probably the best way to measure any time
variability of the fundamental `constants'. Finally, the nature of dark energy and
the physics of inflation can be empirically probed, according to our
current knowledge, only in the largest laboratory available -- the
universe itself.
\item
{\it What are the appropriate methods with which these key questions
can be answered?}\\[0.3ex]
Studies of the dark energy equation of
state can profit from four different methods: the large-scale
structure of the three-dimensional galaxy distribution, clusters of
galaxies, weak lensing, and distant supernovae. An attempt was made to
judge the relative strengths of these methods, though all of them will
require a substantial increase of measurement accuracies compared to
current results, so that unanticipated systematic limits may
become a problem in future work.  For this reason, and given the
central importance of this key question for cosmology and fundamental
physics, pursuing only a single method bears an unacceptable
risk. \\[1.0ex]
The physics of inflation can be studied by three main
methods: the B-mode polarization signal of the CMB; the direct
detection of gravitational waves from the inflationary epoch; and a
precise measurement of the density fluctuation spectrum over a very
large range of length scales, to determine the slope (tilt) and possibly
the curvature (running) of the power spectrum. These parameters are bounded by CMB
measurements on the largest scales, and by weak lensing and Ly$\alpha$
forest studies on the smallest scales.
\item
{\it Which of these methods appear promising for realization within
Europe, or with strong European participation, over
the next $\sim 15$ years?}\\[0.3ex]
This issue is subject to considerable uncertainty, as it depends on
the funding situation as much as on international developments, in
particular when it comes to cooperation with partners outside
Europe. Nevertheless, much work has been invested in planning for
potential future projects, so in many cases there is a strong
basis on which to pick the best future prospects. Certainly,
there is no shortage of input, and it is a sign of the scientific
vitality of European cosmology that there are unfortunately more
attractive ideas than can feasibly be funded.
We have paid particular attention to the findings of the ESA advisory
structure, summarized in the Cosmic Vision 2015-2025 document.
Given the interagency nature of this WG, we have naturally chosen
to emphasise particularly timely opportunities for collaboration between these
two major players in European astronomy.
\item
{\it Which of these methods has a broad range of applications and a
high degree of versatility even outside the field of fundamental
cosmology?}\\[0.3ex]
Given that the next major steps towards answering the key cosmological questions
will in any case require substantial resources,
it is desirable that the projects to be pursued should lead to datasets
of general applicability.
Whereas the cosmological issues are the prime
science drivers of these projects, and determine their specifications,
a broad range of applications will increase the scientific value of
the investments, and boost their level of support in the community.
\end{itemize}
Based on these considerations, our recommendation are as follows:
\begin{enumerate}
\item
{\it Wide-field optical and near-IR imaging survey.}\\[0.3ex]
ESA and ESO have the opportunity to collaborate in executing an
imaging survey across a major fraction of the sky by constructing a
space-borne high-resolution wide-field optical and near-IR imager and
providing the essential optical multi-colour photometry from the
ground. The VST KIDS and VISTA VIKING projects will be essential
pathfinders for this sort of data, but substantial increases in grasp
and improvements in image quality will be needed in order to match or
exceed global efforts in this area. Near-IR photometry is extremely important
for obtaining reliable photometric redshifts, in particular for
galaxies beyond redshift unity, but also to minimize the fraction of
outliers at lower redshifts. VISTA will be able to perform this role
to some extent with regard to KIDS. However, imaging in space offers
huge advantages in the near-IR via the low background, and this is the
only feasible route to quasi all-sky surveys in this band that match
the depth of optical surveys. Therefore,
\begin{itemize}

\item
ESA should give the highest immediate priority in its
astronomy programme to a satellite that offers this
high-resolution optical imaging, preferably combined with
near-IR photometry, and in parallel,

\item
ESO should give high priority to expanding its wide-field optical
imaging capabilities to provide the required multi-band photometric
data.

\item
Furthermore, since the calibration of photo-z's is key to the
success of this plan, ESO should aim to conduct large
spectroscopic surveys spread sparsely over $\sim 10,000$~deg$^2$,
involving $>100,000$ redshifts.  This will require the initiation
of a large Key Programme with the VLT, integrated with the imaging survey.
\end{itemize}
This project will be an essential asset for several of the cosmological
probes discussed in this report. It will provide the necessary data for weak lensing
and large-scale structure studies of the dark energy component in the
Universe. Furthermore, it will provide an indispensable dataset for
statistical studies of dark energy using galaxy clusters, yielding the
means to determine redshifts and optical luminosity of X-ray and
SZ-selected clusters, as provided by, e.g., eROSITA and Planck.
In addition, such a project (essentially an SDSS imaging survey 4 magnitudes
deeper and with $\sim 3$ times larger area, plus
2MASS with a 7 magnitude increase in depth), together with highly
accurate photometric redshifts for galaxies and quasars, would
be a profound resource for astronomy in general, a legacy comparable
in value to the Palomar surveys some 50 years ago. Among the numerous
applications of such a dataset, we mention the selection of targets
for deep spectroscopic studies, either for the VLT, the JWST and
finally an ELT.

\item
The existence of major future imaging surveys presents a
challenge for spectroscopic follow-up. For some applications,
such as weak gravitational lensing, photometric redshifts with
few \% precision are sufficient. But some science questions
need true spectroscopy, and this presents of problem of
grasp.  A capability
for massive multiplexed deep spectroscopy (at the level of
several thousand simultaneous spectra over a field of order
one degree) is required for this. Such a facility would permit
surveys of $>10^6$
redshifts needed to probe dark energy using the galaxy power
spectrum as a standard ruler, and there are a number
of international plans for instruments of this sort.
ESO should secure access to such an instrument, either through the
development of such a facility for the VLT, or as a collaborative
arrangement with an external project, perhaps in conjunction with
sharing some of Europe's proposed imaging data.

\item
A powerful multi-colour imaging capability can also carry out a
supernova survey extending existing samples of $z=0.5-1$ SNe by an
order of magnitude,
although an imager of 4m class is required if
this work is to be pursued from the ground.
In order to exploit the supernova technique fully, an improved local
sample is also required. The VST could provide this, provided that time is
not required for other cosmological surveys, in particular lensing.

\item
Whereas the WG sees the main science drivers for a European Extremely
Large Telescope (ELT) as lying in other fields of astronomy,
we recommend that the following applications in fundamental cosmology
should be regarded as forming an essential part of the ELT capability:
\begin{itemize}
\item
Supernova surveys need to be backed up with spectroscopy to assure
the classification for at least a significant subsample and to check
for evolutionary effects. The spectroscopy requires access to
the largest possible telescopes, and an ELT will be essential for the
study of distant supernovae with redshifts $z>1$.
\item
A European ELT will also be important in fundamental
cosmology via the study of the intergalactic medium. Detailed
quasar spectroscopy can limit the nature of dark matter by
searching for a small-scale coherence length in the mass
distribution. These studies can also measure directly the
acceleration of the universe, by looking at the time dependence
of the cosmological redshift. Furthermore, by providing information of
the density fluctuation power spectrum at the smallest scales, the
Lyman-$\alpha$ forest provides the biggest lever arm on the shape of
the power spectrum, and thus on its tilt and its potentially running
spectral index.
\item
ELT quasar spectroscopy also offers the possibility of better
constraints on any time variation of dimensionless
atomic parameters such as the fine-structure constant $\alpha$ and the
proton-to-electron mass ratio. There presently exist controversial
claims of evidence for variations in $\alpha$, which potentially
relate to the dynamics of dark energy. It is essential to validate
these claims with a wider range of targets and atomic tracers.
\end{itemize}

\item
In the domain of CMB research, Europe is well positioned with the
imminent arrival of Planck. The next steps are (1) to deal with the
effects of foreground gravitational lensing of the CMB and (2) to
measure the `B-mode' polarization signal, which is the prime indicator
of primordial gravitational waves from inflation.  The former effect
is aided by the optical/near-IR imaging experiments discussed
earlier. The latter effect is potentially detectable by Planck, since
simple inflation models combined with data from the WMAP CMB satellite
predict a tensor-to-scalar ratio of $r\simeq 0.15$.  A next-generation
polarization experiment would offer the chance to probe this signature
in detail, providing a direct test of the physics of inflation and
thus of the fundamental physical laws at energies $\sim 10^{12}$ times
higher than achievable in Earth-bound accelerators.  For reasons of
stability, such studies are best done from space; we thus recommend
such a CMB satellite as a strong future priority for ESA,
together with the
support of corresponding technological developments.
\item
An alternative means of probing the earliest phases of cosmology is to
look for primordial gravitational waves at much shorter wavelengths. LISA
has the potential to detect this signature by direct observation of a
background in some models, and even upper limits would be of extreme
importance, given the vast lever arm in scales between direct studies
and the information from the CMB.  We thus endorse space-borne
gravity-wave studies as an essential current and future priority for
ESA.
\end{enumerate}

\ssec{The longer-term outlook for cosmology}

Beyond the time-frame considered here (up to about 2020),
the power of cosmological facilities in astronomy will
inevitably increase still further. One of the most exciting
prospects will be the Square Kilometre Array, which will be
capable of detecting redshifted neutral hydrogen throughout the
visible universe. The general science programme of the SKA 
and its impact on all areas of extragalactic astronomy is
described in Carilli \& Rawlings (2004). For the present purpose,
the most obvious application is that the SKA will obtain
redshifts for $10^8 - 10^9$ galaxies (depending on the
configuration chosen), thus pushing baryon-oscillation
studies of the dark energy into new regimes.
This project will operate on a longer timescale, and is a natural
successor to the studies described above.
A strong European participation in the SKA is therefore essential,
although of course the justification for this lies as much
in the domain of astrophysics as in fundamental cosmology.

In this report, we have shown that there are many exciting opportunities
for ESA and ESO to act in concert to achieve great advances in
our knowledge of cosmology. Where we will stand in 2020 is
impossible to predict, but some speculation is irresistible.
There is a good case that the current evidence for tilt places
us half-way to a proof of inflation, and the great hope must be be
that primordial gravitational waves will be detected in the CMB to
complete the picture. If we were also to
have found by this time that the
dark energy is more than a cosmological constant, then we would
have two completely new windows into previously unstudied physics.
Continuing this optimistic view, the simplest models for dark
matter suggest that the WIMP responsible for this phenomenon
will have been seen both directly in underground experiments
and at the LHC by 2020. Of course, it is possible that none of these
developments will come to pass -- but then future cosmological
research will be steered into new and equally interesting
directions.

\def\mnras{MNRAS}
\def\aj{AJ}
\def\apj{ApJ}
\def\apjs{ApJ Suppl.}
\def\pra{Phys. Rev. A}
\def\prd{Phys. Rev. D}
\def\prl{Phys. Rev. Lett.}
\def\araa{ARAA}
\def\pasp{PASP}
\def\aap{A\&A}

\def\ok{}

\def\japref#1#2{\parskip=0pt\par\noindent\hangindent\parindent
\parskip =2ex plus .2ex minus .1ex}

\addcontentsline{toc}{section}{Bibliography}

%\begin{thebibliography}{}

\section*{Bibliography}

\vglue 0.5em

\japref{[Allen et al.(2001)]}{Allen01} Allen, S.W., Ettori, S., Fabian, A.C., 2001, MNRAS, 324, 877
\japref{[Allen et al.(2004)]}{Allen04} Allen, S.W., Schmidt, R.W., Ebeling, H., Fabian, A.C., van Speybroeck, L., 2004, \mnras, 353, 457 \ok
\japref{[Angulo et al.(2005)]}{Angulo05} Angulo, R., Baugh, C.M., Frenk, C.S., Bower, R.G., Jenkins, A., Morris, S.L., 2005, \mnras, 362, L25 \ok
\japref{[Annis et al.(2005)]}{Annis05} Annis, J., et al., 2005, astro-ph/0510195 \ok
\japref{[Arnett (1982)]}{Arnett82} Arnett, W.D., 1982, \apj, 253, 785 \ok
\japref{[Astier et al.(2006)]}{Astier06} Astier, P., et al., 2006, \aap, 447, 31 \ok
\japref{[Bacon et al.(2003)]}{Bacon03} Bacon, D.J., Massey, R.J., R\'efr\'egier, A.R., Ellis, R.S., 2003, \mnras, 344, 673 \ok
\japref{[Bacon et al.(2005)]}{Bacon05} Bacon, D.J., et al., 2005, \mnras, 363, 723 \ok
\japref{[Bartolo et al.(2004)]}{Bart04} Bartolo, N., Komatsu, E., Matarrese, S., Riotto, A., 2004, Physics Reports, 402, 103
\japref{[Bean et al.(2006)]}{Bean06} Bean, R., Dunkley, J., Pierpaoli, E., 2006,  astro-ph/0606685 \ok
\japref{[Bekenstein(2004)]}{Bekenstein04} Bekenstein, J.D., 2004, \prd, 70, 3509
\japref{[Bell et al.(2004)]}{Bell04} Bell, E.F., et al., 2004, \apj, 608, 752 \ok
\japref{[Bento et al.(2004)]}{Bento04} Bento, M.C., Bertolami, O., Santos, N.M., 2004, \prd, 70, 107304 \ok
\japref{[Bernardeau(1998)]}{Bernardeau98} Bernardeau, F., 1998, \aap, 338, 375 \ok
\japref{[Bernardeau et al.(2002)]}{Bernardeau02} Bernardeau, F., Mellier, Y., van Waerbeke, L., 2002, \aap, 389, L28 \ok
\japref{[Bernardeau et al.(1997)]}{Bernardeau97} Bernardeau, F., van Waerbeke, L., Mellier, Y., 1997, \aap, 322, 1 \ok
\japref{[Blake, Bridle(2005)]}{Blake05} Blake, C., Bridle, S., 2005, \mnras, 363, 1329 \ok
\japref{[Blake, Glazebrook(2003)]}{Blake03} Blake, C., Glazebrook, K., 2003, \apj, 594, 665 \ok
\japref{[Blandford et al.(1991)]}{Blandford91} Blandford, R.D., Saust, A.B., Brainerd, T.G., Villumsen, J.V., 1991, \mnras, 251, 600 \ok
\japref{[Blinnikov et al.(2006)]}{Blinnikov06} Blinnikov, S., et al., 2006, \aap, 543, 229 \ok
\japref{[Bock et al.(2006)]}{Bock06} Bock, J., et al., 2006, astro-ph/0604101 \ok
\japref{[Bohlin et al.(1983)]}{Bohlin83} Bohlin, R.C., Jenkins, E.B., Spitzer, L., Jr., York, D.G., Hill, J.K., Savage, B.D., Snow, T.P., Jr., 1983, \apjs, 51, 277 \ok
\japref{[Brown et al.(2003)]}{Brown03} Brown, M.L., Taylor, A.N., Bacon, D.J., Gray, M.E., Dye, S., Meisenheimer, K., Wolf, C. 2003, \mnras, 341, 100 \ok
\japref{[Carilli, Rawlings(2004)]}{Carilli04} Carilli, C.L., Rawlings, S. (eds), 2004, New Astronomy Review, 48 \ok
\japref{[Catelan et al.(2000)]}{Catelan00} Catelan, P., Porciani, C., Kamionkowski, M., 2000, \mnras, 318, L39 \ok
\japref{[Carroll et al.(1992)]}{Carroll92} Carroll, S.M., Press, W.H., Turner, E.L., 1992, \araa, 30, 499 \ok
\japref{[Chand et al.(2004)]}{Chand04} Chand, H., Srianand, R., Petitjean, P., Aracil, B., 2004, \aap, 417, 853 \ok
\japref{[Chang et al.(2004)]}{Chang04} Chang, T.-C., R\'efr\'egier, A., Helfand, D.J., 2004, \apj, 617, 794 \ok
\japref{[Clowe et al.(2006)]}{Clowe06} Clowe, D., Bradac, M., Gonzalez, A.H., Markevitch, M., Randall, S.W., Jones, C., Zaritsky, D., 2006,  astro-ph/0608407 \ok
\japref{[Coil et al.(2006)]}{Coil06} Coil, A.L., Newman, J.A., Cooper, M.C., Davis, M., Faber, S.M., Koo, D.C., Willmer, C.N.A., 2006, \apj, 644, 671 \ok  %  astro-ph/0512233
\japref{[Cole et al.(2005)]}{Cole05} Cole, S., et al., (The 2dFGRS Team), , 2005, \mnras, 362, 505 \ok
\japref{[Collister, Lahav(2004)]}{Collister04} Collister, A.A., Lahav, O., 2004, \pasp, 116, 345 \ok
\japref{[Contaldi et al.(2003)]}{Contaldi03} Contaldi, C.R., Hoekstra, H., Lewis, A., 2003, Physical Review Letters, 90, 221303 \ok
\japref{[Cooray, Sheth(2002)]}{Cooray02} Cooray, A., Sheth, R., 2002, Physics Reports, 372, 1 \ok
\japref{[Cowie, Songaila(1995)]}{Cowie95} Cowie, L.L., Songaila, A., 1995, \apj, 453, 596 \ok
\japref{[Croft et al.(2002)]}{Croft02} Croft, R.A.C., Weinberg, D.H., Bolte, M., Burles, S., Hernquist, L., Katz, N., Kirkman, D., Tytler, D., 2002, \apj, 581, 20 \ok
\japref{[Desjacques, Nusser(2005)]}{Desjacques05} Desjacques, V., Nusser, A., 2005, \mnras, 361, 1257 \ok
\japref{[Dine et al.(2003)]}{Dine03} Dine, M., Nir, Y., Raz, G., Volansky, T., 2003, \prd, 67, 015009 \ok
\japref{[Dvali et al.(2000)]}{Dvali00} Dvali, G., Gabadadze, G., Porrati, M., 2000, Physics Letters B, 485, 208 \ok
\japref{[Dzuba et al.(2002)]}{Dzuba02} Dzuba, V.A., Flambaum, V.V., Kozlov, M.G., Marchenko, M., 2002, \pra, 66, 022501 \ok
\japref{[Dzuba et al.(1999)]}{Dzuba99} Dzuba, V.A., Flambaum, V.V., Webb, J.K., 1999, \pra, 59, 230 \ok
\japref{[Efstathiou et al.(1990)]}{Efstathiou90} Efstathiou, G., Sutherland, W., Maddox, S.J., 1990, Nature, 348, 705 \ok
\japref{[Eisenstein et al.(2005)]}{Eisenstein05} Eisenstein, D.J., et al., 2005, \apj, 633, 560 \ok
\japref{[Elgaroy(2005)]}{Elgaroy05} Elgaroy, O., Lahav, O., 2005, New Journal of Physics, 7, 61 \ok
\japref{[Elias-Rosa et al.(2006)]}{Elias06} Elias-Rosa, N., et al., 2006, \mnras, 369, 1880 \ok
\japref{[Feldman et al.(1994)]}{Feldman94} Feldman, H.A., Kaiser, N., Peacock, J.A., 1994, \apj, 426, 23 \ok
\japref{[Flambaum et al.(2004)]}{Flambaum04} Flambaum, V.V., Leinweber, D.B., Thomas, A.W., Young, R.D., 2004, \prd, 69, 115006 \ok
\japref{[]}{Fu06} Fu et al., 2006, in preparation\ok
\japref{[Fujii(2005)]}{Fujii05} Fujii, Y., 2005, Physics Letters B, 616, 141 \ok  %  astro-ph/0502191
\japref{[Goldhaber et al.(2001)]}{Goldhaber01} Goldhaber, G., et al., 2001, \apj, 558, 359 \ok
\japref{[Goobar et al.(2006)]}{Goo06} Goobar, A., Hannestad, S., M\"{o}rtsel, E., Tu, H., 2006, JCAP, 06, 019 \ok
\japref{[Guth(1981)]}{guth81} Guth, A.H., 1981, \prd,  23, 347
\japref{[Guy et al.(2005)]}{Guy05} Guy, J., et al., 2005, \aap, 443, 781 \ok
\japref{[Haiman et al.(2005)]}{Haiman05} Haiman, Z., et al., 2005,  astro-ph/0507013 \ok
\japref{[Hamana et al.(2002)]}{Hamana02} Hamana, T., Colombi, S., Thion, A., Devriendt, J., Mellier, Y., Bernardeau, F., 2002, MNRAS, 330, 365
\japref{[Hamana et al.(2003)]}{Hamana03} Hamana, T., et al., 2003, \apj, 597, 98 \ok
\japref{[Hamuy et al.(2003)]}{Hamuy03} Hamuy, M., et al., 2003, Nature, 424, 651 \ok
\japref{[Heavens(2003)]}{Heavens03} Heavens, A.F., 2003, \mnras, 343, 1327 \ok
\japref{[Heavens et al.(2006)]}{Heavens06} Heavens, A.F, Kitching, T.D., Taylor, A.N., 2006, astro-ph/0606568 \ok
\japref{[Hennawi, Spergel(2005)]}{Hennawi05} Hennawi, J.F., Spergel, D.N., 2005, \apj, 624, 59 \ok
\japref{[Hetterscheidt et al.(2006)]}{Hetterscheidt06} Hetterscheidt, M., Simon, P., Schirmer, M., Hildebrandt, H., Schrabback, T., Erben, T., Schneider, P., 2006,  astro-ph/0606571 \ok
\japref{[Heymans et al.(2004)]}{Heymans04} Heymans, C., Brown, M., Heavens, A., Meisenheimer, K., Taylor, A., Wolf, C., 2004, \mnras, 347, 895 \ok
\japref{[Heymans et al.(2005)]}{Heymans05} Heymans, C., et al., 2005, \mnras, 361, 160 \ok\ok
\japref{[Heymans, Heavens(2003)]}{Heymans03} Heymans, C., Heavens, A., 2003, \mnras, 339, 711 \ok
\japref{[Heymans et al.(2006)]}{Heymans06a} Heymans, C., et al., 2006a, \mnras, 368, 1323 \ok
\japref{[Heymans et al.(2006)]}{Heymans06b} Heymans, C., White, M., Heavens, A., Vale, C., Van Waerbeke, L., 2006b, \mnras, 371, 750 \ok
%\japref{[Hildebrandt et al.(2006)]}{Hildebrandt06} Hildebrandt, H., Erben, T., Schneider, P., Eifler, T., Pielorz, J., Simon, P., Dietrich, J.P., 2006,  astro-ph/0606578 \ok
\japref{[Hildebrandt et al.(2006)]}{Hildebrandt06} Hildebrandt, H., et al. in preparation \ok
\japref{[Hillebrandt \& Niemeyer(2000)]}{Hillebrandt00} Hillebrandt, W., Niemeyer, J.C., 2000, \araa, 38, 191 \ok
\japref{[Hirata, Seljak(2004)]}{Hirata04} Hirata, C.M., Seljak, U., 2004, \prd, 70, 063526 \ok
\japref{[Hoekstra et al.(2006)]}{Hoekstra05} Hoekstra, H., et al., 2006, \apj, 647, 116 \ok %astro-ph/0511089\ok
\japref{[Hoekstra et al.(2003)]}{Hoekstra03} Hoekstra, H., Franx, M., Kuijken, K., Carlberg, R.G., Yee, H.K.C., 2003, \mnras, 340, 609 \ok
\japref{[Hoekstra et al.(2002)]}{Hoekstra02} Hoekstra, H., Yee, H.K.C., Gladders, M.D., 2002a, \apj, 577, 595 \ok
\japref{[Hoekstra et al.(2002)]}{Hoekstra02a} Hoekstra, H., van Waerbeke, L., Gladders, M.D., Mellier, Y., Yee, H.K.C., 2002b, \apj, 577, 604 \ok
\japref{[Holder et al.(2001)]}{Holder01} Holder, G., Haiman, Z., Mohr, J.J., 2001, \apj, 560, L111 \ok
\japref{[Hook et al.(2005)]}{Hook05} Hook, I., et al., 2005, \aj, 130, 2788 \ok
\japref{[Howell et al.(2005)]}{Howell05} Howell, D.A., et al., 2005, \aj, 131, 2216 \ok
\japref{[Hubble(1934)]}{hubble34} Hubble, E., 1934, \apj, 79, 8 \ok
\japref{[Huterer, Takada(2005)]}{Huterer05} Huterer, D., Takada, M., 2005, Astroparticle Physics, 23, 369 \ok
\japref{[Huterer et al.(2006)]}{Huterer06} Huterer, D., Takada, M., Bernstein, G., Jain, B., 2006, \mnras, 366, 101 \ok
\japref{[Huterer, White(2005)]}{Huterer05a} Huterer, D., White, M., 2005, \prd, 72, 043002 \ok
\japref{[Hu, Haiman(2003)]}{Hu03} Hu, W., Haiman, Z. 2003, \prd, 68, 063004 \ok
\japref{[Ilbert et al.(2006)]}{Ilbert06} Ilbert, O., et al., 2006, astro-ph/0603217 \ok
\japref{[Ivanchik et al.(2005)]}{Ivanchik05} Ivanchik, A., Petitjean, P., Varshalovich, D., Aracil, B., Srianand, R., Chand, H., Ledoux, C., Boiss{\'e}, P., 2005, \aap, 440, 45 \ok\ok
\japref{[Jain, Seljak(1997)]}{Jain97} Jain, B., Seljak, U., 1997, \apj, 484, 560 \ok
\japref{[Jain, Taylor(2003)]}{Jain02} Jain, B., Taylor, A., 2003, Physical Review Letters, 91, 141302 \ok
\japref{[Jarvis et al.(2003)]}{Jarvis03} Jarvis, M., Bernstein, G.M., Fischer, P., Smith, D., Jain, B., Tyson, J.A., Wittman, D. 2003, AJ, 125, 1014 \ok
\japref{[Jarvis et al.(2004)]}{Jarvis04} Jarvis, M., Bernstein, G., Jain, B., 2004, \mnras, 352, 338 \ok % not wrong page number
\japref{[Jarvis et al.(2006)]}{Jarvis06} Jarvis, M., Jain, B., Bernstein, G., Dolney, D., 2006, \apj, 644, 71 \ok
\japref{[Jha(2002)]}{Jha02} Jha, S., 2002, PhD Thesis, Harvard University \ok
\japref{[Jing et al.(2006)]}{Jing06} Jing, Y.P., Zhang, P., Lin, W.P., Gao, L., Springel, V., 2006, \apj, 640, L119 \ok
\japref{[Jones et al.(2005)]}{Jones05} Jones, D.H., Saunders, W., Read, M., Colless, M., 2005, Publications of the Astronomical Society of Australia, 22, 277 \ok  %  astro-ph/0505068
\japref{[Kaiser(1986)]}{Kaiser86} Kaiser, N., 1986, \mnras, 219, 785 \ok
\japref{[Kaiser(1992)]}{Kaiser92} Kaiser, N., 1992, \apj, 388, 272 \ok
\japref{[Kamionkowski et al.(1997)]}{Kametal97} Kamionkowski, M., Kosowsky, A., Stebbins, A., 1997, \prd, 55, 7368 \ok
\japref{[Kilbinger, Schneider(2005)]}{Kilbinger05} Kilbinger, M., Schneider, P., 2005, \aap, 442, 69 \ok
\japref{[Kim et al.(2004)]}{Kim04} Kim, T.-S., Viel, M., Haehnelt, M.G., Carswell, B., Cristiani, S., 2004a, \mnras, 351, 1471 \ok
\japref{[Kim et al.(2004)]}{Kim04a} Kim, T.-S., Viel, M., Haehnelt, M.G., Carswell, R.F., Cristiani, S., 2004b, \mnras, 347, 355 \ok
\japref{[King(2005)]}{King05} King, L.J., 2005, \aap, 441, 47 \ok
\japref{[King, Schneider(2003)]}{King03} King, L.J., Schneider, P., 2003, \aap, 398, 23 \ok
\japref{[Kohler, Gnedin(2006)]}{KohlerG06} Kohler, K., Gnedin, N.Y., 2006,  astro-ph/0605032 \ok
\japref{[]}{}Kolb, R., et al.,  2006,  DoE/NASA/NSF interagency task force on Dark Energy \ok
\japref{[Kravtsov et al.(2006)]}{Kravtsov06} Kravtsov, A.V., Vikhlinin, A., Nagai, D., 2006,  astro-ph/0603205 \ok
\japref{[Krisciunas et al.(2000)]}{Krisciunas00} Krisciunas, K., et al., 2000, \apj, 539, 658 \ok
\japref{[Krisciunas et al.(2004)]}{Krisciunas04} Krisciunas, K., Phillips, M.M., Suntzeff, B.N., 2004, \apj, 602, L81 \ok
\japref{[Lamoreaux, Torgerson(2004)]}{Lamoreaux04} Lamoreaux, S.K., Torgerson, J.R., 2004, \prd, 69, 121701 \ok
\japref{[Le F{\`e}vre et al.(2004)]}{le04} Le F{\`e}vre, O., et al., 2004, \aap, 428, 1043 \ok
\japref{[Leibundgut(2001)]}{Leibundgut01} Leibundgut, B., 2001, \araa, 39, 67 \ok
\japref{[Levine et al.(2002)]}{Levine02} Levine, E.S., Schulz, A.E., White, M., 2002, \apj, 577, 569 \ok
\japref{[Levshakov et al.(2005)]}{Levshakov05}Levshakov S.A., Centuri\'on, M., Molaro, P., D'Odorico, S., 2005, \aap, 434, 827 \ok
\japref{[Levshakov et al.(2006)]}{Levshakov06} Levshakov S.A., Centuri\'on, M., Molaro, P., Kostina, M.V., 2006, \aap, 447, L21 \ok
\japref{[Levshakov(2004)]}{Levshakov04} Levshakov, S.A., 2004, LNPVol.648: Astrophysics, Clocks and Fundamental Constants, 648, 151 \ok
\japref{[Levshakov et al.(2002)]}{Levshakov02} Levshakov, S.A., Dessauges-Zavadsky, M., D'Odorico, S., Molaro, P., 2002, \mnras, 333, 373 \ok
\japref{[Liddle, Scherrer(1999)]}{Liddle99} Liddle, A.R., Scherrer, R.J., 1999, \prd, 59, 023509 \ok
\japref{[Lidman et al.(2005)]}{Lidman05} Lidman, C., et al., 2005, \aap, 430, 843 \ok
\japref{[Lima, Hu(2004)]}{Lima04} Lima, M., Hu, W., 2004, \prd, 70, 043504 \ok
\japref{[Linder(2003)]}{Linder03} Linder, E.V., 2003, \prd, 68, 083503 \ok
\japref{[Linder \& Huterer(2003)]}{Linder03a} Linder, E.V., Huterer, D., 2003, \prd, 67, 1303 \ok 
\japref{[Livio(2000)]}{Livio00} Livio, M., 2000, in Type Ia Supernovae: Theory and Cosmology, eds. J. C. Niemeyer and J. W. Truran, Cambridge, Cambridge University Press, 33 \ok
\japref{[Majumdar, Mohr(2003)]}{Majumdar03} Majumdar, S., Mohr, J.J., 2003, \apj, 585, 603 \ok
\japref{[Majumdar, Mohr(2004)]}{Majumdar04} Majumdar, S., Mohr, J.J., 2004, \apj, 613, 41 \ok
\japref{[Maoli et al.(2001)]}{Maoli01} Maoli, R., Van Waerbeke, L., Mellier, Y., Schneider, P., Jain, B., Bernardeau, F., Erben, T., Fort, B., 2001, \aap, 368, 766 \ok
\japref{[Marciano(1984)]}{Marciano84} Marciano, W.J., 1984, Physical Review Letters, 52, 489 \ok
\japref{[Massey et al.(2005)]}{Massey05} Massey, R., R\'efr\'egier, A., Bacon, D.J., Ellis, R., Brown, M.L., 2005, \mnras, 359, 1277 \ok
\japref{[McDonald et al.(2000)]}{McDonald00} McDonald P., Miralda-Escud\'e J., Rauch, M., Sargent, W.L.W., Barlow, T.A., Cen, R., Ostriker, J.P., 2000, \apj, 543, 1 \ok
\japref{[McDonald et al.(2005)]}{McDonald05} McDonald, P., et al., 2005, \apj, 635, 761
\japref{[Matheson et al.(2005)]}{Matheson05} Matheson, T., et al., 2005, \aj, 129, 2352 \ok
\japref{[Meiksin et al.(1999)]}{Meiksin99} Meiksin, A., White, M., Peacock, J.A., 1999, \mnras, 304, 851 \ok
\japref{[Miralda-Escude(1991)]}{Miralda-Escude91} Miralda-Escud\'e, J., 1991, \apj, 380, 1 \ok
\japref{[Mohr, Taylor(2005)]}{Mohr05} Mohr, P.J., Taylor, B.N., 2005, Reviews of Modern Physics, 77, 1 \ok
\japref{[Molnar et al.(2004)]}{Molnar04} Molnar, S.M., Haiman, Z., Birkinshaw, M., Mushotzky, R.F., 2004, \apj, 601, 22 \ok
\japref{[Mullis et al.(2005)]}{Mullis05} Mullis, C.R., Rosati, P., Lamer, G., B{\"o}hringer, H., Schwope, A., Schuecker, P., Fassbender, R., 2005, \apj, 623, L85 \ok
\japref{[Murphy et al.(2004)]}{Murphy04} Murphy, M.T., Flambaum, V.V., Webb, J.K., et al., 2004, LNP Vol.648: Astrophysics, Clocks and Fundamental Constants, 648, 131 \ok
\japref{[Nugent et al.(2002)]}{Nugent02} Nugent, P., Kim, A., Perlmutter, S., 2002, \pasp, 114, 803 \ok
\japref{[Peacock, Dodds(1996)]}{Peacock96} Peacock, J.A., Dodds, S.J., 1996, \mnras, 280, L19 \ok
\japref{[Peacock et al.(2001)]}{Peacock01} Peacock, J.A., et al. (The 2dFGRS Team), 2001, Nature, 410, 169 \ok
\japref{[Pen et al.(2003a)]}{Pen03} Pen, U.-L., Zhang, T., van Waerbeke, L., Mellier, Y., Zhang, P.,  Dubinski, J., 2003a, ApJ 592, 664
\japref{[Pen et al.(2003b)]}{Pen03b} Pen, U.-L., Lu, T., van Waerbeke, L., Mellier, Y., 2003b, MNRAS, 346, 994
\japref{[Perlmutter et al.(1999)]}{Perlmutter98} Perlmutter, S., et al. 1998, \apj, 517, 656 \ok
\japref{[Phillips et al.(1999)]}{Phillips99} Phillips, M.M., et al., 1999, \aj, 118, 1766 \ok
\japref{[Popesso et al.(2005)]}{Popesso05} Popesso, P., Biviano, A., B{\"o}hringer, H., Romaniello, M., Voges, W., 2005, \aap, 433, 431 \ok
\japref{[Quast et al.(2004)]}{Quast04} Quast, R., Reimers, D., Levshakov, S.A., 2004, \aap, 415, L7 \ok
\japref{[Refregier et al.(2002)]}{Refregier02} R\'efr\'egier, A., Rhodes, J., Groth, E.J., 2002, \apj, 572, L131 \ok
\japref{[Reinhold et al.(2006)]}{Reinhold06} Reinhold, E., Buning, R., Hollenstein, U., Ivanchik, A., Petitjean, P., Ubachs, W., 2006, Physical Review Letters, 96, 151101 \ok
\japref{[Reiprich, B\"ohringer(2002)]}{Reiprich02} Reiprich, T.H., B\"ohringer, H., 2002, \apj, 567, 716 \ok
\japref{[Rhodes et al.(2001)]}{Rhodes01} Rhodes, J., R\'efr\'egier, A., Groth, E.J., 2001, \apj, 552, L85 \ok
\japref{[Riess et al.(1998)]}{Riess98} Riess, A.G., et al., 1998, \aj, 116, 1009 \ok
\japref{[Riess et al.(2004)]}{Riess04} Riess, A.G., et al., 2004, \apj, 607, 665 \ok
\japref{[Riess et al.(1996)]}{Riess96} Riess, A.G., et al., 1996, \apj, 473, 88 \ok
\japref{[Riotto, Trodden(1999)]}{Riotto99} Riotto, A., Trodden, M., 1999, Annual Review of Nuclear and Particle Science, 49, 35 \ok
\japref{[R\"{o}pke et al.(2006)]}{Ropke06} R\"{o}pke, F.K., et al., 2006, \aap, 453, 203 \ok
\japref{[Sawicki, Carroll(2005)]}{Sawicki05} Sawicki, I., Carroll, S.M., 2005,  astro-ph/0510364 \ok
\japref{[Seljak, Zaldarriaga(1998)]}{SelZal98} Seljak, U., Zaldarriaga, M., 1998, astro-ph/9805010 \ok
\japref{[Schmid et al.(2006)]}{Schmid06} Schmid, C., et al., 2006,  astro-ph/0603158 \ok
\japref{[Schirmer et al.(2006)]}{Schirmer06} Schirmer, M., Erben, T., Hetterscheidt, M., Schneider, P., 2006, astro-ph/0607022
\japref{[Schlegel et al.(1998)]}{Schlegel98} Schlegel, D., Finkbeiner, D.B., Davis, M., 1998, \apj, 500, 525 \ok
\japref{[Schneider et al.(2002)]}{Schneider02} Schneider, P., van Waerbeke, L., Mellier, Y., 2002, \aap, 389, 729 \ok
\japref{[Schrabback et al.(2006)]}{Schrabback06} Schrabback, T., et al., 2006,  astro-ph/0606611 \ok
\japref{[Schuecker et al.(2003)]}{Schuecker03a} Schuecker, P., B{\"o}hringer, H., Collins, C.A., Guzzo, L., 2003a, \aap, 398, 867 \ok
\japref{[Schuecker et al.(2003)]}{Schuecker03b} Schuecker, P., Caldwell, R.R., B{\"o}hringer, H., Collins, C.A., Guzzo, L., Weinberg, N.N., 2003b, \aap, 402, 53
\japref{[]}{Scoville06} Scoville N.Z. et al., (the COSMOS Team), 2006, ApJ, submitted \ok
\japref{[Seljak et al.(2003)]}{Seljak03} Seljak, U., McDonald, P., Makarov, A., 2003, \mnras, 342, L79 \ok
\japref{[Seljak et al.(2005b)]}{Seljak05a} Seljak, U., et al., 2005a, \prd, 71, 043511 \ok
\japref{[Seljak et al.(2005b)]}{Seljak05b} Seljak, U., et al., 2005b, \prd, 71, 103515 \ok
\japref{[Seljak et al.(2006)]}{Seljak06} Seljak, U., Makarov, A., McDonald, P., Trac, H., 2006,  astro-ph/0602430 \ok
\japref{[Semboloni et al.(2006)]}{Semboloni06} Semboloni, E., et al., 2006a, \aap, 452, 51 \ok
\japref{[Semboloni et al.(2006)]}{Semboloni06a} Semboloni, E., van Waerbeke, L., Heymans, C., Hamana, T., Colombi, S., White, M., Mellier, Y., 2006b,  astro-ph/0606648 \ok
\japref{[Seo, Eisenstein(2003)]}{Seo03} Seo, H.-J., Eisenstein, D.J., 2003, \apj, 598, 720
\japref{[Simpson, Bridle(2006)]}{Simpson06} Simpson, F., Bridle, S., 2006, \prd, 73, 083001 \ok
\japref{[Skordis et al.(2006)]}{Skordis06} Skordis C., Mota D.F., Ferreira P.G., Boehm, C., 2006, Physical Review Letters, 96, 011301 \ok
\japref{[Smith et al.(2003)]}{Smith03} Smith, R.E., et al., 2003, \mnras, 341, 1311 \ok
\japref{[Spergel et al.(2006)]}{Spergel06} Spergel, D.N., et al., 2006,  astro-ph/0603449 \ok
\japref{[Spergel et al.(2003)]}{Spergel03} Spergel, D.N., et al., 2003, \apjs, 148, 175 \ok
\japref{[Stebbins(1996)]}{Stebbins96} Stebbins, R.T., 1996, astro-ph/9609149 \ok % no refereed publication
\japref{[Springel et al.(2005)]}{Springel05} Springel, V., et al., 2005, Nature, 435, 629 \ok
\japref{[Starobinski{\v i}(1985)]}{Starobinsky85} Starobinski{\v i}, A.A., 1985, Sov. Astr. Lett., 11, 133 \ok
\japref{[Strolger et al.(2004)]}{Strolger04} Strolger, L.-G., et al., 2004, \apj, 613, 200 \ok
\japref{[Stritzinger \& Leibundgut(2005)]}{Stritzinger05} Stritzinger, M., Leibundgut, B., 2005, \aap, 431, 423 \ok
\japref{[Stritzinger et al.(2002)]}{Stritzinger02} Stritzinger, M., et al., 2002, \aj, 124, 2100 \ok
\japref{[Stritzinger et al.(2006)]}{Stritzinger06} Stritzinger, M., Leibundgut, B., Walch, S., Contardo, G., 2006, \aap, 450, 241 \ok
\japref{[Sullivan et al.(2006)]}{Sullivan06} Sullivan, M., et al., 2006, \aj, 131, 960 \ok
\japref{[Susskind (2003)]}{Susskind03} Susskind, L., 2003, hep-th/0302219 \ok
\japref{[Takada, Jain(2002)]}{Takada02} Takada, M., Jain, B., 2002, \mnras, 337, 875 \ok
\japref{[Tegmark et al.(2004)]}{Tegmark04} Tegmark, M., et al., 2004, \apj, 606, 702 \ok
\japref{[Tonry et al.(2003)]}{Tonry03} Tonry, J.L., et al., 2003, \apj, 594, 1 \ok
\japref{[Trotta(2006)]}{Trotta05} Trotta, R., 2006,  astro-ph/0607496 \ok
\japref{[Tytler et al.(2004)]}{Tytler04} Tytler, D., et al., 2004, \apj, 617, 1 \ok
\japref{[Ubachs, Reinhold(2004)]}{Ubachs04} Ubachs, W., Reinhold, E., 2004, Physical Review Letters, 92, 101302 \ok
\japref{[Uzan(2003)]}{Uzan03} Uzan, J.-P., 2003, Reviews of Modern Physics, 75, 403 \ok
\japref{[Uzan, Bernardeau (2001)]}{UzanBern01} Uzan, J.-P., Bernardeau, F.,  2001, \prd, 64, 3501 \ok
\japref{[Van Waerbeke et al.(2000)]}{Van00} Van Waerbeke, L., et al., 2000, \aap, 358, 30 \ok
\japref{[Van Waerbeke et al.(2002)]}{Van02} Van Waerbeke, L., Mellier, Y., Pell\'o, R., Pen, U.-L., McCracken, H.J., Jain, B. 2002, \aap, 393, 369 \ok\ok
\japref{[Van Waerbeke et al.(2005)]}{Van05} Van Waerbeke, L., Mellier, Y., Hoekstra, H., 2005, \aap, 429, 75 \ok
\japref{[Varshalovich, Levshakov(1993)]}{1993JETPL..58..237V}Varshalovich, D.A., Levshakov, S.A., 1993, Journal of Experimental and Theoretical Physics Letters, 58, 237 \ok
\japref{[Verde et al.(2002)]}{Verde02} Verde, L., et al., 2002, \mnras, 335, 432 \ok
\japref{[Viel et al.(2004)]}{Viel04a} Viel, M., Haehnelt, M.G., Springel, V., 2004, \mnras, 354, 684 \ok
\japref{[Viel et al.(2005)]}{Viel05} Viel, M., Lesgourgues, J., Haehnelt, M.G., Matarrese, S., Riotto, A., 2005, \prd, 71, 063534
\japref{[Viel et al.(2006)]}{Viel06a} Viel, M., Haehnelt, M.G., Lewis, A., 2006a, \mnras, 370, L51  %  astro-ph/0604310 \ok
\japref{[Viel et al.(2006)]}{Viel06b} Viel, M., Lesgourgues, J., Haehnelt, M.G., Matarrese, S., Riotto, A., 2006b,  astro-ph/0605706 \ok
\japref{[Vikhlinin et al.(2005)]}{Vikhlinin05} Vikhlinin, A., Markevitch, M., Murray, S.S., Jones, C., Forman, W., Van Speybroeck, L., 2005, \apj, 628, 655 \ok
\japref{[Vikhlinin et al.(2003)]}{Vikhlinin03} Vikhlinin, A., et al., 2003, \apj, 590, 15 \ok
\japref{[Voit(2005)]}{Voit05} Voit, G.M., 2005, Reviews of Modern Physics, 77, 207
\japref{[]}{Wang88} Wang, L., Steinhardt, P.J., 1998, ApJ, 508, 483\ok
\japref{[Wang et al.(2004)]}{Wang04} Wang, S., Khoury, J., Haiman, Z., May, M., 2004, \prd, 70, 123008 \ok
\japref{[Webb et al.(1999)]}{Webb99} Webb, J.K., Flambaum, V.V., Churchill, C.W., Drinkwater, M.J., Barrow, J.D., 1999, Physical Review Letters, 82, 884 \ok
\japref{[]}{WFMOS}WFMOS study, 2005,  \url{http://www.gemini.edu/files/docman/science/aspen/}\phantom{xxxxx}  \url{WFMOS_feasibility_report_public.pdf}
\japref{[Wetterich (2003)]}{Wetterich03} Wetterich, C., 2003, Phys. Lett. B, 561, 10 % hep-ph/0301261 \ok
\japref{[White, Kochanek(2001)]}{White01} White, M., Kochanek, C.S., 2001, \apj, 574, 24 \ok
\japref{[Wittman(2006)]}{Wittman06} Wittman, D., Dell'Antonio, I.P., Hughes, J.P., Margoniner, V.E., Tyson, J.A., Cohen, J.G., Norman, D., 2006. ApJ, 643, 128
\japref{[Zaldarriaga, Seljak(1997)]}{ZalSel97} Zaldarriaga, M., Seljak, U., 1997, \prd, 55, 1830
\japref{[Zaroubi et al.(2006)]}{Zaroubi06} Zaroubi, S., Viel, M., Nusser, A., Haehnelt, M., Kim, T.-S., 2006, \mnras, 369, 734 \ok  %  astro-ph/0509563
\japref{[Zehavi et al.(1998)]}{Zehavi98} Zehavi, I., Riess, A.G., Kirshner, R.P., Dekel, A., 1998, \apj, 503, 483 \ok
\japref{[Zhang et al.(2005)]}{Zhang05} Zhang, J., Hui, L., Stebbins, A., 2005, \apj, 635, 806 \ok
\japref{[Zhao et al.(2006)]}{Zhao06} Zhao, H., Bacon, D.J., Taylor, A.N., Horne, K., 2006, \mnras, 368, 171 \ok
\japref{[Zlatev et al.(1999)]}{Zlatev99} Zlatev, I., Wang, L., Steinhardt, P.J., 1999, Physical Review Letters, 82, 896 \ok

%\end{thebibliography}

\def\japlistitem{\parskip=0pt\par\noindent\hangindent\parindent
\parskip =2ex plus .2ex minus .1ex}

\addcontentsline{toc}{section}{List of abbreviations}

\section*{List of abbreviations}

\vglue 0.5em

%\begin{enumerate}

\japlistitem 2dF: Two-degree Field
\japlistitem 2dFGRS: The Two-degree Field Galaxy Redshift Survey
\japlistitem 2MASS: Two Micron All Sky Survey
\japlistitem 6dFGS: Six-degree Field Galaxy Survey
\japlistitem AAOmega: the upgraded two degree field (2dF) on the Anglo-Australian 4m telescope
\japlistitem ACES: Atomic Clock Ensemble in Space
\japlistitem ACS:  Advanced Camera for Surveys
\japlistitem ACT:  Atacama Cosmology Telescope
\japlistitem AD: alkali-doublets
\japlistitem ADEPT: Advanced Dark Energy Physics Telescope
\japlistitem AGB: Asymptotic Giant Branch
\japlistitem ALMA: Atacama Large Millimetre Array
\japlistitem ALPACA: Advanced Liquid-mirror Probe for Astrophysics, Cosmology and Asteroids
\japlistitem AMI: Arcminute Micro-Kelvin Imager
\japlistitem APEX: Atacama Pathfinder Experiment
\japlistitem APO: Apache Point Observatory
\japlistitem AU: Astronomical Unit
\japlistitem BAO: Baryon Acoustic Oscillations
\japlistitem BBN: Big Bang Nucleosynthesis
\japlistitem BBO: Big Bang Observer
\japlistitem BaBar: CP-violation experiment located at the Stanford Linear Accelerator Center
\japlistitem BELLE: CP-violation experiment located at KEK, Japan
\japlistitem CBI: Cosmic Background Imager, a radio telescope located in the Chilean Andes
\japlistitem CDM: Cold Dark Matter
\japlistitem CFHT: Canada-France-Hawaii Telescope
\japlistitem CFHTLS:  Canada-France-Hawaii Telescope Legacy Survey
\japlistitem CMB: Cosmic Microwave Background
\japlistitem CMBPOL: CMB POLarision mission
\japlistitem COBE: COsmic Background Explorer
\japlistitem COMBO-17: Classifying Objects by Medium-Band Observations - a spectrophotometric 17-filter survey
\japlistitem COSMOS: Cosmic Evolution Survey
\japlistitem CP: charge-parity conjugation
\japlistitem CTIO: Cerro Tololo Inter-American Observatory
\japlistitem DECam: Camera for Dark Energy Survey
\japlistitem DES: Dark Energy Survey
\japlistitem DEEP2: Deep Extragalactic Evolution Probe, a spectroscopic survey carried out with Keck
\japlistitem DENIS: Deep Near-Infrared Sky Survey
\japlistitem DESTINY: A space-borne near-IR grism survey instrument for probing dark energy with supernovae
\japlistitem DETF: Dark Energy Task Force
\japlistitem DGP: Dvali, Gabadadze \& Porrati extra-dimensional cosmological model
\japlistitem DUNE: Dark Universe Explorer
\japlistitem DarkCam: optical camera proposed for the VISTA telescope
\japlistitem Descart: Dark matter from Ellipticity Sources CARTography, ESO/CFHT imaging programme
\japlistitem DoE: Department of Energy
\japlistitem ELT: Extremely Large Telescope
\japlistitem ESA: European Space Agency
\japlistitem ESO: European Southern Observatory
\japlistitem ESSENCE: Equation of State: SupErNovae trace Cosmic Expansion project
\japlistitem FIRST: Faint Images of the Radio Sky at Twenty-centimetres survey
\japlistitem FLAMES: Fibre Large Array Multi Element Spectrograph at the VLT
\japlistitem FMOS: Fiber Multi-Object Spectrograph. An instrument at Subaru
\japlistitem GAIA: Galactic Astrophysics through Imaging and Astrometry. An  ESA mission
\japlistitem GALEX : GALaxy Evolution EXplorer
\japlistitem GEMS: Galaxy Evolution From Morphology And SEDs survey
\japlistitem GOODS: The Great Observatories Origins Deep Survey
\japlistitem GUT: Grand Unified Theory
\japlistitem GaBoDS: The Garching-Bonn Deep Survey
\japlistitem Gadget-II: hydrodynamical N-body code
\japlistitem HETDEX: Hobby-Eberly Telescope Dark Energy Experiment
\japlistitem HFI: High-Frequency Instrument for Planck
\japlistitem HIRES: High Resolution Echelle Spectrometer on Keck
\japlistitem HST: Hubble Space Telescope
\japlistitem HyperCam: Camera on Subaru
\japlistitem HyperSuprimeCam: Camera on Subaru
\japlistitem IGM: Intergalactic Medium
\japlistitem ILC: International Linear Collider
\japlistitem IR: Infra Red
\japlistitem ISO:  Infrared Space Observatory
\japlistitem JDEM: Joint Dark Energy Mission
\japlistitem JEDI: Joint Efficient Dark-energy Investigation
\japlistitem JHK: 1--2 micron filter bands
\japlistitem JWST: James Webb Space Telescope
\japlistitem KIDS: 1400-square degree gravitational shear survey with OmegaCAM
\japlistitem LHC: Large Hadron Collider
\japlistitem LIGO: Laser Interferometry Gravitational-Wave Observatory
\japlistitem LISA: Laser Interferometry Satellite Antenna gravitational-wave observatory
\japlistitem LOTOSS: Lick Observatory and Tenagra Observatory Supernova Search
\japlistitem LSS: Large Scale Structure
\japlistitem LSST: Large Synoptic Survey Telescope
\japlistitem MDS: Medium Deep Survey with HST
\japlistitem MM: Many-Multiplet method
\japlistitem MMT: Multiple Miror Telescope
\japlistitem MOND: Modified Newtonian Dynamics
\japlistitem NASA: National Aeronautics and Space Administration
\japlistitem NEAT: Near Earth Asteroid Tracking program
\japlistitem NSF: National Science Foundation
\japlistitem PAENS/SHOES: HST supernova survey
\japlistitem POSS: Palomar Observatory Sky Survey
\japlistitem PSF: Point Spread Function
\japlistitem PanSTARRS: Panoramic Survey Telescope \& Rapid Response System
\japlistitem QCD: Quantum ChromoDynamics
\japlistitem QSO: Quasi-Stellar Object
\japlistitem QUIET: Q/U Imaging ExperimenT for CMB polarization
\japlistitem RCS:  Red-sequence Cluster Survey
\japlistitem ROSAT: R\"ontgen SATellite
\japlistitem ROSITA: R\"ontgen Survey with an Imaging Telescope Array
\japlistitem SDSS: Sloan Digital Sky Survey
\japlistitem SED:  Spectral Energy Distribution
\japlistitem SKA: Square Kilometre Array
\japlistitem SMBH: SuperMassive Black Hole
\japlistitem SN: SuperNova
\japlistitem SNAP: Supernova Acceleration Probe
\japlistitem SNIFS: Supernova Integral Field Spectrograph
\japlistitem SNIa: SuperNova type Ia
\japlistitem SNLS:  SuperNova Legacy Survey
\japlistitem SPIDER: CalTech balloon-borne experiment to search for cosmic gravitational wave background
\japlistitem SPT: South Pole Telescope
\japlistitem STEP: Shear TEsting Program
\japlistitem SUPRIME-33: Suprime-Cam weak lensing cluster survey
\japlistitem SZ: Sunyaev-Zeldovich effect
\japlistitem SuprimeCam: Subaru Prime Focus Camera
\japlistitem TeVeS: tensor, vector and scalar field theory of modified gravity, due to Bekenstein
\japlistitem UKIDSS:  UKIRT Infrared Deep Sky Survey
\japlistitem UKIRT: United Kingdom Infra-Red Telescope
\japlistitem UKST: United Kingdom Schmidt Telescope
\japlistitem UV: Ultraviolet
\japlistitem UVES: Ultraviolet and Visual Echelle Spectrograph
\japlistitem VIKING: VISTA Kilo-degree Infrared Galaxy survey
\japlistitem VIMOS: VIsible MultiObject Spectrograph
\japlistitem VISTA: Visible and Infrared Survey Telescope
\japlistitem VLT: The Very Large Telescope
\japlistitem VST: VLT Survey Telescope
\japlistitem VVDS: VIRMOS-VLT Deep Survey
\japlistitem VIRMOS: VLT Visible and InfraRed Multi-Object Spectrograph
\japlistitem WDM: Warm Dark Matter
\japlistitem WFCam: Wide Field Infrared Camera For UKIRT
\japlistitem WFMOS: Wide-Field Multi-Object Spectrograph
\japlistitem WFPC2: Wide Field and Planetary Camera 2
\japlistitem WG: Working Group
\japlistitem WHT: William Herschel Telescope
\japlistitem WIMP: Weakly Interacting Massive Particle
\japlistitem WIRCam: Wide Field IR Camera - wide-field imaging facility at CFHT
\japlistitem WMAP: Wilkinson Microwave Anisotropy Probe
\japlistitem WMAP1: WMAP first year survey
\japlistitem WMAP3: WMAP three year survey
\japlistitem XEUS: X-ray Evolving Universe Spectroscopy mission
\japlistitem XMM: X-ray Spectroscopy Multi-Mirror Mission

%\end{enumerate}

\end{document}